# GENERAL EXPRESSIONS FOR THE COEFFICIENTS OF CHERN FORMS UP TO THE 13$^{th}$ ORDER IN CURVATURE

C. C. Briggs
*Center for Academic Computing, Penn State University, University Park, PA 16802*
Friday, April 16, 1999

**Abstract.** General expressions are given for the coefficients of Chern forms up to the 13$^{th}$ order in curvature in terms of the Riemann-Christoffel curvature tensor and some of its concomitants (e.g., Pontrjagin's characteristic tensors) for $n$-dimensional differentiable manifolds having a general linear connection.

This paper presents general expressions for the coefficients of Chern forms up to the 13$^{th}$ order in curvature in terms of the Riemann-Christoffel curvature tensor and some of its concomitants (e.g., Pontrjagin's characteristic tensors) for $n$-dimensional differentiable manifolds having a general linear connection.

Figuratively speaking, the $p^{th}$ Chern forms[1] $c_{(p)}$ representing the corresponding $p^{th}$ Chern classes of such a manifold $M$ can be defined by[2-3]

$$c_{(p)} \equiv \begin{cases} 1, & \text{if } p = 0 \\ \frac{i^p}{2^p \pi^p} \Omega_{[i_1}{}^{i_1} \wedge \Omega_{i_2}{}^{i_2} \wedge \ldots \wedge \Omega_{i_p]}{}^{i_p}, & \text{if } p > 0 \end{cases}, \quad (1)$$

where $\Omega_a{}^b$ is the curvature 2-form of $M$ defined by

$$\Omega_a{}^b \equiv \frac{1}{2} R_{cda}{}^b \, \omega^c \wedge \omega^d, \quad (2)$$

where $R_{abc}{}^d$ is the Riemann-Christoffel curvature tensor of $M$ defined by[4]

$$R_{abc}{}^d \equiv 2 \, (\partial_{[a} \Gamma_{b]}{}^d{}_c + \Gamma_{[a|e|}{}^d \Gamma_{b]}{}^e{}_c + \Omega_{a\ b}{}^e \Gamma_e{}^d{}_c), \quad (3)$$

where $\Gamma_a{}^b{}_c$ is the connection coefficient, $\Omega_{a\ c}{}^b$ the object of anholonomity, and $\omega^a$ the basis 1-form of $M$. Thus, the 2$p$-forms $c_{(p)}$ for $p \geq 1$ are given by

$$c_{(p)} = c_{(p)i_1 i_2 \ldots i_{2p}} \omega^{i_1} \wedge \omega^{i_2} \wedge \ldots \wedge \omega^{i_{2p}}, \quad (4)$$

where the coefficients $c_{(p)i_1 i_2 \ldots i_{2p}}$ of $c_{(p)}$ for $p \geq 1$ are given by

$$c_{(p)i_1 i_2 \ldots i_{2p}} = \frac{1}{(2p)!} \langle e_{i_1} \wedge e_{i_2} \wedge \ldots \wedge e_{i_{2p}}, c_{(p)} \rangle \quad (5)$$

$$= \frac{1}{(2p)!} c_{(p)j_1 j_2 \ldots j_{2p}} \langle e_{i_1} \wedge e_{i_2} \wedge \ldots \wedge e_{i_{2p}}, \omega^{j_1} \wedge \omega^{j_2} \wedge \ldots \wedge \omega^{j_{2p}} \rangle$$

$$= \frac{1}{(2p)!} \delta_{i_1 i_2 \ldots i_{2p}}^{j_1 j_2 \ldots j_{2p}} c_{(p)j_1 j_2 \ldots j_{2p}}$$

$$= c_{(p)[i_1 i_2 \ldots i_{2p}]}$$

$$= \frac{i^p}{2^p \pi^p (2p)!} \langle e_{i_1} \wedge e_{i_2} \wedge \ldots \wedge e_{i_{2p}}, \Omega_{[j_1}{}^{j_1} \wedge \Omega_{j_2}{}^{j_2} \wedge \ldots \wedge \Omega_{j_p]}{}^{j_p} \rangle$$

$$= \frac{i^p}{2^{2p} \pi^p (2p)!} R_{k_1 k_2 j_1}{}^{[j_1} R_{k_3 k_4 j_2}{}^{j_2} \ldots R_{k_{2p-1} k_{2p} j_p}{}^{j_p]} \langle e_{i_1} \wedge e_{i_2} \wedge \ldots \wedge e_{i_{2p}}, \omega^{k_1} \wedge \omega^{k_2} \wedge \ldots \wedge \omega^{k_{2p}} \rangle$$

$$= \frac{i^p}{2^{2p} \pi^p (2p)!} \delta_{i_1 i_2 \ldots i_{2p}}^{k_1 k_2 \ldots k_{2p}} R_{k_1 k_2 j_1}{}^{[j_1} R_{k_3 k_4 j_2}{}^{j_2} \ldots R_{k_{2p-1} k_{2p} j_p}{}^{j_p]}$$

$$= \frac{i^p}{2^{2p} \pi^p} R_{[i_1 i_2|j_1|}{}^{[j_1} R_{i_3 i_4|j_2|}{}^{j_2} \ldots R_{i_{2p-1} i_{2p}|j_p|}{}^{j_p]},$$

where $e_a$ is the basis tangent vector of $M$ dual to $\omega^a$, i.e. $\langle e_b, \omega^a \rangle = \delta_b^a$, where $\delta_b^a$ is the Kronecker delta, and

$$\langle e_{i_1} \wedge e_{i_2} \wedge \ldots \wedge e_{i_{2p}}, \omega^{j_1} \wedge \omega^{j_2} \wedge \ldots \wedge \omega^{j_{2p}} \rangle = \delta_{i_1 i_2 \ldots i_{2p}}^{j_1 j_2 \ldots j_{2p}}, \quad (6)$$

where $\delta_{i_1 i_2 \ldots i_{2p}}^{j_1 j_2 \ldots j_{2p}}$ is the generalized Kronecker delta.

Some numerical properties of $c_{(p)i_1 i_2 \ldots i_{2p}}$ for $0 \leq p \leq 13$ appear in Table 1. General expressions for $c_{(p)i_1 i_2 \ldots i_{2p}}$ for $0 \leq p \leq 13$ appear in Eqs. (9) through (22), the expressions being given (1$^{st}$) in terms of $R_{abc}{}^d$, (2$^{nd}$) in terms of $R_{abc}{}^d$ and Schouten's tensor $V_{ab}$ defined by[5]

$$V_{ab} \equiv R_{abc}{}^c = \nabla_{[a} Q_{b]c}{}^c + S_{ab}{}^d Q_{dc}{}^c, \quad (7)$$

where $S_{ac}{}^b$ is the torsion tensor defined by $S_{ac}{}^b \equiv \Gamma_{[a\ c]}{}^b + \Omega_a{}^b{}_c$ and $Q_a{}^{bc}$ the non-metricity tensor defined by $Q_a{}^{bc} \equiv \nabla_a g^{bc}$, where $g^{ab}$ is the contravariant metric tensor, and (3$^{rd}$) in terms of Pontrjagin's characteristic tensors (from the Russian "характеристические тензоры") $P^{(2p)}{}_{i_1 i_2 \ldots i_{2p}}$ defined by[6-7]

$$P^{(2p)}{}_{i_1 i_2 \ldots i_{2p}} \equiv \frac{1}{2^p} R_{[i_1 i_2|j_2|}{}^{j_1} R_{i_3 i_4|j_3|}{}^{j_2} \ldots \quad (8)$$

$$\ldots R_{i_{2p-3} i_{2p-2}|j_p|}{}^{j_{p-1}} R_{i_{2p-1} i_{2p}|j_1|}{}^{j_p}.$$

TABLE 1. SOME NUMERICAL PROPERTIES OF $c_{(p)i_1 i_2 \ldots i_{2p}}$ FOR $0 \leq p \leq 13$

| QUANTITY | ORDER | CURVATURE DEPENDENCE | MINIMUM NUMBER OF DIMENSIONS | NUMBER OF TERMS | 1$^{st}$ & 2$^{nd}$ OVERALL NUMERICAL FACTORS | 3$^{rd}$ OVERALL NUMERICAL FACTOR |
|---|---|---|---|---|---|---|
| $c_{(p)i_1 i_2 \ldots i_{2p}}$ | $p$ | — | $2p$ | — | $\frac{i^p}{2^{2p} \pi^p p!}$ | $\frac{i^p}{2^p \pi^p p!}$ |
| $c_{(0)}$ | 0 | Zero | 0 | **1** | $+1$ | $+1$ |
| $c_{(1)ab}$ | 1 | Linear | 2 | **1** | $+\frac{1}{4\pi} i$ | $+\frac{1}{2\pi} i$ |
| $c_{(2)abcd}$ | 2 | Quadratic | 4 | **2** | $-\frac{1}{32\pi^2}$ | $-\frac{1}{8\pi^2}$ |
| $c_{(3)abcdef}$ | 3 | Cubic | 6 | **3** | $-\frac{1}{384\pi^3} i$ | $-\frac{1}{48\pi^3} i$ |
| $c_{(4)abcdefgh}$ | 4 | Quartic | 8 | **5** | $+\frac{1}{6144\pi^4}$ | $+\frac{1}{384\pi^4}$ |
| $c_{(5)abcdefghij}$ | 5 | Quintic | 10 | **7** | $+\frac{1}{122,880\pi^5} i$ | $+\frac{1}{3840\pi^5} i$ |
| $c_{(6)abcdefghijkl}$ | 6 | Sextic | 12 | **11** | $-\frac{1}{2,949,120\pi^6}$ | $-\frac{1}{46,080\pi^6}$ |
| $c_{(7)abcdefghijklmn}$ | 7 | Septic | 14 | **15** | $-\frac{1}{82,575,360\pi^7} i$ | $-\frac{1}{645,120\pi^7} i$ |
| $c_{(8)abcdefghijklmnop}$ | 8 | Octic | 16 | **22** | $+\frac{1}{2,642,411,520\pi^8}$ | $+\frac{1}{10,321,920\pi^8}$ |
| $c_{(9)abcdefghijklmnopqr}$ | 9 | Nonic | 18 | **30** | $+\frac{1}{95,126,814,720\pi^9} i$ | $+\frac{1}{185,794,560\pi^9} i$ |
| $c_{(10)abcdefghijklmnopqrst}$ | 10 | Decic | 20 | **42** | $-\frac{1}{3,805,072,588,800\pi^{10}}$ | $-\frac{1}{3,715,891,200\pi^{10}}$ |
| $c_{(11)abcdefghijklmnopqrstuv}$ | 11 | Undecic | 22 | **56** | $-\frac{1}{167,423,193,907,200\pi^{11}} i$ | $-\frac{1}{81,749,606,400\pi^{11}} i$ |
| $c_{(12)abcdefghijklmnopqrstuvwx}$ | 12 | Duodecic | 24 | **77** | $+\frac{1}{8,036,313,307,545,600\pi^{12}}$ | $+\frac{1}{1,961,990,553,600\pi^{12}}$ |
| $c_{(13)abcdefghijklmnopqrstuvwxyz}$ | 13 | Tredecic | 26 | **101** | $+\frac{1}{417,888,291,992,371,200\pi^{13}} i$ | $+\frac{1}{51,011,754,393,600\pi^{13}} i$ |

### COEFFICIENT OF THE $0^{th}$ CHERN FORM

$$c_{(0)} = \frac{1}{0!}\langle 1, c_{(0)}\rangle = \frac{i^0}{2^0 \pi^0}(+1) \qquad (9)$$

### COEFFICIENT OF THE $1^{st}$ CHERN FORM

$$c_{(1)i_1 i_2} = \frac{1}{2!}\langle \mathbf{e}_{i_1} \wedge \mathbf{e}_{i_2}, c_{(1)}\rangle \qquad (10)$$

$$= \frac{i^1}{2^2 \pi^1 1!}(+R_{i_1 i_2 a}{}^a) = \frac{i^1}{2^2 \pi^1 1!}(+V_{i_1 i_2}) = \frac{i^1}{2^1 \pi^1 1!}(+P^{(2)}{}_{i_1 i_2})$$

### COEFFICIENT OF THE $2^{nd}$ CHERN FORM

$$c_{(2)i_1 i_2 i_3 i_4} = \frac{1}{4!}\langle \mathbf{e}_{i_1} \wedge \mathbf{e}_{i_2} \wedge \mathbf{e}_{i_3} \wedge \mathbf{e}_{i_4}, c_{(2)}\rangle \qquad (11)$$

$$= \frac{i^2}{2^4 \pi^2 2!}(-R_{[i_1 i_2|b|}{}^a R_{i_3 i_4]a}{}^b + R_{[i_1 i_2|a|}{}^a R_{i_3 i_4]b}{}^b)$$

$$= \frac{i^2}{2^4 \pi^2 2!}(-R_{[i_1 i_2|b|}{}^a R_{i_3 i_4]a}{}^b + V_{[i_1 i_2} V_{i_3 i_4]}) =$$

$$= \frac{i^2}{2^2 \pi^2 2!}(-P^{(4)}{}_{i_1 i_2 i_3 i_4} + P^{(2)}{}_{[i_1 i_2} P^{(2)}{}_{i_3 i_4]})$$

### COEFFICIENT OF THE $3^{rd}$ CHERN FORM

$$c_{(3)i_1 i_2 i_3 i_4 i_5 i_6} = \qquad (12)$$

$$= \frac{1}{6!}\langle \mathbf{e}_{i_1} \wedge \mathbf{e}_{i_2} \wedge \mathbf{e}_{i_3} \wedge \mathbf{e}_{i_4} \wedge \mathbf{e}_{i_5} \wedge \mathbf{e}_{i_6}, c_{(3)}\rangle$$

$$= \frac{i^3}{2^6 \pi^3 3!}(+2R_{[i_1 i_2|c|}{}^a R_{i_3 i_4|a|}{}^b R_{i_5 i_6]b}{}^c - 3R_{[i_1 i_2|b|}{}^a R_{i_3 i_4|a|}{}^b R_{i_5 i_6]c}{}^c +$$

$$+ R_{[i_1 i_2|a|}{}^a R_{i_3 i_4|b|}{}^b R_{i_5 i_6]c}{}^c)$$

$$= \frac{i^3}{2^6 \pi^3 3!}(+2R_{[i_1 i_2|c|}{}^a R_{i_3 i_4|a|}{}^b R_{i_5 i_6]b}{}^c - 3R_{[i_1 i_2|b|}{}^a R_{i_3 i_4|a|}{}^b V_{i_5 i_6]} +$$

$$+ V_{[i_1 i_2} V_{i_3 i_4} V_{i_5 i_6]})$$

$$= \frac{i^3}{2^3 \pi^3 3!}(+2P^{(6)}{}_{i_1 i_2 i_3 i_4 i_5 i_6} - 3P^{(4)}{}_{[i_1 i_2 i_3 i_4} P^{(2)}{}_{i_5 i_6]} +$$

$$+ P^{(2)}{}_{[i_1 i_2} P^{(2)}{}_{i_3 i_4} P^{(2)}{}_{i_5 i_6]})$$

### COEFFICIENT OF THE $4^{th}$ CHERN FORM

$$c_{(4)i_1 i_2 i_3 i_4 i_5 i_6 i_7 i_8} = \qquad (13)$$

$$= \frac{1}{8!}\langle \mathbf{e}_{i_1} \wedge \mathbf{e}_{i_2} \wedge \mathbf{e}_{i_3} \wedge \mathbf{e}_{i_4} \wedge \mathbf{e}_{i_5} \wedge \mathbf{e}_{i_6} \wedge \mathbf{e}_{i_7} \wedge \mathbf{e}_{i_8}, c_{(4)}\rangle$$

$$= \frac{i^4}{2^8 \pi^4 4!}(-6R_{[i_1 i_2|d|}{}^a R_{i_3 i_4|a|}{}^b R_{i_5 i_6|b|}{}^c R_{i_7 i_8|c|}{}^d + 8R_{[i_1 i_2|c|}{}^a R_{i_3 i_4|a|}{}^b R_{i_5 i_6|b|}{}^c R_{i_7 i_8]d}{}^d + 3R_{[i_1 i_2|b|}{}^a R_{i_3 i_4|a|}{}^b R_{i_5 i_6|d|}{}^c R_{i_7 i_8]c}{}^d -$$

$$- 6R_{[i_1 i_2|b|}{}^a R_{i_3 i_4|a|}{}^b R_{i_5 i_6|c|}{}^c R_{i_7 i_8]d}{}^d + R_{[i_1 i_2|a|}{}^a R_{i_3 i_4|b|}{}^b R_{i_5 i_6|c|}{}^c R_{i_7 i_8]d}{}^d)$$

$$= \frac{i^4}{2^8 \pi^4 4!}(-6R_{[i_1 i_2|d|}{}^a R_{i_3 i_4|a|}{}^b R_{i_5 i_6|b|}{}^c R_{i_7 i_8]c}{}^d + 8R_{[i_1 i_2|c|}{}^a R_{i_3 i_4|a|}{}^b R_{i_5 i_6|b|}{}^c V_{i_7 i_8]} + 3R_{[i_1 i_2|b|}{}^a R_{i_3 i_4|a|}{}^b R_{i_5 i_6|d|}{}^c R_{i_7 i_8]c}{}^d -$$

$$- 6R_{[i_1 i_2|b|}{}^a R_{i_3 i_4|a|}{}^b V_{i_5 i_6} V_{i_7 i_8]} + V_{[i_1 i_2} V_{i_3 i_4} V_{i_5 i_6} V_{i_7 i_8]})$$

$$= \frac{i^4}{2^4 \pi^4 4!}(-6P^{(8)}{}_{i_1 i_2 i_3 i_4 i_5 i_6 i_7 i_8} + 8P^{(6)}{}_{[i_1 i_2 i_3 i_4 i_5 i_6} P^{(2)}{}_{i_7 i_8]} + 3P^{(4)}{}_{[i_1 i_2 i_3 i_4} P^{(4)}{}_{i_5 i_6 i_7 i_8]} - 6P^{(4)}{}_{[i_1 i_2 i_3 i_4} P^{(2)}{}_{i_5 i_6} P^{(2)}{}_{i_7 i_8]} + P^{(2)}{}_{[i_1 i_2} P^{(2)}{}_{i_3 i_4} P^{(2)}{}_{i_5 i_6} P^{(2)}{}_{i_7 i_8]})$$

### COEFFICIENT OF THE $5^{th}$ CHERN FORM

$$c_{(5)i_1 i_2 i_3 i_4 i_5 i_6 i_7 i_8 i_9 i_{10}} = \qquad (14)$$

$$= \frac{1}{10!}\langle \mathbf{e}_{i_1} \wedge \mathbf{e}_{i_2} \wedge \mathbf{e}_{i_3} \wedge \mathbf{e}_{i_4} \wedge \mathbf{e}_{i_5} \wedge \mathbf{e}_{i_6} \wedge \mathbf{e}_{i_7} \wedge \mathbf{e}_{i_8} \wedge \mathbf{e}_{i_9} \wedge \mathbf{e}_{i_{10}}, c_{(5)}\rangle$$

$$= \frac{i^5}{2^{10} \pi^5 5!}(+24R_{[i_1 i_2|e|}{}^a R_{i_3 i_4|a|}{}^b R_{i_5 i_6|b|}{}^c R_{i_7 i_8|c|}{}^d R_{i_9 i_{10}]d}{}^e - 30R_{[i_1 i_2|d|}{}^a R_{i_3 i_4|a|}{}^b R_{i_5 i_6|b|}{}^c R_{i_7 i_8|c|}{}^d R_{i_9 i_{10}]e}{}^e - 20R_{[i_1 i_2|c|}{}^a R_{i_3 i_4|a|}{}^b R_{\mathbf{i_5 i_6}|b|}{}^c R_{i_7 i_8|e|}{}^d R_{i_9 i_{10}]d}{}^e +$$

$$+ 20R_{[i_1 i_2|c|}{}^a R_{i_3 i_4|a|}{}^b R_{i_5 i_6|b|}{}^c R_{i_7 i_8|d|}{}^d R_{i_9 i_{10}]e}{}^e + 15R_{[i_1 i_2|b|}{}^a R_{i_3 i_4|a|}{}^b R_{i_5 i_6|d|}{}^c R_{i_7 i_8|c|}{}^d R_{i_9 i_{10}]e}{}^e - 10R_{[i_1 i_2|b|}{}^a R_{i_3 i_4|a|}{}^b R_{i_5 i_6|c|}{}^c R_{i_7 i_8|d|}{}^d R_{i_9 i_{10}]e}{}^e +$$

$$+ R_{[i_1 i_2|a|}{}^a R_{i_3 i_4|b|}{}^b R_{i_5 i_6|c|}{}^c R_{i_7 i_8|d|}{}^d R_{i_9 i_{10}]e}{}^e)$$

$$= \frac{i^5}{2^{10} \pi^5 5!}(+24R_{[i_1 i_2|e|}{}^a R_{i_3 i_4|a|}{}^b R_{i_5 i_6|b|}{}^c R_{i_7 i_8|c|}{}^d R_{i_9 i_{10}]d}{}^e - 30R_{[i_1 i_2|d|}{}^a R_{i_3 i_4|a|}{}^b R_{i_5 i_6|b|}{}^c R_{i_7 i_8|c|}{}^d V_{i_9 i_{10}]} - 20R_{[i_1 i_2|c|}{}^a R_{i_3 i_4|a|}{}^b R_{i_5 i_6|b|}{}^c R_{i_7 i_8|e|}{}^d R_{i_9 i_{10}]d}{}^e +$$

$$+ 20R_{[i_1 i_2|c|}{}^a R_{i_3 i_4|a|}{}^b R_{i_5 i_6|b|}{}^c V_{i_7 i_8} V_{i_9 i_{10}]} + 15R_{[i_1 i_2|b|}{}^a R_{i_3 i_4|a|}{}^b R_{i_5 i_6|d|}{}^c R_{i_7 i_8|c|}{}^d V_{i_9 i_{10}]} - 10R_{[i_1 i_2|b|}{}^a R_{i_3 i_4|a|}{}^b V_{i_5 i_6} V_{i_7 i_8} V_{i_9 i_{10}]} +$$

$$+ V_{[i_1 i_2} V_{i_3 i_4} V_{i_5 i_6} V_{i_7 i_8} V_{i_9 i_{10}]})$$

$$= \frac{i^5}{2^5 \pi^5 5!}(+24P^{(10)}{}_{i_1 i_2 i_3 i_4 i_5 i_6 i_7 i_8 i_9 i_{10}} - 30P^{(8)}{}_{[i_1 i_2 i_3 i_4 i_5 i_6 i_7 i_8} P^{(2)}{}_{i_9 i_{10}]} - 20P^{(6)}{}_{[i_1 i_2 i_3 i_4 i_5 i_6} P^{(4)}{}_{i_7 i_8 i_9 i_{10}]} + 20P^{(6)}{}_{[i_1 i_2 i_3 i_4 i_5 i_6} P^{(2)}{}_{i_7 i_8} P^{(2)}{}_{i_9 i_{10}]} +$$

$$+ 15P^{(4)}{}_{[i_1 i_2 i_3 i_4} P^{(4)}{}_{i_5 i_6 i_7 i_8} P^{(2)}{}_{i_9 i_{10}]} - 10P^{(4)}{}_{[i_1 i_2 i_3 i_4} P^{(2)}{}_{i_5 i_6} P^{(2)}{}_{i_7 i_8} P^{(2)}{}_{i_9 i_{10}]} + P^{(2)}{}_{[i_1 i_2} P^{(2)}{}_{i_3 i_4} P^{(2)}{}_{i_5 i_6} P^{(2)}{}_{i_7 i_8} P^{(2)}{}_{i_9 i_{10}]})$$

### COEFFICIENT OF THE $6^{th}$ CHERN FORM

$$c_{(6)i_1 i_2 i_3 i_4 i_5 i_6 i_7 i_8 i_9 i_{10} i_{11} i_{12}} = \qquad (15)$$

$$= \frac{1}{12!}\langle \mathbf{e}_{i_1} \wedge \mathbf{e}_{i_2} \wedge \mathbf{e}_{i_3} \wedge \mathbf{e}_{i_4} \wedge \mathbf{e}_{i_5} \wedge \mathbf{e}_{i_6} \wedge \mathbf{e}_{i_7} \wedge \mathbf{e}_{i_8} \wedge \mathbf{e}_{i_9} \wedge \mathbf{e}_{i_{10}} \wedge \mathbf{e}_{i_{11}} \wedge \mathbf{e}_{i_{12}}, c_{(6)}\rangle$$

$$= \frac{i^6}{2^{12} \pi^6 6!}(-120R_{[i_1 i_2|f|}{}^a R_{i_3 i_4|a|}{}^b R_{i_5 i_6|b|}{}^c R_{i_7 i_8|c|}{}^d R_{i_9 i_{10}|d|}{}^e R_{i_{11} i_{12}]e}{}^f + 144R_{[i_1 i_2|e|}{}^a R_{i_3 i_4|a|}{}^b R_{i_5 i_6|b|}{}^c R_{i_7 i_8|c|}{}^d R_{i_9 i_{10}|d|}{}^e R_{i_{11} i_{12}]f}{}^f +$$

$$+ 90R_{[i_1 i_2|d|}{}^a R_{i_3 i_4|a|}{}^b R_{i_5 i_6|b|}{}^c R_{i_7 i_8|c|}{}^d R_{i_9 i_{10}|f|}{}^e R_{i_{11} i_{12}]e}{}^f - 90R_{[i_1 i_2|d|}{}^a R_{i_3 i_4|a|}{}^b R_{i_5 i_6|b|}{}^c R_{i_7 i_8|c|}{}^d R_{i_9 i_{10}|e|}{}^e R_{i_{11} i_{12}]f}{}^f +$$

$$+ 40R_{[i_1 i_2|c|}{}^a R_{i_3 i_4|a|}{}^b R_{i_5 i_6|b|}{}^c R_{i_7 i_8|f|}{}^d R_{i_9 i_{10}|d|}{}^e R_{i_{11} i_{12}]e}{}^f - 120R_{[i_1 i_2|c|}{}^a R_{i_3 i_4|a|}{}^b R_{i_5 i_6|b|}{}^c R_{i_7 i_8|e|}{}^d R_{i_9 i_{10}|d|}{}^e R_{i_{11} i_{12}]f}{}^f +$$

$$+ 40R_{[i_1 i_2|c|}{}^a R_{i_3 i_4|a|}{}^b R_{i_5 i_6|b|}{}^c R_{i_7 i_8|d|}{}^d R_{i_9 i_{10}|e|}{}^e R_{i_{11} i_{12}]f}{}^f - 15R_{[i_1 i_2|b|}{}^a R_{i_3 i_4|a|}{}^b R_{i_5 i_6|d|}{}^c R_{i_7 i_8|c|}{}^d R_{i_9 i_{10}|f|}{}^e R_{i_{11} i_{12}]e}{}^f +$$

$$+ 45R_{[i_1 i_2|b|}{}^a R_{i_3 i_4|a|}{}^b R_{i_5 i_6|d|}{}^c R_{i_7 i_8|c|}{}^d R_{i_9 i_{10}|e|}{}^e R_{i_{11} i_{12}]f}{}^f - 15R_{[i_1 i_2|b|}{}^a R_{i_3 i_4|a|}{}^b R_{i_5 i_6|c|}{}^c R_{i_7 i_8|d|}{}^d R_{i_9 i_{10}|e|}{}^e R_{i_{11} i_{12}]f}{}^f +$$

$$+ R_{[i_1 i_2|a|}{}^a R_{i_3 i_4|b|}{}^b R_{i_5 i_6|c|}{}^c R_{i_7 i_8|d|}{}^d R_{i_9 i_{10}|e|}{}^e R_{i_{11} i_{12}]f}{}^f)$$

$$= \frac{i^6}{2^{12} \pi^6 6!}(-120R_{[i_1 i_2|f|}{}^a R_{i_3 i_4|a|}{}^b R_{i_5 i_6|b|}{}^c R_{i_7 i_8|c|}{}^d R_{i_9 i_{10}|d|}{}^e R_{i_{11} i_{12}]e}{}^f + 144R_{[i_1 i_2|e|}{}^a R_{i_3 i_4|a|}{}^b R_{i_5 i_6|b|}{}^c R_{i_7 i_8|c|}{}^d R_{i_9 i_{10}|d|}{}^e V_{i_{11} i_{12}]} +$$

$$+ 90R_{[i_1 i_2|d|}{}^a R_{i_3 i_4|a|}{}^b R_{i_5 i_6|b|}{}^c R_{i_7 i_8|c|}{}^d R_{i_9 i_{10}|f|}{}^e R_{i_{11} i_{12}]e}{}^f - 90R_{[i_1 i_2|d|}{}^a R_{i_3 i_4|a|}{}^b R_{i_5 i_6|b|}{}^c R_{i_7 i_8|c|}{}^d V_{i_9 i_{10}} V_{i_{11} i_{12}]} +$$

$$+ 40R_{[i_1 i_2|c|}{}^a R_{i_3 i_4|a|}{}^b R_{i_5 i_6|b|}{}^c R_{i_7 i_8|f|}{}^d R_{i_9 i_{10}|d|}{}^e R_{i_{11} i_{12}]e}{}^f - 120R_{[i_1 i_2|c|}{}^a R_{i_3 i_4|a|}{}^b R_{i_5 i_6|b|}{}^c R_{i_7 i_8|e|}{}^d R_{i_9 i_{10}|d|}{}^e V_{i_{11} i_{12}]} +$$

$$+ 40\, R_{[i_1i_2|c|}{}^a R_{i_3i_4|a|}{}^b R_{i_5i_6|b|}{}^c V_{i_7i_8} V_{i_9i_{10}} V_{i_{11}i_{12}]} - 15\, R_{[i_1i_2|b|}{}^a R_{i_3i_4|a|}{}^b R_{i_5i_6|d|}{}^c R_{i_7i_8|c|}{}^d R_{i_9i_{10}|f|}{}^e R_{i_{11}i_{12}]}{}^f +$$

$$+ 45\, R_{[i_1i_2|b|}{}^a R_{i_3i_4|a|}{}^b R_{i_5i_6|d|}{}^c R_{i_7i_8|c|}{}^d V_{i_9i_{10}} V_{i_{11}i_{12}]} - 15\, R_{[i_1i_2|b|}{}^a R_{i_3i_4|a|}{}^b V_{i_5i_6} V_{i_7i_8} V_{i_9i_{10}} V_{i_{11}i_{12}]} + V_{[i_1i_2} V_{i_3i_4} V_{i_5i_6} V_{i_7i_8} V_{i_9i_{10}} V_{i_{11}i_{12}]})$$

$$+ V_{[i_1i_2} V_{i_3i_4} V_{i_5i_6} V_{i_7i_8} V_{i_9i_{10}} V_{i_{11}i_{12}]})$$

$$= \frac{i^6}{2^6 \pi^6 6!}\bigl(-120\, P^{(12)}{}_{i_1i_2i_3i_4i_5i_6i_7i_8i_9i_{10}i_{11}i_{12}} + 144\, P^{(10)}{}_{[i_1i_2i_3i_4i_5i_6i_7i_8i_9i_{10}} P^{(2)}{}_{i_{11}i_{12}]} + 90\, P^{(8)}{}_{[i_1i_2i_3i_4i_5i_6i_7i_8} P^{(4)}{}_{i_9i_{10}i_{11}i_{12}]} -$$

$$- 90\, P^{(8)}{}_{[i_1i_2i_3i_4i_5i_6i_7i_8} P^{(2)}{}_{i_9i_{10}} P^{(2)}{}_{i_{11}i_{12}]} + 40\, P^{(6)}{}_{[i_1i_2i_3i_4i_5i_6} P^{(6)}{}_{i_7i_8i_9i_{10}i_{11}i_{12}]} - 120\, P^{(6)}{}_{[i_1i_2i_3i_4i_5i_6} P^{(4)}{}_{i_7i_8i_9i_{10}} P^{(2)}{}_{i_{11}i_{12}]} +$$

$$+ 40\, P^{(6)}{}_{[i_1i_2i_3i_4i_5i_6} P^{(2)}{}_{i_7i_8} P^{(2)}{}_{i_9i_{10}} P^{(2)}{}_{i_{11}i_{12}]} - 15\, P^{(4)}{}_{[i_1i_2i_3i_4} P^{(4)}{}_{i_5i_6i_7i_8} P^{(4)}{}_{i_9i_{10}i_{11}i_{12}]} + 45\, P^{(4)}{}_{[i_1i_2i_3i_4} P^{(4)}{}_{i_5i_6i_7i_8} P^{(2)}{}_{i_9i_{10}} P^{(2)}{}_{i_{11}i_{12}]} -$$

$$- 15\, P^{(4)}{}_{[i_1i_2i_3i_4} P^{(2)}{}_{i_5i_6} P^{(2)}{}_{i_7i_8} P^{(2)}{}_{i_9i_{10}} P^{(2)}{}_{i_{11}i_{12}]} + P^{(2)}{}_{[i_1i_2} P^{(2)}{}_{i_3i_4} P^{(2)}{}_{i_5i_6} P^{(2)}{}_{i_7i_8} P^{(2)}{}_{i_9i_{10}} P^{(2)}{}_{i_{11}i_{12}]}\bigr)$$

COEFFICIENT OF THE 7$^{th}$ CHERN FORM

$$c_{(7)i_1i_2i_3i_4i_5i_6i_7i_8i_9i_{10}i_{11}i_{12}i_{13}i_{14}} = \tag{16}$$

$$= \frac{1}{14!}\langle \mathbf{e}_{i_1} \wedge \mathbf{e}_{i_2} \wedge \mathbf{e}_{i_3} \wedge \mathbf{e}_{i_4} \wedge \mathbf{e}_{i_5} \wedge \mathbf{e}_{i_6} \wedge \mathbf{e}_{i_7} \wedge \mathbf{e}_{i_8} \wedge \mathbf{e}_{i_9} \wedge \mathbf{e}_{i_{10}} \wedge \mathbf{e}_{i_{11}} \wedge \mathbf{e}_{i_{12}} \wedge \mathbf{e}_{i_{13}} \wedge \mathbf{e}_{i_{14}}, c_{(7)}\rangle$$

$$= \frac{i^7}{2^{14}\pi^7 7!}\bigl(+ 720\, R_{[i_1i_2|g|}{}^a R_{i_3i_4|a|}{}^b R_{i_5i_6|b|}{}^c R_{i_7i_8|c|}{}^d R_{i_9i_{10}|d|}{}^e R_{i_{11}i_{12}|e|}{}^f R_{i_{13}i_{14}]f}{}^g - 840\, R_{[i_1i_2|f|}{}^a R_{i_3i_4|a|}{}^b R_{i_5i_6|b|}{}^c R_{i_7i_8|c|}{}^d R_{i_9i_{10}|d|}{}^e R_{i_{11}i_{12}|e|}{}^f R_{i_{13}i_{14}]g}{}^g -$$

$$- 504\, R_{[i_1i_2|e|}{}^a R_{i_3i_4|a|}{}^b R_{i_5i_6|b|}{}^c R_{i_7i_8|c|}{}^d R_{i_9i_{10}|d|}{}^e R_{i_{11}i_{12}|f|}{}^f R_{i_{13}i_{14}]f}{}^g + 504\, R_{[i_1i_2|e|}{}^a R_{i_3i_4|a|}{}^b R_{i_5i_6|b|}{}^c R_{i_7i_8|c|}{}^d R_{i_9i_{10}|d|}{}^e R_{i_{11}i_{12}|f|}{}^f R_{i_{13}i_{14}]g}{}^g -$$

$$- 420\, R_{[i_1i_2|d|}{}^a R_{i_3i_4|a|}{}^b R_{i_5i_6|b|}{}^c R_{i_7i_8|c|}{}^d R_{i_9i_{10}|g|}{}^e R_{i_{11}i_{12}|e|}{}^f R_{i_{13}i_{14}]f}{}^g + 630\, R_{[i_1i_2|d|}{}^a R_{i_3i_4|a|}{}^b R_{i_5i_6|b|}{}^c R_{i_7i_8|c|}{}^d R_{i_9i_{10}|e|}{}^e R_{i_{11}i_{12}|e|}{}^f R_{i_{13}i_{14}]f}{}^g -$$

$$- 210\, R_{[i_1i_2|d|}{}^a R_{i_3i_4|a|}{}^b R_{i_5i_6|b|}{}^c R_{i_7i_8|c|}{}^d R_{i_9i_{10}|e|}{}^e R_{i_{11}i_{12}|f|}{}^f R_{i_{13}i_{14}]g}{}^g + 280\, R_{[i_1i_2|c|}{}^a R_{i_3i_4|a|}{}^b R_{i_5i_6|b|}{}^c R_{i_7i_8|f|}{}^d R_{i_9i_{10}|d|}{}^e R_{i_{11}i_{12}|e|}{}^f R_{i_{13}i_{14}]f}{}^g +$$

$$+ 210\, R_{[i_1i_2|c|}{}^a R_{i_3i_4|a|}{}^b R_{i_5i_6|b|}{}^c R_{i_7i_8|e|}{}^d R_{i_9i_{10}|d|}{}^e R_{i_{11}i_{12}|e|}{}^f R_{i_{13}i_{14}]f}{}^g - 420\, R_{[i_1i_2|c|}{}^a R_{i_3i_4|a|}{}^b R_{i_5i_6|b|}{}^c R_{i_7i_8|e|}{}^d R_{i_9i_{10}|d|}{}^e R_{i_{11}i_{12}|f|}{}^f R_{i_{13}i_{14}]g}{}^g +$$

$$+ 70\, R_{[i_1i_2|c|}{}^a R_{i_3i_4|a|}{}^b R_{i_5i_6|b|}{}^c R_{i_7i_8|d|}{}^d R_{i_9i_{10}|e|}{}^e R_{i_{11}i_{12}|f|}{}^f R_{i_{13}i_{14}]g}{}^g - 105\, R_{[i_1i_2|b|}{}^a R_{i_3i_4|a|}{}^b R_{i_5i_6|d|}{}^c R_{i_7i_8|c|}{}^d R_{i_9i_{10}|f|}{}^e R_{i_{11}i_{12}|e|}{}^f R_{i_{13}i_{14}]g}{}^g +$$

$$+ 105\, R_{[i_1i_2|b|}{}^a R_{i_3i_4|a|}{}^b R_{i_5i_6|d|}{}^c R_{i_7i_8|c|}{}^d R_{i_9i_{10}|e|}{}^e R_{i_{11}i_{12}|f|}{}^f R_{i_{13}i_{14}]g}{}^g - 21\, R_{[i_1i_2|b|}{}^a R_{i_3i_4|a|}{}^b R_{i_5i_6|c|}{}^c R_{i_7i_8|d|}{}^d R_{i_9i_{10}|e|}{}^e R_{i_{11}i_{12}|f|}{}^f R_{i_{13}i_{14}]g}{}^g +$$

$$+ R_{[i_1i_2|a|}{}^a R_{i_3i_4|b|}{}^b R_{i_5i_6|c|}{}^c R_{i_7i_8|d|}{}^d R_{i_9i_{10}|e|}{}^e R_{i_{11}i_{12}|f|}{}^f R_{i_{13}i_{14}]g}{}^g\bigr)$$

$$= \frac{i^7}{2^{14}\pi^7 7!}\bigl(+ 720\, R_{[i_1i_2|g|}{}^a R_{i_3i_4|a|}{}^b R_{i_5i_6|b|}{}^c R_{i_7i_8|c|}{}^d R_{i_9i_{10}|d|}{}^e R_{i_{11}i_{12}|e|}{}^f R_{i_{13}i_{14}]f}{}^g - 840\, R_{[i_1i_2|f|}{}^a R_{i_3i_4|a|}{}^b R_{i_5i_6|b|}{}^c R_{i_7i_8|c|}{}^d R_{i_9i_{10}|d|}{}^e R_{i_{11}i_{12}|e|}{}^f V_{i_{13}i_{14}]} -$$

$$- 504\, R_{[i_1i_2|e|}{}^a R_{i_3i_4|a|}{}^b R_{i_5i_6|b|}{}^c R_{i_7i_8|c|}{}^d R_{i_9i_{10}|d|}{}^e R_{i_{11}i_{12}|e|}{}^f R_{i_{13}i_{14}]f}{}^g + 504\, R_{[i_1i_2|e|}{}^a R_{i_3i_4|a|}{}^b R_{i_5i_6|b|}{}^c R_{i_7i_8|c|}{}^d R_{i_9i_{10}|d|}{}^e V_{i_{11}i_{12}} V_{i_{13}i_{14}]} -$$

$$- 420\, R_{[i_1i_2|d|}{}^a R_{i_3i_4|a|}{}^b R_{i_5i_6|b|}{}^c R_{i_7i_8|c|}{}^d R_{i_9i_{10}|e|}{}^e R_{i_{11}i_{12}|e|}{}^f R_{i_{13}i_{14}]f}{}^g + 630\, R_{[i_1i_2|d|}{}^a R_{i_3i_4|a|}{}^b R_{i_5i_6|b|}{}^c R_{i_7i_8|c|}{}^d R_{i_9i_{10}|e|}{}^e R_{i_{11}i_{12}|e|}{}^f V_{i_{13}i_{14}]} -$$

$$- 210\, R_{[i_1i_2|d|}{}^a R_{i_3i_4|a|}{}^b R_{i_5i_6|b|}{}^c R_{i_7i_8|c|}{}^d V_{i_9i_{10}} V_{i_{11}i_{12}} V_{i_{13}i_{14}]} + 280\, R_{[i_1i_2|c|}{}^a R_{i_3i_4|a|}{}^b R_{i_5i_6|b|}{}^c R_{i_7i_8|f|}{}^d R_{i_9i_{10}|d|}{}^e R_{i_{11}i_{12}|e|}{}^f V_{i_{13}i_{14}]} +$$

$$+ 210\, R_{[i_1i_2|c|}{}^a R_{i_3i_4|a|}{}^b R_{i_5i_6|b|}{}^c R_{i_7i_8|e|}{}^d R_{i_9i_{10}|d|}{}^e R_{i_{11}i_{12}|e|}{}^f R_{i_{13}i_{14}]f}{}^g - 420\, R_{[i_1i_2|c|}{}^a R_{i_3i_4|a|}{}^b R_{i_5i_6|b|}{}^c R_{i_7i_8|e|}{}^d R_{i_9i_{10}|d|}{}^e V_{i_{11}i_{12}} V_{i_{13}i_{14}]} +$$

$$+ 70\, R_{[i_1i_2|c|}{}^a R_{i_3i_4|a|}{}^b R_{i_5i_6|b|}{}^c V_{i_7i_8} V_{i_9i_{10}} V_{i_{11}i_{12}} V_{i_{13}i_{14}]} - 105\, R_{[i_1i_2|b|}{}^a R_{i_3i_4|a|}{}^b R_{i_5i_6|d|}{}^c R_{i_7i_8|c|}{}^d R_{i_9i_{10}|f|}{}^e R_{i_{11}i_{12}|e|}{}^f V_{i_{13}i_{14}]} +$$

$$+ 105\, R_{[i_1i_2|b|}{}^a R_{i_3i_4|a|}{}^b R_{i_5i_6|d|}{}^c R_{i_7i_8|c|}{}^d V_{i_9i_{10}} V_{i_{11}i_{12}} V_{i_{13}i_{14}]} - 21\, R_{[i_1i_2|b|}{}^a R_{i_3i_4|a|}{}^b V_{i_5i_6} V_{i_7i_8} V_{i_9i_{10}} V_{i_{11}i_{12}} V_{i_{13}i_{14}]} +$$

$$+ V_{[i_1i_2} V_{i_3i_4} V_{i_5i_6} V_{i_7i_8} V_{i_9i_{10}} V_{i_{11}i_{12}} V_{i_{13}i_{14}]}\bigr)$$

$$= \frac{i^7}{2^7 \pi^7 7!}\bigl(+ 720\, P^{(14)}{}_{i_1i_2i_3i_4i_5i_6i_7i_8i_9i_{10}i_{11}i_{12}i_{13}i_{14}} - 840\, P^{(12)}{}_{[i_1i_2i_3i_4i_5i_6i_7i_8i_9i_{10}i_{11}i_{12}} P^{(2)}{}_{i_{13}i_{14}]} - 504\, P^{(10)}{}_{[i_1i_2i_3i_4i_5i_6i_7i_8i_9i_{10}} P^{(4)}{}_{i_{11}i_{12}i_{13}i_{14}]} +$$

$$+ 504\, P^{(10)}{}_{[i_1i_2i_3i_4i_5i_6i_7i_8i_9i_{10}} P^{(2)}{}_{i_{11}i_{12}} P^{(2)}{}_{i_{13}i_{14}]} - 420\, P^{(8)}{}_{[i_1i_2i_3i_4i_5i_6i_7i_8} P^{(6)}{}_{i_9i_{10}i_{11}i_{12}i_{13}i_{14}]} + 630\, P^{(8)}{}_{[i_1i_2i_3i_4i_5i_6i_7i_8} P^{(4)}{}_{i_9i_{10}i_{11}i_{12}} P^{(2)}{}_{i_{13}i_{14}]} -$$

$$- 210\, P^{(8)}{}_{[i_1i_2i_3i_4i_5i_6i_7i_8} P^{(2)}{}_{i_9i_{10}} P^{(2)}{}_{i_{11}i_{12}} P^{(2)}{}_{i_{13}i_{14}]} + 280\, P^{(6)}{}_{[i_1i_2i_3i_4i_5i_6} P^{(6)}{}_{i_7i_8i_9i_{10}i_{11}i_{12}} P^{(2)}{}_{i_{13}i_{14}]} + 210\, P^{(6)}{}_{[i_1i_2i_3i_4i_5i_6} P^{(4)}{}_{i_7i_8i_9i_{10}} P^{(4)}{}_{i_{11}i_{12}i_{13}i_{14}]} -$$

$$- 420\, P^{(6)}{}_{[i_1i_2i_3i_4i_5i_6} P^{(4)}{}_{i_7i_8i_9i_{10}} P^{(2)}{}_{i_{11}i_{12}} P^{(2)}{}_{i_{13}i_{14}]} + 70\, P^{(6)}{}_{[i_1i_2i_3i_4i_5i_6} P^{(2)}{}_{i_7i_8} P^{(2)}{}_{i_9i_{10}} P^{(2)}{}_{i_{11}i_{12}} P^{(2)}{}_{i_{13}i_{14}]} -$$

$$- 105\, P^{(4)}{}_{[i_1i_2i_3i_4} P^{(4)}{}_{i_5i_6i_7i_8} P^{(4)}{}_{i_9i_{10}i_{11}i_{12}} P^{(2)}{}_{i_{13}i_{14}]} + 105\, P^{(4)}{}_{[i_1i_2i_3i_4} P^{(4)}{}_{i_5i_6i_7i_8} P^{(2)}{}_{i_9i_{10}} P^{(2)}{}_{i_{11}i_{12}} P^{(2)}{}_{i_{13}i_{14}]} -$$

$$- 21\, P^{(4)}{}_{[i_1i_2i_3i_4} P^{(2)}{}_{i_5i_6} P^{(2)}{}_{i_7i_8} P^{(2)}{}_{i_9i_{10}} P^{(2)}{}_{i_{11}i_{12}} P^{(2)}{}_{i_{13}i_{14}]} + P^{(2)}{}_{[i_1i_2} P^{(2)}{}_{i_3i_4} P^{(2)}{}_{i_5i_6} P^{(2)}{}_{i_7i_8} P^{(2)}{}_{i_9i_{10}} P^{(2)}{}_{i_{11}i_{12}} P^{(2)}{}_{i_{13}i_{14}]}\bigr)$$

COEFFICIENT OF THE 8$^{th}$ CHERN FORM

$$c_{(8)i_1i_2i_3i_4i_5i_6i_7i_8i_9i_{10}i_{11}i_{12}i_{13}i_{14}i_{15}i_{16}} = \frac{1}{16!}\langle \mathbf{e}_{i_1} \wedge \mathbf{e}_{i_2} \wedge \mathbf{e}_{i_3} \wedge \mathbf{e}_{i_4} \wedge \mathbf{e}_{i_5} \wedge \mathbf{e}_{i_6} \wedge \mathbf{e}_{i_7} \wedge \mathbf{e}_{i_8} \wedge \mathbf{e}_{i_9} \wedge \mathbf{e}_{i_{10}} \wedge \mathbf{e}_{i_{11}} \wedge \mathbf{e}_{i_{12}} \wedge \mathbf{e}_{i_{13}} \wedge \mathbf{e}_{i_{14}} \wedge \mathbf{e}_{i_{15}} \wedge \mathbf{e}_{i_{16}}, c_{(8)}\rangle \tag{17}$$

$$= \frac{i^8}{2^{16}\pi^8 8!}\bigl(- 5040\, R_{[i_1i_2|h|}{}^a R_{i_3i_4|a|}{}^b R_{i_5i_6|b|}{}^c R_{i_7i_8|c|}{}^d R_{i_9i_{10}|d|}{}^e R_{i_{11}i_{12}|e|}{}^f R_{i_{13}i_{14}|f|}{}^g R_{i_{15}i_{16}]g}{}^h +$$

$$+ 5760\, R_{[i_1i_2|g|}{}^a R_{i_3i_4|a|}{}^b R_{i_5i_6|b|}{}^c R_{i_7i_8|c|}{}^d R_{i_9i_{10}|d|}{}^e R_{i_{11}i_{12}|e|}{}^f R_{i_{13}i_{14}|f|}{}^g R_{i_{15}i_{16}]h}{}^h +$$

$$+ 3360\, R_{[i_1i_2|f|}{}^a R_{i_3i_4|a|}{}^b R_{i_5i_6|b|}{}^c R_{i_7i_8|c|}{}^d R_{i_9i_{10}|d|}{}^e R_{i_{11}i_{12}|e|}{}^f R_{i_{13}i_{14}|h|}{}^g R_{i_{15}i_{16}]g}{}^h -$$

$$- 3360\, R_{[i_1i_2|f|}{}^a R_{i_3i_4|a|}{}^b R_{i_5i_6|b|}{}^c R_{i_7i_8|c|}{}^d R_{i_9i_{10}|d|}{}^e R_{i_{11}i_{12}|e|}{}^f R_{i_{13}i_{14}|g|}{}^g R_{i_{15}i_{16}]h}{}^h +$$

$$+ 2688\, R_{[i_1i_2|e|}{}^a R_{i_3i_4|a|}{}^b R_{i_5i_6|b|}{}^c R_{i_7i_8|c|}{}^d R_{i_9i_{10}|d|}{}^e R_{i_{11}i_{12}|h|}{}^f R_{i_{13}i_{14}|f|}{}^g R_{i_{15}i_{16}]g}{}^h -$$

$$- 4032\, R_{[i_1i_2|e|}{}^a R_{i_3i_4|a|}{}^b R_{i_5i_6|b|}{}^c R_{i_7i_8|c|}{}^d R_{i_9i_{10}|d|}{}^e R_{i_{11}i_{12}|g|}{}^f R_{i_{13}i_{14}|f|}{}^g R_{i_{15}i_{16}]h}{}^h +$$

$$+ 1344\, R_{[i_1i_2|e|}{}^a R_{i_3i_4|a|}{}^b R_{i_5i_6|b|}{}^c R_{i_7i_8|c|}{}^d R_{i_9i_{10}|d|}{}^e R_{i_{11}i_{12}|f|}{}^f R_{i_{13}i_{14}|g|}{}^g R_{i_{15}i_{16}]h}{}^h +$$

$$+ 1260\, R_{[i_1 i_2|d|}{}^a R_{i_3 i_4|a|}{}^b R_{i_5 i_6|b|}{}^c R_{i_7 i_8|c|}{}^d R_{i_9 i_{10}|h|}{}^e R_{i_{11} i_{12}|e|}{}^f R_{i_{13} i_{14}|f|}{}^g R_{i_{15} i_{16}]g}{}^h -$$
$$- 3360\, R_{[i_1 i_2|d|}{}^a R_{i_3 i_4|a|}{}^b R_{i_5 i_6|b|}{}^c R_{i_7 i_8|c|}{}^d R_{i_9 i_{10}|g|}{}^e R_{i_{11} i_{12}|e|}{}^f R_{i_{13} i_{14}|f|}{}^g R_{i_{15} i_{16}]h}{}^h -$$
$$- 1260\, R_{[i_1 i_2|d|}{}^a R_{i_3 i_4|a|}{}^b R_{i_5 i_6|b|}{}^c R_{i_7 i_8|c|}{}^d R_{i_9 i_{10}|f|}{}^e R_{i_{11} i_{12}|e|}{}^f R_{i_{13} i_{14}|h|}{}^g R_{i_{15} i_{16}]h}{}^h +$$
$$+ 2520\, R_{[i_1 i_2|d|}{}^a R_{i_3 i_4|a|}{}^b R_{i_5 i_6|b|}{}^c R_{i_7 i_8|c|}{}^d R_{i_9 i_{10}|e|}{}^e R_{i_{11} i_{12}|e|}{}^f R_{i_{13} i_{14}|f|}{}^g R_{i_{15} i_{16}]h}{}^h -$$
$$- 420\, R_{[i_1 i_2|d|}{}^a R_{i_3 i_4|a|}{}^b R_{i_5 i_6|b|}{}^c R_{i_7 i_8|c|}{}^d R_{i_9 i_{10}|e|}{}^e R_{i_{11} i_{12}|f|}{}^f R_{i_{13} i_{14}|g|}{}^g R_{i_{15} i_{16}]h}{}^h -$$
$$- 1120\, R_{[i_1 i_2|c|}{}^a R_{i_3 i_4|a|}{}^b R_{i_5 i_6|b|}{}^c R_{i_7 i_8|f|}{}^d R_{i_9 i_{10}|d|}{}^e R_{i_{11} i_{12}|e|}{}^f R_{i_{13} i_{14}|h|}{}^g R_{i_{15} i_{16}]g}{}^h +$$
$$+ 1120\, R_{[i_1 i_2|c|}{}^a R_{i_3 i_4|a|}{}^b R_{i_5 i_6|b|}{}^c R_{i_7 i_8|f|}{}^d R_{i_9 i_{10}|d|}{}^e R_{i_{11} i_{12}|e|}{}^f R_{i_{13} i_{14}|g|}{}^g R_{i_{15} i_{16}]h}{}^h +$$
$$+ 1680\, R_{[i_1 i_2|c|}{}^a R_{i_3 i_4|a|}{}^b R_{i_5 i_6|b|}{}^c R_{i_7 i_8|e|}{}^d R_{i_9 i_{10}|d|}{}^e R_{i_{11} i_{12}|g|}{}^f R_{i_{13} i_{14}|f|}{}^g R_{i_{15} i_{16}]h}{}^h -$$
$$- 1120\, R_{[i_1 i_2|c|}{}^a R_{i_3 i_4|a|}{}^b R_{i_5 i_6|b|}{}^c R_{i_7 i_8|e|}{}^d R_{i_9 i_{10}|d|}{}^e R_{i_{11} i_{12}|f|}{}^f R_{i_{13} i_{14}|g|}{}^g R_{i_{15} i_{16}]h}{}^h +$$
$$+ 112\, R_{[i_1 i_2|c|}{}^a R_{i_3 i_4|a|}{}^b R_{i_5 i_6|b|}{}^c R_{i_7 i_8|d|}{}^d R_{i_9 i_{10}|e|}{}^e R_{i_{11} i_{12}|f|}{}^f R_{i_{13} i_{14}|g|}{}^g R_{i_{15} i_{16}]h}{}^h +$$
$$+ 105\, R_{[i_1 i_2|b|}{}^a R_{i_3 i_4|a|}{}^b R_{i_5 i_6|d|}{}^c R_{i_7 i_8|c|}{}^d R_{i_9 i_{10}|f|}{}^e R_{i_{11} i_{12}|e|}{}^f R_{i_{13} i_{14}|h|}{}^g R_{i_{15} i_{16}]g}{}^h -$$
$$- 420\, R_{[i_1 i_2|b|}{}^a R_{i_3 i_4|a|}{}^b R_{i_5 i_6|d|}{}^c R_{i_7 i_8|c|}{}^d R_{i_9 i_{10}|e|}{}^e R_{i_{11} i_{12}|e|}{}^f R_{i_{13} i_{14}|g|}{}^g R_{i_{15} i_{16}]h}{}^h +$$
$$+ 210\, R_{[i_1 i_2|b|}{}^a R_{i_3 i_4|a|}{}^b R_{i_5 i_6|d|}{}^c R_{i_7 i_8|c|}{}^d R_{i_9 i_{10}|e|}{}^e R_{i_{11} i_{12}|f|}{}^f R_{i_{13} i_{14}|g|}{}^g R_{i_{15} i_{16}]h}{}^h -$$
$$- 28\, R_{[i_1 i_2|b|}{}^a R_{i_3 i_4|a|}{}^b R_{i_5 i_6|c|}{}^c R_{i_7 i_8|d|}{}^d R_{i_9 i_{10}|e|}{}^e R_{i_{11} i_{12}|f|}{}^f R_{i_{13} i_{14}|g|}{}^g R_{i_{15} i_{16}]h}{}^h +$$
$$+ R_{[i_1 i_2|a|}{}^a R_{i_3 i_4|b|}{}^b R_{i_5 i_6|c|}{}^c R_{i_7 i_8|d|}{}^d R_{i_9 i_{10}|e|}{}^e R_{i_{11} i_{12}|f|}{}^f R_{i_{13} i_{14}|g|}{}^g R_{i_{15} i_{16}]h}{}^h )$$

$$= \frac{i^8}{2^{16} \pi^8 8!} \Big( -5040\, R_{[i_1 i_2|h|}{}^a R_{i_3 i_4|a|}{}^b R_{i_5 i_6|b|}{}^c R_{i_7 i_8|c|}{}^d R_{i_9 i_{10}|d|}{}^e R_{i_{11} i_{12}|e|}{}^f R_{i_{13} i_{14}|f|}{}^g R_{i_{15} i_{16}]g}{}^h +$$
$$+ 5760\, R_{[i_1 i_2|g|}{}^a R_{i_3 i_4|a|}{}^b R_{i_5 i_6|b|}{}^c R_{i_7 i_8|c|}{}^d R_{i_9 i_{10}|d|}{}^e R_{i_{11} i_{12}|e|}{}^f R_{i_{13} i_{14}|f|}{}^g V_{i_{15} i_{16}]} +$$
$$+ 3360\, R_{[i_1 i_2|f|}{}^a R_{i_3 i_4|a|}{}^b R_{i_5 i_6|b|}{}^c R_{i_7 i_8|c|}{}^d R_{i_9 i_{10}|d|}{}^e R_{i_{11} i_{12}|e|}{}^f R_{i_{13} i_{14}|h|}{}^g R_{i_{15} i_{16}]g}{}^h -$$
$$- 3360\, R_{[i_1 i_2|f|}{}^a R_{i_3 i_4|a|}{}^b R_{i_5 i_6|b|}{}^c R_{i_7 i_8|c|}{}^d R_{i_9 i_{10}|d|}{}^e R_{i_{11} i_{12}|e|}{}^f V_{i_{13} i_{14}} V_{i_{15} i_{16}]} +$$
$$+ 2688\, R_{[i_1 i_2|e|}{}^a R_{i_3 i_4|a|}{}^b R_{i_5 i_6|b|}{}^c R_{i_7 i_8|c|}{}^d R_{i_9 i_{10}|d|}{}^e R_{i_{11} i_{12}|h|}{}^f R_{i_{13} i_{14}|f|}{}^g R_{i_{15} i_{16}]g}{}^h -$$
$$- 4032\, R_{[i_1 i_2|e|}{}^a R_{i_3 i_4|a|}{}^b R_{i_5 i_6|b|}{}^c R_{i_7 i_8|c|}{}^d R_{i_9 i_{10}|d|}{}^e R_{i_{11} i_{12}|g|}{}^f R_{i_{13} i_{14}|f|}{}^g V_{i_{15} i_{16}]} +$$
$$+ 1344\, R_{[i_1 i_2|e|}{}^a R_{i_3 i_4|a|}{}^b R_{i_5 i_6|b|}{}^c R_{i_7 i_8|c|}{}^d R_{i_9 i_{10}|d|}{}^e V_{i_{11} i_{12}} V_{i_{13} i_{14}} V_{i_{15} i_{16}]} +$$
$$+ 1260\, R_{[i_1 i_2|d|}{}^a R_{i_3 i_4|a|}{}^b R_{i_5 i_6|b|}{}^c R_{i_7 i_8|c|}{}^d R_{i_9 i_{10}|h|}{}^e R_{i_{11} i_{12}|e|}{}^f R_{i_{13} i_{14}|f|}{}^g R_{i_{15} i_{16}]g}{}^h -$$
$$- 3360\, R_{[i_1 i_2|d|}{}^a R_{i_3 i_4|a|}{}^b R_{i_5 i_6|b|}{}^c R_{i_7 i_8|c|}{}^d R_{i_9 i_{10}|g|}{}^e R_{i_{11} i_{12}|e|}{}^f R_{i_{13} i_{14}|f|}{}^g V_{i_{15} i_{16}]} -$$
$$- 1260\, R_{[i_1 i_2|d|}{}^a R_{i_3 i_4|a|}{}^b R_{i_5 i_6|b|}{}^c R_{i_7 i_8|c|}{}^d R_{i_9 i_{10}|f|}{}^e R_{i_{11} i_{12}|e|}{}^f R_{i_{13} i_{14}|h|}{}^g R_{i_{15} i_{16}]g}{}^h +$$
$$+ 2520\, R_{[i_1 i_2|d|}{}^a R_{i_3 i_4|a|}{}^b R_{i_5 i_6|b|}{}^c R_{i_7 i_8|c|}{}^d R_{i_9 i_{10}|f|}{}^e R_{i_{11} i_{12}|e|}{}^f V_{i_{13} i_{14}} V_{i_{15} i_{16}]} -$$
$$- 420\, R_{[i_1 i_2|d|}{}^a R_{i_3 i_4|a|}{}^b R_{i_5 i_6|b|}{}^c R_{i_7 i_8|c|}{}^d V_{i_9 i_{10}} V_{i_{11} i_{12}} V_{i_{13} i_{14}} V_{i_{15} i_{16}]} -$$
$$- 1120\, R_{[i_1 i_2|c|}{}^a R_{i_3 i_4|a|}{}^b R_{i_5 i_6|b|}{}^c R_{i_7 i_8|f|}{}^d R_{i_9 i_{10}|d|}{}^e R_{i_{11} i_{12}|e|}{}^f R_{i_{13} i_{14}|h|}{}^g R_{i_{15} i_{16}]g}{}^h +$$
$$+ 1120\, R_{[i_1 i_2|c|}{}^a R_{i_3 i_4|a|}{}^b R_{i_5 i_6|b|}{}^c R_{i_7 i_8|f|}{}^d R_{i_9 i_{10}|d|}{}^e R_{i_{11} i_{12}|e|}{}^f V_{i_{13} i_{14}} V_{i_{15} i_{16}]} +$$
$$+ 1680\, R_{[i_1 i_2|c|}{}^a R_{i_3 i_4|a|}{}^b R_{i_5 i_6|b|}{}^c R_{i_7 i_8|e|}{}^d R_{i_9 i_{10}|d|}{}^e R_{i_{11} i_{12}|g|}{}^f R_{i_{13} i_{14}|f|}{}^g V_{i_{15} i_{16}]} -$$
$$- 1120\, R_{[i_1 i_2|c|}{}^a R_{i_3 i_4|a|}{}^b R_{i_5 i_6|b|}{}^c R_{i_7 i_8|e|}{}^d R_{i_9 i_{10}|d|}{}^e V_{i_{11} i_{12}} V_{i_{13} i_{14}} V_{i_{15} i_{16}]} +$$
$$+ 112\, R_{[i_1 i_2|c|}{}^a R_{i_3 i_4|a|}{}^b R_{i_5 i_6|b|}{}^c V_{i_7 i_8} V_{i_9 i_{10}} V_{i_{11} i_{12}} V_{i_{13} i_{14}} V_{i_{15} i_{16}]} +$$
$$+ 105\, R_{[i_1 i_2|b|}{}^a R_{i_3 i_4|a|}{}^b R_{i_5 i_6|d|}{}^c R_{i_7 i_8|c|}{}^d R_{i_9 i_{10}|f|}{}^e R_{i_{11} i_{12}|e|}{}^f R_{i_{13} i_{14}|h|}{}^g R_{i_{15} i_{16}]g}{}^h -$$
$$- 420\, R_{[i_1 i_2|b|}{}^a R_{i_3 i_4|a|}{}^b R_{i_5 i_6|d|}{}^c R_{i_7 i_8|c|}{}^d R_{i_9 i_{10}|f|}{}^e R_{i_{11} i_{12}|e|}{}^f V_{i_{13} i_{14}} V_{i_{15} i_{16}]} +$$
$$+ 210\, R_{[i_1 i_2|b|}{}^a R_{i_3 i_4|a|}{}^b R_{i_5 i_6|d|}{}^c R_{i_7 i_8|c|}{}^d V_{i_9 i_{10}} V_{i_{11} i_{12}} V_{i_{13} i_{14}} V_{i_{15} i_{16}]} -$$
$$- 28\, R_{[i_1 i_2|b|}{}^a R_{i_3 i_4|a|}{}^b V_{i_5 i_6} V_{i_7 i_8} V_{i_9 i_{10}} V_{i_{11} i_{12}} V_{i_{13} i_{14}} V_{i_{15} i_{16}]} +$$
$$+ V_{[i_1 i_2} V_{i_3 i_4} V_{i_5 i_6} V_{i_7 i_8} V_{i_9 i_{10}} V_{i_{11} i_{12}} V_{i_{13} i_{14}} V_{i_{15} i_{16}]} \Big)$$

$$= \frac{i^8}{2^8 \pi^8 8!} \Big( -5040\, P^{(16)}{}_{i_1 i_2 i_3 i_4 i_5 i_6 i_7 i_8 i_9 i_{10} i_{11} i_{12} i_{13} i_{14} i_{15} i_{16}} + 5760\, P^{(14)}{}_{[i_1 i_2 i_3 i_4 i_5 i_6 i_7 i_8 i_9 i_{10} i_{11} i_{12} i_{13} i_{14}} P^{(2)}{}_{i_{15} i_{16}]} +$$
$$+ 3360\, P^{(12)}{}_{[i_1 i_2 i_3 i_4 i_5 i_6 i_7 i_8 i_9 i_{10} i_{11} i_{12}} P^{(4)}{}_{i_{13} i_{14} i_{15} i_{16}]} - 3360\, P^{(12)}{}_{[i_1 i_2 i_3 i_4 i_5 i_6 i_7 i_8 i_9 i_{10} i_{11} i_{12}} P^{(2)}{}_{i_{13} i_{14}} P^{(2)}{}_{i_{15} i_{16}]} +$$
$$+ 2688\, P^{(10)}{}_{[i_1 i_2 i_3 i_4 i_5 i_6 i_7 i_8 i_9 i_{10}} P^{(6)}{}_{i_{11} i_{12} i_{13} i_{14} i_{15} i_{16}]} - 4032\, P^{(10)}{}_{[i_1 i_2 i_3 i_4 i_5 i_6 i_7 i_8 i_9 i_{10}} P^{(4)}{}_{i_{11} i_{12} i_{13} i_{14}} P^{(2)}{}_{i_{15} i_{16}]} +$$
$$+ 1344\, P^{(10)}{}_{[i_1 i_2 i_3 i_4 i_5 i_6 i_7 i_8 i_9 i_{10}} P^{(2)}{}_{i_{11} i_{12}} P^{(2)}{}_{i_{13} i_{14}} P^{(2)}{}_{i_{15} i_{16}]} + 1260\, P^{(8)}{}_{[i_1 i_2 i_3 i_4 i_5 i_6 i_7 i_8} P^{(8)}{}_{i_9 i_{10} i_{11} i_{12} i_{13} i_{14} i_{15} i_{16}]} -$$

$$- 3360 \, P^{(8)}_{[i_1 i_2 i_3 i_4 i_5 i_6 i_7 i_8} P^{(6)}_{i_9 i_{10} i_{11} i_{12} i_{13} i_{14}} P^{(2)}_{i_{15} i_{16}]} - 1260 \, P^{(8)}_{[i_1 i_2 i_3 i_4 i_5 i_6 i_7 i_8} P^{(4)}_{i_9 i_{10} i_{11} i_{12}} P^{(4)}_{i_{13} i_{14} i_{15} i_{16}]} +$$

$$+ 2520 \, P^{(8)}_{[i_1 i_2 i_3 i_4 i_5 i_6 i_7 i_8} P^{(4)}_{i_9 i_{10} i_{11} i_{12}} P^{(2)}_{i_{13} i_{14}} P^{(2)}_{i_{15} i_{16}]} -$$

$$- 420 \, P^{(8)}_{[i_1 i_2 i_3 i_4 i_5 i_6 i_7 i_8} P^{(2)}_{i_9 i_{10}} P^{(2)}_{i_{11} i_{12}} P^{(2)}_{i_{13} i_{14}} P^{(2)}_{i_{15} i_{16}]} -$$

$$- 1120 \, P^{(6)}_{[i_1 i_2 i_3 i_4 i_5 i_6} P^{(6)}_{i_7 i_8 i_9 i_{10} i_{11} i_{12}} P^{(4)}_{i_{13} i_{14} i_{15} i_{16}]} +$$

$$+ 1120 \, P^{(6)}_{[i_1 i_2 i_3 i_4 i_5 i_6} P^{(6)}_{i_7 i_8 i_9 i_{10} i_{11} i_{12}} P^{(2)}_{i_{13} i_{14}} P^{(2)}_{i_{15} i_{16}]} +$$

$$+ 1680 \, P^{(6)}_{[i_1 i_2 i_3 i_4 i_5 i_6} P^{(4)}_{i_7 i_8 i_9 i_{10}} P^{(4)}_{i_{11} i_{12} i_{13} i_{14}} P^{(2)}_{i_{15} i_{16}]} -$$

$$- 1120 \, P^{(6)}_{[i_1 i_2 i_3 i_4 i_5 i_6} P^{(4)}_{i_7 i_8 i_9 i_{10}} P^{(2)}_{i_{11} i_{12}} P^{(2)}_{i_{13} i_{14}} P^{(2)}_{i_{15} i_{16}]} +$$

$$+ 112 \, P^{(6)}_{[i_1 i_2 i_3 i_4 i_5 i_6} P^{(2)}_{i_7 i_8} P^{(2)}_{i_9 i_{10}} P^{(2)}_{i_{11} i_{12}} P^{(2)}_{i_{13} i_{14}} P^{(2)}_{i_{15} i_{16}]} +$$

$$+ 105 \, P^{(4)}_{[i_1 i_2 i_3 i_4} P^{(4)}_{i_5 i_6 i_7 i_8} P^{(4)}_{i_9 i_{10} i_{11} i_{12}} P^{(4)}_{i_{13} i_{14} i_{15} i_{16}]} -$$

$$- 420 \, P^{(4)}_{[i_1 i_2 i_3 i_4} P^{(4)}_{i_5 i_6 i_7 i_8} P^{(4)}_{i_9 i_{10} i_{11} i_{12}} P^{(2)}_{i_{13} i_{14}} P^{(2)}_{i_{15} i_{16}]} +$$

$$+ 210 \, P^{(4)}_{[i_1 i_2 i_3 i_4} P^{(4)}_{i_5 i_6 i_7 i_8} P^{(2)}_{i_9 i_{10}} P^{(2)}_{i_{11} i_{12}} P^{(2)}_{i_{13} i_{14}} P^{(2)}_{i_{15} i_{16}]} -$$

$$- 28 \, P^{(4)}_{[i_1 i_2 i_3 i_4} P^{(2)}_{i_5 i_6} P^{(2)}_{i_7 i_8} P^{(2)}_{i_9 i_{10}} P^{(2)}_{i_{11} i_{12}} P^{(2)}_{i_{13} i_{14}} P^{(2)}_{i_{15} i_{16}]} +$$

$$+ P^{(2)}_{[i_1 i_2} P^{(2)}_{i_3 i_4} P^{(2)}_{i_5 i_6} P^{(2)}_{i_7 i_8} P^{(2)}_{i_9 i_{10}} P^{(2)}_{i_{11} i_{12}} P^{(2)}_{i_{13} i_{14}} P^{(2)}_{i_{15} i_{16}]} )$$

### COEFFICIENT OF THE 9$^{th}$ CHERN FORM

$$c_{(9) i_1 i_2 i_3 i_4 i_5 i_6 i_7 i_8 i_9 i_{10} i_{11} i_{12} i_{13} i_{14} i_{15} i_{16} i_{17} i_{18}} = \tag{18}$$

$$= \frac{1}{18!} \langle \mathbf{e}_{i_1} \wedge \mathbf{e}_{i_2} \wedge \mathbf{e}_{i_3} \wedge \mathbf{e}_{i_4} \wedge \mathbf{e}_{i_5} \wedge \mathbf{e}_{i_6} \wedge \mathbf{e}_{i_7} \wedge \mathbf{e}_{i_8} \wedge \mathbf{e}_{i_9} \wedge \mathbf{e}_{i_{10}} \wedge \mathbf{e}_{i_{11}} \wedge \mathbf{e}_{i_{12}} \wedge \mathbf{e}_{i_{13}} \wedge \mathbf{e}_{i_{14}} \wedge \mathbf{e}_{i_{15}} \wedge \mathbf{e}_{i_{16}} \wedge \mathbf{e}_{i_{17}} \wedge \mathbf{e}_{i_{18}}, c_{(9)} \rangle$$

$$= \frac{i^9}{2^{18} \pi^9 9!} (+ 40{,}320 \, R_{[i_1 i_2 | i |}{}^a R_{i_3 i_4 | a |}{}^b R_{i_5 i_6 | b |}{}^c R_{i_7 i_8 | c |}{}^d R_{i_9 i_{10} | d |}{}^e R_{i_{11} i_{12} | e |}{}^f R_{i_{13} i_{14} | f |}{}^g R_{i_{15} i_{16} | g |}{}^h R_{i_{17} i_{18} | h |}{}^i -$$

$$- 45{,}360 \, R_{[i_1 i_2 | h |}{}^a R_{i_3 i_4 | a |}{}^b R_{i_5 i_6 | b |}{}^c R_{i_7 i_8 | c |}{}^d R_{i_9 i_{10} | d |}{}^e R_{i_{11} i_{12} | e |}{}^f R_{i_{13} i_{14} | f |}{}^g R_{i_{15} i_{16} | g |}{}^h R_{i_{17} i_{18} | i |}{}^i -$$

$$- 25{,}920 \, R_{[i_1 i_2 | g |}{}^a R_{i_3 i_4 | a |}{}^b R_{i_5 i_6 | b |}{}^c R_{i_7 i_8 | c |}{}^d R_{i_9 i_{10} | d |}{}^e R_{i_{11} i_{12} | e |}{}^f R_{i_{13} i_{14} | f |}{}^g R_{i_{15} i_{16} | i |}{}^h R_{i_{17} i_{18} | h |}{}^i +$$

$$+ 25{,}920 \, R_{[i_1 i_2 | g |}{}^a R_{i_3 i_4 | a |}{}^b R_{i_5 i_6 | b |}{}^c R_{i_7 i_8 | c |}{}^d R_{i_9 i_{10} | d |}{}^e R_{i_{11} i_{12} | e |}{}^f R_{i_{13} i_{14} | f |}{}^g R_{i_{15} i_{16} | h |}{}^h R_{i_{17} i_{18} | i |}{}^i -$$

$$- 20{,}160 \, R_{[i_1 i_2 | f |}{}^a R_{i_3 i_4 | a |}{}^b R_{i_5 i_6 | b |}{}^c R_{i_7 i_8 | c |}{}^d R_{i_9 i_{10} | d |}{}^e R_{i_{11} i_{12} | e |}{}^f R_{i_{13} i_{14} | f |}{}^g R_{i_{15} i_{16} | g |}{}^h R_{i_{17} i_{18} | h |}{}^i +$$

$$+ 30{,}240 \, R_{[i_1 i_2 | f |}{}^a R_{i_3 i_4 | a |}{}^b R_{i_5 i_6 | b |}{}^c R_{i_7 i_8 | c |}{}^d R_{i_9 i_{10} | d |}{}^e R_{i_{11} i_{12} | e |}{}^f R_{i_{13} i_{14} | h |}{}^g R_{i_{15} i_{16} | g |}{}^h R_{i_{17} i_{18} | i |}{}^i -$$

$$- 10{,}080 \, R_{[i_1 i_2 | f |}{}^a R_{i_3 i_4 | a |}{}^b R_{i_5 i_6 | b |}{}^c R_{i_7 i_8 | c |}{}^d R_{i_9 i_{10} | d |}{}^e R_{i_{11} i_{12} | e |}{}^f R_{i_{13} i_{14} | g |}{}^g R_{i_{15} i_{16} | h |}{}^h R_{i_{17} i_{18} | i |}{}^i -$$

$$- 18{,}144 \, R_{[i_1 i_2 | e |}{}^a R_{i_3 i_4 | a |}{}^b R_{i_5 i_6 | b |}{}^c R_{i_7 i_8 | c |}{}^d R_{i_9 i_{10} | d |}{}^e R_{i_{11} i_{12} | i |}{}^f R_{i_{13} i_{14} | f |}{}^g R_{i_{15} i_{16} | g |}{}^h R_{i_{17} i_{18} | h |}{}^i +$$

$$+ 24{,}192 \, R_{[i_1 i_2 | e |}{}^a R_{i_3 i_4 | a |}{}^b R_{i_5 i_6 | b |}{}^c R_{i_7 i_8 | c |}{}^d R_{i_9 i_{10} | d |}{}^e R_{i_{11} i_{12} | h |}{}^f R_{i_{13} i_{14} | f |}{}^g R_{i_{15} i_{16} | g |}{}^h R_{i_{17} i_{18} | i |}{}^i +$$

$$+ 9072 \, R_{[i_1 i_2 | e |}{}^a R_{i_3 i_4 | a |}{}^b R_{i_5 i_6 | b |}{}^c R_{i_7 i_8 | c |}{}^d R_{i_9 i_{10} | d |}{}^e R_{i_{11} i_{12} | g |}{}^f R_{i_{13} i_{14} | f |}{}^g R_{i_{15} i_{16} | i |}{}^h R_{i_{17} i_{18} | h |}{}^i -$$

$$- 18{,}144 \, R_{[i_1 i_2 | e |}{}^a R_{i_3 i_4 | a |}{}^b R_{i_5 i_6 | b |}{}^c R_{i_7 i_8 | c |}{}^d R_{i_9 i_{10} | d |}{}^e R_{i_{11} i_{12} | g |}{}^f R_{i_{13} i_{14} | f |}{}^g R_{i_{15} i_{16} | h |}{}^h R_{i_{17} i_{18} | i |}{}^i +$$

$$+ 3024 \, R_{[i_1 i_2 | e |}{}^a R_{i_3 i_4 | a |}{}^b R_{i_5 i_6 | b |}{}^c R_{i_7 i_8 | c |}{}^d R_{i_9 i_{10} | d |}{}^e R_{i_{11} i_{12} | f |}{}^f R_{i_{13} i_{14} | g |}{}^g R_{i_{15} i_{16} | h |}{}^h R_{i_{17} i_{18} | i |}{}^i +$$

$$+ 11{,}340 \, R_{[i_1 i_2 | d |}{}^a R_{i_3 i_4 | a |}{}^b R_{i_5 i_6 | b |}{}^c R_{i_7 i_8 | c |}{}^d R_{i_9 i_{10} | h |}{}^e R_{i_{11} i_{12} | e |}{}^f R_{i_{13} i_{14} | f |}{}^g R_{i_{15} i_{16} | g |}{}^h R_{i_{17} i_{18} | i |}{}^i +$$

$$+ 15{,}120 \, R_{[i_1 i_2 | d |}{}^a R_{i_3 i_4 | a |}{}^b R_{i_5 i_6 | b |}{}^c R_{i_7 i_8 | c |}{}^d R_{i_9 i_{10} | g |}{}^e R_{i_{11} i_{12} | e |}{}^f R_{i_{13} i_{14} | f |}{}^g R_{i_{15} i_{16} | i |}{}^h R_{i_{17} i_{18} | h |}{}^i -$$

$$- 15{,}120 \, R_{[i_1 i_2 | d |}{}^a R_{i_3 i_4 | a |}{}^b R_{i_5 i_6 | b |}{}^c R_{i_7 i_8 | c |}{}^d R_{i_9 i_{10} | g |}{}^e R_{i_{11} i_{12} | e |}{}^f R_{i_{13} i_{14} | f |}{}^g R_{i_{15} i_{16} | h |}{}^h R_{i_{17} i_{18} | i |}{}^i -$$

$$- 11{,}340 \, R_{[i_1 i_2 | d |}{}^a R_{i_3 i_4 | a |}{}^b R_{i_5 i_6 | b |}{}^c R_{i_7 i_8 | c |}{}^d R_{i_9 i_{10} | f |}{}^e R_{i_{11} i_{12} | e |}{}^f R_{i_{13} i_{14} | g |}{}^g R_{i_{15} i_{16} | h |}{}^h R_{i_{17} i_{18} | i |}{}^i +$$

$$+ 7560 \, R_{[i_1 i_2 | d |}{}^a R_{i_3 i_4 | a |}{}^b R_{i_5 i_6 | b |}{}^c R_{i_7 i_8 | c |}{}^d R_{i_9 i_{10} | f |}{}^e R_{i_{11} i_{12} | e |}{}^f R_{i_{13} i_{14} | g |}{}^g R_{i_{15} i_{16} | h |}{}^h R_{i_{17} i_{18} | i |}{}^i -$$

$$- 756 \, R_{[i_1 i_2 | d |}{}^a R_{i_3 i_4 | a |}{}^b R_{i_5 i_6 | b |}{}^c R_{i_7 i_8 | c |}{}^d R_{i_9 i_{10} | e |}{}^e R_{i_{11} i_{12} | f |}{}^f R_{i_{13} i_{14} | g |}{}^g R_{i_{15} i_{16} | h |}{}^h R_{i_{17} i_{18} | i |}{}^i +$$

$$+ 2240 \, R_{[i_1 i_2 | c |}{}^a R_{i_3 i_4 | a |}{}^b R_{i_5 i_6 | b |}{}^c R_{i_7 i_8 | f |}{}^d R_{i_9 i_{10} | d |}{}^e R_{i_{11} i_{12} | e |}{}^f R_{i_{13} i_{14} | i |}{}^g R_{i_{15} i_{16} | g |}{}^h R_{i_{17} i_{18} | h |}{}^i -$$

$$- 10{,}080 \, R_{[i_1 i_2 | c |}{}^a R_{i_3 i_4 | a |}{}^b R_{i_5 i_6 | b |}{}^c R_{i_7 i_8 | f |}{}^d R_{i_9 i_{10} | d |}{}^e R_{i_{11} i_{12} | e |}{}^f R_{i_{13} i_{14} | f |}{}^g R_{i_{15} i_{16} | g |}{}^h R_{i_{17} i_{18} | h |}{}^i +$$

$$+ 3360 \, R_{[i_1 i_2 | c |}{}^a R_{i_3 i_4 | a |}{}^b R_{i_5 i_6 | b |}{}^c R_{i_7 i_8 | f |}{}^d R_{i_9 i_{10} | d |}{}^e R_{i_{11} i_{12} | e |}{}^f R_{i_{13} i_{14} | g |}{}^g R_{i_{15} i_{16} | h |}{}^h R_{i_{17} i_{18} | i |}{}^i -$$

$$- 2520 \, R_{[i_1 i_2 | c |}{}^a R_{i_3 i_4 | a |}{}^b R_{i_5 i_6 | b |}{}^c R_{i_7 i_8 | e |}{}^d R_{i_9 i_{10} | d |}{}^e R_{i_{11} i_{12} | f |}{}^f R_{i_{13} i_{14} | f |}{}^g R_{i_{15} i_{16} | i |}{}^h R_{i_{17} i_{18} | h |}{}^i +$$

$$+ 7560 \, R_{[i_1 i_2 | c |}{}^a R_{i_3 i_4 | a |}{}^b R_{i_5 i_6 | b |}{}^c R_{i_7 i_8 | e |}{}^d R_{i_9 i_{10} | d |}{}^e R_{i_{11} i_{12} | g |}{}^f R_{i_{13} i_{14} | f |}{}^g R_{i_{15} i_{16} | h |}{}^h R_{i_{17} i_{18} | i |}{}^i -$$

$$- 2520 \, R_{[i_1 i_2 | c |}{}^a R_{i_3 i_4 | a |}{}^b R_{i_5 i_6 | b |}{}^c R_{i_7 i_8 | e |}{}^d R_{i_9 i_{10} | d |}{}^e R_{i_{11} i_{12} | f |}{}^f R_{i_{13} i_{14} | g |}{}^g R_{i_{15} i_{16} | h |}{}^h R_{i_{17} i_{18} | i |}{}^i +$$

$$+ 168 \, R_{[i_1 i_2 | c |}{}^a R_{i_3 i_4 | a |}{}^b R_{i_5 i_6 | b |}{}^c R_{i_7 i_8 | d |}{}^d R_{i_9 i_{10} | e |}{}^e R_{i_{11} i_{12} | f |}{}^f R_{i_{13} i_{14} | g |}{}^g R_{i_{15} i_{16} | h |}{}^h R_{i_{17} i_{18} | i |}{}^i +$$

$$+ 945\, R_{[i_1i_2|b|}{}^a\, R_{i_3i_4|a|}{}^b\, R_{i_5i_6|d|}{}^c\, R_{i_7i_8|c|}{}^d\, R_{i_9i_{10}|f|}{}^e\, R_{i_{11}i_{12}|e|}{}^f\, R_{i_{13}i_{14}|h|}{}^g\, R_{i_{15}i_{16}|g|}{}^h\, R_{i_{17}i_{18}]i}{}^i\, -$$

$$- 1260\, R_{[i_1i_2|b|}{}^a\, R_{i_3i_4|a|}{}^b\, R_{i_5i_6|d|}{}^c\, R_{i_7i_8|c|}{}^d\, R_{i_9i_{10}|f|}{}^e\, R_{i_{11}i_{12}|e|}{}^f\, R_{i_{13}i_{14}|g|}{}^g\, R_{i_{15}i_{16}|h|}{}^h\, R_{i_{17}i_{18}]i}{}^i\, +$$

$$+ 378\, R_{[i_1i_2|b|}{}^a\, R_{i_3i_4|a|}{}^b\, R_{i_5i_6|d|}{}^c\, R_{i_7i_8|c|}{}^d\, R_{i_9i_{10}|e|}{}^e\, R_{i_{11}i_{12}|f|}{}^f\, R_{i_{13}i_{14}|g|}{}^g\, R_{i_{15}i_{16}|h|}{}^h\, R_{i_{17}i_{18}]i}{}^i\, -$$

$$- 36\, R_{[i_1i_2|b|}{}^a\, R_{i_3i_4|a|}{}^b\, R_{i_5i_6|c|}{}^c\, R_{i_7i_8|d|}{}^d\, R_{i_9i_{10}|e|}{}^e\, R_{i_{11}i_{12}|f|}{}^f\, R_{i_{13}i_{14}|g|}{}^g\, R_{i_{15}i_{16}|h|}{}^h\, R_{i_{17}i_{18}]i}{}^i\, +$$

$$+ R_{[i_1i_2|a|}{}^a\, R_{i_3i_4|b|}{}^b\, R_{i_5i_6|c|}{}^c\, R_{i_7i_8|d|}{}^d\, R_{i_9i_{10}|e|}{}^e\, R_{i_{11}i_{12}|f|}{}^f\, R_{i_{13}i_{14}|g|}{}^g\, R_{i_{15}i_{16}|h|}{}^h\, R_{i_{17}i_{18}]i}{}^i\, )$$

$$= \frac{i^9}{2^{18}\pi^9 9!}\,(+\, 40{,}320\, R_{[i_1i_2|i|}{}^a\, R_{i_3i_4|a|}{}^b\, R_{i_5i_6|b|}{}^c\, R_{i_7i_8|c|}{}^d\, R_{i_9i_{10}|d|}{}^e\, R_{i_{11}i_{12}|e|}{}^f\, R_{i_{13}i_{14}|f|}{}^g\, R_{i_{15}i_{16}|g|}{}^h\, R_{i_{17}i_{18}]h}{}^i\, -$$

$$- 45{,}360\, R_{[i_1i_2|h|}{}^a\, R_{i_3i_4|a|}{}^b\, R_{i_5i_6|b|}{}^c\, R_{i_7i_8|c|}{}^d\, R_{i_9i_{10}|d|}{}^e\, R_{i_{11}i_{12}|e|}{}^f\, R_{i_{13}i_{14}|f|}{}^g\, R_{i_{15}i_{16}]g}{}^h\, V_{i_{17}i_{18}]}\, -$$

$$- 25{,}920\, R_{[i_1i_2|g|}{}^a\, R_{i_3i_4|a|}{}^b\, R_{i_5i_6|b|}{}^c\, R_{i_7i_8|c|}{}^d\, R_{i_9i_{10}|d|}{}^e\, R_{i_{11}i_{12}|e|}{}^f\, R_{i_{13}i_{14}|f|}{}^g\, R_{i_{15}i_{16}|i|}{}^h\, R_{i_{17}i_{18}]h}{}^i\, +$$

$$+ 25{,}920\, R_{[i_1i_2|g|}{}^a\, R_{i_3i_4|a|}{}^b\, R_{i_5i_6|b|}{}^c\, R_{i_7i_8|c|}{}^d\, R_{i_9i_{10}|d|}{}^e\, R_{i_{11}i_{12}|e|}{}^f\, R_{i_{13}i_{14}]f}{}^g\, V_{i_{15}i_{16}}\, V_{i_{17}i_{18}]}\, -$$

$$- 20{,}160\, R_{[i_1i_2|f|}{}^a\, R_{i_3i_4|a|}{}^b\, R_{i_5i_6|b|}{}^c\, R_{i_7i_8|c|}{}^d\, R_{i_9i_{10}|d|}{}^e\, R_{i_{11}i_{12}|e|}{}^f\, R_{i_{13}i_{14}|i|}{}^g\, R_{i_{15}i_{16}|g|}{}^h\, R_{i_{17}i_{18}]h}{}^i\, +$$

$$+ 30{,}240\, R_{[i_1i_2|f|}{}^a\, R_{i_3i_4|a|}{}^b\, R_{i_5i_6|b|}{}^c\, R_{i_7i_8|c|}{}^d\, R_{i_9i_{10}|d|}{}^e\, R_{i_{11}i_{12}|e|}{}^f\, R_{i_{13}i_{14}|h|}{}^g\, R_{i_{15}i_{16}]g}{}^h\, V_{i_{17}i_{18}]}\, -$$

$$- 10{,}080\, R_{[i_1i_2|f|}{}^a\, R_{i_3i_4|a|}{}^b\, R_{i_5i_6|b|}{}^c\, R_{i_7i_8|c|}{}^d\, R_{i_9i_{10}|d|}{}^e\, R_{i_{11}i_{12}|e|}{}^f\, V_{i_{13}i_{14}}\, V_{i_{15}i_{16}}\, V_{i_{17}i_{18}]}\, -$$

$$- 18{,}144\, R_{[i_1i_2|e|}{}^a\, R_{i_3i_4|a|}{}^b\, R_{i_5i_6|b|}{}^c\, R_{i_7i_8|c|}{}^d\, R_{i_9i_{10}|d|}{}^e\, R_{i_{11}i_{12}|i|}{}^f\, R_{i_{13}i_{14}|f|}{}^g\, R_{i_{15}i_{16}|g|}{}^h\, R_{i_{17}i_{18}]h}{}^i\, +$$

$$+ 24{,}192\, R_{[i_1i_2|e|}{}^a\, R_{i_3i_4|a|}{}^b\, R_{i_5i_6|b|}{}^c\, R_{i_7i_8|c|}{}^d\, R_{i_9i_{10}|d|}{}^e\, R_{i_{11}i_{12}|h|}{}^f\, R_{i_{13}i_{14}|f|}{}^g\, R_{i_{15}i_{16}]g}{}^h\, V_{i_{17}i_{18}]}\, +$$

$$+ 9072\, R_{[i_1i_2|e|}{}^a\, R_{i_3i_4|a|}{}^b\, R_{i_5i_6|b|}{}^c\, R_{i_7i_8|c|}{}^d\, R_{i_9i_{10}|d|}{}^e\, R_{i_{11}i_{12}|g|}{}^f\, R_{i_{13}i_{14}|f|}{}^g\, R_{i_{15}i_{16}|i|}{}^h\, R_{i_{17}i_{18}]h}{}^i\, -$$

$$- 18{,}144\, R_{[i_1i_2|e|}{}^a\, R_{i_3i_4|a|}{}^b\, R_{i_5i_6|b|}{}^c\, R_{i_7i_8|c|}{}^d\, R_{i_9i_{10}|d|}{}^e\, R_{i_{11}i_{12}|g|}{}^f\, R_{i_{13}i_{14}]f}{}^g\, V_{i_{15}i_{16}}\, V_{i_{17}i_{18}]}\, +$$

$$+ 3024\, R_{[i_1i_2|e|}{}^a\, R_{i_3i_4|a|}{}^b\, R_{i_5i_6|b|}{}^c\, R_{i_7i_8|c|}{}^d\, R_{i_9i_{10}]d}{}^e\, V_{i_{11}i_{12}}\, V_{i_{13}i_{14}}\, V_{i_{15}i_{16}}\, V_{i_{17}i_{18}]}\, +$$

$$+ 11{,}340\, R_{[i_1i_2|d|}{}^a\, R_{i_3i_4|a|}{}^b\, R_{i_5i_6|b|}{}^c\, R_{i_7i_8|c|}{}^d\, R_{i_9i_{10}|h|}{}^e\, R_{i_{11}i_{12}|e|}{}^f\, R_{i_{13}i_{14}|f|}{}^g\, R_{i_{15}i_{16}]g}{}^h\, V_{i_{17}i_{18}]}\, +$$

$$+ 15{,}120\, R_{[i_1i_2|d|}{}^a\, R_{i_3i_4|a|}{}^b\, R_{i_5i_6|b|}{}^c\, R_{i_7i_8|c|}{}^d\, R_{i_9i_{10}|g|}{}^e\, R_{i_{11}i_{12}|e|}{}^f\, R_{i_{13}i_{14}|f|}{}^g\, R_{i_{15}i_{16}|i|}{}^h\, R_{i_{17}i_{18}]h}{}^i\, -$$

$$- 15{,}120\, R_{[i_1i_2|d|}{}^a\, R_{i_3i_4|a|}{}^b\, R_{i_5i_6|b|}{}^c\, R_{i_7i_8|c|}{}^d\, R_{i_9i_{10}|g|}{}^e\, R_{i_{11}i_{12}|e|}{}^f\, R_{i_{13}i_{14}]f}{}^g\, V_{i_{15}i_{16}}\, V_{i_{17}i_{18}]}\, -$$

$$- 11{,}340\, R_{[i_1i_2|d|}{}^a\, R_{i_3i_4|a|}{}^b\, R_{i_5i_6|b|}{}^c\, R_{i_7i_8|c|}{}^d\, R_{i_9i_{10}|f|}{}^e\, R_{i_{11}i_{12}|e|}{}^f\, R_{i_{13}i_{14}|h|}{}^g\, R_{i_{15}i_{16}]g}{}^h\, V_{i_{17}i_{18}]}\, +$$

$$+ 7560\, R_{[i_1i_2|d|}{}^a\, R_{i_3i_4|a|}{}^b\, R_{i_5i_6|b|}{}^c\, R_{i_7i_8|c|}{}^d\, R_{i_9i_{10}|f|}{}^e\, R_{i_{11}i_{12}]e}{}^f\, V_{i_{13}i_{14}}\, V_{i_{15}i_{16}}\, V_{i_{17}i_{18}]}\, -$$

$$- 756\, R_{[i_1i_2|d|}{}^a\, R_{i_3i_4|a|}{}^b\, R_{i_5i_6|b|}{}^c\, R_{i_7i_8]c}{}^d\, V_{i_9i_{10}}\, V_{i_{11}i_{12}}\, V_{i_{13}i_{14}}\, V_{i_{15}i_{16}}\, V_{i_{17}i_{18}]}\, +$$

$$+ 2240\, R_{[i_1i_2|c|}{}^a\, R_{i_3i_4|a|}{}^b\, R_{i_5i_6|b|}{}^c\, R_{i_7i_8|f|}{}^d\, R_{i_9i_{10}|d|}{}^e\, R_{i_{11}i_{12}|e|}{}^f\, R_{i_{13}i_{14}|i|}{}^g\, R_{i_{15}i_{16}|g|}{}^h\, R_{i_{17}i_{18}]h}{}^i\, -$$

$$- 10{,}080\, R_{[i_1i_2|c|}{}^a\, R_{i_3i_4|a|}{}^b\, R_{i_5i_6|b|}{}^c\, R_{i_7i_8|f|}{}^d\, R_{i_9i_{10}|d|}{}^e\, R_{i_{11}i_{12}|e|}{}^f\, R_{i_{13}i_{14}|h|}{}^g\, R_{i_{15}i_{16}]g}{}^h\, V_{i_{17}i_{18}]}\, +$$

$$+ 3360\, R_{[i_1i_2|c|}{}^a\, R_{i_3i_4|a|}{}^b\, R_{i_5i_6|b|}{}^c\, R_{i_7i_8|f|}{}^d\, R_{i_9i_{10}|d|}{}^e\, R_{i_{11}i_{12}]e}{}^f\, V_{i_{13}i_{14}}\, V_{i_{15}i_{16}}\, V_{i_{17}i_{18}]}\, -$$

$$- 2520\, R_{[i_1i_2|c|}{}^a\, R_{i_3i_4|a|}{}^b\, R_{i_5i_6|b|}{}^c\, R_{i_7i_8|e|}{}^d\, R_{i_9i_{10}|d|}{}^e\, R_{i_{11}i_{12}|g|}{}^f\, R_{i_{13}i_{14}|f|}{}^g\, R_{i_{15}i_{16}|i|}{}^h\, R_{i_{17}i_{18}]h}{}^i\, +$$

$$+ 7560\, R_{[i_1i_2|c|}{}^a\, R_{i_3i_4|a|}{}^b\, R_{i_5i_6|b|}{}^c\, R_{i_7i_8|e|}{}^d\, R_{i_9i_{10}|d|}{}^e\, R_{i_{11}i_{12}|g|}{}^f\, R_{i_{13}i_{14}]f}{}^g\, V_{i_{15}i_{16}}\, V_{i_{17}i_{18}]}\, -$$

$$- 2520\, R_{[i_1i_2|c|}{}^a\, R_{i_3i_4|a|}{}^b\, R_{i_5i_6|b|}{}^c\, R_{i_7i_8|e|}{}^d\, R_{i_9i_{10}]d}{}^e\, V_{i_{11}i_{12}}\, V_{i_{13}i_{14}}\, V_{i_{15}i_{16}}\, V_{i_{17}i_{18}]}\, +$$

$$+ 168\, R_{[i_1i_2|c|}{}^a\, R_{i_3i_4|a|}{}^b\, R_{i_5i_6]b}{}^c\, V_{i_7i_8}\, V_{i_9i_{10}}\, V_{i_{11}i_{12}}\, V_{i_{13}i_{14}}\, V_{i_{15}i_{16}}\, V_{i_{17}i_{18}]}\, +$$

$$+ 945\, R_{[i_1i_2|b|}{}^a\, R_{i_3i_4|a|}{}^b\, R_{i_5i_6|d|}{}^c\, R_{i_7i_8|c|}{}^d\, R_{i_9i_{10}|f|}{}^e\, R_{i_{11}i_{12}|e|}{}^f\, R_{i_{13}i_{14}|h|}{}^g\, R_{i_{15}i_{16}]g}{}^h\, V_{i_{17}i_{18}]}\, -$$

$$- 1260\, R_{[i_1i_2|b|}{}^a\, R_{i_3i_4|a|}{}^b\, R_{i_5i_6|d|}{}^c\, R_{i_7i_8|c|}{}^d\, R_{i_9i_{10}|f|}{}^e\, R_{i_{11}i_{12}]e}{}^f\, V_{i_{13}i_{14}}\, V_{i_{15}i_{16}}\, V_{i_{17}i_{18}]}\, +$$

$$+ 378\, R_{[i_1i_2|b|}{}^a\, R_{i_3i_4|a|}{}^b\, R_{i_5i_6|d|}{}^c\, R_{i_7i_8]c}{}^d\, V_{i_9i_{10}}\, V_{i_{11}i_{12}}\, V_{i_{13}i_{14}}\, V_{i_{15}i_{16}}\, V_{i_{17}i_{18}]}\, -$$

$$- 36\, R_{[i_1i_2|b|}{}^a\, R_{i_3i_4|a|}{}^b\, V_{i_5i_6}\, V_{i_7i_8}\, V_{i_9i_{10}}\, V_{i_{11}i_{12}}\, V_{i_{13}i_{14}}\, V_{i_{15}i_{16}}\, V_{i_{17}i_{18}]}\, +$$

$$+ V_{[i_1i_2}\, V_{i_3i_4}\, V_{i_5i_6}\, V_{i_7i_8}\, V_{i_9i_{10}}\, V_{i_{11}i_{12}}\, V_{i_{13}i_{14}}\, V_{i_{15}i_{16}}\, V_{i_{17}i_{18}]})$$

$$= \frac{i^9}{2^9\pi^9 9!}\,(+\, 40{,}320\, P^{(18)}{}_{i_1i_2i_3i_4i_5i_6i_7i_8i_9i_{10}i_{11}i_{12}i_{13}i_{14}i_{15}i_{16}i_{17}i_{18}} - 45{,}360\, P^{(16)}{}_{[i_1i_2i_3i_4i_5i_6i_7i_8i_9i_{10}i_{11}i_{12}i_{13}i_{14}i_{15}i_{16}}\, P^{(2)}{}_{i_{17}i_{18}]}\, -$$

$$- 25{,}920\, P^{(14)}{}_{[i_1i_2i_3i_4i_5i_6i_7i_8i_9i_{10}i_{11}i_{12}i_{13}i_{14}}\, P^{(4)}{}_{i_{15}i_{16}i_{17}i_{18}]} + 25{,}920\, P^{(14)}{}_{[i_1i_2i_3i_4i_5i_6i_7i_8i_9i_{10}i_{11}i_{12}i_{13}i_{14}}\, P^{(2)}{}_{i_{15}i_{16}}\, P^{(2)}{}_{i_{17}i_{18}]}\, -$$

$$- 20{,}160\, P^{(12)}{}_{[i_1i_2i_3i_4i_5i_6i_7i_8i_9i_{10}i_{11}i_{12}}\, P^{(6)}{}_{i_{13}i_{14}i_{15}i_{16}i_{17}i_{18}]} + 30{,}240\, P^{(12)}{}_{[i_1i_2i_3i_4i_5i_6i_7i_8i_9i_{10}i_{11}i_{12}}\, P^{(4)}{}_{i_{13}i_{14}i_{15}i_{16}}\, P^{(2)}{}_{i_{17}i_{18}]}\, -$$

$$- 10{,}080\, P^{(12)}{}_{[i_1i_2i_3i_4i_5i_6i_7i_8i_9i_{10}i_{11}i_{12}}\, P^{(2)}{}_{i_{13}i_{14}}\, P^{(2)}{}_{i_{15}i_{16}}\, P^{(2)}{}_{i_{17}i_{18}]} - 18{,}144\, P^{(10)}{}_{[i_1i_2i_3i_4i_5i_6i_7i_8i_9i_{10}}\, P^{(8)}{}_{i_{11}i_{12}i_{13}i_{14}i_{15}i_{16}i_{17}i_{18}]}\, +$$

$$+ 24{,}192\, P^{(10)}{}_{[i_1i_2i_3i_4i_5i_6i_7i_8i_9i_{10}}\, P^{(6)}{}_{i_{11}i_{12}i_{13}i_{14}i_{15}i_{16}}\, P^{(2)}{}_{i_{17}i_{18}]} + 9{,}072\, P^{(10)}{}_{[i_1i_2i_3i_4i_5i_6i_7i_8i_9i_{10}}\, P^{(4)}{}_{i_{11}i_{12}i_{13}i_{14}}\, P^{(4)}{}_{i_{15}i_{16}i_{17}i_{18}]}\, -$$

$$- 18{,}144\, P^{(10)}{}_{[i_1i_2i_3i_4i_5i_6i_7i_8i_9i_{10}}\, P^{(4)}{}_{i_{11}i_{12}i_{13}i_{14}}\, P^{(2)}{}_{i_{15}i_{16}}\, P^{(2)}{}_{i_{17}i_{18}]} + 3{,}024\, P^{(10)}{}_{[i_1i_2i_3i_4i_5i_6i_7i_8i_9i_{10}}\, P^{(2)}{}_{i_{11}i_{12}}\, P^{(2)}{}_{i_{13}i_{14}}\, P^{(2)}{}_{i_{15}i_{16}}\, P^{(2)}{}_{i_{17}i_{18}]}\, +$$

$$+ 11{,}340\, P^{(8)}_{[i_1i_2i_3i_4i_5i_6i_7i_8}\, P^{(8)}_{i_9i_{10}i_{11}i_{12}i_{13}i_{14}i_{15}i_{16}}\, P^{(2)}_{i_{17}i_{18}]} + 15{,}120\, P^{(8)}_{[i_1i_2i_3i_4i_5i_6i_7i_8}\, P^{(6)}_{i_9i_{10}i_{11}i_{12}i_{13}i_{14}}\, P^{(4)}_{i_{15}i_{16}i_{17}i_{18}]} -$$

$$- 15{,}120\, P^{(8)}_{[i_1i_2i_3i_4i_5i_6i_7i_8}\, P^{(6)}_{i_9i_{10}i_{11}i_{12}i_{13}i_{14}}\, P^{(2)}_{i_{15}i_{16}}\, P^{(2)}_{i_{17}i_{18}]} - 11{,}340\, P^{(8)}_{[i_1i_2i_3i_4i_5i_6i_7i_8}\, P^{(4)}_{i_9i_{10}i_{11}i_{12}}\, P^{(4)}_{i_{13}i_{14}i_{15}i_{16}}\, P^{(2)}_{i_{17}i_{18}]} +$$

$$+ 7560\, P^{(8)}_{[i_1i_2i_3i_4i_5i_6i_7i_8}\, P^{(4)}_{i_9i_{10}i_{11}i_{12}}\, P^{(2)}_{i_{13}i_{14}}\, P^{(2)}_{i_{15}i_{16}}\, P^{(2)}_{i_{17}i_{18}]} - 756\, P^{(8)}_{[i_1i_2i_3i_4i_5i_6i_7i_8}\, P^{(2)}_{i_9i_{10}}\, P^{(2)}_{i_{11}i_{12}}\, P^{(2)}_{i_{13}i_{14}}\, P^{(2)}_{i_{15}i_{16}}\, P^{(2)}_{i_{17}i_{18}]} +$$

$$+ 2240\, P^{(6)}_{[i_1i_2i_3i_4i_5i_6}\, P^{(6)}_{i_7i_8i_9i_{10}i_{11}i_{12}}\, P^{(6)}_{i_{13}i_{14}i_{15}i_{16}i_{17}i_{18}]} - 10{,}080\, P^{(6)}_{[i_1i_2i_3i_4i_5i_6}\, P^{(6)}_{i_7i_8i_9i_{10}i_{11}i_{12}}\, P^{(4)}_{i_{13}i_{14}i_{15}i_{16}}\, P^{(2)}_{i_{17}i_{18}]} +$$

$$+ 3360\, P^{(6)}_{[i_1i_2i_3i_4i_5i_6}\, P^{(6)}_{i_7i_8i_9i_{10}i_{11}i_{12}}\, P^{(2)}_{i_{13}i_{14}}\, P^{(2)}_{i_{15}i_{16}}\, P^{(2)}_{i_{17}i_{18}]} - 2520\, P^{(6)}_{[i_1i_2i_3i_4i_5i_6}\, P^{(4)}_{i_7i_8i_9i_{10}}\, P^{(4)}_{i_{11}i_{12}i_{13}i_{14}}\, P^{(4)}_{i_{15}i_{16}i_{17}i_{18}]} +$$

$$+ 7560\, P^{(6)}_{[i_1i_2i_3i_4i_5i_6}\, P^{(4)}_{i_7i_8i_9i_{10}}\, P^{(4)}_{i_{11}i_{12}i_{13}i_{14}}\, P^{(2)}_{i_{15}i_{16}}\, P^{(2)}_{i_{17}i_{18}]} - 2520\, P^{(6)}_{[i_1i_2i_3i_4i_5i_6}\, P^{(4)}_{i_7i_8i_9i_{10}}\, P^{(2)}_{i_{11}i_{12}}\, P^{(2)}_{i_{13}i_{14}}\, P^{(2)}_{i_{15}i_{16}}\, P^{(2)}_{i_{17}i_{18}]} +$$

$$+ 168\, P^{(6)}_{[i_1i_2i_3i_4i_5i_6}\, P^{(2)}_{i_7i_8}\, P^{(2)}_{i_9i_{10}}\, P^{(2)}_{i_{11}i_{12}}\, P^{(2)}_{i_{13}i_{14}}\, P^{(2)}_{i_{15}i_{16}}\, P^{(2)}_{i_{17}i_{18}]} + 945\, P^{(4)}_{[i_1i_2i_3i_4}\, P^{(4)}_{i_5i_6i_7i_8}\, P^{(4)}_{i_9i_{10}i_{11}i_{12}}\, P^{(4)}_{i_{13}i_{14}i_{15}i_{16}}\, P^{(2)}_{i_{17}i_{18}]} -$$

$$- 1260\, P^{(4)}_{[i_1i_2i_3i_4}\, P^{(4)}_{i_5i_6i_7i_8}\, P^{(4)}_{i_9i_{10}i_{11}i_{12}}\, P^{(2)}_{i_{13}i_{14}}\, P^{(2)}_{i_{15}i_{16}}\, P^{(2)}_{i_{17}i_{18}]} + 378\, P^{(4)}_{[i_1i_2i_3i_4}\, P^{(4)}_{i_5i_6i_7i_8}\, P^{(2)}_{i_9i_{10}}\, P^{(2)}_{i_{11}i_{12}}\, P^{(2)}_{i_{13}i_{14}}\, P^{(2)}_{i_{15}i_{16}}\, P^{(2)}_{i_{17}i_{18}]} -$$

$$- 36\, P^{(4)}_{[i_1i_2i_3i_4}\, P^{(2)}_{i_5i_6}\, P^{(2)}_{i_7i_8}\, P^{(2)}_{i_9i_{10}}\, P^{(2)}_{i_{11}i_{12}}\, P^{(2)}_{i_{13}i_{14}}\, P^{(2)}_{i_{15}i_{16}}\, P^{(2)}_{i_{17}i_{18}]} +$$

$$+ P^{(2)}_{[i_1i_2}\, P^{(2)}_{i_3i_4}\, P^{(2)}_{i_5i_6}\, P^{(2)}_{i_7i_8}\, P^{(2)}_{i_9i_{10}}\, P^{(2)}_{i_{11}i_{12}}\, P^{(2)}_{i_{13}i_{14}}\, P^{(2)}_{i_{15}i_{16}}\, P^{(2)}_{i_{17}i_{18}]} \Big)$$

### COEFFICIENT OF THE 10$^{th}$ CHERN FORM

$$c_{(10)\, i_1i_2i_3i_4i_5i_6i_7i_8i_9i_{10}i_{11}i_{12}i_{13}i_{14}i_{15}i_{16}i_{17}i_{18}i_{19}i_{20}} = \qquad (19)$$

$$= \frac{1}{20!}\, \langle \mathbf{e}_{i_1}\wedge \mathbf{e}_{i_2}\wedge \mathbf{e}_{i_3}\wedge \mathbf{e}_{i_4}\wedge \mathbf{e}_{i_5}\wedge \mathbf{e}_{i_6}\wedge \mathbf{e}_{i_7}\wedge \mathbf{e}_{i_8}\wedge \mathbf{e}_{i_9}\wedge \mathbf{e}_{i_{10}}\wedge \mathbf{e}_{i_{11}}\wedge \mathbf{e}_{i_{12}}\wedge \mathbf{e}_{i_{13}}\wedge \mathbf{e}_{i_{14}}\wedge \mathbf{e}_{i_{15}}\wedge \mathbf{e}_{i_{16}}\wedge \mathbf{e}_{i_{17}}\wedge \mathbf{e}_{i_{18}}\wedge \mathbf{e}_{i_{19}}\wedge \mathbf{e}_{i_{20}},\, c_{(10)}\rangle$$

$$= \frac{i^{10}}{2^{20}\pi^{10} 10!}\, \big(-\, 362{,}880\, R_{[i_1i_2|j|}{}^a\, R_{i_3i_4|a|}{}^b\, R_{i_5i_6|b|}{}^c\, R_{i_7i_8|c|}{}^d\, R_{i_9i_{10}|d|}{}^e\, R_{i_{11}i_{12}|e|}{}^f\, R_{i_{13}i_{14}|f|}{}^g\, R_{i_{15}i_{16}|g|}{}^h\, R_{i_{17}i_{18}|h|}{}^i\, R_{i_{19}i_{20}]i}{}^j +$$

$$+ 403{,}200\, R_{[i_1i_2|i|}{}^a\, R_{i_3i_4|a|}{}^b\, R_{i_5i_6|b|}{}^c\, R_{i_7i_8|c|}{}^d\, R_{i_9i_{10}|d|}{}^e\, R_{i_{11}i_{12}|e|}{}^f\, R_{i_{13}i_{14}|f|}{}^g\, R_{i_{15}i_{16}|g|}{}^h\, R_{i_{17}i_{18}|h|}{}^j\, R_{i_{19}i_{20}]j}{}^i +$$

$$+ 226{,}800\, R_{[i_1i_2|h|}{}^a\, R_{i_3i_4|a|}{}^b\, R_{i_5i_6|b|}{}^c\, R_{i_7i_8|c|}{}^d\, R_{i_9i_{10}|d|}{}^e\, R_{i_{11}i_{12}|e|}{}^f\, R_{i_{13}i_{14}|f|}{}^g\, R_{i_{15}i_{16}|g|}{}^h\, R_{i_{17}i_{18}|j|}{}^i\, R_{i_{19}i_{20}]i}{}^j -$$

$$- 226{,}800\, R_{[i_1i_2|h|}{}^a\, R_{i_3i_4|a|}{}^b\, R_{i_5i_6|b|}{}^c\, R_{i_7i_8|c|}{}^d\, R_{i_9i_{10}|d|}{}^e\, R_{i_{11}i_{12}|e|}{}^f\, R_{i_{13}i_{14}|f|}{}^g\, R_{i_{15}i_{16}|g|}{}^h\, R_{i_{17}i_{18}|i|}{}^i\, R_{i_{19}i_{20}]i}{}^j +$$

$$+ 172{,}800\, R_{[i_1i_2|g|}{}^a\, R_{i_3i_4|a|}{}^b\, R_{i_5i_6|b|}{}^c\, R_{i_7i_8|c|}{}^d\, R_{i_9i_{10}|d|}{}^e\, R_{i_{11}i_{12}|e|}{}^f\, R_{i_{13}i_{14}|f|}{}^g\, R_{i_{15}i_{16}|j|}{}^h\, R_{i_{17}i_{18}|h|}{}^i\, R_{i_{19}i_{20}]i}{}^j -$$

$$- 259{,}200\, R_{[i_1i_2|g|}{}^a\, R_{i_3i_4|a|}{}^b\, R_{i_5i_6|b|}{}^c\, R_{i_7i_8|c|}{}^d\, R_{i_9i_{10}|d|}{}^e\, R_{i_{11}i_{12}|e|}{}^f\, R_{i_{13}i_{14}|f|}{}^g\, R_{i_{15}i_{16}|j|}{}^h\, R_{i_{17}i_{18}|i|}{}^i\, R_{i_{19}i_{20}]j}{}^j +$$

$$+ 86{,}400\, R_{[i_1i_2|g|}{}^a\, R_{i_3i_4|a|}{}^b\, R_{i_5i_6|b|}{}^c\, R_{i_7i_8|c|}{}^d\, R_{i_9i_{10}|d|}{}^e\, R_{i_{11}i_{12}|e|}{}^f\, R_{i_{13}i_{14}|f|}{}^g\, R_{i_{15}i_{16}|h|}{}^h\, R_{i_{17}i_{18}|i|}{}^i\, R_{i_{19}i_{20}]j}{}^j +$$

$$+ 151{,}200\, R_{[i_1i_2|f|}{}^a\, R_{i_3i_4|a|}{}^b\, R_{i_5i_6|b|}{}^c\, R_{i_7i_8|c|}{}^d\, R_{i_9i_{10}|d|}{}^e\, R_{i_{11}i_{12}|e|}{}^f\, R_{i_{13}i_{14}|j|}{}^g\, R_{i_{15}i_{16}|g|}{}^h\, R_{i_{17}i_{18}|h|}{}^i\, R_{i_{19}i_{20}]i}{}^j -$$

$$- 201{,}600\, R_{[i_1i_2|f|}{}^a\, R_{i_3i_4|a|}{}^b\, R_{i_5i_6|b|}{}^c\, R_{i_7i_8|c|}{}^d\, R_{i_9i_{10}|d|}{}^e\, R_{i_{11}i_{12}|e|}{}^f\, R_{i_{13}i_{14}|i|}{}^g\, R_{i_{15}i_{16}|g|}{}^h\, R_{i_{17}i_{18}|h|}{}^i\, R_{i_{19}i_{20}]j}{}^j -$$

$$- 75{,}600\, R_{[i_1i_2|f|}{}^a\, R_{i_3i_4|a|}{}^b\, R_{i_5i_6|b|}{}^c\, R_{i_7i_8|c|}{}^d\, R_{i_9i_{10}|d|}{}^e\, R_{i_{11}i_{12}|e|}{}^f\, R_{i_{13}i_{14}|h|}{}^g\, R_{i_{15}i_{16}|g|}{}^h\, R_{i_{17}i_{18}|j|}{}^i\, R_{i_{19}i_{20}]i}{}^j +$$

$$+ 151{,}200\, R_{[i_1i_2|f|}{}^a\, R_{i_3i_4|a|}{}^b\, R_{i_5i_6|b|}{}^c\, R_{i_7i_8|c|}{}^d\, R_{i_9i_{10}|d|}{}^e\, R_{i_{11}i_{12}|e|}{}^f\, R_{i_{13}i_{14}|h|}{}^g\, R_{i_{15}i_{16}|i|}{}^h\, R_{i_{17}i_{18}|i|}{}^i\, R_{i_{19}i_{20}]j}{}^j -$$

$$- 25{,}200\, R_{[i_1i_2|f|}{}^a\, R_{i_3i_4|a|}{}^b\, R_{i_5i_6|b|}{}^c\, R_{i_7i_8|c|}{}^d\, R_{i_9i_{10}|d|}{}^e\, R_{i_{11}i_{12}|e|}{}^f\, R_{i_{13}i_{14}|g|}{}^g\, R_{i_{15}i_{16}|h|}{}^h\, R_{i_{17}i_{18}|i|}{}^i\, R_{i_{19}i_{20}]j}{}^j +$$

$$+ 72{,}576\, R_{[i_1i_2|e|}{}^a\, R_{i_3i_4|a|}{}^b\, R_{i_5i_6|b|}{}^c\, R_{i_7i_8|c|}{}^d\, R_{i_9i_{10}|d|}{}^e\, R_{i_{11}i_{12}|j|}{}^f\, R_{i_{13}i_{14}|f|}{}^g\, R_{i_{15}i_{16}|g|}{}^h\, R_{i_{17}i_{18}|h|}{}^i\, R_{i_{19}i_{20}]i}{}^j -$$

$$- 181{,}440\, R_{[i_1i_2|e|}{}^a\, R_{i_3i_4|a|}{}^b\, R_{i_5i_6|b|}{}^c\, R_{i_7i_8|c|}{}^d\, R_{i_9i_{10}|d|}{}^e\, R_{i_{11}i_{12}|i|}{}^f\, R_{i_{13}i_{14}|f|}{}^g\, R_{i_{15}i_{16}|g|}{}^h\, R_{i_{17}i_{18}|h|}{}^i\, R_{i_{19}i_{20}]j}{}^j -$$

$$- 120{,}960\, R_{[i_1i_2|e|}{}^a\, R_{i_3i_4|a|}{}^b\, R_{i_5i_6|b|}{}^c\, R_{i_7i_8|c|}{}^d\, R_{i_9i_{10}|d|}{}^e\, R_{i_{11}i_{12}|h|}{}^f\, R_{i_{13}i_{14}|f|}{}^g\, R_{i_{15}i_{16}|g|}{}^h\, R_{i_{17}i_{18}|j|}{}^i\, R_{i_{19}i_{20}]i}{}^j +$$

$$+ 120{,}960\, R_{[i_1i_2|e|}{}^a\, R_{i_3i_4|a|}{}^b\, R_{i_5i_6|b|}{}^c\, R_{i_7i_8|c|}{}^d\, R_{i_9i_{10}|d|}{}^e\, R_{i_{11}i_{12}|h|}{}^f\, R_{i_{13}i_{14}|f|}{}^g\, R_{i_{15}i_{16}|g|}{}^h\, R_{i_{17}i_{18}|i|}{}^i\, R_{i_{19}i_{20}]j}{}^j +$$

$$+ 90{,}720\, R_{[i_1i_2|e|}{}^a\, R_{i_3i_4|a|}{}^b\, R_{i_5i_6|b|}{}^c\, R_{i_7i_8|c|}{}^d\, R_{i_9i_{10}|d|}{}^e\, R_{i_{11}i_{12}|g|}{}^f\, R_{i_{13}i_{14}|f|}{}^g\, R_{i_{15}i_{16}|i|}{}^h\, R_{i_{17}i_{18}|h|}{}^i\, R_{i_{19}i_{20}]j}{}^j -$$

$$- 60{,}480\, R_{[i_1i_2|e|}{}^a\, R_{i_3i_4|a|}{}^b\, R_{i_5i_6|b|}{}^c\, R_{i_7i_8|c|}{}^d\, R_{i_9i_{10}|d|}{}^e\, R_{i_{11}i_{12}|g|}{}^f\, R_{i_{13}i_{14}|f|}{}^g\, R_{i_{15}i_{16}|h|}{}^h\, R_{i_{17}i_{18}|i|}{}^i\, R_{i_{19}i_{20}]j}{}^j +$$

$$+ 6048\, R_{[i_1i_2|e|}{}^a\, R_{i_3i_4|a|}{}^b\, R_{i_5i_6|b|}{}^c\, R_{i_7i_8|c|}{}^d\, R_{i_9i_{10}|d|}{}^e\, R_{i_{11}i_{12}|f|}{}^f\, R_{i_{13}i_{14}|g|}{}^g\, R_{i_{15}i_{16}|h|}{}^h\, R_{i_{17}i_{18}|i|}{}^i\, R_{i_{19}i_{20}]j}{}^j -$$

$$- 56{,}700\, R_{[i_1i_2|d|}{}^a\, R_{i_3i_4|a|}{}^b\, R_{i_5i_6|b|}{}^c\, R_{i_7i_8|c|}{}^d\, R_{i_9i_{10}|h|}{}^e\, R_{i_{11}i_{12}|e|}{}^f\, R_{i_{13}i_{14}|f|}{}^g\, R_{i_{15}i_{16}|g|}{}^h\, R_{i_{17}i_{18}|j|}{}^i\, R_{i_{19}i_{20}]i}{}^j +$$

$$+ 56{,}700\, R_{[i_1i_2|d|}{}^a\, R_{i_3i_4|a|}{}^b\, R_{i_5i_6|b|}{}^c\, R_{i_7i_8|c|}{}^d\, R_{i_9i_{10}|h|}{}^e\, R_{i_{11}i_{12}|e|}{}^f\, R_{i_{13}i_{14}|f|}{}^g\, R_{i_{15}i_{16}|g|}{}^h\, R_{i_{17}i_{18}|i|}{}^i\, R_{i_{19}i_{20}]j}{}^j -$$

$$- 50{,}400\, R_{[i_1i_2|d|}{}^a\, R_{i_3i_4|a|}{}^b\, R_{i_5i_6|b|}{}^c\, R_{i_7i_8|c|}{}^d\, R_{i_9i_{10}|g|}{}^e\, R_{i_{11}i_{12}|e|}{}^f\, R_{i_{13}i_{14}|f|}{}^g\, R_{i_{15}i_{16}|j|}{}^h\, R_{i_{17}i_{18}|h|}{}^i\, R_{i_{19}i_{20}]i}{}^j +$$

$$+ 151{,}200\, R_{[i_1i_2|d|}{}^a\, R_{i_3i_4|a|}{}^b\, R_{i_5i_6|b|}{}^c\, R_{i_7i_8|c|}{}^d\, R_{i_9i_{10}|g|}{}^e\, R_{i_{11}i_{12}|e|}{}^f\, R_{i_{13}i_{14}|f|}{}^g\, R_{i_{15}i_{16}|i|}{}^h\, R_{i_{17}i_{18}|h|}{}^i\, R_{i_{19}i_{20}]j}{}^j -$$

$$- 50{,}400\, R_{[i_1i_2|d|}{}^a\, R_{i_3i_4|a|}{}^b\, R_{i_5i_6|b|}{}^c\, R_{i_7i_8|c|}{}^d\, R_{i_9i_{10}|g|}{}^e\, R_{i_{11}i_{12}|e|}{}^f\, R_{i_{13}i_{14}|f|}{}^g\, R_{i_{15}i_{16}|h|}{}^h\, R_{i_{17}i_{18}|i|}{}^i\, R_{i_{19}i_{20}]j}{}^j +$$

$$+ 18{,}900\, R_{[i_1i_2|d|}{}^a\, R_{i_3i_4|a|}{}^b\, R_{i_5i_6|b|}{}^c\, R_{i_7i_8|c|}{}^d\, R_{i_9i_{10}|f|}{}^e\, R_{i_{11}i_{12}|e|}{}^f\, R_{i_{13}i_{14}|h|}{}^g\, R_{i_{15}i_{16}|g|}{}^h\, R_{i_{17}i_{18}|j|}{}^i\, R_{i_{19}i_{20}]i}{}^j -$$

$$- 56{,}700\, R_{[i_1i_2|d|}{}^a\, R_{i_3i_4|a|}{}^b\, R_{i_5i_6|b|}{}^c\, R_{i_7i_8|c|}{}^d\, R_{i_9i_{10}|f|}{}^e\, R_{i_{11}i_{12}|e|}{}^f\, R_{i_{13}i_{14}|h|}{}^g\, R_{i_{15}i_{16}|g|}{}^h\, R_{i_{17}i_{18}|i|}{}^i\, R_{i_{19}i_{20}]j}{}^j +$$

$$+ 18{,}900\, R_{[i_1i_2|d|}{}^a\, R_{i_3i_4|a|}{}^b\, R_{i_5i_6|b|}{}^c\, R_{i_7i_8|c|}{}^d\, R_{i_9i_{10}|f|}{}^e\, R_{i_{11}i_{12}|e|}{}^f\, R_{i_{13}i_{14}|g|}{}^g\, R_{i_{15}i_{16}|h|}{}^h\, R_{i_{17}i_{18}|i|}{}^i\, R_{i_{19}i_{20}]j}{}^j -$$

$$- 1260\, R_{[i_1i_2|d|}{}^a\, R_{i_3i_4|a|}{}^b\, R_{i_5i_6|b|}{}^c\, R_{i_7i_8|c|}{}^d\, R_{i_9i_{10}|e|}{}^e\, R_{i_{11}i_{12}|f|}{}^f\, R_{i_{13}i_{14}|g|}{}^g\, R_{i_{15}i_{16}|h|}{}^h\, R_{i_{17}i_{18}|i|}{}^i\, R_{i_{19}i_{20}]j}{}^j +$$

$$+ 22{,}400\, R_{[i_1 i_2|c|}{}^a R_{i_3 i_4|a|}{}^b R_{i_5 i_6|b|}{}^c R_{i_7 i_8|f|}{}^d R_{i_9 i_{10}|d|}{}^e R_{i_{11} i_{12}|e|}{}^f R_{i_{13} i_{14}|i|}{}^g R_{i_{15} i_{16}|g|}{}^h R_{i_{17} i_{18}|h|}{}^i R_{i_{19} i_{20}]}{}^j +$$

$$+ 25{,}200\, R_{[i_1 i_2|c|}{}^a R_{i_3 i_4|a|}{}^b R_{i_5 i_6|b|}{}^c R_{i_7 i_8|f|}{}^d R_{i_9 i_{10}|d|}{}^e R_{i_{11} i_{12}|e|}{}^f R_{i_{13} i_{14}|h|}{}^g R_{i_{15} i_{16}|g|}{}^h R_{i_{17} i_{18}|j|}{}^i R_{i_{19} i_{20}]}{}^j -$$

$$- 50{,}400\, R_{[i_1 i_2|c|}{}^a R_{i_3 i_4|a|}{}^b R_{i_5 i_6|b|}{}^c R_{i_7 i_8|f|}{}^d R_{i_9 i_{10}|d|}{}^e R_{i_{11} i_{12}|e|}{}^f R_{i_{13} i_{14}|h|}{}^g R_{i_{15} i_{16}]}{}^h R_{i_{17} i_{18}|i|}{}^i R_{i_{19} i_{20}]}{}^j +$$

$$+ 8400\, R_{[i_1 i_2|c|}{}^a R_{i_3 i_4|a|}{}^b R_{i_5 i_6|b|}{}^c R_{i_7 i_8|f|}{}^d R_{i_9 i_{10}|d|}{}^e R_{i_{11} i_{12}|}{}^f R_{i_{13} i_{14}|}{}^g R_{i_{15} i_{16}|}{}^h R_{i_{17} i_{18}|}{}^i R_{i_{19} i_{20}]}{}^j -$$

$$- 25{,}200\, R_{[i_1 i_2|c|}{}^a R_{i_3 i_4|a|}{}^b R_{i_5 i_6|b|}{}^c R_{i_7 i_8|e|}{}^d R_{i_9 i_{10}|d|}{}^e R_{i_{11} i_{12}|g|}{}^f R_{i_{13} i_{14}|f|}{}^g R_{i_{15} i_{16}|i|}{}^h R_{i_{17} i_{18}|h|}{}^i R_{i_{19} i_{20}]}{}^j +$$

$$+ 25{,}200\, R_{[i_1 i_2|c|}{}^a R_{i_3 i_4|a|}{}^b R_{i_5 i_6|b|}{}^c R_{i_7 i_8|e|}{}^d R_{i_9 i_{10}|d|}{}^e R_{i_{11} i_{12}|g|}{}^f R_{i_{13} i_{14}|f|}{}^g R_{i_{15} i_{16}|}{}^h R_{i_{17} i_{18}|}{}^i R_{i_{19} i_{20}]}{}^j -$$

$$- 5040\, R_{[i_1 i_2|c|}{}^a R_{i_3 i_4|a|}{}^b R_{i_5 i_6|b|}{}^c R_{i_7 i_8|e|}{}^d R_{i_9 i_{10}|d|}{}^e R_{i_{11} i_{12}|f|}{}^f R_{i_{13} i_{14}|}{}^g R_{i_{15} i_{16}|}{}^h R_{i_{17} i_{18}|}{}^i R_{i_{19} i_{20}]}{}^j +$$

$$+ 240\, R_{[i_1 i_2|c|}{}^a R_{i_3 i_4|a|}{}^b R_{i_5 i_6|b|}{}^c R_{i_7 i_8|d|}{}^d R_{i_9 i_{10}|e|}{}^e R_{i_{11} i_{12}|f|}{}^f R_{i_{13} i_{14}|}{}^g R_{i_{15} i_{16}|}{}^h R_{i_{17} i_{18}|}{}^i R_{i_{19} i_{20}]}{}^j -$$

$$- 945\, R_{[i_1 i_2|b|}{}^a R_{i_3 i_4|a|}{}^b R_{i_5 i_6|d|}{}^c R_{i_7 i_8|c|}{}^d R_{i_9 i_{10}|f|}{}^e R_{i_{11} i_{12}|e|}{}^f R_{i_{13} i_{14}|h|}{}^g R_{i_{15} i_{16}|g|}{}^h R_{i_{17} i_{18}|j|}{}^i R_{i_{19} i_{20}]}{}^j +$$

$$+ 4725\, R_{[i_1 i_2|b|}{}^a R_{i_3 i_4|a|}{}^b R_{i_5 i_6|d|}{}^c R_{i_7 i_8|c|}{}^d R_{i_9 i_{10}|f|}{}^e R_{i_{11} i_{12}|e|}{}^f R_{i_{13} i_{14}|}{}^g R_{i_{15} i_{16}|}{}^h R_{i_{17} i_{18}|}{}^i R_{i_{19} i_{20}]}{}^j -$$

$$- 3150\, R_{[i_1 i_2|b|}{}^a R_{i_3 i_4|a|}{}^b R_{i_5 i_6|d|}{}^c R_{i_7 i_8|c|}{}^d R_{i_9 i_{10}|f|}{}^e R_{i_{11} i_{12}|e|}{}^f R_{i_{13} i_{14}|}{}^g R_{i_{15} i_{16}|}{}^h R_{i_{17} i_{18}|}{}^i R_{i_{19} i_{20}]}{}^j +$$

$$+ 630\, R_{[i_1 i_2|b|}{}^a R_{i_3 i_4|a|}{}^b R_{i_5 i_6|d|}{}^c R_{i_7 i_8|c|}{}^d R_{i_9 i_{10}|e|}{}^e R_{i_{11} i_{12}|f|}{}^f R_{i_{13} i_{14}|}{}^g R_{i_{15} i_{16}|}{}^h R_{i_{17} i_{18}|}{}^i R_{i_{19} i_{20}]}{}^j -$$

$$- 45\, R_{[i_1 i_2|b|}{}^a R_{i_3 i_4|a|}{}^b R_{i_5 i_6|c|}{}^c R_{i_7 i_8|d|}{}^d R_{i_9 i_{10}|e|}{}^e R_{i_{11} i_{12}|f|}{}^f R_{i_{13} i_{14}|}{}^g R_{i_{15} i_{16}|}{}^h R_{i_{17} i_{18}|}{}^i R_{i_{19} i_{20}]}{}^j +$$

$$+ R_{[i_1 i_2|a|}{}^a R_{i_3 i_4|b|}{}^b R_{i_5 i_6|c|}{}^c R_{i_7 i_8|d|}{}^d R_{i_9 i_{10}|e|}{}^e R_{i_{11} i_{12}|f|}{}^f R_{i_{13} i_{14}|}{}^g R_{i_{15} i_{16}|}{}^h R_{i_{17} i_{18}|}{}^i R_{i_{19} i_{20}]}{}^j )$$

$$= \frac{i^{10}}{2^{20} \pi^{10} 10!} \Big( -362{,}880\, R_{[i_1 i_2|j|}{}^a R_{i_3 i_4|a|}{}^b R_{i_5 i_6|b|}{}^c R_{i_7 i_8|c|}{}^d R_{i_9 i_{10}|d|}{}^e R_{i_{11} i_{12}|e|}{}^f R_{i_{13} i_{14}|f|}{}^g R_{i_{15} i_{16}|g|}{}^h R_{i_{17} i_{18}|h|}{}^i R_{i_{19} i_{20}]}{}^j +$$

$$+ 403{,}200\, R_{[i_1 i_2|i|}{}^a R_{i_3 i_4|a|}{}^b R_{i_5 i_6|b|}{}^c R_{i_7 i_8|c|}{}^d R_{i_9 i_{10}|d|}{}^e R_{i_{11} i_{12}|e|}{}^f R_{i_{13} i_{14}|f|}{}^g R_{i_{15} i_{16}|g|}{}^h R_{i_{17} i_{18}|h|}{}^i V_{i_{19} i_{20}]} +$$

$$+ 226{,}800\, R_{[i_1 i_2|h|}{}^a R_{i_3 i_4|a|}{}^b R_{i_5 i_6|b|}{}^c R_{i_7 i_8|c|}{}^d R_{i_9 i_{10}|d|}{}^e R_{i_{11} i_{12}|e|}{}^f R_{i_{13} i_{14}|f|}{}^g R_{i_{15} i_{16}|g|}{}^h R_{i_{17} i_{18}|j|}{}^i R_{i_{19} i_{20}]}{}^i -$$

$$- 226{,}800\, R_{[i_1 i_2|h|}{}^a R_{i_3 i_4|a|}{}^b R_{i_5 i_6|b|}{}^c R_{i_7 i_8|c|}{}^d R_{i_9 i_{10}|d|}{}^e R_{i_{11} i_{12}|e|}{}^f R_{i_{13} i_{14}|f|}{}^g R_{i_{15} i_{16}|g|}{}^h V_{i_{17} i_{18}} V_{i_{19} i_{20}]} +$$

$$+ 172{,}800\, R_{[i_1 i_2|g|}{}^a R_{i_3 i_4|a|}{}^b R_{i_5 i_6|b|}{}^c R_{i_7 i_8|c|}{}^d R_{i_9 i_{10}|d|}{}^e R_{i_{11} i_{12}|e|}{}^f R_{i_{13} i_{14}|f|}{}^g R_{i_{15} i_{16}|j|}{}^h R_{i_{17} i_{18}|h|}{}^i R_{i_{19} i_{20}]}{}^i -$$

$$- 259{,}200\, R_{[i_1 i_2|g|}{}^a R_{i_3 i_4|a|}{}^b R_{i_5 i_6|b|}{}^c R_{i_7 i_8|c|}{}^d R_{i_9 i_{10}|d|}{}^e R_{i_{11} i_{12}|e|}{}^f R_{i_{13} i_{14}|f|}{}^g R_{i_{15} i_{16}|}{}^h R_{i_{17} i_{18}|h|}{}^i V_{i_{19} i_{20}]} +$$

$$+ 86{,}400\, R_{[i_1 i_2|g|}{}^a R_{i_3 i_4|a|}{}^b R_{i_5 i_6|b|}{}^c R_{i_7 i_8|c|}{}^d R_{i_9 i_{10}|d|}{}^e R_{i_{11} i_{12}|e|}{}^f R_{i_{13} i_{14}|f|}{}^g V_{i_{15} i_{16}} V_{i_{17} i_{18}} V_{i_{19} i_{20}]} +$$

$$+ 151{,}200\, R_{[i_1 i_2|f|}{}^a R_{i_3 i_4|a|}{}^b R_{i_5 i_6|b|}{}^c R_{i_7 i_8|c|}{}^d R_{i_9 i_{10}|d|}{}^e R_{i_{11} i_{12}|e|}{}^f R_{i_{13} i_{14}|j|}{}^g R_{i_{15} i_{16}|g|}{}^h R_{i_{17} i_{18}|h|}{}^i R_{i_{19} i_{20}]}{}^i -$$

$$- 201{,}600\, R_{[i_1 i_2|f|}{}^a R_{i_3 i_4|a|}{}^b R_{i_5 i_6|b|}{}^c R_{i_7 i_8|c|}{}^d R_{i_9 i_{10}|d|}{}^e R_{i_{11} i_{12}|e|}{}^f R_{i_{13} i_{14}|i|}{}^g R_{i_{15} i_{16}|g|}{}^h R_{i_{17} i_{18}|h|}{}^i V_{i_{19} i_{20}]} -$$

$$- 75{,}600\, R_{[i_1 i_2|f|}{}^a R_{i_3 i_4|a|}{}^b R_{i_5 i_6|b|}{}^c R_{i_7 i_8|c|}{}^d R_{i_9 i_{10}|d|}{}^e R_{i_{11} i_{12}|e|}{}^f R_{i_{13} i_{14}|h|}{}^g R_{i_{15} i_{16}|g|}{}^h R_{i_{17} i_{18}|j|}{}^i R_{i_{19} i_{20}]}{}^i +$$

$$+ 151{,}200\, R_{[i_1 i_2|f|}{}^a R_{i_3 i_4|a|}{}^b R_{i_5 i_6|b|}{}^c R_{i_7 i_8|c|}{}^d R_{i_9 i_{10}|d|}{}^e R_{i_{11} i_{12}|e|}{}^f R_{i_{13} i_{14}|}{}^g R_{i_{15} i_{16}|}{}^h V_{i_{17} i_{18}} V_{i_{19} i_{20}]} -$$

$$- 25{,}200\, R_{[i_1 i_2|f|}{}^a R_{i_3 i_4|a|}{}^b R_{i_5 i_6|b|}{}^c R_{i_7 i_8|c|}{}^d R_{i_9 i_{10}|d|}{}^e R_{i_{11} i_{12}|e|}{}^f V_{i_{13} i_{14}} V_{i_{15} i_{16}} V_{i_{17} i_{18}} V_{i_{19} i_{20}]} +$$

$$+ 72{,}576\, R_{[i_1 i_2|e|}{}^a R_{i_3 i_4|a|}{}^b R_{i_5 i_6|b|}{}^c R_{i_7 i_8|c|}{}^d R_{i_9 i_{10}|d|}{}^e R_{i_{11} i_{12}|j|}{}^f R_{i_{13} i_{14}|f|}{}^g R_{i_{15} i_{16}|g|}{}^h R_{i_{17} i_{18}|h|}{}^i R_{i_{19} i_{20}]}{}^i -$$

$$- 181{,}440\, R_{[i_1 i_2|e|}{}^a R_{i_3 i_4|a|}{}^b R_{i_5 i_6|b|}{}^c R_{i_7 i_8|c|}{}^d R_{i_9 i_{10}|d|}{}^e R_{i_{11} i_{12}|i|}{}^f R_{i_{13} i_{14}|f|}{}^g R_{i_{15} i_{16}|g|}{}^h R_{i_{17} i_{18}|h|}{}^i V_{i_{19} i_{20}]} -$$

$$- 120{,}960\, R_{[i_1 i_2|e|}{}^a R_{i_3 i_4|a|}{}^b R_{i_5 i_6|b|}{}^c R_{i_7 i_8|c|}{}^d R_{i_9 i_{10}|d|}{}^e R_{i_{11} i_{12}|f|}{}^f R_{i_{13} i_{14}|}{}^g R_{i_{15} i_{16}|}{}^h R_{i_{17} i_{18}|j|}{}^i R_{i_{19} i_{20}]}{}^i +$$

$$+ 120{,}960\, R_{[i_1 i_2|e|}{}^a R_{i_3 i_4|a|}{}^b R_{i_5 i_6|b|}{}^c R_{i_7 i_8|c|}{}^d R_{i_9 i_{10}|d|}{}^e R_{i_{11} i_{12}|h|}{}^f R_{i_{13} i_{14}|f|}{}^g R_{i_{15} i_{16}|}{}^h V_{i_{17} i_{18}} V_{i_{19} i_{20}]} +$$

$$+ 90{,}720\, R_{[i_1 i_2|e|}{}^a R_{i_3 i_4|a|}{}^b R_{i_5 i_6|b|}{}^c R_{i_7 i_8|c|}{}^d R_{i_9 i_{10}|d|}{}^e R_{i_{11} i_{12}|g|}{}^f R_{i_{13} i_{14}|f|}{}^g R_{i_{15} i_{16}|i|}{}^h R_{i_{17} i_{18}|h|}{}^i V_{i_{19} i_{20}]} -$$

$$- 60{,}480\, R_{[i_1 i_2|e|}{}^a R_{i_3 i_4|a|}{}^b R_{i_5 i_6|b|}{}^c R_{i_7 i_8|c|}{}^d R_{i_9 i_{10}|d|}{}^e R_{i_{11} i_{12}|g|}{}^f R_{i_{13} i_{14}|f|}{}^g V_{i_{15} i_{16}} V_{i_{17} i_{18}} V_{i_{19} i_{20}]} +$$

$$+ 6048\, R_{[i_1 i_2|e|}{}^a R_{i_3 i_4|a|}{}^b R_{i_5 i_6|b|}{}^c R_{i_7 i_8|c|}{}^d R_{i_9 i_{10}|d|}{}^e V_{i_{11} i_{12}} V_{i_{13} i_{14}} V_{i_{15} i_{16}} V_{i_{17} i_{18}} V_{i_{19} i_{20}]} -$$

$$- 56{,}700\, R_{[i_1 i_2|d|}{}^a R_{i_3 i_4|a|}{}^b R_{i_5 i_6|b|}{}^c R_{i_7 i_8|c|}{}^d R_{i_9 i_{10}|h|}{}^e R_{i_{11} i_{12}|e|}{}^f R_{i_{13} i_{14}|f|}{}^g R_{i_{15} i_{16}|g|}{}^h R_{i_{17} i_{18}|j|}{}^i R_{i_{19} i_{20}]}{}^i +$$

$$+ 56{,}700\, R_{[i_1 i_2|d|}{}^a R_{i_3 i_4|a|}{}^b R_{i_5 i_6|b|}{}^c R_{i_7 i_8|c|}{}^d R_{i_9 i_{10}|h|}{}^e R_{i_{11} i_{12}|e|}{}^f R_{i_{13} i_{14}|f|}{}^g R_{i_{15} i_{16}|g|}{}^h V_{i_{17} i_{18}} V_{i_{19} i_{20}]} -$$

$$- 50{,}400\, R_{[i_1 i_2|d|}{}^a R_{i_3 i_4|a|}{}^b R_{i_5 i_6|b|}{}^c R_{i_7 i_8|c|}{}^d R_{i_9 i_{10}|g|}{}^e R_{i_{11} i_{12}|e|}{}^f R_{i_{13} i_{14}|f|}{}^g R_{i_{15} i_{16}|j|}{}^h R_{i_{17} i_{18}|h|}{}^i R_{i_{19} i_{20}]}{}^i +$$

$$+ 151{,}200\, R_{[i_1 i_2|d|}{}^a R_{i_3 i_4|a|}{}^b R_{i_5 i_6|b|}{}^c R_{i_7 i_8|c|}{}^d R_{i_9 i_{10}|g|}{}^e R_{i_{11} i_{12}|e|}{}^f R_{i_{13} i_{14}|f|}{}^g R_{i_{15} i_{16}|}{}^h R_{i_{17} i_{18}|h|}{}^i V_{i_{19} i_{20}]} -$$

$$- 50{,}400\, R_{[i_1 i_2|d|}{}^a R_{i_3 i_4|a|}{}^b R_{i_5 i_6|b|}{}^c R_{i_7 i_8|c|}{}^d R_{i_9 i_{10}|g|}{}^e R_{i_{11} i_{12}|e|}{}^f R_{i_{13} i_{14}|f|}{}^g V_{i_{15} i_{16}} V_{i_{17} i_{18}} V_{i_{19} i_{20}]} +$$

$$+ 18{,}900\, R_{[i_1 i_2|d|}{}^a R_{i_3 i_4|a|}{}^b R_{i_5 i_6|b|}{}^c R_{i_7 i_8|c|}{}^d R_{i_9 i_{10}|f|}{}^e R_{i_{11} i_{12}|e|}{}^f R_{i_{13} i_{14}|h|}{}^g R_{i_{15} i_{16}|g|}{}^h R_{i_{17} i_{18}|j|}{}^i R_{i_{19} i_{20}]}{}^i -$$

$$- 56{,}700\, R_{[i_1 i_2|d|}{}^a R_{i_3 i_4|a|}{}^b R_{i_5 i_6|b|}{}^c R_{i_7 i_8|c|}{}^d R_{i_9 i_{10}|f|}{}^e R_{i_{11} i_{12}|e|}{}^f R_{i_{13} i_{14}|h|}{}^g R_{i_{15} i_{16}|g|}{}^h V_{i_{17} i_{18}} V_{i_{19} i_{20}]} +$$

$$+ 18{,}900\, R_{[i_1 i_2|d|}{}^a R_{i_3 i_4|a|}{}^b R_{i_5 i_6|b|}{}^c R_{i_7 i_8|c|}{}^d R_{i_9 i_{10}|f|}{}^e R_{i_{11} i_{12}|e|}{}^f V_{i_{13} i_{14}} V_{i_{15} i_{16}} V_{i_{17} i_{18}} V_{i_{19} i_{20}]} -$$

$$\begin{aligned}
&- 1260\, R_{[i_1i_2|d|}{}^a R_{i_3i_4|a|}{}^b R_{i_5i_6|b|}{}^c R_{i_7i_8|c|}{}^d V_{i_9i_{10}} V_{i_{11}i_{12}} V_{i_{13}i_{14}} V_{i_{15}i_{16}} V_{i_{17}i_{18}} V_{i_{19}i_{20}]} + \\
&+ 22{,}400\, R_{[i_1i_2|c|}{}^a R_{i_3i_4|a|}{}^b R_{i_5i_6|b|}{}^c R_{i_7i_8|f|}{}^d R_{i_9i_{10}|d|}{}^e R_{i_{11}i_{12}|e|}{}^f R_{i_{13}i_{14}|i|}{}^g R_{i_{15}i_{16}|g|}{}^h R_{i_{17}i_{18}|h|}{}^i V_{i_{19}i_{20}]} + \\
&+ 25{,}200\, R_{[i_1i_2|c|}{}^a R_{i_3i_4|a|}{}^b R_{i_5i_6|b|}{}^c R_{i_7i_8|f|}{}^d R_{i_9i_{10}|d|}{}^e R_{i_{11}i_{12}|e|}{}^f R_{i_{13}i_{14}|h|}{}^g R_{i_{15}i_{16}|g|}{}^h R_{i_{17}i_{18}|j|}{}^i R_{i_{19}i_{20}]i}{}^j - \\
&- 50{,}400\, R_{[i_1i_2|c|}{}^a R_{i_3i_4|a|}{}^b R_{i_5i_6|b|}{}^c R_{i_7i_8|f|}{}^d R_{i_9i_{10}|d|}{}^e R_{i_{11}i_{12}|e|}{}^f R_{i_{13}i_{14}|h|}{}^g R_{i_{15}i_{16}|g|}{}^h V_{i_{17}i_{18}} V_{i_{19}i_{20}]} + \\
&+ 8400\, R_{[i_1i_2|c|}{}^a R_{i_3i_4|a|}{}^b R_{i_5i_6|b|}{}^c R_{i_7i_8|f|}{}^d R_{i_9i_{10}|d|}{}^e R_{i_{11}i_{12}|e|}{}^f V_{i_{13}i_{14}} V_{i_{15}i_{16}} V_{i_{17}i_{18}} V_{i_{19}i_{20}]} - \\
&- 25{,}200\, R_{[i_1i_2|c|}{}^a R_{i_3i_4|a|}{}^b R_{i_5i_6|b|}{}^c R_{i_7i_8|e|}{}^d R_{i_9i_{10}|d|}{}^e R_{i_{11}i_{12}|g|}{}^f R_{i_{13}i_{14}|f|}{}^g R_{i_{15}i_{16}|i|}{}^h R_{i_{17}i_{18}|h|}{}^i V_{i_{19}i_{20}]} + \\
&+ 25{,}200\, R_{[i_1i_2|c|}{}^a R_{i_3i_4|a|}{}^b R_{i_5i_6|b|}{}^c R_{i_7i_8|e|}{}^d R_{i_9i_{10}|d|}{}^e R_{i_{11}i_{12}|g|}{}^f R_{i_{13}i_{14}|f|}{}^g V_{i_{15}i_{16}} V_{i_{17}i_{18}} V_{i_{19}i_{20}]} - \\
&- 5040\, R_{[i_1i_2|c|}{}^a R_{i_3i_4|a|}{}^b R_{i_5i_6|b|}{}^c R_{i_7i_8|e|}{}^d R_{i_9i_{10}|d|}{}^e V_{i_{11}i_{12}} V_{i_{13}i_{14}} V_{i_{15}i_{16}} V_{i_{17}i_{18}} V_{i_{19}i_{20}]} + \\
&+ 240\, R_{[i_1i_2|c|}{}^a R_{i_3i_4|a|}{}^b R_{i_5i_6|b|}{}^c V_{i_7i_8} V_{i_9i_{10}} V_{i_{11}i_{12}} V_{i_{13}i_{14}} V_{i_{15}i_{16}} V_{i_{17}i_{18}} V_{i_{19}i_{20}]} - \\
&- 945\, R_{[i_1i_2|b|}{}^a R_{i_3i_4|a|}{}^b R_{i_5i_6|d|}{}^c R_{i_7i_8|c|}{}^d R_{i_9i_{10}|f|}{}^e R_{i_{11}i_{12}|e|}{}^f R_{i_{13}i_{14}|h|}{}^g R_{i_{15}i_{16}|g|}{}^h R_{i_{17}i_{18}|j|}{}^i R_{i_{19}i_{20}]i}{}^j + \\
&+ 4725\, R_{[i_1i_2|b|}{}^a R_{i_3i_4|a|}{}^b R_{i_5i_6|d|}{}^c R_{i_7i_8|c|}{}^d R_{i_9i_{10}|f|}{}^e R_{i_{11}i_{12}|e|}{}^f R_{i_{13}i_{14}|h|}{}^g R_{i_{15}i_{16}|g|}{}^h V_{i_{17}i_{18}} V_{i_{19}i_{20}]} - \\
&- 3150\, R_{[i_1i_2|b|}{}^a R_{i_3i_4|a|}{}^b R_{i_5i_6|d|}{}^c R_{i_7i_8|c|}{}^d R_{i_9i_{10}|f|}{}^e R_{i_{11}i_{12}|e|}{}^f V_{i_{13}i_{14}} V_{i_{15}i_{16}} V_{i_{17}i_{18}} V_{i_{19}i_{20}]} + \\
&+ 630\, R_{[i_1i_2|b|}{}^a R_{i_3i_4|a|}{}^b R_{i_5i_6|d|}{}^c R_{i_7i_8|c|}{}^d V_{i_9i_{10}} V_{i_{11}i_{12}} V_{i_{13}i_{14}} V_{i_{15}i_{16}} V_{i_{17}i_{18}} V_{i_{19}i_{20}]} - \\
&- 45\, R_{[i_1i_2|b|}{}^a R_{i_3i_4|a|}{}^b V_{i_5i_6} V_{i_7i_8} V_{i_9i_{10}} V_{i_{11}i_{12}} V_{i_{13}i_{14}} V_{i_{15}i_{16}} V_{i_{17}i_{18}} V_{i_{19}i_{20}]} + \\
&+ V_{[i_1i_2} V_{i_3i_4} V_{i_5i_6} V_{i_7i_8} V_{i_9i_{10}} V_{i_{11}i_{12}} V_{i_{13}i_{14}} V_{i_{15}i_{16}} V_{i_{17}i_{18}} V_{i_{19}i_{20}]}\big)
\end{aligned}$$

$$\begin{aligned}
&= \frac{i^{10}}{2^{10}\pi^{10} 10!}\Big(- 362{,}880\, P^{(20)}{}_{i_1i_2i_3i_4i_5i_6i_7i_8i_9i_{10}i_{11}i_{12}i_{13}i_{14}i_{15}i_{16}i_{17}i_{18}i_{19}i_{20}} + 403{,}200\, P^{(18)}{}_{[i_1i_2i_3i_4i_5i_6i_7i_8i_9i_{10}i_{11}i_{12}i_{13}i_{14}i_{15}i_{16}i_{17}i_{18}} P^{(2)}{}_{i_{19}i_{20}]} + \\
&+ 226{,}800\, P^{(16)}{}_{[i_1i_2i_3i_4i_5i_6i_7i_8i_9i_{10}i_{11}i_{12}i_{13}i_{14}i_{15}i_{16}} P^{(4)}{}_{i_{17}i_{18}i_{19}i_{20}]} - 226{,}800\, P^{(16)}{}_{[i_1i_2i_3i_4i_5i_6i_7i_8i_9i_{10}i_{11}i_{12}i_{13}i_{14}i_{15}i_{16}} P^{(2)}{}_{i_{17}i_{18}} P^{(2)}{}_{i_{19}i_{20}]} + \\
&+ 172{,}800\, P^{(14)}{}_{[i_1i_2i_3i_4i_5i_6i_7i_8i_9i_{10}i_{11}i_{12}i_{13}i_{14}} P^{(6)}{}_{i_{15}i_{16}i_{17}i_{18}i_{19}i_{20}]} - 259{,}200\, P^{(14)}{}_{[i_1i_2i_3i_4i_5i_6i_7i_8i_9i_{10}i_{11}i_{12}i_{13}i_{14}} P^{(4)}{}_{i_{15}i_{16}i_{17}i_{18}} P^{(2)}{}_{i_{19}i_{20}]} + \\
&+ 86{,}400\, P^{(14)}{}_{[i_1i_2i_3i_4i_5i_6i_7i_8i_9i_{10}i_{11}i_{12}i_{13}i_{14}} P^{(2)}{}_{i_{15}i_{16}} P^{(2)}{}_{i_{17}i_{18}} P^{(2)}{}_{i_{19}i_{20}]} + 151{,}200\, P^{(12)}{}_{[i_1i_2i_3i_4i_5i_6i_7i_8i_9i_{10}i_{11}i_{12}} P^{(8)}{}_{i_{13}i_{14}i_{15}i_{16}i_{17}i_{18}i_{19}i_{20}]} - \\
&- 201{,}600\, P^{(12)}{}_{[i_1i_2i_3i_4i_5i_6i_7i_8i_9i_{10}i_{11}i_{12}} P^{(6)}{}_{i_{13}i_{14}i_{15}i_{16}i_{17}i_{18}} P^{(2)}{}_{i_{19}i_{20}]} - 75{,}600\, P^{(12)}{}_{[i_1i_2i_3i_4i_5i_6i_7i_8i_9i_{10}i_{11}i_{12}} P^{(4)}{}_{i_{13}i_{14}i_{15}i_{16}} P^{(4)}{}_{i_{17}i_{18}i_{19}i_{20}]} + \\
&+ 151{,}200\, P^{(12)}{}_{[i_1i_2i_3i_4i_5i_6i_7i_8i_9i_{10}i_{11}i_{12}} P^{(4)}{}_{i_{13}i_{14}i_{15}i_{16}} P^{(2)}{}_{i_{17}i_{18}} P^{(2)}{}_{i_{19}i_{20}]} - 25{,}200\, P^{(12)}{}_{[i_1i_2i_3i_4i_5i_6i_7i_8i_9i_{10}i_{11}i_{12}} P^{(2)}{}_{i_{13}i_{14}} P^{(2)}{}_{i_{15}i_{16}} P^{(2)}{}_{i_{17}i_{18}} P^{(2)}{}_{i_{19}i_{20}]} + \\
&+ 72{,}576\, P^{(10)}{}_{[i_1i_2i_3i_4i_5i_6i_7i_8i_9i_{10}} P^{(10)}{}_{i_{11}i_{12}i_{13}i_{14}i_{15}i_{16}i_{17}i_{18}i_{19}i_{20}]} - 181{,}440\, P^{(10)}{}_{[i_1i_2i_3i_4i_5i_6i_7i_8i_9i_{10}} P^{(8)}{}_{i_{11}i_{12}i_{13}i_{14}i_{15}i_{16}i_{17}i_{18}} P^{(2)}{}_{i_{19}i_{20}]} - \\
&- 120{,}960\, P^{(10)}{}_{[i_1i_2i_3i_4i_5i_6i_7i_8i_9i_{10}} P^{(6)}{}_{i_{11}i_{12}i_{13}i_{14}i_{15}i_{16}} P^{(4)}{}_{i_{17}i_{18}i_{19}i_{20}]} + 120{,}960\, P^{(10)}{}_{[i_1i_2i_3i_4i_5i_6i_7i_8i_9i_{10}} P^{(6)}{}_{i_{11}i_{12}i_{13}i_{14}i_{15}i_{16}} P^{(2)}{}_{i_{17}i_{18}} P^{(2)}{}_{i_{19}i_{20}]} + \\
&+ 90{,}720\, P^{(10)}{}_{[i_1i_2i_3i_4i_5i_6i_7i_8i_9i_{10}} P^{(4)}{}_{i_{11}i_{12}i_{13}i_{14}} P^{(4)}{}_{i_{15}i_{16}i_{17}i_{18}} P^{(2)}{}_{i_{19}i_{20}]} - 60{,}480\, P^{(10)}{}_{[i_1i_2i_3i_4i_5i_6i_7i_8i_9i_{10}} P^{(4)}{}_{i_{11}i_{12}i_{13}i_{14}} P^{(2)}{}_{i_{15}i_{16}} P^{(2)}{}_{i_{17}i_{18}} P^{(2)}{}_{i_{19}i_{20}]} + \\
&+ 6048\, P^{(10)}{}_{[i_1i_2i_3i_4i_5i_6i_7i_8i_9i_{10}} P^{(2)}{}_{i_{11}i_{12}} P^{(2)}{}_{i_{13}i_{14}} P^{(2)}{}_{i_{15}i_{16}} P^{(2)}{}_{i_{17}i_{18}} P^{(2)}{}_{i_{19}i_{20}]} - 56{,}700\, P^{(8)}{}_{[i_1i_2i_3i_4i_5i_6i_7i_8} P^{(8)}{}_{i_9i_{10}i_{11}i_{12}i_{13}i_{14}i_{15}i_{16}} P^{(4)}{}_{i_{17}i_{18}i_{19}i_{20}]} + \\
&+ 56{,}700\, P^{(8)}{}_{[i_1i_2i_3i_4i_5i_6i_7i_8} P^{(8)}{}_{i_9i_{10}i_{11}i_{12}i_{13}i_{14}i_{15}i_{16}} P^{(2)}{}_{i_{17}i_{18}} P^{(2)}{}_{i_{19}i_{20}]} - 50{,}400\, P^{(8)}{}_{[i_1i_2i_3i_4i_5i_6i_7i_8} P^{(6)}{}_{i_9i_{10}i_{11}i_{12}i_{13}i_{14}} P^{(6)}{}_{i_{15}i_{16}i_{17}i_{18}i_{19}i_{20}]} + \\
&+ 151{,}200\, P^{(8)}{}_{[i_1i_2i_3i_4i_5i_6i_7i_8} P^{(6)}{}_{i_9i_{10}i_{11}i_{12}i_{13}i_{14}} P^{(4)}{}_{i_{15}i_{16}i_{17}i_{18}} P^{(2)}{}_{i_{19}i_{20}]} - 50{,}400\, P^{(8)}{}_{[i_1i_2i_3i_4i_5i_6i_7i_8} P^{(6)}{}_{i_9i_{10}i_{11}i_{12}i_{13}i_{14}} P^{(2)}{}_{i_{15}i_{16}} P^{(2)}{}_{i_{17}i_{18}} P^{(2)}{}_{i_{19}i_{20}]} + \\
&+ 18{,}900\, P^{(8)}{}_{[i_1i_2i_3i_4i_5i_6i_7i_8} P^{(4)}{}_{i_9i_{10}i_{11}i_{12}} P^{(4)}{}_{i_{13}i_{14}i_{15}i_{16}} P^{(4)}{}_{i_{17}i_{18}i_{19}i_{20}]} - 56{,}700\, P^{(8)}{}_{[i_1i_2i_3i_4i_5i_6i_7i_8} P^{(4)}{}_{i_9i_{10}i_{11}i_{12}} P^{(4)}{}_{i_{13}i_{14}i_{15}i_{16}} P^{(2)}{}_{i_{17}i_{18}} P^{(2)}{}_{i_{19}i_{20}]} + \\
&+ 18{,}900\, P^{(8)}{}_{[i_1i_2i_3i_4i_5i_6i_7i_8} P^{(4)}{}_{i_9i_{10}i_{11}i_{12}} P^{(2)}{}_{i_{13}i_{14}} P^{(2)}{}_{i_{15}i_{16}} P^{(2)}{}_{i_{17}i_{18}} P^{(2)}{}_{i_{19}i_{20}]} - \\
&- 1260\, P^{(8)}{}_{[i_1i_2i_3i_4i_5i_6i_7i_8} P^{(2)}{}_{i_9i_{10}} P^{(2)}{}_{i_{11}i_{12}} P^{(2)}{}_{i_{13}i_{14}} P^{(2)}{}_{i_{15}i_{16}} P^{(2)}{}_{i_{17}i_{18}} P^{(2)}{}_{i_{19}i_{20}]} + \\
&+ 22{,}400\, P^{(6)}{}_{[i_1i_2i_3i_4i_5i_6} P^{(6)}{}_{i_7i_8i_9i_{10}i_{11}i_{12}} P^{(6)}{}_{i_{13}i_{14}i_{15}i_{16}i_{17}i_{18}} P^{(2)}{}_{i_{19}i_{20}]} + \\
&+ 25{,}200\, P^{(6)}{}_{[i_1i_2i_3i_4i_5i_6} P^{(6)}{}_{i_7i_8i_9i_{10}i_{11}i_{12}} P^{(4)}{}_{i_{13}i_{14}i_{15}i_{16}} P^{(4)}{}_{i_{17}i_{18}i_{19}i_{20}]} - \\
&- 50{,}400\, P^{(6)}{}_{[i_1i_2i_3i_4i_5i_6} P^{(6)}{}_{i_7i_8i_9i_{10}i_{11}i_{12}} P^{(4)}{}_{i_{13}i_{14}i_{15}i_{16}} P^{(2)}{}_{i_{17}i_{18}} P^{(2)}{}_{i_{19}i_{20}]} + \\
&+ 8400\, P^{(6)}{}_{[i_1i_2i_3i_4i_5i_6} P^{(6)}{}_{i_7i_8i_9i_{10}i_{11}i_{12}} P^{(2)}{}_{i_{13}i_{14}} P^{(2)}{}_{i_{15}i_{16}} P^{(2)}{}_{i_{17}i_{18}} P^{(2)}{}_{i_{19}i_{20}]} - \\
&- 25{,}200\, P^{(6)}{}_{[i_1i_2i_3i_4i_5i_6} P^{(4)}{}_{i_7i_8i_9i_{10}} P^{(4)}{}_{i_{11}i_{12}i_{13}i_{14}} P^{(4)}{}_{i_{15}i_{16}i_{17}i_{18}} P^{(2)}{}_{i_{19}i_{20}]} + \\
&+ 25{,}200\, P^{(6)}{}_{[i_1i_2i_3i_4i_5i_6} P^{(4)}{}_{i_7i_8i_9i_{10}} P^{(4)}{}_{i_{11}i_{12}i_{13}i_{14}} P^{(2)}{}_{i_{15}i_{16}} P^{(2)}{}_{i_{17}i_{18}} P^{(2)}{}_{i_{19}i_{20}]} - \\
&- 5040\, P^{(6)}{}_{[i_1i_2i_3i_4i_5i_6} P^{(4)}{}_{i_7i_8i_9i_{10}} P^{(2)}{}_{i_{11}i_{12}} P^{(2)}{}_{i_{13}i_{14}} P^{(2)}{}_{i_{15}i_{16}} P^{(2)}{}_{i_{17}i_{18}} P^{(2)}{}_{i_{19}i_{20}]} + \\
&+ 240\, P^{(6)}{}_{[i_1i_2i_3i_4i_5i_6} P^{(2)}{}_{i_7i_8} P^{(2)}{}_{i_9i_{10}} P^{(2)}{}_{i_{11}i_{12}} P^{(2)}{}_{i_{13}i_{14}} P^{(2)}{}_{i_{15}i_{16}} P^{(2)}{}_{i_{17}i_{18}} P^{(2)}{}_{i_{19}i_{20}]} - \\
&- 945\, P^{(4)}{}_{[i_1i_2i_3i_4} P^{(4)}{}_{i_5i_6i_7i_8} P^{(4)}{}_{i_9i_{10}i_{11}i_{12}} P^{(4)}{}_{i_{13}i_{14}i_{15}i_{16}} P^{(4)}{}_{i_{17}i_{18}i_{19}i_{20}]} + \\
&+ 4725\, P^{(4)}{}_{[i_1i_2i_3i_4} P^{(4)}{}_{i_5i_6i_7i_8} P^{(4)}{}_{i_9i_{10}i_{11}i_{12}} P^{(4)}{}_{i_{13}i_{14}i_{15}i_{16}} P^{(2)}{}_{i_{17}i_{18}} P^{(2)}{}_{i_{19}i_{20}]} - \\
&- 3150\, P^{(4)}{}_{[i_1i_2i_3i_4} P^{(4)}{}_{i_5i_6i_7i_8} P^{(4)}{}_{i_9i_{10}i_{11}i_{12}} P^{(2)}{}_{i_{13}i_{14}} P^{(2)}{}_{i_{15}i_{16}} P^{(2)}{}_{i_{17}i_{18}} P^{(2)}{}_{i_{19}i_{20}]} +
\end{aligned}$$

$$+ 630\, P^{(4)}{}_{[i_1i_2i_3i_4}\, P^{(4)}{}_{i_5i_6i_7i_8}\, P^{(2)}{}_{i_9i_{10}}\, P^{(2)}{}_{i_{11}i_{12}}\, P^{(2)}{}_{i_{13}i_{14}}\, P^{(2)}{}_{i_{15}i_{16}}\, P^{(2)}{}_{i_{17}i_{18}}\, P^{(2)}{}_{i_{19}i_{20}]} -$$

$$- 45\, P^{(4)}{}_{[i_1i_2i_3i_4}\, P^{(2)}{}_{i_5i_6}\, P^{(2)}{}_{i_7i_8}\, P^{(2)}{}_{i_9i_{10}}\, P^{(2)}{}_{i_{11}i_{12}}\, P^{(2)}{}_{i_{13}i_{14}}\, P^{(2)}{}_{i_{15}i_{16}}\, P^{(2)}{}_{i_{17}i_{18}}\, P^{(2)}{}_{i_{19}i_{20}]} +$$

$$+ P^{(2)}{}_{[i_1i_2}\, P^{(2)}{}_{i_3i_4}\, P^{(2)}{}_{i_5i_6}\, P^{(2)}{}_{i_7i_8}\, P^{(2)}{}_{i_9i_{10}}\, P^{(2)}{}_{i_{11}i_{12}}\, P^{(2)}{}_{i_{13}i_{14}}\, P^{(2)}{}_{i_{15}i_{16}}\, P^{(2)}{}_{i_{17}i_{18}}\, P^{(2)}{}_{i_{19}i_{20}]})$$

### COEFFICIENT OF THE 11$^{th}$ CHERN FORM

$$c_{(11)\,i_1i_2i_3i_4i_5i_6i_7i_8i_9i_{10}i_{11}i_{12}i_{13}i_{14}i_{15}i_{16}i_{17}i_{18}i_{19}i_{20}i_{21}i_{22}} = \tag{20}$$

$$= \frac{1}{22!} \langle \mathbf{e}_{i_1} \wedge \mathbf{e}_{i_2} \wedge \mathbf{e}_{i_3} \wedge \mathbf{e}_{i_4} \wedge \mathbf{e}_{i_5} \wedge \mathbf{e}_{i_6} \wedge \mathbf{e}_{i_7} \wedge \mathbf{e}_{i_8} \wedge \mathbf{e}_{i_9} \wedge \mathbf{e}_{i_{10}} \wedge \mathbf{e}_{i_{11}} \wedge \mathbf{e}_{i_{12}} \wedge \mathbf{e}_{i_{13}} \wedge \mathbf{e}_{i_{14}} \wedge \mathbf{e}_{i_{15}} \wedge \mathbf{e}_{i_{16}} \wedge \mathbf{e}_{i_{17}} \wedge \mathbf{e}_{i_{18}} \wedge \mathbf{e}_{i_{19}} \wedge \mathbf{e}_{i_{20}} \wedge \mathbf{e}_{i_{21}} \wedge \mathbf{e}_{i_{22}},\, c_{(11)} \rangle$$

$$= \frac{i^{11}}{2^{22}\pi^{11} 11!} (+\, 3{,}628{,}800\, R_{[i_1i_2|k|}{}^{a}\, R_{i_3i_4|a|}{}^{b}\, R_{i_5i_6|b|}{}^{c}\, R_{i_7i_8|c|}{}^{d}\, R_{i_9i_{10}|d|}{}^{e}\, R_{i_{11}i_{12}|e|}{}^{f}\, R_{i_{13}i_{14}|f|}{}^{g}\, R_{i_{15}i_{16}|g|}{}^{h}\, R_{i_{17}i_{18}|h|}{}^{i}\, R_{i_{19}i_{20}|i|}{}^{j}\, R_{i_{21}i_{22}]j}{}^{k} -$$

$$-\, 3{,}991{,}680\, R_{[i_1i_2|j|}{}^{a}\, R_{i_3i_4|a|}{}^{b}\, R_{i_5i_6|b|}{}^{c}\, R_{i_7i_8|c|}{}^{d}\, R_{i_9i_{10}|d|}{}^{e}\, R_{i_{11}i_{12}|e|}{}^{f}\, R_{i_{13}i_{14}|f|}{}^{g}\, R_{i_{15}i_{16}|g|}{}^{h}\, R_{i_{17}i_{18}|h|}{}^{i}\, R_{i_{19}i_{20}|i|}{}^{j}\, R_{i_{21}i_{22}]k}{}^{k} -$$

$$-\, 2{,}217{,}600\, R_{[i_1i_2|i|}{}^{a}\, R_{i_3i_4|a|}{}^{b}\, R_{i_5i_6|b|}{}^{c}\, R_{i_7i_8|c|}{}^{d}\, R_{i_9i_{10}|d|}{}^{e}\, R_{i_{11}i_{12}|e|}{}^{f}\, R_{i_{13}i_{14}|f|}{}^{g}\, R_{i_{15}i_{16}|g|}{}^{h}\, R_{i_{17}i_{18}|h|}{}^{i}\, R_{i_{19}i_{20}|k|}{}^{j}\, R_{i_{21}i_{22}]j}{}^{k} +$$

$$+\, 2{,}217{,}600\, R_{[i_1i_2|i|}{}^{a}\, R_{i_3i_4|a|}{}^{b}\, R_{i_5i_6|b|}{}^{c}\, R_{i_7i_8|c|}{}^{d}\, R_{i_9i_{10}|d|}{}^{e}\, R_{i_{11}i_{12}|e|}{}^{f}\, R_{i_{13}i_{14}|f|}{}^{g}\, R_{i_{15}i_{16}|g|}{}^{h}\, R_{i_{17}i_{18}|h|}{}^{i}\, R_{i_{19}i_{20}|j|}{}^{j}\, R_{i_{21}i_{22}]k}{}^{k} -$$

$$-\, 1{,}663{,}200\, R_{[i_1i_2|h|}{}^{a}\, R_{i_3i_4|a|}{}^{b}\, R_{i_5i_6|b|}{}^{c}\, R_{i_7i_8|c|}{}^{d}\, R_{i_9i_{10}|d|}{}^{e}\, R_{i_{11}i_{12}|e|}{}^{f}\, R_{i_{13}i_{14}|f|}{}^{g}\, R_{i_{15}i_{16}|g|}{}^{h}\, R_{i_{17}i_{18}|h|}{}^{i}\, R_{i_{19}i_{20}|i|}{}^{j}\, R_{i_{21}i_{22}]k}{}^{k} +$$

$$+\, 2{,}494{,}800\, R_{[i_1i_2|h|}{}^{a}\, R_{i_3i_4|a|}{}^{b}\, R_{i_5i_6|b|}{}^{c}\, R_{i_7i_8|c|}{}^{d}\, R_{i_9i_{10}|d|}{}^{e}\, R_{i_{11}i_{12}|e|}{}^{f}\, R_{i_{13}i_{14}|f|}{}^{g}\, R_{i_{15}i_{16}|g|}{}^{h}\, R_{i_{17}i_{18}|j|}{}^{i}\, R_{i_{19}i_{20}|i|}{}^{j}\, R_{i_{21}i_{22}]k}{}^{k} -$$

$$-\, 831{,}600\, R_{[i_1i_2|h|}{}^{a}\, R_{i_3i_4|a|}{}^{b}\, R_{i_5i_6|b|}{}^{c}\, R_{i_7i_8|c|}{}^{d}\, R_{i_9i_{10}|d|}{}^{e}\, R_{i_{11}i_{12}|e|}{}^{f}\, R_{i_{13}i_{14}|f|}{}^{g}\, R_{i_{15}i_{16}|g|}{}^{h}\, R_{i_{17}i_{18}|j|}{}^{i}\, R_{i_{19}i_{20}|i|}{}^{j}\, R_{i_{21}i_{22}]k}{}^{k} -$$

$$-\, 1{,}425{,}600\, R_{[i_1i_2|g|}{}^{a}\, R_{i_3i_4|a|}{}^{b}\, R_{i_5i_6|b|}{}^{c}\, R_{i_7i_8|c|}{}^{d}\, R_{i_9i_{10}|d|}{}^{e}\, R_{i_{11}i_{12}|e|}{}^{f}\, R_{i_{13}i_{14}|f|}{}^{g}\, R_{i_{15}i_{16}|k|}{}^{h}\, R_{i_{17}i_{18}|h|}{}^{i}\, R_{i_{19}i_{20}|i|}{}^{j}\, R_{i_{21}i_{22}]j}{}^{k} +$$

$$+\, 1{,}900{,}800\, R_{[i_1i_2|g|}{}^{a}\, R_{i_3i_4|a|}{}^{b}\, R_{i_5i_6|b|}{}^{c}\, R_{i_7i_8|c|}{}^{d}\, R_{i_9i_{10}|d|}{}^{e}\, R_{i_{11}i_{12}|e|}{}^{f}\, R_{i_{13}i_{14}|f|}{}^{g}\, R_{i_{15}i_{16}|j|}{}^{h}\, R_{i_{17}i_{18}|h|}{}^{i}\, R_{i_{19}i_{20}|i|}{}^{j}\, R_{i_{21}i_{22}]k}{}^{k} +$$

$$+\, 712{,}800\, R_{[i_1i_2|g|}{}^{a}\, R_{i_3i_4|a|}{}^{b}\, R_{i_5i_6|b|}{}^{c}\, R_{i_7i_8|c|}{}^{d}\, R_{i_9i_{10}|d|}{}^{e}\, R_{i_{11}i_{12}|e|}{}^{f}\, R_{i_{13}i_{14}|f|}{}^{g}\, R_{i_{15}i_{16}|i|}{}^{h}\, R_{i_{17}i_{18}|h|}{}^{i}\, R_{i_{19}i_{20}|k|}{}^{j}\, R_{i_{21}i_{22}]j}{}^{k} -$$

$$-\, 1{,}425{,}600\, R_{[i_1i_2|g|}{}^{a}\, R_{i_3i_4|a|}{}^{b}\, R_{i_5i_6|b|}{}^{c}\, R_{i_7i_8|c|}{}^{d}\, R_{i_9i_{10}|d|}{}^{e}\, R_{i_{11}i_{12}|e|}{}^{f}\, R_{i_{13}i_{14}|f|}{}^{g}\, R_{i_{15}i_{16}|i|}{}^{h}\, R_{i_{17}i_{18}|h|}{}^{i}\, R_{i_{19}i_{20}|j|}{}^{j}\, R_{i_{21}i_{22}]k}{}^{k} +$$

$$+\, 237{,}600\, R_{[i_1i_2|g|}{}^{a}\, R_{i_3i_4|a|}{}^{b}\, R_{i_5i_6|b|}{}^{c}\, R_{i_7i_8|c|}{}^{d}\, R_{i_9i_{10}|d|}{}^{e}\, R_{i_{11}i_{12}|e|}{}^{f}\, R_{i_{13}i_{14}|f|}{}^{g}\, R_{i_{15}i_{16}|h|}{}^{h}\, R_{i_{17}i_{18}|i|}{}^{i}\, R_{i_{19}i_{20}|j|}{}^{j}\, R_{i_{21}i_{22}]k}{}^{k} -$$

$$-\, 1{,}330{,}560\, R_{[i_1i_2|f|}{}^{a}\, R_{i_3i_4|a|}{}^{b}\, R_{i_5i_6|b|}{}^{c}\, R_{i_7i_8|c|}{}^{d}\, R_{i_9i_{10}|d|}{}^{e}\, R_{i_{11}i_{12}|e|}{}^{f}\, R_{i_{13}i_{14}|k|}{}^{g}\, R_{i_{15}i_{16}|g|}{}^{h}\, R_{i_{17}i_{18}|h|}{}^{i}\, R_{i_{19}i_{20}|i|}{}^{j}\, R_{i_{21}i_{22}]j}{}^{k} +$$

$$+\, 1{,}663{,}200\, R_{[i_1i_2|f|}{}^{a}\, R_{i_3i_4|a|}{}^{b}\, R_{i_5i_6|b|}{}^{c}\, R_{i_7i_8|c|}{}^{d}\, R_{i_9i_{10}|d|}{}^{e}\, R_{i_{11}i_{12}|e|}{}^{f}\, R_{i_{13}i_{14}|j|}{}^{g}\, R_{i_{15}i_{16}|g|}{}^{h}\, R_{i_{17}i_{18}|h|}{}^{i}\, R_{i_{19}i_{20}|i|}{}^{j}\, R_{i_{21}i_{22}]k}{}^{k} +$$

$$+\, 1{,}108{,}800\, R_{[i_1i_2|f|}{}^{a}\, R_{i_3i_4|a|}{}^{b}\, R_{i_5i_6|b|}{}^{c}\, R_{i_7i_8|c|}{}^{d}\, R_{i_9i_{10}|d|}{}^{e}\, R_{i_{11}i_{12}|e|}{}^{f}\, R_{i_{13}i_{14}|i|}{}^{g}\, R_{i_{15}i_{16}|g|}{}^{h}\, R_{i_{17}i_{18}|h|}{}^{i}\, R_{i_{19}i_{20}|k|}{}^{j}\, R_{i_{21}i_{22}]j}{}^{k} -$$

$$-\, 1{,}108{,}800\, R_{[i_1i_2|f|}{}^{a}\, R_{i_3i_4|a|}{}^{b}\, R_{i_5i_6|b|}{}^{c}\, R_{i_7i_8|c|}{}^{d}\, R_{i_9i_{10}|d|}{}^{e}\, R_{i_{11}i_{12}|e|}{}^{f}\, R_{i_{13}i_{14}|i|}{}^{g}\, R_{i_{15}i_{16}|g|}{}^{h}\, R_{i_{17}i_{18}|h|}{}^{i}\, R_{i_{19}i_{20}|j|}{}^{j}\, R_{i_{21}i_{22}]k}{}^{k} -$$

$$-\, 831{,}600\, R_{[i_1i_2|f|}{}^{a}\, R_{i_3i_4|a|}{}^{b}\, R_{i_5i_6|b|}{}^{c}\, R_{i_7i_8|c|}{}^{d}\, R_{i_9i_{10}|d|}{}^{e}\, R_{i_{11}i_{12}|e|}{}^{f}\, R_{i_{13}i_{14}|h|}{}^{g}\, R_{i_{15}i_{16}|g|}{}^{h}\, R_{i_{17}i_{18}|j|}{}^{i}\, R_{i_{19}i_{20}|i|}{}^{j}\, R_{i_{21}i_{22}]k}{}^{k} +$$

$$+\, 554{,}400\, R_{[i_1i_2|f|}{}^{a}\, R_{i_3i_4|a|}{}^{b}\, R_{i_5i_6|b|}{}^{c}\, R_{i_7i_8|c|}{}^{d}\, R_{i_9i_{10}|d|}{}^{e}\, R_{i_{11}i_{12}|e|}{}^{f}\, R_{i_{13}i_{14}|h|}{}^{g}\, R_{i_{15}i_{16}|g|}{}^{h}\, R_{i_{17}i_{18}|i|}{}^{i}\, R_{i_{19}i_{20}|j|}{}^{j}\, R_{i_{21}i_{22}]k}{}^{k} -$$

$$-\, 55{,}440\, R_{[i_1i_2|f|}{}^{a}\, R_{i_3i_4|a|}{}^{b}\, R_{i_5i_6|b|}{}^{c}\, R_{i_7i_8|c|}{}^{d}\, R_{i_9i_{10}|d|}{}^{e}\, R_{i_{11}i_{12}|e|}{}^{f}\, R_{i_{13}i_{14}|g|}{}^{g}\, R_{i_{15}i_{16}|h|}{}^{h}\, R_{i_{17}i_{18}|i|}{}^{i}\, R_{i_{19}i_{20}|j|}{}^{j}\, R_{i_{21}i_{22}]k}{}^{k} +$$

$$+\, 798{,}336\, R_{[i_1i_2|e|}{}^{a}\, R_{i_3i_4|a|}{}^{b}\, R_{i_5i_6|b|}{}^{c}\, R_{i_7i_8|c|}{}^{d}\, R_{i_9i_{10}|d|}{}^{e}\, R_{i_{11}i_{12}|j|}{}^{f}\, R_{i_{13}i_{14}|f|}{}^{g}\, R_{i_{15}i_{16}|g|}{}^{h}\, R_{i_{17}i_{18}|h|}{}^{i}\, R_{i_{19}i_{20}|i|}{}^{j}\, R_{i_{21}i_{22}]k}{}^{k} +$$

$$+\, 997{,}920\, R_{[i_1i_2|e|}{}^{a}\, R_{i_3i_4|a|}{}^{b}\, R_{i_5i_6|b|}{}^{c}\, R_{i_7i_8|c|}{}^{d}\, R_{i_9i_{10}|d|}{}^{e}\, R_{i_{11}i_{12}|i|}{}^{f}\, R_{i_{13}i_{14}|f|}{}^{g}\, R_{i_{15}i_{16}|g|}{}^{h}\, R_{i_{17}i_{18}|h|}{}^{i}\, R_{i_{19}i_{20}|k|}{}^{j}\, R_{i_{21}i_{22}]j}{}^{k} -$$

$$-\, 997{,}920\, R_{[i_1i_2|e|}{}^{a}\, R_{i_3i_4|a|}{}^{b}\, R_{i_5i_6|b|}{}^{c}\, R_{i_7i_8|c|}{}^{d}\, R_{i_9i_{10}|d|}{}^{e}\, R_{i_{11}i_{12}|i|}{}^{f}\, R_{i_{13}i_{14}|f|}{}^{g}\, R_{i_{15}i_{16}|g|}{}^{h}\, R_{i_{17}i_{18}|h|}{}^{i}\, R_{i_{19}i_{20}|j|}{}^{j}\, R_{i_{21}i_{22}]k}{}^{k} +$$

$$+\, 443{,}520\, R_{[i_1i_2|e|}{}^{a}\, R_{i_3i_4|a|}{}^{b}\, R_{i_5i_6|b|}{}^{c}\, R_{i_7i_8|c|}{}^{d}\, R_{i_9i_{10}|d|}{}^{e}\, R_{i_{11}i_{12}|h|}{}^{f}\, R_{i_{13}i_{14}|f|}{}^{g}\, R_{i_{15}i_{16}|g|}{}^{h}\, R_{i_{17}i_{18}|k|}{}^{i}\, R_{i_{19}i_{20}|i|}{}^{j}\, R_{i_{21}i_{22}]j}{}^{k} -$$

$$-\, 1{,}330{,}560\, R_{[i_1i_2|e|}{}^{a}\, R_{i_3i_4|a|}{}^{b}\, R_{i_5i_6|b|}{}^{c}\, R_{i_7i_8|c|}{}^{d}\, R_{i_9i_{10}|d|}{}^{e}\, R_{i_{11}i_{12}|h|}{}^{f}\, R_{i_{13}i_{14}|f|}{}^{g}\, R_{i_{15}i_{16}|g|}{}^{h}\, R_{i_{17}i_{18}|j|}{}^{i}\, R_{i_{19}i_{20}|i|}{}^{j}\, R_{i_{21}i_{22}]k}{}^{k} +$$

$$+\, 443{,}520\, R_{[i_1i_2|e|}{}^{a}\, R_{i_3i_4|a|}{}^{b}\, R_{i_5i_6|b|}{}^{c}\, R_{i_7i_8|c|}{}^{d}\, R_{i_9i_{10}|d|}{}^{e}\, R_{i_{11}i_{12}|h|}{}^{f}\, R_{i_{13}i_{14}|f|}{}^{g}\, R_{i_{15}i_{16}|g|}{}^{h}\, R_{i_{17}i_{18}|i|}{}^{i}\, R_{i_{19}i_{20}|j|}{}^{j}\, R_{i_{21}i_{22}]k}{}^{k} -$$

$$-\, 166{,}320\, R_{[i_1i_2|e|}{}^{a}\, R_{i_3i_4|a|}{}^{b}\, R_{i_5i_6|b|}{}^{c}\, R_{i_7i_8|c|}{}^{d}\, R_{i_9i_{10}|d|}{}^{e}\, R_{i_{11}i_{12}|g|}{}^{f}\, R_{i_{13}i_{14}|f|}{}^{g}\, R_{i_{15}i_{16}|h|}{}^{h}\, R_{i_{17}i_{18}|i|}{}^{i}\, R_{i_{19}i_{20}|k|}{}^{j}\, R_{i_{21}i_{22}]j}{}^{k} +$$

$$+\, 498{,}960\, R_{[i_1i_2|e|}{}^{a}\, R_{i_3i_4|a|}{}^{b}\, R_{i_5i_6|b|}{}^{c}\, R_{i_7i_8|c|}{}^{d}\, R_{i_9i_{10}|d|}{}^{e}\, R_{i_{11}i_{12}|g|}{}^{f}\, R_{i_{13}i_{14}|f|}{}^{g}\, R_{i_{15}i_{16}|i|}{}^{h}\, R_{i_{17}i_{18}|h|}{}^{i}\, R_{i_{19}i_{20}|j|}{}^{j}\, R_{i_{21}i_{22}]k}{}^{k} -$$

$$-\, 166{,}320\, R_{[i_1i_2|e|}{}^{a}\, R_{i_3i_4|a|}{}^{b}\, R_{i_5i_6|b|}{}^{c}\, R_{i_7i_8|c|}{}^{d}\, R_{i_9i_{10}|d|}{}^{e}\, R_{i_{11}i_{12}|g|}{}^{f}\, R_{i_{13}i_{14}|f|}{}^{g}\, R_{i_{15}i_{16}|h|}{}^{h}\, R_{i_{17}i_{18}|i|}{}^{i}\, R_{i_{19}i_{20}|j|}{}^{j}\, R_{i_{21}i_{22}]k}{}^{k} +$$

$$+\, 11{,}088\, R_{[i_1i_2|e|}{}^{a}\, R_{i_3i_4|a|}{}^{b}\, R_{i_5i_6|b|}{}^{c}\, R_{i_7i_8|c|}{}^{d}\, R_{i_9i_{10}|d|}{}^{e}\, R_{i_{11}i_{12}|f|}{}^{f}\, R_{i_{13}i_{14}|g|}{}^{g}\, R_{i_{15}i_{16}|h|}{}^{h}\, R_{i_{17}i_{18}|i|}{}^{i}\, R_{i_{19}i_{20}|j|}{}^{j}\, R_{i_{21}i_{22}]k}{}^{k} +$$

$$+\, 415{,}800\, R_{[i_1i_2|d|}{}^{a}\, R_{i_3i_4|a|}{}^{b}\, R_{i_5i_6|b|}{}^{c}\, R_{i_7i_8|c|}{}^{d}\, R_{i_9i_{10}|h|}{}^{e}\, R_{i_{11}i_{12}|e|}{}^{f}\, R_{i_{13}i_{14}|f|}{}^{g}\, R_{i_{15}i_{16}|g|}{}^{h}\, R_{i_{17}i_{18}|k|}{}^{i}\, R_{i_{19}i_{20}|i|}{}^{j}\, R_{i_{21}i_{22}]j}{}^{k} -$$

$$-\, 623{,}700\, R_{[i_1i_2|d|}{}^{a}\, R_{i_3i_4|a|}{}^{b}\, R_{i_5i_6|b|}{}^{c}\, R_{i_7i_8|c|}{}^{d}\, R_{i_9i_{10}|h|}{}^{e}\, R_{i_{11}i_{12}|e|}{}^{f}\, R_{i_{13}i_{14}|f|}{}^{g}\, R_{i_{15}i_{16}|g|}{}^{h}\, R_{i_{17}i_{18}|j|}{}^{i}\, R_{i_{19}i_{20}|i|}{}^{j}\, R_{i_{21}i_{22}]k}{}^{k} +$$

$$+\, 207{,}900\, R_{[i_1i_2|d|}{}^{a}\, R_{i_3i_4|a|}{}^{b}\, R_{i_5i_6|b|}{}^{c}\, R_{i_7i_8|c|}{}^{d}\, R_{i_9i_{10}|h|}{}^{e}\, R_{i_{11}i_{12}|e|}{}^{f}\, R_{i_{13}i_{14}|f|}{}^{g}\, R_{i_{15}i_{16}|g|}{}^{h}\, R_{i_{17}i_{18}|i|}{}^{i}\, R_{i_{19}i_{20}|j|}{}^{j}\, R_{i_{21}i_{22}]k}{}^{k} -$$

$$-\, 554{,}400\, R_{[i_1i_2|d|}{}^{a}\, R_{i_3i_4|a|}{}^{b}\, R_{i_5i_6|b|}{}^{c}\, R_{i_7i_8|c|}{}^{d}\, R_{i_9i_{10}|g|}{}^{e}\, R_{i_{11}i_{12}|e|}{}^{f}\, R_{i_{13}i_{14}|f|}{}^{g}\, R_{i_{15}i_{16}|h|}{}^{h}\, R_{i_{17}i_{18}|h|}{}^{i}\, R_{i_{19}i_{20}|i|}{}^{j}\, R_{i_{21}i_{22}]k}{}^{k} -$$

$$-\, 415{,}800\, R_{[i_1i_2|d|}{}^{a}\, R_{i_3i_4|a|}{}^{b}\, R_{i_5i_6|b|}{}^{c}\, R_{i_7i_8|c|}{}^{d}\, R_{i_9i_{10}|g|}{}^{e}\, R_{i_{11}i_{12}|e|}{}^{f}\, R_{i_{13}i_{14}|f|}{}^{g}\, R_{i_{15}i_{16}|i|}{}^{h}\, R_{i_{17}i_{18}|h|}{}^{i}\, R_{i_{19}i_{20}|k|}{}^{j}\, R_{i_{21}i_{22}]j}{}^{k} +$$

$$+\, 831{,}600\, R_{[i_1i_2|d|}{}^{a}\, R_{i_3i_4|a|}{}^{b}\, R_{i_5i_6|b|}{}^{c}\, R_{i_7i_8|c|}{}^{d}\, R_{i_9i_{10}|g|}{}^{e}\, R_{i_{11}i_{12}|e|}{}^{f}\, R_{i_{13}i_{14}|f|}{}^{g}\, R_{i_{15}i_{16}|i|}{}^{h}\, R_{i_{17}i_{18}|h|}{}^{i}\, R_{i_{19}i_{20}|j|}{}^{j}\, R_{i_{21}i_{22}]k}{}^{k} -$$

$$\begin{aligned}
&- 138{,}600\, R_{[i_1i_2|d|}{}^a R_{i_3i_4|a|}{}^b R_{i_5i_6|b|}{}^c R_{i_7i_8|c|}{}^d R_{i_9i_{10}|g|}{}^e R_{i_{11}i_{12}|e|}{}^f R_{i_{13}i_{14}|f|}{}^g R_{i_{15}i_{16}|h|}{}^h R_{i_{17}i_{18}|i|}{}^i R_{i_{19}i_{20}|j|}{}^j R_{i_{21}i_{22}]}{}^k \\
&+ 207{,}900\, R_{[i_1i_2|d|}{}^a R_{i_3i_4|a|}{}^b R_{i_5i_6|b|}{}^c R_{i_7i_8|c|}{}^d R_{i_9i_{10}|f|}{}^e R_{i_{11}i_{12}|e|}{}^f R_{i_{13}i_{14}|h|}{}^g R_{i_{15}i_{16}|g|}{}^h R_{i_{17}i_{18}|j|}{}^i R_{i_{19}i_{20}|i|}{}^j R_{i_{21}i_{22}]}{}^k \\
&- 207{,}900\, R_{[i_1i_2|d|}{}^a R_{i_3i_4|a|}{}^b R_{i_5i_6|b|}{}^c R_{i_7i_8|c|}{}^d R_{i_9i_{10}|e|}{}^e R_{i_{11}i_{12}|e|}{}^f R_{i_{13}i_{14}|h|}{}^g R_{i_{15}i_{16}|g|}{}^h R_{i_{17}i_{18}|i|}{}^i R_{i_{19}i_{20}|j|}{}^j R_{i_{21}i_{22}]}{}^k \\
&+ 41{,}580\, R_{[i_1i_2|d|}{}^a R_{i_3i_4|a|}{}^b R_{i_5i_6|b|}{}^c R_{i_7i_8|c|}{}^d R_{i_9i_{10}|e|}{}^e R_{i_{11}i_{12}|e|}{}^f R_{i_{13}i_{14}|g|}{}^g R_{i_{15}i_{16}|h|}{}^h R_{i_{17}i_{18}|i|}{}^i R_{i_{19}i_{20}|j|}{}^j R_{i_{21}i_{22}]}{}^k \\
&- 1980\, R_{[i_1i_2|d|}{}^a R_{i_3i_4|a|}{}^b R_{i_5i_6|b|}{}^c R_{i_7i_8|c|}{}^d R_{i_9i_{10}|e|}{}^e R_{i_{11}i_{12}|f|}{}^f R_{i_{13}i_{14}|g|}{}^g R_{i_{15}i_{16}|h|}{}^h R_{i_{17}i_{18}|i|}{}^i R_{i_{19}i_{20}|j|}{}^j R_{i_{21}i_{22}]}{}^k \\
&- 123{,}200\, R_{[i_1i_2|c|}{}^a R_{i_3i_4|a|}{}^b R_{i_5i_6|b|}{}^c R_{i_7i_8|f|}{}^d R_{i_9i_{10}|d|}{}^e R_{i_{11}i_{12}|e|}{}^f R_{i_{13}i_{14}|i|}{}^g R_{i_{15}i_{16}|g|}{}^h R_{i_{17}i_{18}|h|}{}^i R_{i_{19}i_{20}|k|}{}^j R_{i_{21}i_{22}]}{}^k \\
&+ 123{,}200\, R_{[i_1i_2|c|}{}^a R_{i_3i_4|a|}{}^b R_{i_5i_6|b|}{}^c R_{i_7i_8|f|}{}^d R_{i_9i_{10}|d|}{}^e R_{i_{11}i_{12}|e|}{}^f R_{i_{13}i_{14}|g|}{}^g R_{i_{15}i_{16}|h|}{}^h R_{i_{17}i_{18}|h|}{}^i R_{i_{19}i_{20}|j|}{}^j R_{i_{21}i_{22}]}{}^k \\
&+ 277{,}200\, R_{[i_1i_2|c|}{}^a R_{i_3i_4|a|}{}^b R_{i_5i_6|b|}{}^c R_{i_7i_8|f|}{}^d R_{i_9i_{10}|d|}{}^e R_{i_{11}i_{12}|e|}{}^f R_{i_{13}i_{14}|h|}{}^g R_{i_{15}i_{16}|g|}{}^h R_{i_{17}i_{18}|j|}{}^i R_{i_{19}i_{20}|i|}{}^j R_{i_{21}i_{22}]}{}^k \\
&- 184{,}800\, R_{[i_1i_2|c|}{}^a R_{i_3i_4|a|}{}^b R_{i_5i_6|b|}{}^c R_{i_7i_8|f|}{}^d R_{i_9i_{10}|d|}{}^e R_{i_{11}i_{12}|e|}{}^f R_{i_{13}i_{14}|h|}{}^g R_{i_{15}i_{16}|g|}{}^h R_{i_{17}i_{18}|i|}{}^i R_{i_{19}i_{20}|j|}{}^j R_{i_{21}i_{22}]}{}^k \\
&+ 18{,}480\, R_{[i_1i_2|c|}{}^a R_{i_3i_4|a|}{}^b R_{i_5i_6|b|}{}^c R_{i_7i_8|f|}{}^d R_{i_9i_{10}|d|}{}^e R_{i_{11}i_{12}|e|}{}^f R_{i_{13}i_{14}|g|}{}^g R_{i_{15}i_{16}|h|}{}^h R_{i_{17}i_{18}|i|}{}^i R_{i_{19}i_{20}|j|}{}^j R_{i_{21}i_{22}]}{}^k \\
&+ 34{,}650\, R_{[i_1i_2|c|}{}^a R_{i_3i_4|a|}{}^b R_{i_5i_6|b|}{}^c R_{i_7i_8|e|}{}^d R_{i_9i_{10}|d|}{}^e R_{i_{11}i_{12}|g|}{}^f R_{i_{13}i_{14}|f|}{}^g R_{i_{15}i_{16}|i|}{}^h R_{i_{17}i_{18}|h|}{}^i R_{i_{19}i_{20}|k|}{}^j R_{i_{21}i_{22}]j}{}^k \\
&- 138{,}600\, R_{[i_1i_2|c|}{}^a R_{i_3i_4|a|}{}^b R_{i_5i_6|b|}{}^c R_{i_7i_8|e|}{}^d R_{i_9i_{10}|d|}{}^e R_{i_{11}i_{12}|g|}{}^f R_{i_{13}i_{14}|f|}{}^g R_{i_{15}i_{16}|i|}{}^h R_{i_{17}i_{18}|h|}{}^i R_{i_{19}i_{20}|j|}{}^j R_{i_{21}i_{22}]}{}^k \\
&+ 69{,}300\, R_{[i_1i_2|c|}{}^a R_{i_3i_4|a|}{}^b R_{i_5i_6|b|}{}^c R_{i_7i_8|e|}{}^d R_{i_9i_{10}|d|}{}^e R_{i_{11}i_{12}|g|}{}^f R_{i_{13}i_{14}|f|}{}^g R_{i_{15}i_{16}|h|}{}^h R_{i_{17}i_{18}|i|}{}^i R_{i_{19}i_{20}|j|}{}^j R_{i_{21}i_{22}]}{}^k \\
&- 9240\, R_{[i_1i_2|c|}{}^a R_{i_3i_4|a|}{}^b R_{i_5i_6|b|}{}^c R_{i_7i_8|e|}{}^d R_{i_9i_{10}|d|}{}^e R_{i_{11}i_{12}|f|}{}^f R_{i_{13}i_{14}|g|}{}^g R_{i_{15}i_{16}|h|}{}^h R_{i_{17}i_{18}|i|}{}^i R_{i_{19}i_{20}|j|}{}^j R_{i_{21}i_{22}]}{}^k \\
&+ 330\, R_{[i_1i_2|c|}{}^a R_{i_3i_4|a|}{}^b R_{i_5i_6|b|}{}^c R_{i_7i_8|d|}{}^d R_{i_9i_{10}|e|}{}^e R_{i_{11}i_{12}|f|}{}^f R_{i_{13}i_{14}|g|}{}^g R_{i_{15}i_{16}|h|}{}^h R_{i_{17}i_{18}|i|}{}^i R_{i_{19}i_{20}|j|}{}^j R_{i_{21}i_{22}]}{}^k \\
&- 10{,}395\, R_{[i_1i_2|b|}{}^a R_{i_3i_4|a|}{}^b R_{i_5i_6|d|}{}^c R_{i_7i_8|c|}{}^d R_{i_9i_{10}|e|}{}^e R_{i_{11}i_{12}|f|}{}^f R_{i_{13}i_{14}|h|}{}^g R_{i_{15}i_{16}|g|}{}^h R_{i_{17}i_{18}|j|}{}^i R_{i_{19}i_{20}|i|}{}^j R_{i_{21}i_{22}]}{}^k \\
&+ 17{,}325\, R_{[i_1i_2|b|}{}^a R_{i_3i_4|a|}{}^b R_{i_5i_6|d|}{}^c R_{i_7i_8|c|}{}^d R_{i_9i_{10}|e|}{}^e R_{i_{11}i_{12}|e|}{}^f R_{i_{13}i_{14}|h|}{}^g R_{i_{15}i_{16}|g|}{}^h R_{i_{17}i_{18}|i|}{}^i R_{i_{19}i_{20}|j|}{}^j R_{i_{21}i_{22}]}{}^k \\
&- 6930\, R_{[i_1i_2|b|}{}^a R_{i_3i_4|a|}{}^b R_{i_5i_6|d|}{}^c R_{i_7i_8|c|}{}^d R_{i_9i_{10}|f|}{}^e R_{i_{11}i_{12}|e|}{}^f R_{i_{13}i_{14}|g|}{}^g R_{i_{15}i_{16}|h|}{}^h R_{i_{17}i_{18}|i|}{}^i R_{i_{19}i_{20}|j|}{}^j R_{i_{21}i_{22}]}{}^k \\
&+ 990\, R_{[i_1i_2|b|}{}^a R_{i_3i_4|a|}{}^b R_{i_5i_6|d|}{}^c R_{i_7i_8|c|}{}^d R_{i_9i_{10}|e|}{}^e R_{i_{11}i_{12}|f|}{}^f R_{i_{13}i_{14}|g|}{}^g R_{i_{15}i_{16}|h|}{}^h R_{i_{17}i_{18}|i|}{}^i R_{i_{19}i_{20}|j|}{}^j R_{i_{21}i_{22}]}{}^k \\
&- 55\, R_{[i_1i_2|b|}{}^a R_{i_3i_4|a|}{}^b R_{i_5i_6|c|}{}^c R_{i_7i_8|d|}{}^d R_{i_9i_{10}|e|}{}^e R_{i_{11}i_{12}|f|}{}^f R_{i_{13}i_{14}|g|}{}^g R_{i_{15}i_{16}|h|}{}^h R_{i_{17}i_{18}|i|}{}^i R_{i_{19}i_{20}|j|}{}^j R_{i_{21}i_{22}]}{}^k \\
&+ R_{[i_1i_2|a|}{}^a R_{i_3i_4|b|}{}^b R_{i_5i_6|c|}{}^c R_{i_7i_8|d|}{}^d R_{i_9i_{10}|e|}{}^e R_{i_{11}i_{12}|f|}{}^f R_{i_{13}i_{14}|g|}{}^g R_{i_{15}i_{16}|h|}{}^h R_{i_{17}i_{18}|i|}{}^i R_{i_{19}i_{20}|j|}{}^j R_{i_{21}i_{22}]}{}^k \big)
\end{aligned}$$

$$= \frac{i^{11}}{2^{22}\pi^{11}11!}\Big(+ 3{,}628{,}800\, R_{[i_1i_2|k|}{}^a R_{i_3i_4|a|}{}^b R_{i_5i_6|b|}{}^c R_{i_7i_8|c|}{}^d R_{i_9i_{10}|d|}{}^e R_{i_{11}i_{12}|e|}{}^f R_{i_{13}i_{14}|f|}{}^g R_{i_{15}i_{16}|g|}{}^h R_{i_{17}i_{18}|h|}{}^i R_{i_{19}i_{20}|i|}{}^j R_{i_{21}i_{22}]j}{}$$

$$- 3{,}991{,}680\, R_{[i_1i_2|j|}{}^a R_{i_3i_4|a|}{}^b R_{i_5i_6|b|}{}^c R_{i_7i_8|c|}{}^d R_{i_9i_{10}|d|}{}^e R_{i_{11}i_{12}|e|}{}^f R_{i_{13}i_{14}|f|}{}^g R_{i_{15}i_{16}|g|}{}^h R_{i_{17}i_{18}|h|}{}^i R_{i_{19}i_{20}|i|}{}^j V_{i_{21}i_{22}]}$$

$$- 2{,}217{,}600\, R_{[i_1i_2|i|}{}^a R_{i_3i_4|a|}{}^b R_{i_5i_6|b|}{}^c R_{i_7i_8|c|}{}^d R_{i_9i_{10}|d|}{}^e R_{i_{11}i_{12}|e|}{}^f R_{i_{13}i_{14}|f|}{}^g R_{i_{15}i_{16}|g|}{}^h R_{i_{17}i_{18}|h|}{}^i R_{i_{19}i_{20}|k|}{}^j R_{i_{21}i_{22}]j}{}^k$$

$$+ 2{,}217{,}600\, R_{[i_1i_2|i|}{}^a R_{i_3i_4|a|}{}^b R_{i_5i_6|b|}{}^c R_{i_7i_8|c|}{}^d R_{i_9i_{10}|d|}{}^e R_{i_{11}i_{12}|e|}{}^f R_{i_{13}i_{14}|f|}{}^g R_{i_{15}i_{16}|g|}{}^h R_{i_{17}i_{18}|h|}{}^i V_{i_{19}i_{20}} V_{i_{21}i_{22}]}$$

$$- 1{,}663{,}200\, R_{[i_1i_2|h|}{}^a R_{i_3i_4|a|}{}^b R_{i_5i_6|b|}{}^c R_{i_7i_8|c|}{}^d R_{i_9i_{10}|d|}{}^e R_{i_{11}i_{12}|e|}{}^f R_{i_{13}i_{14}|f|}{}^g R_{i_{15}i_{16}|g|}{}^h R_{i_{17}i_{18}|h|}{}^i R_{i_{19}i_{20}|i|}{}^j R_{i_{21}i_{22}]j}{}^k$$

$$+ 2{,}494{,}800\, R_{[i_1i_2|h|}{}^a R_{i_3i_4|a|}{}^b R_{i_5i_6|b|}{}^c R_{i_7i_8|c|}{}^d R_{i_9i_{10}|d|}{}^e R_{i_{11}i_{12}|e|}{}^f R_{i_{13}i_{14}|f|}{}^g R_{i_{15}i_{16}|g|}{}^h R_{i_{17}i_{18}|j|}{}^i R_{i_{19}i_{20}|i|}{}^j V_{i_{21}i_{22}]}$$

$$- 831{,}600\, R_{[i_1i_2|h|}{}^a R_{i_3i_4|a|}{}^b R_{i_5i_6|b|}{}^c R_{i_7i_8|c|}{}^d R_{i_9i_{10}|d|}{}^e R_{i_{11}i_{12}|e|}{}^f R_{i_{13}i_{14}|f|}{}^g R_{i_{15}i_{16}|g|}{}^h V_{i_{17}i_{18}} V_{i_{19}i_{20}} V_{i_{21}i_{22}]}$$

$$- 1{,}425{,}600\, R_{[i_1i_2|g|}{}^a R_{i_3i_4|a|}{}^b R_{i_5i_6|b|}{}^c R_{i_7i_8|c|}{}^d R_{i_9i_{10}|d|}{}^e R_{i_{11}i_{12}|e|}{}^f R_{i_{13}i_{14}|f|}{}^g R_{i_{15}i_{16}|k|}{}^h R_{i_{17}i_{18}|h|}{}^i R_{i_{19}i_{20}|i|}{}^j R_{i_{21}i_{22}]j}{}$$

$$+ 1{,}900{,}800\, R_{[i_1i_2|g|}{}^a R_{i_3i_4|a|}{}^b R_{i_5i_6|b|}{}^c R_{i_7i_8|c|}{}^d R_{i_9i_{10}|d|}{}^e R_{i_{11}i_{12}|e|}{}^f R_{i_{13}i_{14}|f|}{}^g R_{i_{15}i_{16}|j|}{}^h R_{i_{17}i_{18}|h|}{}^i R_{i_{19}i_{20}|i|}{}^j V_{i_{21}i_{22}]}$$

$$+ 712{,}800\, R_{[i_1i_2|g|}{}^a R_{i_3i_4|a|}{}^b R_{i_5i_6|b|}{}^c R_{i_7i_8|c|}{}^d R_{i_9i_{10}|d|}{}^e R_{i_{11}i_{12}|e|}{}^f R_{i_{13}i_{14}|f|}{}^g R_{i_{15}i_{16}}{}^h R_{i_{17}i_{18}|i|}{}^i R_{i_{19}i_{20}|k|}{}^j R_{i_{21}i_{22}]j}{}$$

$$- 1{,}425{,}600\, R_{[i_1i_2|g|}{}^a R_{i_3i_4|a|}{}^b R_{i_5i_6|b|}{}^c R_{i_7i_8|c|}{}^d R_{i_9i_{10}|d|}{}^e R_{i_{11}i_{12}|e|}{}^f R_{i_{13}i_{14}|f|}{}^g R_{i_{15}i_{16}|i|}{}^h R_{i_{17}i_{18}|h|}{}^i V_{i_{19}i_{20}} V_{i_{21}i_{22}]}$$

$$+ 237{,}600\, R_{[i_1i_2|g|}{}^a R_{i_3i_4|a|}{}^b R_{i_5i_6|b|}{}^c R_{i_7i_8|c|}{}^d R_{i_9i_{10}|d|}{}^e R_{i_{11}i_{12}|e|}{}^f R_{i_{13}i_{14}}{}^g V_{i_{15}i_{16}} V_{i_{17}i_{18}} V_{i_{19}i_{20}} V_{i_{21}i_{22}]}$$

$$- 1{,}330{,}560\, R_{[i_1i_2|f|}{}^a R_{i_3i_4|a|}{}^b R_{i_5i_6|b|}{}^c R_{i_7i_8|c|}{}^d R_{i_9i_{10}|d|}{}^e R_{i_{11}i_{12}|e|}{}^f R_{i_{13}i_{14}|k|}{}^g R_{i_{15}i_{16}|g|}{}^h R_{i_{17}i_{18}|h|}{}^i R_{i_{19}i_{20}|i|}{}^j R_{i_{21}i_{22}]j}{}$$

$$+ 1{,}663{,}200\, R_{[i_1i_2|f|}{}^a R_{i_3i_4|a|}{}^b R_{i_5i_6|b|}{}^c R_{i_7i_8|c|}{}^d R_{i_9i_{10}|d|}{}^e R_{i_{11}i_{12}|e|}{}^f R_{i_{13}i_{14}|j|}{}^g R_{i_{15}i_{16}|g|}{}^h R_{i_{17}i_{18}|h|}{}^i R_{i_{19}i_{20}|i|}{}^j V_{i_{21}i_{22}]}$$

$$+ 1{,}108{,}800\, R_{[i_1i_2|f|}{}^a R_{i_3i_4|a|}{}^b R_{i_5i_6|b|}{}^c R_{i_7i_8|c|}{}^d R_{i_9i_{10}|d|}{}^e R_{i_{11}i_{12}|e|}{}^f R_{i_{13}i_{14}|i|}{}^g R_{i_{15}i_{16}|g|}{}^h R_{i_{17}i_{18}|h|}{}^i R_{i_{19}i_{20}|k|}{}^j R_{i_{21}i_{22}]j}{}$$

$$- 1{,}108{,}800\, R_{[i_1i_2|f|}{}^a R_{i_3i_4|a|}{}^b R_{i_5i_6|b|}{}^c R_{i_7i_8|c|}{}^d R_{i_9i_{10}|d|}{}^e R_{i_{11}i_{12}|e|}{}^f R_{i_{13}i_{14}|i|}{}^g R_{i_{15}i_{16}|g|}{}^h R_{i_{17}i_{18}|h|}{}^i V_{i_{19}i_{20}} V_{i_{21}i_{22}]}$$

$$- 831{,}600\, R_{[i_1i_2|f|}{}^a R_{i_3i_4|a|}{}^b R_{i_5i_6|b|}{}^c R_{i_7i_8|c|}{}^d R_{i_9i_{10}|d|}{}^e R_{i_{11}i_{12}|e|}{}^f R_{i_{13}i_{14}|h|}{}^g R_{i_{15}i_{16}|g|}{}^h R_{i_{17}i_{18}|i|}{}^i R_{i_{19}i_{20}|j|}{}^j V_{i_{21}i_{22}]}$$

$$+ 554{,}400\, R_{[i_1i_2|f|}{}^a R_{i_3i_4|a|}{}^b R_{i_5i_6|b|}{}^c R_{i_7i_8|c|}{}^d R_{i_9i_{10}|d|}{}^e R_{i_{11}i_{12}|e|}{}^f R_{i_{13}i_{14}|h|}{}^g R_{i_{15}i_{16}}{}^h V_{i_{17}i_{18}} V_{i_{19}i_{20}} V_{i_{21}i_{22}]}$$

$$- 55{,}440\, R_{[i_1i_2|f|}{}^a R_{i_3i_4|a|}{}^b R_{i_5i_6|b|}{}^c R_{i_7i_8|c|}{}^d R_{i_9i_{10}|d|}{}^e R_{i_{11}i_{12}|e|}{}^f V_{i_{13}i_{14}} V_{i_{15}i_{16}} V_{i_{17}i_{18}} V_{i_{19}i_{20}} V_{i_{21}i_{22}]}$$

$$+ 798{,}336\, R_{[i_1i_2|e|}{}^a R_{i_3i_4|a|}{}^b R_{i_5i_6|b|}{}^c R_{i_7i_8|c|}{}^d R_{i_9i_{10}|d|}{}^e R_{i_{11}i_{12}|j|}{}^f R_{i_{13}i_{14}|f|}{}^g R_{i_{15}i_{16}|g|}{}^h R_{i_{17}i_{18}|h|}{}^i R_{i_{19}i_{20}|i|}{}^j V_{i_{21}i_{22}]}$$

$$+ 997{,}920\, R_{[i_1 i_2|e|}{}^a\, R_{i_3 i_4|a|}{}^b\, R_{i_5 i_6|b|}{}^c\, R_{i_7 i_8|c|}{}^d\, R_{i_9 i_{10}|d|}{}^e\, R_{i_{11} i_{12}|i|}{}^f\, R_{i_{13} i_{14}|f|}{}^g\, R_{i_{15} i_{16}|g|}{}^h\, R_{i_{17} i_{18}|h|}{}^i\, R_{i_{19} i_{20}|k|}{}^j\, R_{i_{21} i_{22}]j}{}^k -$$

$$- 997{,}920\, R_{[i_1 i_2|e|}{}^a\, R_{i_3 i_4|a|}{}^b\, R_{i_5 i_6|b|}{}^c\, R_{i_7 i_8|c|}{}^d\, R_{i_9 i_{10}|d|}{}^e\, R_{i_{11} i_{12}|i|}{}^f\, R_{i_{13} i_{14}|f|}{}^g\, R_{i_{15} i_{16}|g|}{}^h\, R_{i_{17} i_{18}|h|}{}^i\, V_{i_{19} i_{20}}\, V_{i_{21} i_{22}]} +$$

$$+ 443{,}520\, R_{[i_1 i_2|e|}{}^a\, R_{i_3 i_4|a|}{}^b\, R_{i_5 i_6|b|}{}^c\, R_{i_7 i_8|c|}{}^d\, R_{i_9 i_{10}|d|}{}^e\, R_{i_{11} i_{12}|h|}{}^f\, R_{i_{13} i_{14}|f|}{}^g\, R_{i_{15} i_{16}|g|}{}^h\, R_{i_{17} i_{18}|k|}{}^i\, R_{i_{19} i_{20}|i|}{}^j\, R_{i_{21} i_{22}]j}{}^k -$$

$$- 1{,}330{,}560\, R_{[i_1 i_2|e|}{}^a\, R_{i_3 i_4|a|}{}^b\, R_{i_5 i_6|b|}{}^c\, R_{i_7 i_8|c|}{}^d\, R_{i_9 i_{10}|d|}{}^e\, R_{i_{11} i_{12}|h|}{}^f\, R_{i_{13} i_{14}|f|}{}^g\, R_{i_{15} i_{16}|g|}{}^h\, R_{i_{17} i_{18}|i|}{}^i\, R_{i_{19} i_{20}|i|}{}^j\, V_{i_{21} i_{22}]} +$$

$$+ 443{,}520\, R_{[i_1 i_2|e|}{}^a\, R_{i_3 i_4|a|}{}^b\, R_{i_5 i_6|b|}{}^c\, R_{i_7 i_8|c|}{}^d\, R_{i_9 i_{10}|d|}{}^e\, R_{i_{11} i_{12}|h|}{}^f\, R_{i_{13} i_{14}|f|}{}^g\, R_{i_{15} i_{16}|g|}{}^h\, V_{i_{17} i_{18}}\, V_{i_{19} i_{20}}\, V_{i_{21} i_{22}]} -$$

$$- 166{,}320\, R_{[i_1 i_2|e|}{}^a\, R_{i_3 i_4|a|}{}^b\, R_{i_5 i_6|b|}{}^c\, R_{i_7 i_8|c|}{}^d\, R_{i_9 i_{10}|d|}{}^e\, R_{i_{11} i_{12}|g|}{}^f\, R_{i_{13} i_{14}|f|}{}^g\, R_{i_{15} i_{16}|i|}{}^h\, R_{i_{17} i_{18}|h|}{}^i\, R_{i_{19} i_{20}|k|}{}^j\, R_{i_{21} i_{22}]j}{}^k +$$

$$+ 498{,}960\, R_{[i_1 i_2|e|}{}^a\, R_{i_3 i_4|a|}{}^b\, R_{i_5 i_6|b|}{}^c\, R_{i_7 i_8|c|}{}^d\, R_{i_9 i_{10}|d|}{}^e\, R_{i_{11} i_{12}|g|}{}^f\, R_{i_{13} i_{14}|f|}{}^g\, R_{i_{15} i_{16}|i|}{}^h\, R_{i_{17} i_{18}|h|}{}^i\, V_{i_{19} i_{20}}\, V_{i_{21} i_{22}]} -$$

$$- 166{,}320\, R_{[i_1 i_2|e|}{}^a\, R_{i_3 i_4|a|}{}^b\, R_{i_5 i_6|b|}{}^c\, R_{i_7 i_8|c|}{}^d\, R_{i_9 i_{10}|d|}{}^e\, R_{i_{11} i_{12}|g|}{}^f\, R_{i_{13} i_{14}|f|}{}^g\, V_{i_{15} i_{16}}\, V_{i_{17} i_{18}}\, V_{i_{19} i_{20}}\, V_{i_{21} i_{22}]} +$$

$$+ 11{,}088\, R_{[i_1 i_2|e|}{}^a\, R_{i_3 i_4|a|}{}^b\, R_{i_5 i_6|b|}{}^c\, R_{i_7 i_8|c|}{}^d\, R_{i_9 i_{10}|d|}{}^e\, V_{i_{11} i_{12}}\, V_{i_{13} i_{14}}\, V_{i_{15} i_{16}}\, V_{i_{17} i_{18}}\, V_{i_{19} i_{20}}\, V_{i_{21} i_{22}]} +$$

$$+ 415{,}800\, R_{[i_1 i_2|d|}{}^a\, R_{i_3 i_4|a|}{}^b\, R_{i_5 i_6|b|}{}^c\, R_{i_7 i_8|c|}{}^d\, R_{i_9 i_{10}|h|}{}^e\, R_{i_{11} i_{12}|e|}{}^f\, R_{i_{13} i_{14}|f|}{}^g\, R_{i_{15} i_{16}|g|}{}^h\, R_{i_{17} i_{18}|k|}{}^i\, R_{i_{19} i_{20}|i|}{}^j\, R_{i_{21} i_{22}]j}{}^k -$$

$$- 623{,}700\, R_{[i_1 i_2|d|}{}^a\, R_{i_3 i_4|a|}{}^b\, R_{i_5 i_6|b|}{}^c\, R_{i_7 i_8|c|}{}^d\, R_{i_9 i_{10}|h|}{}^e\, R_{i_{11} i_{12}|e|}{}^f\, R_{i_{13} i_{14}|f|}{}^g\, R_{i_{15} i_{16}|g|}{}^h\, R_{i_{17} i_{18}|j|}{}^i\, R_{i_{19} i_{20}|i|}{}^j\, V_{i_{21} i_{22}]} +$$

$$+ 207{,}900\, R_{[i_1 i_2|d|}{}^a\, R_{i_3 i_4|a|}{}^b\, R_{i_5 i_6|b|}{}^c\, R_{i_7 i_8|c|}{}^d\, R_{i_9 i_{10}|h|}{}^e\, R_{i_{11} i_{12}|e|}{}^f\, R_{i_{13} i_{14}|f|}{}^g\, R_{i_{15} i_{16}|g|}{}^h\, V_{i_{17} i_{18}}\, V_{i_{19} i_{20}}\, V_{i_{21} i_{22}]} -$$

$$- 554{,}400\, R_{[i_1 i_2|d|}{}^a\, R_{i_3 i_4|a|}{}^b\, R_{i_5 i_6|b|}{}^c\, R_{i_7 i_8|c|}{}^d\, R_{i_9 i_{10}|g|}{}^e\, R_{i_{11} i_{12}|e|}{}^f\, R_{i_{13} i_{14}|f|}{}^g\, R_{i_{15} i_{16}|j|}{}^h\, R_{i_{17} i_{18}|h|}{}^i\, R_{i_{19} i_{20}|i|}{}^j\, V_{i_{21} i_{22}]} -$$

$$- 415{,}800\, R_{[i_1 i_2|d|}{}^a\, R_{i_3 i_4|a|}{}^b\, R_{i_5 i_6|b|}{}^c\, R_{i_7 i_8|c|}{}^d\, R_{i_9 i_{10}|g|}{}^e\, R_{i_{11} i_{12}|e|}{}^f\, R_{i_{13} i_{14}|f|}{}^g\, R_{i_{15} i_{16}|i|}{}^h\, R_{i_{17} i_{18}|h|}{}^i\, R_{i_{19} i_{20}|k|}{}^j\, R_{i_{21} i_{22}]j}{}^k +$$

$$+ 831{,}600\, R_{[i_1 i_2|d|}{}^a\, R_{i_3 i_4|a|}{}^b\, R_{i_5 i_6|b|}{}^c\, R_{i_7 i_8|c|}{}^d\, R_{i_9 i_{10}|g|}{}^e\, R_{i_{11} i_{12}|e|}{}^f\, R_{i_{13} i_{14}|f|}{}^g\, R_{i_{15} i_{16}|i|}{}^h\, R_{i_{17} i_{18}|h|}{}^i\, V_{i_{19} i_{20}}\, V_{i_{21} i_{22}]} -$$

$$- 138{,}600\, R_{[i_1 i_2|d|}{}^a\, R_{i_3 i_4|a|}{}^b\, R_{i_5 i_6|b|}{}^c\, R_{i_7 i_8|c|}{}^d\, R_{i_9 i_{10}|g|}{}^e\, R_{i_{11} i_{12}|e|}{}^f\, R_{i_{13} i_{14}|f|}{}^g\, V_{i_{15} i_{16}}\, V_{i_{17} i_{18}}\, V_{i_{19} i_{20}}\, V_{i_{21} i_{22}]} +$$

$$+ 207{,}900\, R_{[i_1 i_2|d|}{}^a\, R_{i_3 i_4|a|}{}^b\, R_{i_5 i_6|b|}{}^c\, R_{i_7 i_8|c|}{}^d\, R_{i_9 i_{10}|f|}{}^e\, R_{i_{11} i_{12}|e|}{}^f\, R_{i_{13} i_{14}|h|}{}^g\, R_{i_{15} i_{16}|g|}{}^h\, R_{i_{17} i_{18}|j|}{}^i\, R_{i_{19} i_{20}|i|}{}^j\, V_{i_{21} i_{22}]} -$$

$$- 207{,}900\, R_{[i_1 i_2|d|}{}^a\, R_{i_3 i_4|a|}{}^b\, R_{i_5 i_6|b|}{}^c\, R_{i_7 i_8|c|}{}^d\, R_{i_9 i_{10}|f|}{}^e\, R_{i_{11} i_{12}|e|}{}^f\, R_{i_{13} i_{14}|h|}{}^g\, R_{i_{15} i_{16}|g|}{}^h\, V_{i_{17} i_{18}}\, V_{i_{19} i_{20}}\, V_{i_{21} i_{22}]} +$$

$$+ 41{,}580\, R_{[i_1 i_2|d|}{}^a\, R_{i_3 i_4|a|}{}^b\, R_{i_5 i_6|b|}{}^c\, R_{i_7 i_8|c|}{}^d\, R_{i_9 i_{10}|f|}{}^e\, R_{i_{11} i_{12}|e|}{}^f\, V_{i_{13} i_{14}}\, V_{i_{15} i_{16}}\, V_{i_{17} i_{18}}\, V_{i_{19} i_{20}}\, V_{i_{21} i_{22}]} -$$

$$- 1980\, R_{[i_1 i_2|d|}{}^a\, R_{i_3 i_4|a|}{}^b\, R_{i_5 i_6|b|}{}^c\, R_{i_7 i_8|c|}{}^d\, V_{i_9 i_{10}}\, V_{i_{11} i_{12}}\, V_{i_{13} i_{14}}\, V_{i_{15} i_{16}}\, V_{i_{17} i_{18}}\, V_{i_{19} i_{20}}\, V_{i_{21} i_{22}]} -$$

$$- 123{,}200\, R_{[i_1 i_2|c|}{}^a\, R_{i_3 i_4|a|}{}^b\, R_{i_5 i_6|b|}{}^c\, R_{i_7 i_8|f|}{}^d\, R_{i_9 i_{10}|d|}{}^e\, R_{i_{11} i_{12}|e|}{}^f\, R_{i_{13} i_{14}|i|}{}^g\, R_{i_{15} i_{16}|g|}{}^h\, R_{i_{17} i_{18}|h|}{}^i\, R_{i_{19} i_{20}|k|}{}^j\, R_{i_{21} i_{22}]j}{}^k +$$

$$+ 123{,}200\, R_{[i_1 i_2|c|}{}^a\, R_{i_3 i_4|a|}{}^b\, R_{i_5 i_6|b|}{}^c\, R_{i_7 i_8|f|}{}^d\, R_{i_9 i_{10}|d|}{}^e\, R_{i_{11} i_{12}|e|}{}^f\, R_{i_{13} i_{14}|i|}{}^g\, R_{i_{15} i_{16}|g|}{}^h\, R_{i_{17} i_{18}|h|}{}^i\, V_{i_{19} i_{20}}\, V_{i_{21} i_{22}]} +$$

$$+ 277{,}200\, R_{[i_1 i_2|c|}{}^a\, R_{i_3 i_4|a|}{}^b\, R_{i_5 i_6|b|}{}^c\, R_{i_7 i_8|f|}{}^d\, R_{i_9 i_{10}|d|}{}^e\, R_{i_{11} i_{12}|e|}{}^f\, R_{i_{13} i_{14}|h|}{}^g\, R_{i_{15} i_{16}|g|}{}^h\, R_{i_{17} i_{18}|j|}{}^i\, R_{i_{19} i_{20}|i|}{}^j\, V_{i_{21} i_{22}]} -$$

$$- 184{,}800\, R_{[i_1 i_2|c|}{}^a\, R_{i_3 i_4|a|}{}^b\, R_{i_5 i_6|b|}{}^c\, R_{i_7 i_8|f|}{}^d\, R_{i_9 i_{10}|d|}{}^e\, R_{i_{11} i_{12}|e|}{}^f\, R_{i_{13} i_{14}|h|}{}^g\, R_{i_{15} i_{16}|g|}{}^h\, V_{i_{17} i_{18}}\, V_{i_{19} i_{20}}\, V_{i_{21} i_{22}]} +$$

$$+ 18{,}480\, R_{[i_1 i_2|c|}{}^a\, R_{i_3 i_4|a|}{}^b\, R_{i_5 i_6|b|}{}^c\, R_{i_7 i_8|f|}{}^d\, R_{i_9 i_{10}|d|}{}^e\, R_{i_{11} i_{12}|e|}{}^f\, V_{i_{13} i_{14}}\, V_{i_{15} i_{16}}\, V_{i_{17} i_{18}}\, V_{i_{19} i_{20}}\, V_{i_{21} i_{22}]} +$$

$$+ 34{,}650\, R_{[i_1 i_2|c|}{}^a\, R_{i_3 i_4|a|}{}^b\, R_{i_5 i_6|b|}{}^c\, R_{i_7 i_8|e|}{}^d\, R_{i_9 i_{10}|d|}{}^e\, R_{i_{11} i_{12}|g|}{}^f\, R_{i_{13} i_{14}|f|}{}^g\, R_{i_{15} i_{16}|i|}{}^h\, R_{i_{17} i_{18}|h|}{}^i\, R_{i_{19} i_{20}|k|}{}^j\, R_{i_{21} i_{22}]j}{}^k -$$

$$- 138{,}600\, R_{[i_1 i_2|c|}{}^a\, R_{i_3 i_4|a|}{}^b\, R_{i_5 i_6|b|}{}^c\, R_{i_7 i_8|e|}{}^d\, R_{i_9 i_{10}|d|}{}^e\, R_{i_{11} i_{12}|g|}{}^f\, R_{i_{13} i_{14}|f|}{}^g\, R_{i_{15} i_{16}|i|}{}^h\, R_{i_{17} i_{18}|h|}{}^i\, V_{i_{19} i_{20}}\, V_{i_{21} i_{22}]} +$$

$$+ 69{,}300\, R_{[i_1 i_2|c|}{}^a\, R_{i_3 i_4|a|}{}^b\, R_{i_5 i_6|b|}{}^c\, R_{i_7 i_8|e|}{}^d\, R_{i_9 i_{10}|d|}{}^e\, R_{i_{11} i_{12}|g|}{}^f\, R_{i_{13} i_{14}|f|}{}^g\, V_{i_{15} i_{16}}\, V_{i_{17} i_{18}}\, V_{i_{19} i_{20}}\, V_{i_{21} i_{22}]} -$$

$$- 9240\, R_{[i_1 i_2|c|}{}^a\, R_{i_3 i_4|a|}{}^b\, R_{i_5 i_6|b|}{}^c\, R_{i_7 i_8|e|}{}^d\, R_{i_9 i_{10}|d|}{}^e\, V_{i_{11} i_{12}}\, V_{i_{13} i_{14}}\, V_{i_{15} i_{16}}\, V_{i_{17} i_{18}}\, V_{i_{19} i_{20}}\, V_{i_{21} i_{22}]} +$$

$$+ 330\, R_{[i_1 i_2|c|}{}^a\, R_{i_3 i_4|a|}{}^b\, R_{i_5 i_6|b|}{}^c\, V_{i_7 i_8}\, V_{i_9 i_{10}}\, V_{i_{11} i_{12}}\, V_{i_{13} i_{14}}\, V_{i_{15} i_{16}}\, V_{i_{17} i_{18}}\, V_{i_{19} i_{20}}\, V_{i_{21} i_{22}]} -$$

$$- 10{,}395\, R_{[i_1 i_2|b|}{}^a\, R_{i_3 i_4|a|}{}^b\, R_{i_5 i_6|d|}{}^c\, R_{i_7 i_8|c|}{}^d\, R_{i_9 i_{10}|f|}{}^e\, R_{i_{11} i_{12}|e|}{}^f\, R_{i_{13} i_{14}|h|}{}^g\, R_{i_{15} i_{16}|g|}{}^h\, R_{i_{17} i_{18}|j|}{}^i\, R_{i_{19} i_{20}|i|}{}^j\, V_{i_{21} i_{22}]} +$$

$$+ 17{,}325\, R_{[i_1 i_2|b|}{}^a\, R_{i_3 i_4|a|}{}^b\, R_{i_5 i_6|d|}{}^c\, R_{i_7 i_8|c|}{}^d\, R_{i_9 i_{10}|f|}{}^e\, R_{i_{11} i_{12}|e|}{}^f\, R_{i_{13} i_{14}|h|}{}^g\, R_{i_{15} i_{16}|g|}{}^h\, V_{i_{17} i_{18}}\, V_{i_{19} i_{20}}\, V_{i_{21} i_{22}]} -$$

$$- 6930\, R_{[i_1 i_2|b|}{}^a\, R_{i_3 i_4|a|}{}^b\, R_{i_5 i_6|d|}{}^c\, R_{i_7 i_8|c|}{}^d\, R_{i_9 i_{10}|f|}{}^e\, R_{i_{11} i_{12}|e|}{}^f\, V_{i_{13} i_{14}}\, V_{i_{15} i_{16}}\, V_{i_{17} i_{18}}\, V_{i_{19} i_{20}}\, V_{i_{21} i_{22}]} +$$

$$+ 990\, R_{[i_1 i_2|b|}{}^a\, R_{i_3 i_4|a|}{}^b\, R_{i_5 i_6|d|}{}^c\, R_{i_7 i_8|c|}{}^d\, V_{i_9 i_{10}}\, V_{i_{11} i_{12}}\, V_{i_{13} i_{14}}\, V_{i_{15} i_{16}}\, V_{i_{17} i_{18}}\, V_{i_{19} i_{20}}\, V_{i_{21} i_{22}]} -$$

$$- 55\, R_{[i_1 i_2|b|}{}^a\, R_{i_3 i_4|a|}{}^b\, V_{i_5 i_6}\, V_{i_7 i_8}\, V_{i_9 i_{10}}\, V_{i_{11} i_{12}}\, V_{i_{13} i_{14}}\, V_{i_{15} i_{16}}\, V_{i_{17} i_{18}}\, V_{i_{19} i_{20}}\, V_{i_{21} i_{22}]} +$$

$$+ V_{[i_1 i_2}\, V_{i_3 i_4}\, V_{i_5 i_6}\, V_{i_7 i_8}\, V_{i_9 i_{10}}\, V_{i_{11} i_{12}}\, V_{i_{13} i_{14}}\, V_{i_{15} i_{16}}\, V_{i_{17} i_{18}}\, V_{i_{19} i_{20}}\, V_{i_{21} i_{22}]})$$

$$= \frac{i^{11}}{2^{11} \pi^{11} 11!}\, (+\, 3{,}628{,}800\, P^{(22)}{}_{i_1 i_2 i_3 i_4 i_5 i_6 i_7 i_8 i_9 i_{10} i_{11} i_{12} i_{13} i_{14} i_{15} i_{16} i_{17} i_{18} i_{19} i_{20} i_{21} i_{22}} -$$

$$- 3{,}991{,}680\, P^{(20)}{}_{[i_1 i_2 i_3 i_4 i_5 i_6 i_7 i_8 i_9 i_{10} i_{11} i_{12} i_{13} i_{14} i_{15} i_{16} i_{17} i_{18} i_{19} i_{20}}\, P^{(2)}{}_{i_{21} i_{22}]} -$$

$$- 2{,}217{,}600\, P^{(18)}{}_{[i_1 i_2 i_3 i_4 i_5 i_6 i_7 i_8 i_9 i_{10} i_{11} i_{12} i_{13} i_{14} i_{15} i_{16} i_{17} i_{18}}\, P^{(4)}{}_{i_{19} i_{20} i_{21} i_{22}]} +$$

$$+ 2{,}217{,}600\, P^{(18)}{}_{[i_1 i_2 i_3 i_4 i_5 i_6 i_7 i_8 i_9 i_{10} i_{11} i_{12} i_{13} i_{14} i_{15} i_{16} i_{17} i_{18}}\, P^{(2)}{}_{i_{19} i_{20}}\, P^{(2)}{}_{i_{21} i_{22}]} -$$

$$- 1{,}663{,}200\, P^{(16)}{}_{[i_1 i_2 i_3 i_4 i_5 i_6 i_7 i_8 i_9 i_{10} i_{11} i_{12} i_{13} i_{14} i_{15} i_{16}}\, P^{(6)}{}_{i_{17} i_{18} i_{19} i_{20} i_{21} i_{22}]} +$$

$$+ 2{,}494{,}800 \, P^{(16)}{}_{[i_1 i_2 i_3 i_4 i_5 i_6 i_7 i_8 i_9 i_{10} i_{11} i_{12} i_{13} i_{14} i_{15} i_{16}} P^{(4)}{}_{i_{17} i_{18} i_{19} i_{20}} P^{(2)}{}_{i_{21} i_{22}]} -$$

$$- 831{,}600 \, P^{(16)}{}_{[i_1 i_2 i_3 i_4 i_5 i_6 i_7 i_8 i_9 i_{10} i_{11} i_{12} i_{13} i_{14} i_{15} i_{16}} P^{(2)}{}_{i_{17} i_{18}} P^{(2)}{}_{i_{19} i_{20}} P^{(2)}{}_{i_{21} i_{22}]} -$$

$$- 1{,}425{,}600 \, P^{(14)}{}_{[i_1 i_2 i_3 i_4 i_5 i_6 i_7 i_8 i_9 i_{10} i_{11} i_{12} i_{13} i_{14}} P^{(8)}{}_{i_{15} i_{16} i_{17} i_{18} i_{19} i_{20} i_{21} i_{22}]} +$$

$$+ 1{,}900{,}800 \, P^{(14)}{}_{[i_1 i_2 i_3 i_4 i_5 i_6 i_7 i_8 i_9 i_{10} i_{11} i_{12} i_{13} i_{14}} P^{(6)}{}_{i_{15} i_{16} i_{17} i_{18} i_{19} i_{20}} P^{(2)}{}_{i_{21} i_{22}]} +$$

$$+ 712{,}800 \, P^{(14)}{}_{[i_1 i_2 i_3 i_4 i_5 i_6 i_7 i_8 i_9 i_{10} i_{11} i_{12} i_{13} i_{14}} P^{(4)}{}_{i_{15} i_{16} i_{17} i_{18}} P^{(4)}{}_{i_{19} i_{20} i_{21} i_{22}]} -$$

$$- 1{,}425{,}600 \, P^{(14)}{}_{[i_1 i_2 i_3 i_4 i_5 i_6 i_7 i_8 i_9 i_{10} i_{11} i_{12} i_{13} i_{14}} P^{(4)}{}_{i_{15} i_{16} i_{17} i_{18}} P^{(2)}{}_{i_{19} i_{20}} P^{(2)}{}_{i_{21} i_{22}]} +$$

$$+ 237{,}600 \, P^{(14)}{}_{[i_1 i_2 i_3 i_4 i_5 i_6 i_7 i_8 i_9 i_{10} i_{11} i_{12} i_{13} i_{14}} P^{(2)}{}_{i_{15} i_{16}} P^{(2)}{}_{i_{17} i_{18}} P^{(2)}{}_{i_{19} i_{20}} P^{(2)}{}_{i_{21} i_{22}]} -$$

$$- 1{,}330{,}560 \, P^{(12)}{}_{[i_1 i_2 i_3 i_4 i_5 i_6 i_7 i_8 i_9 i_{10} i_{11} i_{12}} P^{(10)}{}_{i_{13} i_{14} i_{15} i_{16} i_{17} i_{18} i_{19} i_{20} i_{21} i_{22}]} +$$

$$+ 1{,}663{,}200 \, P^{(12)}{}_{[i_1 i_2 i_3 i_4 i_5 i_6 i_7 i_8 i_9 i_{10} i_{11} i_{12}} P^{(8)}{}_{i_{13} i_{14} i_{15} i_{16} i_{17} i_{18} i_{19} i_{20}} P^{(2)}{}_{i_{21} i_{22}]} +$$

$$+ 1{,}108{,}800 \, P^{(12)}{}_{[i_1 i_2 i_3 i_4 i_5 i_6 i_7 i_8 i_9 i_{10} i_{11} i_{12}} P^{(6)}{}_{i_{13} i_{14} i_{15} i_{16} i_{17} i_{18}} P^{(4)}{}_{i_{19} i_{20} i_{21} i_{22}]} -$$

$$- 1{,}108{,}800 \, P^{(12)}{}_{[i_1 i_2 i_3 i_4 i_5 i_6 i_7 i_8 i_9 i_{10} i_{11} i_{12}} P^{(6)}{}_{i_{13} i_{14} i_{15} i_{16} i_{17} i_{18}} P^{(2)}{}_{i_{19} i_{20}} P^{(2)}{}_{i_{21} i_{22}]} -$$

$$- 831{,}600 \, P^{(12)}{}_{[i_1 i_2 i_3 i_4 i_5 i_6 i_7 i_8 i_9 i_{10} i_{11} i_{12}} P^{(4)}{}_{i_{13} i_{14} i_{15} i_{16}} P^{(4)}{}_{i_{17} i_{18} i_{19} i_{20}} P^{(2)}{}_{i_{21} i_{22}]} +$$

$$+ 554{,}400 \, P^{(12)}{}_{[i_1 i_2 i_3 i_4 i_5 i_6 i_7 i_8 i_9 i_{10} i_{11} i_{12}} P^{(4)}{}_{i_{13} i_{14} i_{15} i_{16}} P^{(2)}{}_{i_{17} i_{18}} P^{(2)}{}_{i_{19} i_{20}} P^{(2)}{}_{i_{21} i_{22}]} -$$

$$- 55{,}440 \, P^{(12)}{}_{[i_1 i_2 i_3 i_4 i_5 i_6 i_7 i_8 i_9 i_{10} i_{11} i_{12}} P^{(2)}{}_{i_{13} i_{14}} P^{(2)}{}_{i_{15} i_{16}} P^{(2)}{}_{i_{17} i_{18}} P^{(2)}{}_{i_{19} i_{20}} P^{(2)}{}_{i_{21} i_{22}]} +$$

$$+ 798{,}336 \, P^{(10)}{}_{[i_1 i_2 i_3 i_4 i_5 i_6 i_7 i_8 i_9 i_{10}} P^{(10)}{}_{i_{11} i_{12} i_{13} i_{14} i_{15} i_{16} i_{17} i_{18} i_{19} i_{20}} P^{(2)}{}_{i_{21} i_{22}]} +$$

$$+ 997{,}920 \, P^{(10)}{}_{[i_1 i_2 i_3 i_4 i_5 i_6 i_7 i_8 i_9 i_{10}} P^{(8)}{}_{i_{11} i_{12} i_{13} i_{14} i_{15} i_{16} i_{17} i_{18}} P^{(4)}{}_{i_{19} i_{20} i_{21} i_{22}]} -$$

$$- 997{,}920 \, P^{(10)}{}_{[i_1 i_2 i_3 i_4 i_5 i_6 i_7 i_8 i_9 i_{10}} P^{(8)}{}_{i_{11} i_{12} i_{13} i_{14} i_{15} i_{16} i_{17} i_{18}} P^{(2)}{}_{i_{19} i_{20}} P^{(2)}{}_{i_{21} i_{22}]} +$$

$$+ 443{,}520 \, P^{(10)}{}_{[i_1 i_2 i_3 i_4 i_5 i_6 i_7 i_8 i_9 i_{10}} P^{(6)}{}_{i_{11} i_{12} i_{13} i_{14} i_{15} i_{16}} P^{(6)}{}_{i_{17} i_{18} i_{19} i_{20} i_{21} i_{22}]} -$$

$$- 1{,}330{,}560 \, P^{(10)}{}_{[i_1 i_2 i_3 i_4 i_5 i_6 i_7 i_8 i_9 i_{10}} P^{(6)}{}_{i_{11} i_{12} i_{13} i_{14} i_{15} i_{16}} P^{(4)}{}_{i_{17} i_{18} i_{19} i_{20}} P^{(2)}{}_{i_{21} i_{22}]} +$$

$$+ 443{,}520 \, P^{(10)}{}_{[i_1 i_2 i_3 i_4 i_5 i_6 i_7 i_8 i_9 i_{10}} P^{(6)}{}_{i_{11} i_{12} i_{13} i_{14} i_{15} i_{16}} P^{(2)}{}_{i_{17} i_{18}} P^{(2)}{}_{i_{19} i_{20}} P^{(2)}{}_{i_{21} i_{22}]} -$$

$$- 166{,}320 \, P^{(10)}{}_{[i_1 i_2 i_3 i_4 i_5 i_6 i_7 i_8 i_9 i_{10}} P^{(4)}{}_{i_{11} i_{12} i_{13} i_{14}} P^{(4)}{}_{i_{15} i_{16} i_{17} i_{18}} P^{(4)}{}_{i_{19} i_{20} i_{21} i_{22}]} +$$

$$+ 498{,}960 \, P^{(10)}{}_{[i_1 i_2 i_3 i_4 i_5 i_6 i_7 i_8 i_9 i_{10}} P^{(4)}{}_{i_{11} i_{12} i_{13} i_{14}} P^{(4)}{}_{i_{15} i_{16} i_{17} i_{18}} P^{(2)}{}_{i_{19} i_{20}} P^{(2)}{}_{i_{21} i_{22}]} -$$

$$- 166{,}320 \, P^{(10)}{}_{[i_1 i_2 i_3 i_4 i_5 i_6 i_7 i_8 i_9 i_{10}} P^{(4)}{}_{i_{11} i_{12} i_{13} i_{14}} P^{(2)}{}_{i_{15} i_{16}} P^{(2)}{}_{i_{17} i_{18}} P^{(2)}{}_{i_{19} i_{20}} P^{(2)}{}_{i_{21} i_{22}]} +$$

$$+ 11{,}088 \, P^{(10)}{}_{[i_1 i_2 i_3 i_4 i_5 i_6 i_7 i_8 i_9 i_{10}} P^{(2)}{}_{i_{11} i_{12}} P^{(2)}{}_{i_{13} i_{14}} P^{(2)}{}_{i_{15} i_{16}} P^{(2)}{}_{i_{17} i_{18}} P^{(2)}{}_{i_{19} i_{20}} P^{(2)}{}_{i_{21} i_{22}]} +$$

$$+ 415{,}800 \, P^{(8)}{}_{[i_1 i_2 i_3 i_4 i_5 i_6 i_7 i_8} P^{(8)}{}_{i_9 i_{10} i_{11} i_{12} i_{13} i_{14} i_{15} i_{16}} P^{(6)}{}_{i_{17} i_{18} i_{19} i_{20} i_{21} i_{22}]} -$$

$$- 623{,}700 \, P^{(8)}{}_{[i_1 i_2 i_3 i_4 i_5 i_6 i_7 i_8} P^{(8)}{}_{i_9 i_{10} i_{11} i_{12} i_{13} i_{14} i_{15} i_{16}} P^{(4)}{}_{i_{17} i_{18} i_{19} i_{20}} P^{(2)}{}_{i_{21} i_{22}]} +$$

$$+ 207{,}900 \, P^{(8)}{}_{[i_1 i_2 i_3 i_4 i_5 i_6 i_7 i_8} P^{(8)}{}_{i_9 i_{10} i_{11} i_{12} i_{13} i_{14} i_{15} i_{16}} P^{(2)}{}_{i_{17} i_{18}} P^{(2)}{}_{i_{19} i_{20}} P^{(2)}{}_{i_{21} i_{22}]} -$$

$$- 554{,}400 \, P^{(8)}{}_{[i_1 i_2 i_3 i_4 i_5 i_6 i_7 i_8} P^{(6)}{}_{i_9 i_{10} i_{11} i_{12} i_{13} i_{14}} P^{(6)}{}_{i_{15} i_{16} i_{17} i_{18} i_{19} i_{20}} P^{(2)}{}_{i_{21} i_{22}]} -$$

$$- 415{,}800 \, P^{(8)}{}_{[i_1 i_2 i_3 i_4 i_5 i_6 i_7 i_8} P^{(6)}{}_{i_9 i_{10} i_{11} i_{12} i_{13} i_{14}} P^{(4)}{}_{i_{15} i_{16} i_{17} i_{18}} P^{(4)}{}_{i_{19} i_{20} i_{21} i_{22}]} +$$

$$+ 831{,}600 \, P^{(8)}{}_{[i_1 i_2 i_3 i_4 i_5 i_6 i_7 i_8} P^{(6)}{}_{i_9 i_{10} i_{11} i_{12} i_{13} i_{14}} P^{(4)}{}_{i_{15} i_{16} i_{17} i_{18}} P^{(2)}{}_{i_{19} i_{20}} P^{(2)}{}_{i_{21} i_{22}]} -$$

$$- 138{,}600 \, P^{(8)}{}_{[i_1 i_2 i_3 i_4 i_5 i_6 i_7 i_8} P^{(6)}{}_{i_9 i_{10} i_{11} i_{12} i_{13} i_{14}} P^{(2)}{}_{i_{15} i_{16}} P^{(2)}{}_{i_{17} i_{18}} P^{(2)}{}_{i_{19} i_{20}} P^{(2)}{}_{i_{21} i_{22}]} +$$

$$+ 207{,}900 \, P^{(8)}{}_{[i_1 i_2 i_3 i_4 i_5 i_6 i_7 i_8} P^{(4)}{}_{i_9 i_{10} i_{11} i_{12}} P^{(4)}{}_{i_{13} i_{14} i_{15} i_{16}} P^{(4)}{}_{i_{17} i_{18} i_{19} i_{20}} P^{(2)}{}_{i_{21} i_{22}]} -$$

$$- 207{,}900 \, P^{(8)}{}_{[i_1 i_2 i_3 i_4 i_5 i_6 i_7 i_8} P^{(4)}{}_{i_9 i_{10} i_{11} i_{12}} P^{(4)}{}_{i_{13} i_{14} i_{15} i_{16}} P^{(2)}{}_{i_{17} i_{18}} P^{(2)}{}_{i_{19} i_{20}} P^{(2)}{}_{i_{21} i_{22}]} +$$

$$+ 41{,}580 \, P^{(8)}{}_{[i_1 i_2 i_3 i_4 i_5 i_6 i_7 i_8} P^{(4)}{}_{i_9 i_{10} i_{11} i_{12}} P^{(2)}{}_{i_{13} i_{14}} P^{(2)}{}_{i_{15} i_{16}} P^{(2)}{}_{i_{17} i_{18}} P^{(2)}{}_{i_{19} i_{20}} P^{(2)}{}_{i_{21} i_{22}]} -$$

$$- 1980 \, P^{(8)}{}_{[i_1 i_2 i_3 i_4 i_5 i_6 i_7 i_8} P^{(2)}{}_{i_9 i_{10}} P^{(2)}{}_{i_{11} i_{12}} P^{(2)}{}_{i_{13} i_{14}} P^{(2)}{}_{i_{15} i_{16}} P^{(2)}{}_{i_{17} i_{18}} P^{(2)}{}_{i_{19} i_{20}} P^{(2)}{}_{i_{21} i_{22}]} -$$

$$- 123{,}200 \, P^{(6)}{}_{[i_1 i_2 i_3 i_4 i_5 i_6} P^{(6)}{}_{i_7 i_8 i_9 i_{10} i_{11} i_{12}} P^{(6)}{}_{i_{13} i_{14} i_{15} i_{16} i_{17} i_{18}} P^{(4)}{}_{i_{19} i_{20} i_{21} i_{22}]} +$$

$$+ 123{,}200 \, P^{(6)}{}_{[i_1 i_2 i_3 i_4 i_5 i_6} P^{(6)}{}_{i_7 i_8 i_9 i_{10} i_{11} i_{12}} P^{(6)}{}_{i_{13} i_{14} i_{15} i_{16} i_{17} i_{18}} P^{(2)}{}_{i_{19} i_{20}} P^{(2)}{}_{i_{21} i_{22}]} +$$

$$+ 277{,}200 \, P^{(6)}{}_{[i_1 i_2 i_3 i_4 i_5 i_6} P^{(6)}{}_{i_7 i_8 i_9 i_{10} i_{11} i_{12}} P^{(4)}{}_{i_{13} i_{14} i_{15} i_{16}} P^{(4)}{}_{i_{17} i_{18} i_{19} i_{20}} P^{(2)}{}_{i_{21} i_{22}]} -$$

$$- 184{,}800 \, P^{(6)}{}_{[i_1 i_2 i_3 i_4 i_5 i_6} P^{(6)}{}_{i_7 i_8 i_9 i_{10} i_{11} i_{12}} P^{(4)}{}_{i_{13} i_{14} i_{15} i_{16}} P^{(2)}{}_{i_{17} i_{18}} P^{(2)}{}_{i_{19} i_{20}} P^{(2)}{}_{i_{21} i_{22}]} +$$

$$+ 18{,}480 \, P^{(6)}{}_{[i_1 i_2 i_3 i_4 i_5 i_6} P^{(6)}{}_{i_7 i_8 i_9 i_{10} i_{11} i_{12}} P^{(2)}{}_{i_{13} i_{14}} P^{(2)}{}_{i_{15} i_{16}} P^{(2)}{}_{i_{17} i_{18}} P^{(2)}{}_{i_{19} i_{20}} P^{(2)}{}_{i_{21} i_{22}]} +$$

$$+ 34{,}650 \, P^{(6)}{}_{[i_1 i_2 i_3 i_4 i_5 i_6} P^{(4)}{}_{i_7 i_8 i_9 i_{10}} P^{(4)}{}_{i_{11} i_{12} i_{13} i_{14}} P^{(4)}{}_{i_{15} i_{16} i_{17} i_{18}} P^{(4)}{}_{i_{19} i_{20} i_{21} i_{22}]} -$$

$$- 138{,}600\, P^{(6)}{}_{[i_1i_2i_3i_4i_5i_6}\, P^{(4)}{}_{i_7i_8i_9i_{10}}\, P^{(4)}{}_{i_{11}i_{12}i_{13}i_{14}}\, P^{(4)}{}_{i_{15}i_{16}i_{17}i_{18}}\, P^{(2)}{}_{i_{19}i_{20}}\, P^{(2)}{}_{i_{21}i_{22}]} +$$

$$+ 69{,}300\, P^{(6)}{}_{[i_1i_2i_3i_4i_5i_6}\, P^{(4)}{}_{i_7i_8i_9i_{10}}\, P^{(4)}{}_{i_{11}i_{12}i_{13}i_{14}}\, P^{(2)}{}_{i_{15}i_{16}}\, P^{(2)}{}_{i_{17}i_{18}}\, P^{(2)}{}_{i_{19}i_{20}}\, P^{(2)}{}_{i_{21}i_{22}]} -$$

$$- 9240\, P^{(6)}{}_{[i_1i_2i_3i_4i_5i_6}\, P^{(4)}{}_{i_7i_8i_9i_{10}}\, P^{(2)}{}_{i_{11}i_{12}}\, P^{(2)}{}_{i_{13}i_{14}}\, P^{(2)}{}_{i_{15}i_{16}}\, P^{(2)}{}_{i_{17}i_{18}}\, P^{(2)}{}_{i_{19}i_{20}}\, P^{(2)}{}_{i_{21}i_{22}]} +$$

$$+ 330\, P^{(6)}{}_{[i_1i_2i_3i_4i_5i_6}\, P^{(2)}{}_{i_7i_8}\, P^{(2)}{}_{i_9i_{10}}\, P^{(2)}{}_{i_{11}i_{12}}\, P^{(2)}{}_{i_{13}i_{14}}\, P^{(2)}{}_{i_{15}i_{16}}\, P^{(2)}{}_{i_{17}i_{18}}\, P^{(2)}{}_{i_{19}i_{20}}\, P^{(2)}{}_{i_{21}i_{22}]} -$$

$$- 10{,}395\, P^{(4)}{}_{[i_1i_2i_3i_4}\, P^{(4)}{}_{i_5i_6i_7i_8}\, P^{(4)}{}_{i_9i_{10}i_{11}i_{12}}\, P^{(4)}{}_{i_{13}i_{14}i_{15}i_{16}}\, P^{(4)}{}_{i_{17}i_{18}i_{19}i_{20}}\, P^{(2)}{}_{i_{21}i_{22}]} +$$

$$+ 17{,}325\, P^{(4)}{}_{[i_1i_2i_3i_4}\, P^{(4)}{}_{i_5i_6i_7i_8}\, P^{(4)}{}_{i_9i_{10}i_{11}i_{12}}\, P^{(4)}{}_{i_{13}i_{14}i_{15}i_{16}}\, P^{(2)}{}_{i_{17}i_{18}}\, P^{(2)}{}_{i_{19}i_{20}}\, P^{(2)}{}_{i_{21}i_{22}]} -$$

$$- 6930\, P^{(4)}{}_{[i_1i_2i_3i_4}\, P^{(4)}{}_{i_5i_6i_7i_8}\, P^{(4)}{}_{i_9i_{10}i_{11}i_{12}}\, P^{(2)}{}_{i_{13}i_{14}}\, P^{(2)}{}_{i_{15}i_{16}}\, P^{(2)}{}_{i_{17}i_{18}}\, P^{(2)}{}_{i_{19}i_{20}}\, P^{(2)}{}_{i_{21}i_{22}]} +$$

$$+ 990\, P^{(4)}{}_{[i_1i_2i_3i_4}\, P^{(4)}{}_{i_5i_6i_7i_8}\, P^{(2)}{}_{i_9i_{10}}\, P^{(2)}{}_{i_{11}i_{12}}\, P^{(2)}{}_{i_{13}i_{14}}\, P^{(2)}{}_{i_{15}i_{16}}\, P^{(2)}{}_{i_{17}i_{18}}\, P^{(2)}{}_{i_{19}i_{20}}\, P^{(2)}{}_{i_{21}i_{22}]} -$$

$$- 55\, P^{(4)}{}_{[i_1i_2i_3i_4}\, P^{(2)}{}_{i_5i_6}\, P^{(2)}{}_{i_7i_8}\, P^{(2)}{}_{i_9i_{10}}\, P^{(2)}{}_{i_{11}i_{12}}\, P^{(2)}{}_{i_{13}i_{14}}\, P^{(2)}{}_{i_{15}i_{16}}\, P^{(2)}{}_{i_{17}i_{18}}\, P^{(2)}{}_{i_{19}i_{20}}\, P^{(2)}{}_{i_{21}i_{22}]} +$$

$$+ P^{(2)}{}_{[i_1i_2}\, P^{(2)}{}_{i_3i_4}\, P^{(2)}{}_{i_5i_6}\, P^{(2)}{}_{i_7i_8}\, P^{(2)}{}_{i_9i_{10}}\, P^{(2)}{}_{i_{11}i_{12}}\, P^{(2)}{}_{i_{13}i_{14}}\, P^{(2)}{}_{i_{15}i_{16}}\, P^{(2)}{}_{i_{17}i_{18}}\, P^{(2)}{}_{i_{19}i_{20}}\, P^{(2)}{}_{i_{21}i_{22}]})$$

### COEFFICIENT OF THE 12$^{th}$ CHERN FORM

$$c_{(12)i_1i_2i_3i_4i_5i_6i_7i_8i_9i_{10}i_{11}i_{12}i_{13}i_{14}i_{15}i_{16}i_{17}i_{18}i_{19}i_{20}i_{21}i_{22}i_{23}i_{24}} = \tag{21}$$

$$= \tfrac{1}{24!}\langle \mathbf{e}_{i_1} \wedge \mathbf{e}_{i_2} \wedge \mathbf{e}_{i_3} \wedge \mathbf{e}_{i_4} \wedge \mathbf{e}_{i_5} \wedge \mathbf{e}_{i_6} \wedge \mathbf{e}_{i_7} \wedge \mathbf{e}_{i_8} \wedge \mathbf{e}_{i_9} \wedge \mathbf{e}_{i_{10}} \wedge \mathbf{e}_{i_{11}} \wedge \mathbf{e}_{i_{12}} \wedge \ldots$$

$$\ldots \wedge \mathbf{e}_{i_{13}} \wedge \mathbf{e}_{i_{14}} \wedge \mathbf{e}_{i_{15}} \wedge \mathbf{e}_{i_{16}} \wedge \mathbf{e}_{i_{17}} \wedge \mathbf{e}_{i_{18}} \wedge \mathbf{e}_{i_{19}} \wedge \mathbf{e}_{i_{20}} \wedge \mathbf{e}_{i_{21}} \wedge \mathbf{e}_{i_{22}} \wedge \mathbf{e}_{i_{23}} \wedge \mathbf{e}_{i_{24}}, c_{(12)}\rangle$$

$$= \tfrac{i^{12}}{2^{24}\pi^{12}12!}(- 39{,}916{,}800\, R_{[i_1i_2|l|}{}^a\, R_{i_3i_4|a|}{}^b\, R_{i_5i_6|b|}{}^c\, R_{i_7i_8|c|}{}^d\, R_{i_9i_{10}|d|}{}^e\, R_{i_{11}i_{12}|e|}{}^f\, R_{i_{13}i_{14}|f|}{}^g\, R_{i_{15}i_{16}|g|}{}^h\, R_{i_{17}i_{18}|h|}{}^i\, R_{i_{19}i_{20}|i|}{}^j\, R_{i_{21}i_{22}|j|}{}^k\, R_{i_{23}i_{24}]k}{}^l +$$

$$+ 43{,}545{,}600\, R_{[i_1i_2|k|}{}^a\, R_{i_3i_4|a|}{}^b\, R_{i_5i_6|b|}{}^c\, R_{i_7i_8|c|}{}^d\, R_{i_9i_{10}|d|}{}^e\, R_{i_{11}i_{12}|e|}{}^f\, R_{i_{13}i_{14}|f|}{}^g\, R_{i_{15}i_{16}|g|}{}^h\, R_{i_{17}i_{18}|h|}{}^i\, R_{i_{19}i_{20}|i|}{}^j\, R_{i_{21}i_{22}|j|}{}^k\, R_{i_{23}i_{24}]k}{}^l +$$

$$+ 23{,}950{,}080\, R_{[i_1i_2|j|}{}^a\, R_{i_3i_4|a|}{}^b\, R_{i_5i_6|b|}{}^c\, R_{i_7i_8|c|}{}^d\, R_{i_9i_{10}|d|}{}^e\, R_{i_{11}i_{12}|e|}{}^f\, R_{i_{13}i_{14}|f|}{}^g\, R_{i_{15}i_{16}|g|}{}^h\, R_{i_{17}i_{18}|h|}{}^i\, R_{i_{19}i_{20}|i|}{}^j\, R_{i_{21}i_{22}|l|}{}^k\, R_{i_{23}i_{24}]k}{}^l -$$

$$- 23{,}950{,}080\, R_{[i_1i_2|j|}{}^a\, R_{i_3i_4|a|}{}^b\, R_{i_5i_6|b|}{}^c\, R_{i_7i_8|c|}{}^d\, R_{i_9i_{10}|d|}{}^e\, R_{i_{11}i_{12}|e|}{}^f\, R_{i_{13}i_{14}|f|}{}^g\, R_{i_{15}i_{16}|g|}{}^h\, R_{i_{17}i_{18}|h|}{}^i\, R_{i_{19}i_{20}|i|}{}^j\, R_{i_{21}i_{22}|k|}{}^k\, R_{i_{23}i_{24}]k}{}^l +$$

$$+ 17{,}740{,}800\, R_{[i_1i_2|j|}{}^a\, R_{i_3i_4|a|}{}^b\, R_{i_5i_6|b|}{}^c\, R_{i_7i_8|c|}{}^d\, R_{i_9i_{10}|d|}{}^e\, R_{i_{11}i_{12}|e|}{}^f\, R_{i_{13}i_{14}|f|}{}^g\, R_{i_{15}i_{16}|g|}{}^h\, R_{i_{17}i_{18}|h|}{}^i\, R_{i_{19}i_{20}|i|}{}^j\, R_{i_{21}i_{22}|j|}{}^k\, R_{i_{23}i_{24}]k}{}^l -$$

$$- 26{,}611{,}200\, R_{[i_1i_2|i|}{}^a\, R_{i_3i_4|a|}{}^b\, R_{i_5i_6|b|}{}^c\, R_{i_7i_8|c|}{}^d\, R_{i_9i_{10}|d|}{}^e\, R_{i_{11}i_{12}|e|}{}^f\, R_{i_{13}i_{14}|f|}{}^g\, R_{i_{15}i_{16}|g|}{}^h\, R_{i_{17}i_{18}|h|}{}^i\, R_{i_{19}i_{20}|k|}{}^j\, R_{i_{21}i_{22}|j|}{}^k\, R_{i_{23}i_{24}]k}{}^l +$$

$$+ 8{,}870{,}400\, R_{[i_1i_2|i|}{}^a\, R_{i_3i_4|a|}{}^b\, R_{i_5i_6|b|}{}^c\, R_{i_7i_8|c|}{}^d\, R_{i_9i_{10}|d|}{}^e\, R_{i_{11}i_{12}|e|}{}^f\, R_{i_{13}i_{14}|f|}{}^g\, R_{i_{15}i_{16}|g|}{}^h\, R_{i_{17}i_{18}|l|}{}^i\, R_{i_{19}i_{20}|i|}{}^j\, R_{i_{21}i_{22}|j|}{}^k\, R_{i_{23}i_{24}]k}{}^l +$$

$$+ 14{,}968{,}800\, R_{[i_1i_2|h|}{}^a\, R_{i_3i_4|a|}{}^b\, R_{i_5i_6|b|}{}^c\, R_{i_7i_8|c|}{}^d\, R_{i_9i_{10}|d|}{}^e\, R_{i_{11}i_{12}|e|}{}^f\, R_{i_{13}i_{14}|f|}{}^g\, R_{i_{15}i_{16}|g|}{}^h\, R_{i_{17}i_{18}|l|}{}^i\, R_{i_{19}i_{20}|i|}{}^j\, R_{i_{21}i_{22}|j|}{}^k\, R_{i_{23}i_{24}]k}{}^l -$$

$$- 19{,}958{,}400\, R_{[i_1i_2|h|}{}^a\, R_{i_3i_4|a|}{}^b\, R_{i_5i_6|b|}{}^c\, R_{i_7i_8|c|}{}^d\, R_{i_9i_{10}|d|}{}^e\, R_{i_{11}i_{12}|e|}{}^f\, R_{i_{13}i_{14}|f|}{}^g\, R_{i_{15}i_{16}|g|}{}^h\, R_{i_{17}i_{18}|k|}{}^i\, R_{i_{19}i_{20}|i|}{}^j\, R_{i_{21}i_{22}|j|}{}^k\, R_{i_{23}i_{24}]k}{}^l -$$

$$- 7{,}484{,}400\, R_{[i_1i_2|h|}{}^a\, R_{i_3i_4|a|}{}^b\, R_{i_5i_6|b|}{}^c\, R_{i_7i_8|c|}{}^d\, R_{i_9i_{10}|d|}{}^e\, R_{i_{11}i_{12}|e|}{}^f\, R_{i_{13}i_{14}|f|}{}^g\, R_{i_{15}i_{16}|g|}{}^h\, R_{i_{17}i_{18}|j|}{}^i\, R_{i_{19}i_{20}|i|}{}^j\, R_{i_{21}i_{22}|l|}{}^k\, R_{i_{23}i_{24}]k}{}^l +$$

$$+ 14{,}968{,}800\, R_{[i_1i_2|h|}{}^a\, R_{i_3i_4|a|}{}^b\, R_{i_5i_6|b|}{}^c\, R_{i_7i_8|c|}{}^d\, R_{i_9i_{10}|d|}{}^e\, R_{i_{11}i_{12}|e|}{}^f\, R_{i_{13}i_{14}|f|}{}^g\, R_{i_{15}i_{16}|g|}{}^h\, R_{i_{17}i_{18}|j|}{}^i\, R_{i_{19}i_{20}|i|}{}^j\, R_{i_{21}i_{22}|k|}{}^k\, R_{i_{23}i_{24}]k}{}^l -$$

$$- 2{,}494{,}800\, R_{[i_1i_2|h|}{}^a\, R_{i_3i_4|a|}{}^b\, R_{i_5i_6|b|}{}^c\, R_{i_7i_8|c|}{}^d\, R_{i_9i_{10}|d|}{}^e\, R_{i_{11}i_{12}|e|}{}^f\, R_{i_{13}i_{14}|f|}{}^g\, R_{i_{15}i_{16}|g|}{}^h\, R_{i_{17}i_{18}|i|}{}^i\, R_{i_{19}i_{20}|j|}{}^j\, R_{i_{21}i_{22}|k|}{}^k\, R_{i_{23}i_{24}]k}{}^l +$$

$$+ 13{,}685{,}760\, R_{[i_1i_2|g|}{}^a\, R_{i_3i_4|a|}{}^b\, R_{i_5i_6|b|}{}^c\, R_{i_7i_8|c|}{}^d\, R_{i_9i_{10}|d|}{}^e\, R_{i_{11}i_{12}|e|}{}^f\, R_{i_{13}i_{14}|f|}{}^g\, R_{i_{15}i_{16}|l|}{}^h\, R_{i_{17}i_{18}|h|}{}^i\, R_{i_{19}i_{20}|i|}{}^j\, R_{i_{21}i_{22}|j|}{}^k\, R_{i_{23}i_{24}]k}{}^l -$$

$$- 17{,}107{,}200\, R_{[i_1i_2|g|}{}^a\, R_{i_3i_4|a|}{}^b\, R_{i_5i_6|b|}{}^c\, R_{i_7i_8|c|}{}^d\, R_{i_9i_{10}|d|}{}^e\, R_{i_{11}i_{12}|e|}{}^f\, R_{i_{13}i_{14}|f|}{}^g\, R_{i_{15}i_{16}|k|}{}^h\, R_{i_{17}i_{18}|h|}{}^i\, R_{i_{19}i_{20}|i|}{}^j\, R_{i_{21}i_{22}|j|}{}^k\, R_{i_{23}i_{24}]k}{}^l -$$

$$- 11{,}404{,}800\, R_{[i_1i_2|g|}{}^a\, R_{i_3i_4|a|}{}^b\, R_{i_5i_6|b|}{}^c\, R_{i_7i_8|c|}{}^d\, R_{i_9i_{10}|d|}{}^e\, R_{i_{11}i_{12}|e|}{}^f\, R_{i_{13}i_{14}|f|}{}^g\, R_{i_{15}i_{16}|j|}{}^h\, R_{i_{17}i_{18}|h|}{}^i\, R_{i_{19}i_{20}|i|}{}^j\, R_{i_{21}i_{22}|l|}{}^k\, R_{i_{23}i_{24}]k}{}^l +$$

$$+ 11{,}404{,}800\, R_{[i_1i_2|g|}{}^a\, R_{i_3i_4|a|}{}^b\, R_{i_5i_6|b|}{}^c\, R_{i_7i_8|c|}{}^d\, R_{i_9i_{10}|d|}{}^e\, R_{i_{11}i_{12}|e|}{}^f\, R_{i_{13}i_{14}|f|}{}^g\, R_{i_{15}i_{16}|j|}{}^h\, R_{i_{17}i_{18}|h|}{}^i\, R_{i_{19}i_{20}|i|}{}^j\, R_{i_{21}i_{22}|k|}{}^k\, R_{i_{23}i_{24}]k}{}^l +$$

$$+ 8{,}553{,}600\, R_{[i_1i_2|g|}{}^a\, R_{i_3i_4|a|}{}^b\, R_{i_5i_6|b|}{}^c\, R_{i_7i_8|c|}{}^d\, R_{i_9i_{10}|d|}{}^e\, R_{i_{11}i_{12}|e|}{}^f\, R_{i_{13}i_{14}|f|}{}^g\, R_{i_{15}i_{16}|i|}{}^h\, R_{i_{17}i_{18}|h|}{}^i\, R_{i_{19}i_{20}|k|}{}^j\, R_{i_{21}i_{22}|j|}{}^k\, R_{i_{23}i_{24}]k}{}^l -$$

$$- 5{,}702{,}400\, R_{[i_1i_2|g|}{}^a\, R_{i_3i_4|a|}{}^b\, R_{i_5i_6|b|}{}^c\, R_{i_7i_8|c|}{}^d\, R_{i_9i_{10}|d|}{}^e\, R_{i_{11}i_{12}|e|}{}^f\, R_{i_{13}i_{14}|f|}{}^g\, R_{i_{15}i_{16}|i|}{}^h\, R_{i_{17}i_{18}|h|}{}^i\, R_{i_{19}i_{20}|j|}{}^j\, R_{i_{21}i_{22}|k|}{}^k\, R_{i_{23}i_{24}]k}{}^l +$$

$$+ 570{,}240\, R_{[i_1i_2|g|}{}^a\, R_{i_3i_4|a|}{}^b\, R_{i_5i_6|b|}{}^c\, R_{i_7i_8|c|}{}^d\, R_{i_9i_{10}|d|}{}^e\, R_{i_{11}i_{12}|e|}{}^f\, R_{i_{13}i_{14}|f|}{}^g\, R_{i_{15}i_{16}|h|}{}^h\, R_{i_{17}i_{18}|i|}{}^i\, R_{i_{19}i_{20}|j|}{}^j\, R_{i_{21}i_{22}|k|}{}^k\, R_{i_{23}i_{24}]k}{}^l +$$

$$+ 6{,}652{,}800\, R_{[i_1i_2|f|}{}^a\, R_{i_3i_4|a|}{}^b\, R_{i_5i_6|b|}{}^c\, R_{i_7i_8|c|}{}^d\, R_{i_9i_{10}|d|}{}^e\, R_{i_{11}i_{12}|e|}{}^f\, R_{i_{13}i_{14}|l|}{}^g\, R_{i_{15}i_{16}|g|}{}^h\, R_{i_{17}i_{18}|h|}{}^i\, R_{i_{19}i_{20}|i|}{}^j\, R_{i_{21}i_{22}|j|}{}^k\, R_{i_{23}i_{24}]k}{}^l -$$

$$- 15{,}966{,}720\, R_{[i_1i_2|f|}{}^a\, R_{i_3i_4|a|}{}^b\, R_{i_5i_6|b|}{}^c\, R_{i_7i_8|c|}{}^d\, R_{i_9i_{10}|d|}{}^e\, R_{i_{11}i_{12}|e|}{}^f\, R_{i_{13}i_{14}|k|}{}^g\, R_{i_{15}i_{16}|g|}{}^h\, R_{i_{17}i_{18}|h|}{}^i\, R_{i_{19}i_{20}|i|}{}^j\, R_{i_{21}i_{22}|j|}{}^k\, R_{i_{23}i_{24}]k}{}^l -$$

$$- 9{,}979{,}200\, R_{[i_1i_2|f|}{}^a\, R_{i_3i_4|a|}{}^b\, R_{i_5i_6|b|}{}^c\, R_{i_7i_8|c|}{}^d\, R_{i_9i_{10}|d|}{}^e\, R_{i_{11}i_{12}|e|}{}^f\, R_{i_{13}i_{14}|j|}{}^g\, R_{i_{15}i_{16}|g|}{}^h\, R_{i_{17}i_{18}|h|}{}^i\, R_{i_{19}i_{20}|i|}{}^j\, R_{i_{21}i_{22}|l|}{}^k\, R_{i_{23}i_{24}]k}{}^l +$$

$$+ 9{,}979{,}200\, R_{[i_1i_2|f|}{}^a\, R_{i_3i_4|a|}{}^b\, R_{i_5i_6|b|}{}^c\, R_{i_7i_8|c|}{}^d\, R_{i_9i_{10}|d|}{}^e\, R_{i_{11}i_{12}|e|}{}^f\, R_{i_{13}i_{14}|j|}{}^g\, R_{i_{15}i_{16}|g|}{}^h\, R_{i_{17}i_{18}|h|}{}^i\, R_{i_{19}i_{20}|i|}{}^j\, R_{i_{21}i_{22}|k|}{}^k\, R_{i_{23}i_{24}]k}{}^l -$$

$$- 4{,}435{,}200\, R_{[i_1i_2|f|}{}^a\, R_{i_3i_4|a|}{}^b\, R_{i_5i_6|b|}{}^c\, R_{i_7i_8|c|}{}^d\, R_{i_9i_{10}|d|}{}^e\, R_{i_{11}i_{12}|e|}{}^f\, R_{i_{13}i_{14}|i|}{}^g\, R_{i_{15}i_{16}|g|}{}^h\, R_{i_{17}i_{18}|h|}{}^i\, R_{i_{19}i_{20}|i|}{}^j\, R_{i_{21}i_{22}|j|}{}^k\, R_{i_{23}i_{24}]k}{}^l +$$

$$+ 13{,}305{,}600\, R_{[i_1i_2|f|}{}^a\, R_{i_3i_4|a|}{}^b\, R_{i_5i_6|b|}{}^c\, R_{i_7i_8|c|}{}^d\, R_{i_9i_{10}|d|}{}^e\, R_{i_{11}i_{12}|e|}{}^f\, R_{i_{13}i_{14}|i|}{}^g\, R_{i_{15}i_{16}|g|}{}^h\, R_{i_{17}i_{18}|h|}{}^i\, R_{i_{19}i_{20}|k|}{}^j\, R_{i_{21}i_{22}|j|}{}^k\, R_{i_{23}i_{24}]k}{}^l -$$

$$- 4{,}435{,}200\, R_{[i_1i_2|f|}{}^a\, R_{i_3i_4|a|}{}^b\, R_{i_5i_6|b|}{}^c\, R_{i_7i_8|c|}{}^d\, R_{i_9i_{10}|d|}{}^e\, R_{i_{11}i_{12}|e|}{}^f\, R_{i_{13}i_{14}|i|}{}^g\, R_{i_{15}i_{16}|g|}{}^h\, R_{i_{17}i_{18}|h|}{}^i\, R_{i_{19}i_{20}|j|}{}^j\, R_{i_{21}i_{22}|k|}{}^k\, R_{i_{23}i_{24}]k}{}^l +$$

$$+ 1{,}663{,}200\, R_{[i_1i_2|f|}{}^a\, R_{i_3i_4|a|}{}^b\, R_{i_5i_6|b|}{}^c\, R_{i_7i_8|c|}{}^d\, R_{i_9i_{10}|d|}{}^e\, R_{i_{11}i_{12}|e|}{}^f\, R_{i_{13}i_{14}|h|}{}^g\, R_{i_{15}i_{16}|g|}{}^h\, R_{i_{17}i_{18}|j|}{}^i\, R_{i_{19}i_{20}|i|}{}^j\, R_{i_{21}i_{22}|l|}{}^k\, R_{i_{23}i_{24}]k}{}^l -$$

$$- 4{,}989{,}600\ R_{[i_1i_2|f|}{}^a R_{i_3i_4|a|}{}^b R_{i_5i_6|b|}{}^c R_{i_7i_8|c|}{}^d R_{i_9i_{10}|d|}{}^e R_{i_{11}i_{12}|e|}{}^f R_{i_{13}i_{14}|h|}{}^g R_{i_{15}i_{16}|g|}{}^h R_{i_{17}i_{18}|j|}{}^i R_{i_{19}i_{20}|i|}{}^j R_{i_{21}i_{22}|k|}{}^k R_{i_{23}i_{24}]l}{}^l +$$

$$+ 1{,}663{,}200\ R_{[i_1i_2|f|}{}^a R_{i_3i_4|a|}{}^b R_{i_5i_6|b|}{}^c R_{i_7i_8|c|}{}^d R_{i_9i_{10}|d|}{}^e R_{i_{11}i_{12}|e|}{}^f R_{i_{13}i_{14}|h|}{}^g R_{i_{15}i_{16}|g|}{}^h R_{i_{17}i_{18}|i|}{}^i R_{i_{19}i_{20}|j|}{}^j R_{i_{21}i_{22}|k|}{}^k R_{i_{23}i_{24}]l}{}^l -$$

$$- 110{,}880\ R_{[i_1i_2|f|}{}^a R_{i_3i_4|a|}{}^b R_{i_5i_6|b|}{}^c R_{i_7i_8|c|}{}^d R_{i_9i_{10}|d|}{}^e R_{i_{11}i_{12}|e|}{}^f R_{i_{13}i_{14}|h|}{}^g R_{i_{15}i_{16}|h|}{}^h R_{i_{17}i_{18}|i|}{}^i R_{i_{19}i_{20}|j|}{}^j R_{i_{21}i_{22}|k|}{}^k R_{i_{23}i_{24}]l}{}^l -$$

$$- 4{,}790{,}016\ R_{[i_1i_2|e|}{}^a R_{i_3i_4|a|}{}^b R_{i_5i_6|b|}{}^c R_{i_7i_8|c|}{}^d R_{i_9i_{10}|d|}{}^e R_{i_{11}i_{12}|j|}{}^f R_{i_{13}i_{14}|f|}{}^g R_{i_{15}i_{16}|g|}{}^h R_{i_{17}i_{18}|h|}{}^i R_{i_{19}i_{20}|i|}{}^j R_{i_{21}i_{22}|l|}{}^k R_{i_{23}i_{24}]k}{}^l +$$

$$+ 4{,}790{,}016\ R_{[i_1i_2|e|}{}^a R_{i_3i_4|a|}{}^b R_{i_5i_6|b|}{}^c R_{i_7i_8|c|}{}^d R_{i_9i_{10}|d|}{}^e R_{i_{11}i_{12}|j|}{}^f R_{i_{13}i_{14}|f|}{}^g R_{i_{15}i_{16}|g|}{}^h R_{i_{17}i_{18}|h|}{}^i R_{i_{19}i_{20}|i|}{}^j R_{i_{21}i_{22}|k|}{}^k R_{i_{23}i_{24}]l}{}^l -$$

$$- 7{,}983{,}360\ R_{[i_1i_2|e|}{}^a R_{i_3i_4|a|}{}^b R_{i_5i_6|b|}{}^c R_{i_7i_8|c|}{}^d R_{i_9i_{10}|d|}{}^e R_{i_{11}i_{12}|i|}{}^f R_{i_{13}i_{14}|f|}{}^g R_{i_{15}i_{16}|g|}{}^h R_{i_{17}i_{18}|h|}{}^i R_{i_{19}i_{20}|l|}{}^j R_{i_{21}i_{22}|j|}{}^k R_{i_{23}i_{24}]k}{}^l +$$

$$+ 11{,}975{,}040\ R_{[i_1i_2|e|}{}^a R_{i_3i_4|a|}{}^b R_{i_5i_6|b|}{}^c R_{i_7i_8|c|}{}^d R_{i_9i_{10}|d|}{}^e R_{i_{11}i_{12}|i|}{}^f R_{i_{13}i_{14}|f|}{}^g R_{i_{15}i_{16}|g|}{}^h R_{i_{17}i_{18}|h|}{}^i R_{i_{19}i_{20}|k|}{}^j R_{i_{21}i_{22}|j|}{}^k R_{i_{23}i_{24}]l}{}^l -$$

$$- 3{,}991{,}680\ R_{[i_1i_2|e|}{}^a R_{i_3i_4|a|}{}^b R_{i_5i_6|b|}{}^c R_{i_7i_8|c|}{}^d R_{i_9i_{10}|d|}{}^e R_{i_{11}i_{12}|i|}{}^f R_{i_{13}i_{14}|f|}{}^g R_{i_{15}i_{16}|g|}{}^h R_{i_{17}i_{18}|h|}{}^i R_{i_{19}i_{20}|j|}{}^j R_{i_{21}i_{22}|k|}{}^k R_{i_{23}i_{24}]l}{}^l +$$

$$+ 5{,}322{,}240\ R_{[i_1i_2|e|}{}^a R_{i_3i_4|a|}{}^b R_{i_5i_6|b|}{}^c R_{i_7i_8|c|}{}^d R_{i_9i_{10}|d|}{}^e R_{i_{11}i_{12}|h|}{}^f R_{i_{13}i_{14}|f|}{}^g R_{i_{15}i_{16}|g|}{}^h R_{i_{17}i_{18}|k|}{}^i R_{i_{19}i_{20}|i|}{}^j R_{i_{21}i_{22}|j|}{}^k R_{i_{23}i_{24}]l}{}^l +$$

$$+ 3{,}991{,}680\ R_{[i_1i_2|e|}{}^a R_{i_3i_4|a|}{}^b R_{i_5i_6|b|}{}^c R_{i_7i_8|c|}{}^d R_{i_9i_{10}|d|}{}^e R_{i_{11}i_{12}|h|}{}^f R_{i_{13}i_{14}|f|}{}^g R_{i_{15}i_{16}|g|}{}^h R_{i_{17}i_{18}|j|}{}^i R_{i_{19}i_{20}|i|}{}^j R_{i_{21}i_{22}|l|}{}^k R_{i_{23}i_{24}]k}{}^l -$$

$$- 7{,}983{,}360\ R_{[i_1i_2|e|}{}^a R_{i_3i_4|a|}{}^b R_{i_5i_6|b|}{}^c R_{i_7i_8|c|}{}^d R_{i_9i_{10}|d|}{}^e R_{i_{11}i_{12}|h|}{}^f R_{i_{13}i_{14}|f|}{}^g R_{i_{15}i_{16}|g|}{}^h R_{i_{17}i_{18}|i|}{}^i R_{i_{19}i_{20}|j|}{}^j R_{i_{21}i_{22}|k|}{}^k R_{i_{23}i_{24}]l}{}^l +$$

$$+ 1{,}330{,}560\ R_{[i_1i_2|e|}{}^a R_{i_3i_4|a|}{}^b R_{i_5i_6|b|}{}^c R_{i_7i_8|c|}{}^d R_{i_9i_{10}|d|}{}^e R_{i_{11}i_{12}|h|}{}^f R_{i_{13}i_{14}|f|}{}^g R_{i_{15}i_{16}|g|}{}^h R_{i_{17}i_{18}|i|}{}^i R_{i_{19}i_{20}|j|}{}^j R_{i_{21}i_{22}|k|}{}^k R_{i_{23}i_{24}]l}{}^l -$$

$$- 1{,}995{,}840\ R_{[i_1i_2|e|}{}^a R_{i_3i_4|a|}{}^b R_{i_5i_6|b|}{}^c R_{i_7i_8|c|}{}^d R_{i_9i_{10}|d|}{}^e R_{i_{11}i_{12}|g|}{}^f R_{i_{13}i_{14}|f|}{}^g R_{i_{15}i_{16}|i|}{}^h R_{i_{17}i_{18}|h|}{}^i R_{i_{19}i_{20}|k|}{}^j R_{i_{21}i_{22}|j|}{}^k R_{i_{23}i_{24}]l}{}^l +$$

$$+ 1{,}995{,}840\ R_{[i_1i_2|e|}{}^a R_{i_3i_4|a|}{}^b R_{i_5i_6|b|}{}^c R_{i_7i_8|c|}{}^d R_{i_9i_{10}|d|}{}^e R_{i_{11}i_{12}|g|}{}^f R_{i_{13}i_{14}|f|}{}^g R_{i_{15}i_{16}|i|}{}^h R_{i_{17}i_{18}|h|}{}^i R_{i_{19}i_{20}|j|}{}^j R_{i_{21}i_{22}|k|}{}^k R_{i_{23}i_{24}]l}{}^l -$$

$$- 399{,}168\ R_{[i_1i_2|e|}{}^a R_{i_3i_4|a|}{}^b R_{i_5i_6|b|}{}^c R_{i_7i_8|c|}{}^d R_{i_9i_{10}|d|}{}^e R_{i_{11}i_{12}|g|}{}^f R_{i_{13}i_{14}|f|}{}^g R_{i_{15}i_{16}|h|}{}^h R_{i_{17}i_{18}|i|}{}^i R_{i_{19}i_{20}|j|}{}^j R_{i_{21}i_{22}|k|}{}^k R_{i_{23}i_{24}]l}{}^l +$$

$$+ 19{,}008\ R_{[i_1i_2|e|}{}^a R_{i_3i_4|a|}{}^b R_{i_5i_6|b|}{}^c R_{i_7i_8|c|}{}^d R_{i_9i_{10}|d|}{}^e R_{i_{11}i_{12}|f|}{}^f R_{i_{13}i_{14}|g|}{}^g R_{i_{15}i_{16}|h|}{}^h R_{i_{17}i_{18}|i|}{}^i R_{i_{19}i_{20}|j|}{}^j R_{i_{21}i_{22}|k|}{}^k R_{i_{23}i_{24}]l}{}^l -$$

$$- 1{,}247{,}400\ R_{[i_1i_2|d|}{}^a R_{i_3i_4|a|}{}^b R_{i_5i_6|b|}{}^c R_{i_7i_8|c|}{}^d R_{i_9i_{10}|h|}{}^e R_{i_{11}i_{12}|e|}{}^f R_{i_{13}i_{14}|f|}{}^g R_{i_{15}i_{16}|g|}{}^h R_{i_{17}i_{18}|l|}{}^i R_{i_{19}i_{20}|i|}{}^j R_{i_{21}i_{22}|j|}{}^k R_{i_{23}i_{24}]k}{}^l +$$

$$+ 4{,}989{,}600\ R_{[i_1i_2|d|}{}^a R_{i_3i_4|a|}{}^b R_{i_5i_6|b|}{}^c R_{i_7i_8|c|}{}^d R_{i_9i_{10}|h|}{}^e R_{i_{11}i_{12}|e|}{}^f R_{i_{13}i_{14}|f|}{}^g R_{i_{15}i_{16}|g|}{}^h R_{i_{17}i_{18}|h|}{}^i R_{i_{19}i_{20}|i|}{}^j R_{i_{21}i_{22}|j|}{}^k R_{i_{23}i_{24}]l}{}^l +$$

$$+ 1{,}871{,}100\ R_{[i_1i_2|d|}{}^a R_{i_3i_4|a|}{}^b R_{i_5i_6|b|}{}^c R_{i_7i_8|c|}{}^d R_{i_9i_{10}|h|}{}^e R_{i_{11}i_{12}|e|}{}^f R_{i_{13}i_{14}|f|}{}^g R_{i_{15}i_{16}|g|}{}^h R_{i_{17}i_{18}|j|}{}^i R_{i_{19}i_{20}|i|}{}^j R_{i_{21}i_{22}|l|}{}^k R_{i_{23}i_{24}]k}{}^l -$$

$$- 3{,}742{,}200\ R_{[i_1i_2|d|}{}^a R_{i_3i_4|a|}{}^b R_{i_5i_6|b|}{}^c R_{i_7i_8|c|}{}^d R_{i_9i_{10}|h|}{}^e R_{i_{11}i_{12}|e|}{}^f R_{i_{13}i_{14}|f|}{}^g R_{i_{15}i_{16}|g|}{}^h R_{i_{17}i_{18}|j|}{}^i R_{i_{19}i_{20}|i|}{}^j R_{i_{21}i_{22}|k|}{}^k R_{i_{23}i_{24}]l}{}^l +$$

$$+ 623{,}700\ R_{[i_1i_2|d|}{}^a R_{i_3i_4|a|}{}^b R_{i_5i_6|b|}{}^c R_{i_7i_8|c|}{}^d R_{i_9i_{10}|h|}{}^e R_{i_{11}i_{12}|e|}{}^f R_{i_{13}i_{14}|f|}{}^g R_{i_{15}i_{16}|g|}{}^h R_{i_{17}i_{18}|i|}{}^i R_{i_{19}i_{20}|j|}{}^j R_{i_{21}i_{22}|k|}{}^k R_{i_{23}i_{24}]l}{}^l +$$

$$+ 3{,}326{,}400\ R_{[i_1i_2|d|}{}^a R_{i_3i_4|a|}{}^b R_{i_5i_6|b|}{}^c R_{i_7i_8|c|}{}^d R_{i_9i_{10}|g|}{}^e R_{i_{11}i_{12}|e|}{}^f R_{i_{13}i_{14}|f|}{}^g R_{i_{15}i_{16}|j|}{}^h R_{i_{17}i_{18}|h|}{}^i R_{i_{19}i_{20}|i|}{}^j R_{i_{21}i_{22}|l|}{}^k R_{i_{23}i_{24}]k}{}^l -$$

$$- 3{,}326{,}400\ R_{[i_1i_2|d|}{}^a R_{i_3i_4|a|}{}^b R_{i_5i_6|b|}{}^c R_{i_7i_8|c|}{}^d R_{i_9i_{10}|g|}{}^e R_{i_{11}i_{12}|e|}{}^f R_{i_{13}i_{14}|f|}{}^g R_{i_{15}i_{16}|j|}{}^h R_{i_{17}i_{18}|h|}{}^i R_{i_{19}i_{20}|i|}{}^j R_{i_{21}i_{22}|k|}{}^k R_{i_{23}i_{24}]l}{}^l -$$

$$- 4{,}989{,}600\ R_{[i_1i_2|d|}{}^a R_{i_3i_4|a|}{}^b R_{i_5i_6|b|}{}^c R_{i_7i_8|c|}{}^d R_{i_9i_{10}|g|}{}^e R_{i_{11}i_{12}|e|}{}^f R_{i_{13}i_{14}|f|}{}^g R_{i_{15}i_{16}|i|}{}^h R_{i_{17}i_{18}|h|}{}^i R_{i_{19}i_{20}|k|}{}^j R_{i_{21}i_{22}|j|}{}^k R_{i_{23}i_{24}]l}{}^l +$$

$$+ 3{,}326{,}400\ R_{[i_1i_2|d|}{}^a R_{i_3i_4|a|}{}^b R_{i_5i_6|b|}{}^c R_{i_7i_8|c|}{}^d R_{i_9i_{10}|g|}{}^e R_{i_{11}i_{12}|e|}{}^f R_{i_{13}i_{14}|f|}{}^g R_{i_{15}i_{16}|i|}{}^h R_{i_{17}i_{18}|h|}{}^i R_{i_{19}i_{20}|j|}{}^j R_{i_{21}i_{22}|k|}{}^k R_{i_{23}i_{24}]l}{}^l -$$

$$- 332{,}640\ R_{[i_1i_2|d|}{}^a R_{i_3i_4|a|}{}^b R_{i_5i_6|b|}{}^c R_{i_7i_8|c|}{}^d R_{i_9i_{10}|g|}{}^e R_{i_{11}i_{12}|e|}{}^f R_{i_{13}i_{14}|f|}{}^g R_{i_{15}i_{16}|h|}{}^h R_{i_{17}i_{18}|i|}{}^i R_{i_{19}i_{20}|j|}{}^j R_{i_{21}i_{22}|k|}{}^k R_{i_{23}i_{24}]l}{}^l -$$

$$- 311{,}850\ R_{[i_1i_2|d|}{}^a R_{i_3i_4|a|}{}^b R_{i_5i_6|b|}{}^c R_{i_7i_8|c|}{}^d R_{i_9i_{10}|f|}{}^e R_{i_{11}i_{12}|e|}{}^f R_{i_{13}i_{14}|h|}{}^g R_{i_{15}i_{16}|g|}{}^h R_{i_{17}i_{18}|j|}{}^i R_{i_{19}i_{20}|i|}{}^j R_{i_{21}i_{22}|l|}{}^k R_{i_{23}i_{24}]k}{}^l +$$

$$+ 1{,}247{,}400\ R_{[i_1i_2|d|}{}^a R_{i_3i_4|a|}{}^b R_{i_5i_6|b|}{}^c R_{i_7i_8|c|}{}^d R_{i_9i_{10}|f|}{}^e R_{i_{11}i_{12}|e|}{}^f R_{i_{13}i_{14}|h|}{}^g R_{i_{15}i_{16}|g|}{}^h R_{i_{17}i_{18}|j|}{}^i R_{i_{19}i_{20}|i|}{}^j R_{i_{21}i_{22}|k|}{}^k R_{i_{23}i_{24}]l}{}^l -$$

$$- 623{,}700\ R_{[i_1i_2|d|}{}^a R_{i_3i_4|a|}{}^b R_{i_5i_6|b|}{}^c R_{i_7i_8|c|}{}^d R_{i_9i_{10}|f|}{}^e R_{i_{11}i_{12}|e|}{}^f R_{i_{13}i_{14}|h|}{}^g R_{i_{15}i_{16}|g|}{}^h R_{i_{17}i_{18}|i|}{}^i R_{i_{19}i_{20}|j|}{}^j R_{i_{21}i_{22}|k|}{}^k R_{i_{23}i_{24}]l}{}^l +$$

$$+ 83{,}160\ R_{[i_1i_2|d|}{}^a R_{i_3i_4|a|}{}^b R_{i_5i_6|b|}{}^c R_{i_7i_8|c|}{}^d R_{i_9i_{10}|f|}{}^e R_{i_{11}i_{12}|e|}{}^f R_{i_{13}i_{14}|g|}{}^g R_{i_{15}i_{16}|h|}{}^h R_{i_{17}i_{18}|i|}{}^i R_{i_{19}i_{20}|j|}{}^j R_{i_{21}i_{22}|k|}{}^k R_{i_{23}i_{24}]l}{}^l -$$

$$- 2970\ R_{[i_1i_2|d|}{}^a R_{i_3i_4|a|}{}^b R_{i_5i_6|b|}{}^c R_{i_7i_8|c|}{}^d R_{i_9i_{10}|e|}{}^e R_{i_{11}i_{12}|f|}{}^f R_{i_{13}i_{14}|g|}{}^g R_{i_{15}i_{16}|h|}{}^h R_{i_{17}i_{18}|i|}{}^i R_{i_{19}i_{20}|j|}{}^j R_{i_{21}i_{22}|k|}{}^k R_{i_{23}i_{24}]l}{}^l +$$

$$+ 246{,}400\ R_{[i_1i_2|c|}{}^a R_{i_3i_4|a|}{}^b R_{i_5i_6|b|}{}^c R_{i_7i_8|f|}{}^d R_{i_9i_{10}|d|}{}^e R_{i_{11}i_{12}|e|}{}^f R_{i_{13}i_{14}|i|}{}^g R_{i_{15}i_{16}|g|}{}^h R_{i_{17}i_{18}|h|}{}^i R_{i_{19}i_{20}|l|}{}^j R_{i_{21}i_{22}|j|}{}^k R_{i_{23}i_{24}]k}{}^l -$$

$$- 1{,}478{,}400\ R_{[i_1i_2|c|}{}^a R_{i_3i_4|a|}{}^b R_{i_5i_6|b|}{}^c R_{i_7i_8|f|}{}^d R_{i_9i_{10}|d|}{}^e R_{i_{11}i_{12}|e|}{}^f R_{i_{13}i_{14}|i|}{}^g R_{i_{15}i_{16}|g|}{}^h R_{i_{17}i_{18}|h|}{}^i R_{i_{19}i_{20}|k|}{}^j R_{i_{21}i_{22}|j|}{}^k R_{i_{23}i_{24}]l}{}^l +$$

$$+ 492{,}800\ R_{[i_1i_2|c|}{}^a R_{i_3i_4|a|}{}^b R_{i_5i_6|b|}{}^c R_{i_7i_8|f|}{}^d R_{i_9i_{10}|d|}{}^e R_{i_{11}i_{12}|e|}{}^f R_{i_{13}i_{14}|i|}{}^g R_{i_{15}i_{16}|g|}{}^h R_{i_{17}i_{18}|h|}{}^i R_{i_{19}i_{20}|j|}{}^j R_{i_{21}i_{22}|k|}{}^k R_{i_{23}i_{24}]l}{}^l -$$

$$- 554{,}400\ R_{[i_1i_2|c|}{}^a R_{i_3i_4|a|}{}^b R_{i_5i_6|b|}{}^c R_{i_7i_8|f|}{}^d R_{i_9i_{10}|d|}{}^e R_{i_{11}i_{12}|e|}{}^f R_{i_{13}i_{14}|h|}{}^g R_{i_{15}i_{16}|g|}{}^h R_{i_{17}i_{18}|j|}{}^i R_{i_{19}i_{20}|i|}{}^j R_{i_{21}i_{22}|l|}{}^k R_{i_{23}i_{24}]k}{}^l +$$

$$+ 1{,}663{,}200\ R_{[i_1i_2|c|}{}^a R_{i_3i_4|a|}{}^b R_{i_5i_6|b|}{}^c R_{i_7i_8|f|}{}^d R_{i_9i_{10}|d|}{}^e R_{i_{11}i_{12}|e|}{}^f R_{i_{13}i_{14}|h|}{}^g R_{i_{15}i_{16}|g|}{}^h R_{i_{17}i_{18}|j|}{}^i R_{i_{19}i_{20}|i|}{}^j R_{i_{21}i_{22}|k|}{}^k R_{i_{23}i_{24}]l}{}^l -$$

$$- 554{,}400\ R_{[i_1i_2|c|}{}^a R_{i_3i_4|a|}{}^b R_{i_5i_6|b|}{}^c R_{i_7i_8|f|}{}^d R_{i_9i_{10}|d|}{}^e R_{i_{11}i_{12}|e|}{}^f R_{i_{13}i_{14}|h|}{}^g R_{i_{15}i_{16}|g|}{}^h R_{i_{17}i_{18}|i|}{}^i R_{i_{19}i_{20}|j|}{}^j R_{i_{21}i_{22}|k|}{}^k R_{i_{23}i_{24}]l}{}^l +$$

$$+ 36{,}960\ R_{[i_1i_2|c|}{}^a R_{i_3i_4|a|}{}^b R_{i_5i_6|b|}{}^c R_{i_7i_8|f|}{}^d R_{i_9i_{10}|d|}{}^e R_{i_{11}i_{12}|e|}{}^f R_{i_{13}i_{14}|g|}{}^g R_{i_{15}i_{16}|h|}{}^h R_{i_{17}i_{18}|i|}{}^i R_{i_{19}i_{20}|j|}{}^j R_{i_{21}i_{22}|k|}{}^k R_{i_{23}i_{24}]l}{}^l +$$

$$+ 415{,}800\ R_{[i_1i_2|c|}{}^a R_{i_3i_4|a|}{}^b R_{i_5i_6|b|}{}^c R_{i_7i_8|e|}{}^d R_{i_9i_{10}|d|}{}^e R_{i_{11}i_{12}|f|}{}^f R_{i_{13}i_{14}|h|}{}^g R_{i_{15}i_{16}|i|}{}^h R_{i_{17}i_{18}|h|}{}^i R_{i_{19}i_{20}|k|}{}^j R_{i_{21}i_{22}|j|}{}^k R_{i_{23}i_{24}]l}{}^l -$$

$$- 554{,}400\ R_{[i_1i_2|c|}{}^a R_{i_3i_4|a|}{}^b R_{i_5i_6|b|}{}^c R_{i_7i_8|e|}{}^d R_{i_9i_{10}|d|}{}^e R_{i_{11}i_{12}|f|}{}^f R_{i_{13}i_{14}|f|}{}^g R_{i_{15}i_{16}|g|}{}^h R_{i_{17}i_{18}|h|}{}^i R_{i_{19}i_{20}|i|}{}^j R_{i_{21}i_{22}|k|}{}^k R_{i_{23}i_{24}]l}{}^l +$$

$$+ 166{,}320\ R_{[i_1i_2|c|}{}^a R_{i_3i_4|a|}{}^b R_{i_5i_6|b|}{}^c R_{i_7i_8|e|}{}^d R_{i_9i_{10}|d|}{}^e R_{i_{11}i_{12}|f|}{}^f R_{i_{13}i_{14}|g|}{}^g R_{i_{15}i_{16}|h|}{}^h R_{i_{17}i_{18}|i|}{}^i R_{i_{19}i_{20}|j|}{}^j R_{i_{21}i_{22}|k|}{}^k R_{i_{23}i_{24}]l}{}^l -$$

$$- 15{,}840\, R_{[i_1 i_2|c|}{}^a R_{i_3 i_4|a|}{}^b R_{i_5 i_6|b|}{}^c R_{i_7 i_8|e|}{}^d R_{i_9 i_{10}|d|}{}^e R_{i_{11}i_{12}|f|}{}^f R_{i_{13}i_{14}|g|}{}^g R_{i_{15}i_{16}|h|}{}^h R_{i_{17}i_{18}|i|}{}^i R_{i_{19}i_{20}|j|}{}^j R_{i_{21}i_{22}|k|}{}^k R_{i_{23}i_{24}]}{}^l +$$

$$+ 440\, R_{[i_1 i_2|c|}{}^a R_{i_3 i_4|a|}{}^b R_{i_5 i_6|b|}{}^c R_{i_7 i_8|d|}{}^d R_{i_9 i_{10}|e|}{}^e R_{i_{11}i_{12}|f|}{}^f R_{i_{13}i_{14}|g|}{}^g R_{i_{15}i_{16}|h|}{}^h R_{i_{17}i_{18}|i|}{}^i R_{i_{19}i_{20}|j|}{}^j R_{i_{21}i_{22}|k|}{}^k R_{i_{23}i_{24}]}{}^l +$$

$$+ 10{,}395\, R_{[i_1 i_2|b|}{}^a R_{i_3 i_4|a|}{}^b R_{i_5 i_6|d|}{}^c R_{i_7 i_8|c|}{}^d R_{i_9 i_{10}|e|}{}^e R_{i_{11}i_{12}|f|}{}^f R_{i_{13}i_{14}|h|}{}^g R_{i_{15}i_{16}|g|}{}^h R_{i_{17}i_{18}|i|}{}^i R_{i_{19}i_{20}|j|}{}^j R_{i_{21}i_{22}|l|}{}^k R_{i_{23}i_{24}]}{}^k -$$

$$- 62{,}370\, R_{[i_1 i_2|b|}{}^a R_{i_3 i_4|a|}{}^b R_{i_5 i_6|d|}{}^c R_{i_7 i_8|c|}{}^d R_{i_9 i_{10}|f|}{}^e R_{i_{11}i_{12}|e|}{}^f R_{i_{13}i_{14}|h|}{}^g R_{i_{15}i_{16}|g|}{}^h R_{i_{17}i_{18}|j|}{}^i R_{i_{19}i_{20}|i|}{}^j R_{i_{21}i_{22}|k|}{}^k R_{i_{23}i_{24}]}{}^l +$$

$$+ 51{,}975\, R_{[i_1 i_2|b|}{}^a R_{i_3 i_4|a|}{}^b R_{i_5 i_6|d|}{}^c R_{i_7 i_8|c|}{}^d R_{i_9 i_{10}|f|}{}^e R_{i_{11}i_{12}|e|}{}^f R_{i_{13}i_{14}|h|}{}^g R_{i_{15}i_{16}|g|}{}^h R_{i_{17}i_{18}|i|}{}^i R_{i_{19}i_{20}|j|}{}^j R_{i_{21}i_{22}|k|}{}^k R_{i_{23}i_{24}]}{}^l -$$

$$- 13{,}860\, R_{[i_1 i_2|b|}{}^a R_{i_3 i_4|a|}{}^b R_{i_5 i_6|d|}{}^c R_{i_7 i_8|c|}{}^d R_{i_9 i_{10}|e|}{}^e R_{i_{11}i_{12}|f|}{}^f R_{i_{13}i_{14}|g|}{}^g R_{i_{15}i_{16}|h|}{}^h R_{i_{17}i_{18}|i|}{}^i R_{i_{19}i_{20}|j|}{}^j R_{i_{21}i_{22}|k|}{}^k R_{i_{23}i_{24}]}{}^l +$$

$$+ 1485\, R_{[i_1 i_2|b|}{}^a R_{i_3 i_4|a|}{}^b R_{i_5 i_6|c|}{}^c R_{i_7 i_8|c|}{}^d R_{i_9 i_{10}|e|}{}^e R_{i_{11}i_{12}|f|}{}^f R_{i_{13}i_{14}|g|}{}^g R_{i_{15}i_{16}|h|}{}^h R_{i_{17}i_{18}|i|}{}^i R_{i_{19}i_{20}|j|}{}^j R_{i_{21}i_{22}|k|}{}^k R_{i_{23}i_{24}]}{}^l -$$

$$- 66\, R_{[i_1 i_2|b|}{}^a R_{i_3 i_4|a|}{}^b R_{i_5 i_6|c|}{}^c R_{i_7 i_8|d|}{}^d R_{i_9 i_{10}|e|}{}^e R_{i_{11}i_{12}|f|}{}^f R_{i_{13}i_{14}|g|}{}^g R_{i_{15}i_{16}|h|}{}^h R_{i_{17}i_{18}|i|}{}^i R_{i_{19}i_{20}|j|}{}^j R_{i_{21}i_{22}|k|}{}^k R_{i_{23}i_{24}]}{}^l +$$

$$+ R_{[i_1 i_2|a|}{}^a R_{i_3 i_4|b|}{}^b R_{i_5 i_6|c|}{}^c R_{i_7 i_8|d|}{}^d R_{i_9 i_{10}|e|}{}^e R_{i_{11}i_{12}|f|}{}^f R_{i_{13}i_{14}|g|}{}^g R_{i_{15}i_{16}|h|}{}^h R_{i_{17}i_{18}|i|}{}^i R_{i_{19}i_{20}|j|}{}^j R_{i_{21}i_{22}|k|}{}^k R_{i_{23}i_{24}]}{}^l )$$

$$= \frac{i^{12}}{2^{24}\pi^{12}12!}(- 39{,}916{,}800\, R_{[i_1 i_2|l|}{}^a R_{i_3 i_4|a|}{}^b R_{i_5 i_6|b|}{}^c R_{i_7 i_8|c|}{}^d R_{i_9 i_{10}|d|}{}^e R_{i_{11}i_{12}|e|}{}^f R_{i_{13}i_{14}|f|}{}^g R_{i_{15}i_{16}|g|}{}^h R_{i_{17}i_{18}|h|}{}^i R_{i_{19}i_{20}|i|}{}^j R_{i_{21}i_{22}|j|}{}^k R_{i_{23}i_{24}]}{}^l +$$

$$+ 43{,}545{,}600\, R_{[i_1 i_2|k|}{}^a R_{i_3 i_4|a|}{}^b R_{i_5 i_6|b|}{}^c R_{i_7 i_8|c|}{}^d R_{i_9 i_{10}|d|}{}^e R_{i_{11}i_{12}|e|}{}^f R_{i_{13}i_{14}|f|}{}^g R_{i_{15}i_{16}|g|}{}^h R_{i_{17}i_{18}|h|}{}^i R_{i_{19}i_{20}|i|}{}^j R_{i_{21}i_{22}|j|}{}^k V_{i_{23}i_{24}]} +$$

$$+ 23{,}950{,}080\, R_{[i_1 i_2|j|}{}^a R_{i_3 i_4|a|}{}^b R_{i_5 i_6|b|}{}^c R_{i_7 i_8|c|}{}^d R_{i_9 i_{10}|d|}{}^e R_{i_{11}i_{12}|e|}{}^f R_{i_{13}i_{14}|f|}{}^g R_{i_{15}i_{16}|g|}{}^h R_{i_{17}i_{18}|h|}{}^i R_{i_{19}i_{20}|i|}{}^j R_{i_{21}i_{22}|l|}{}^k R_{i_{23}i_{24}]}{}^l -$$

$$- 23{,}950{,}080\, R_{[i_1 i_2|j|}{}^a R_{i_3 i_4|a|}{}^b R_{i_5 i_6|b|}{}^c R_{i_7 i_8|c|}{}^d R_{i_9 i_{10}|d|}{}^e R_{i_{11}i_{12}|e|}{}^f R_{i_{13}i_{14}|f|}{}^g R_{i_{15}i_{16}|g|}{}^h R_{i_{17}i_{18}|h|}{}^i R_{i_{19}i_{20}|i|}{}^j V_{i_{21}i_{22}} V_{i_{23}i_{24}]} +$$

$$+ 17{,}740{,}800\, R_{[i_1 i_2|i|}{}^a R_{i_3 i_4|a|}{}^b R_{i_5 i_6|b|}{}^c R_{i_7 i_8|c|}{}^d R_{i_9 i_{10}|d|}{}^e R_{i_{11}i_{12}|e|}{}^f R_{i_{13}i_{14}|f|}{}^g R_{i_{15}i_{16}|g|}{}^h R_{i_{17}i_{18}|h|}{}^i R_{i_{19}i_{20}|l|}{}^j R_{i_{21}i_{22}|j|}{}^k R_{i_{23}i_{24}]}{}^l -$$

$$- 26{,}611{,}200\, R_{[i_1 i_2|i|}{}^a R_{i_3 i_4|a|}{}^b R_{i_5 i_6|b|}{}^c R_{i_7 i_8|c|}{}^d R_{i_9 i_{10}|d|}{}^e R_{i_{11}i_{12}|e|}{}^f R_{i_{13}i_{14}|f|}{}^g R_{i_{15}i_{16}|g|}{}^h R_{i_{17}i_{18}|h|}{}^i R_{i_{19}i_{20}|k|}{}^j R_{i_{21}i_{22}|j|}{}^k V_{i_{23}i_{24}]} +$$

$$+ 8{,}870{,}400\, R_{[i_1 i_2|i|}{}^a R_{i_3 i_4|a|}{}^b R_{i_5 i_6|b|}{}^c R_{i_7 i_8|c|}{}^d R_{i_9 i_{10}|d|}{}^e R_{i_{11}i_{12}|e|}{}^f R_{i_{13}i_{14}|f|}{}^g R_{i_{15}i_{16}|g|}{}^h R_{i_{17}i_{18}|h|}{}^i V_{i_{19}i_{20}} V_{i_{21}i_{22}} V_{i_{23}i_{24}]} +$$

$$+ 14{,}968{,}800\, R_{[i_1 i_2|h|}{}^a R_{i_3 i_4|a|}{}^b R_{i_5 i_6|b|}{}^c R_{i_7 i_8|c|}{}^d R_{i_9 i_{10}|d|}{}^e R_{i_{11}i_{12}|e|}{}^f R_{i_{13}i_{14}|f|}{}^g R_{i_{15}i_{16}|g|}{}^h R_{i_{17}i_{18}|l|}{}^i R_{i_{19}i_{20}|i|}{}^j R_{i_{21}i_{22}|j|}{}^k R_{i_{23}i_{24}]}{}^l -$$

$$- 19{,}958{,}400\, R_{[i_1 i_2|h|}{}^a R_{i_3 i_4|a|}{}^b R_{i_5 i_6|b|}{}^c R_{i_7 i_8|c|}{}^d R_{i_9 i_{10}|d|}{}^e R_{i_{11}i_{12}|e|}{}^f R_{i_{13}i_{14}|f|}{}^g R_{i_{15}i_{16}|g|}{}^h R_{i_{17}i_{18}|k|}{}^i R_{i_{19}i_{20}|i|}{}^j R_{i_{21}i_{22}|j|}{}^k V_{i_{23}i_{24}]} -$$

$$- 7{,}484{,}400\, R_{[i_1 i_2|h|}{}^a R_{i_3 i_4|a|}{}^b R_{i_5 i_6|b|}{}^c R_{i_7 i_8|c|}{}^d R_{i_9 i_{10}|d|}{}^e R_{i_{11}i_{12}|e|}{}^f R_{i_{13}i_{14}|f|}{}^g R_{i_{15}i_{16}|g|}{}^h R_{i_{17}i_{18}|j|}{}^i R_{i_{19}i_{20}|i|}{}^j R_{i_{21}i_{22}|l|}{}^k R_{i_{23}i_{24}]}{}^l +$$

$$+ 14{,}968{,}800\, R_{[i_1 i_2|h|}{}^a R_{i_3 i_4|a|}{}^b R_{i_5 i_6|b|}{}^c R_{i_7 i_8|c|}{}^d R_{i_9 i_{10}|d|}{}^e R_{i_{11}i_{12}|e|}{}^f R_{i_{13}i_{14}|f|}{}^g R_{i_{15}i_{16}|g|}{}^h R_{i_{17}i_{18}|j|}{}^i R_{i_{19}i_{20}|i|}{}^j V_{i_{21}i_{22}} V_{i_{23}i_{24}]} -$$

$$- 2{,}494{,}800\, R_{[i_1 i_2|h|}{}^a R_{i_3 i_4|a|}{}^b R_{i_5 i_6|b|}{}^c R_{i_7 i_8|c|}{}^d R_{i_9 i_{10}|d|}{}^e R_{i_{11}i_{12}|e|}{}^f R_{i_{13}i_{14}|f|}{}^g R_{i_{15}i_{16}|g|}{}^h V_{i_{17}i_{18}} V_{i_{19}i_{20}} V_{i_{21}i_{22}} V_{i_{23}i_{24}]} +$$

$$+ 13{,}685{,}760\, R_{[i_1 i_2|g|}{}^a R_{i_3 i_4|a|}{}^b R_{i_5 i_6|b|}{}^c R_{i_7 i_8|c|}{}^d R_{i_9 i_{10}|d|}{}^e R_{i_{11}i_{12}|e|}{}^f R_{i_{13}i_{14}|f|}{}^g R_{i_{15}i_{16}|l|}{}^h R_{i_{17}i_{18}|h|}{}^i R_{i_{19}i_{20}|i|}{}^j R_{i_{21}i_{22}|j|}{}^k R_{i_{23}i_{24}]}{}^l -$$

$$- 17{,}107{,}200\, R_{[i_1 i_2|g|}{}^a R_{i_3 i_4|a|}{}^b R_{i_5 i_6|b|}{}^c R_{i_7 i_8|c|}{}^d R_{i_9 i_{10}|d|}{}^e R_{i_{11}i_{12}|e|}{}^f R_{i_{13}i_{14}|f|}{}^g R_{i_{15}i_{16}|k|}{}^h R_{i_{17}i_{18}|h|}{}^i R_{i_{19}i_{20}|i|}{}^j R_{i_{21}i_{22}|j|}{}^k V_{i_{23}i_{24}]} -$$

$$- 11{,}404{,}800\, R_{[i_1 i_2|g|}{}^a R_{i_3 i_4|a|}{}^b R_{i_5 i_6|b|}{}^c R_{i_7 i_8|c|}{}^d R_{i_9 i_{10}|d|}{}^e R_{i_{11}i_{12}|e|}{}^f R_{i_{13}i_{14}|f|}{}^g R_{i_{15}i_{16}|j|}{}^h R_{i_{17}i_{18}|h|}{}^i R_{i_{19}i_{20}|i|}{}^j R_{i_{21}i_{22}|l|}{}^k R_{i_{23}i_{24}]}{}^l +$$

$$+ 11{,}404{,}800\, R_{[i_1 i_2|g|}{}^a R_{i_3 i_4|a|}{}^b R_{i_5 i_6|b|}{}^c R_{i_7 i_8|c|}{}^d R_{i_9 i_{10}|d|}{}^e R_{i_{11}i_{12}|e|}{}^f R_{i_{13}i_{14}|f|}{}^g R_{i_{15}i_{16}|j|}{}^h R_{i_{17}i_{18}|h|}{}^i R_{i_{19}i_{20}|i|}{}^j V_{i_{21}i_{22}} V_{i_{23}i_{24}]} +$$

$$+ 8{,}553{,}600\, R_{[i_1 i_2|g|}{}^a R_{i_3 i_4|a|}{}^b R_{i_5 i_6|b|}{}^c R_{i_7 i_8|c|}{}^d R_{i_9 i_{10}|d|}{}^e R_{i_{11}i_{12}|e|}{}^f R_{i_{13}i_{14}|f|}{}^g R_{i_{15}i_{16}|i|}{}^h R_{i_{17}i_{18}|h|}{}^i R_{i_{19}i_{20}|k|}{}^j R_{i_{21}i_{22}|j|}{}^k V_{i_{23}i_{24}]} -$$

$$- 5{,}702{,}400\, R_{[i_1 i_2|g|}{}^a R_{i_3 i_4|a|}{}^b R_{i_5 i_6|b|}{}^c R_{i_7 i_8|c|}{}^d R_{i_9 i_{10}|d|}{}^e R_{i_{11}i_{12}|e|}{}^f R_{i_{13}i_{14}|f|}{}^g R_{i_{15}i_{16}|i|}{}^h R_{i_{17}i_{18}|h|}{}^i V_{i_{19}i_{20}} V_{i_{21}i_{22}} V_{i_{23}i_{24}]} +$$

$$+ 570{,}240\, R_{[i_1 i_2|g|}{}^a R_{i_3 i_4|a|}{}^b R_{i_5 i_6|b|}{}^c R_{i_7 i_8|c|}{}^d R_{i_9 i_{10}|d|}{}^e R_{i_{11}i_{12}|e|}{}^f R_{i_{13}i_{14}|f|}{}^g V_{i_{15}i_{16}} V_{i_{17}i_{18}} V_{i_{19}i_{20}} V_{i_{21}i_{22}} V_{i_{23}i_{24}]} +$$

$$+ 6{,}652{,}800\, R_{[i_1 i_2|f|}{}^a R_{i_3 i_4|a|}{}^b R_{i_5 i_6|b|}{}^c R_{i_7 i_8|c|}{}^d R_{i_9 i_{10}|d|}{}^e R_{i_{11}i_{12}|e|}{}^f R_{i_{13}i_{14}|l|}{}^g R_{i_{15}i_{16}|g|}{}^h R_{i_{17}i_{18}|h|}{}^i R_{i_{19}i_{20}|i|}{}^j R_{i_{21}i_{22}|j|}{}^k R_{i_{23}i_{24}]}{}^l -$$

$$- 15{,}966{,}720\, R_{[i_1 i_2|f|}{}^a R_{i_3 i_4|a|}{}^b R_{i_5 i_6|b|}{}^c R_{i_7 i_8|c|}{}^d R_{i_9 i_{10}|d|}{}^e R_{i_{11}i_{12}|e|}{}^f R_{i_{13}i_{14}|k|}{}^g R_{i_{15}i_{16}|g|}{}^h R_{i_{17}i_{18}|h|}{}^i R_{i_{19}i_{20}|i|}{}^j R_{i_{21}i_{22}|j|}{}^k V_{i_{23}i_{24}]} -$$

$$- 9{,}979{,}200\, R_{[i_1 i_2|f|}{}^a R_{i_3 i_4|a|}{}^b R_{i_5 i_6|b|}{}^c R_{i_7 i_8|c|}{}^d R_{i_9 i_{10}|d|}{}^e R_{i_{11}i_{12}|e|}{}^f R_{i_{13}i_{14}|j|}{}^g R_{i_{15}i_{16}|g|}{}^h R_{i_{17}i_{18}|h|}{}^i R_{i_{19}i_{20}|i|}{}^j R_{i_{21}i_{22}|l|}{}^k R_{i_{23}i_{24}]}{}^l +$$

$$+ 9{,}979{,}200\, R_{[i_1 i_2|f|}{}^a R_{i_3 i_4|a|}{}^b R_{i_5 i_6|b|}{}^c R_{i_7 i_8|c|}{}^d R_{i_9 i_{10}|d|}{}^e R_{i_{11}i_{12}|e|}{}^f R_{i_{13}i_{14}|j|}{}^g R_{i_{15}i_{16}|g|}{}^h R_{i_{17}i_{18}|h|}{}^i R_{i_{19}i_{20}|i|}{}^j V_{i_{21}i_{22}} V_{i_{23}i_{24}]} -$$

$$- 4{,}435{,}200\, R_{[i_1 i_2|f|}{}^a R_{i_3 i_4|a|}{}^b R_{i_5 i_6|b|}{}^c R_{i_7 i_8|c|}{}^d R_{i_9 i_{10}|d|}{}^e R_{i_{11}i_{12}|e|}{}^f R_{i_{13}i_{14}|i|}{}^g R_{i_{15}i_{16}|g|}{}^h R_{i_{17}i_{18}|h|}{}^i R_{i_{19}i_{20}|l|}{}^j R_{i_{21}i_{22}|j|}{}^k R_{i_{23}i_{24}]}{}^l +$$

$$+ 13{,}305{,}600\, R_{[i_1 i_2|f|}{}^a R_{i_3 i_4|a|}{}^b R_{i_5 i_6|b|}{}^c R_{i_7 i_8|c|}{}^d R_{i_9 i_{10}|d|}{}^e R_{i_{11}i_{12}|e|}{}^f R_{i_{13}i_{14}|i|}{}^g R_{i_{15}i_{16}|g|}{}^h R_{i_{17}i_{18}|h|}{}^i R_{i_{19}i_{20}|k|}{}^j R_{i_{21}i_{22}|j|}{}^k V_{i_{23}i_{24}]} -$$

$$- 4{,}435{,}200\, R_{[i_1 i_2|f|}{}^a R_{i_3 i_4|a|}{}^b R_{i_5 i_6|b|}{}^c R_{i_7 i_8|c|}{}^d R_{i_9 i_{10}|d|}{}^e R_{i_{11}i_{12}|e|}{}^f R_{i_{13}i_{14}|i|}{}^g R_{i_{15}i_{16}|g|}{}^h R_{i_{17}i_{18}|h|}{}^i V_{i_{19}i_{20}} V_{i_{21}i_{22}} V_{i_{23}i_{24}]} +$$

$$+ 1{,}663{,}200\, R_{[i_1 i_2|f|}{}^a R_{i_3 i_4|a|}{}^b R_{i_5 i_6|b|}{}^c R_{i_7 i_8|c|}{}^d R_{i_9 i_{10}|d|}{}^e R_{i_{11}i_{12}|e|}{}^f R_{i_{13}i_{14}|h|}{}^g R_{i_{15}i_{16}|g|}{}^h R_{i_{17}i_{18}|j|}{}^i R_{i_{19}i_{20}|i|}{}^j R_{i_{21}i_{22}|l|}{}^k R_{i_{23}i_{24}]}{}^l -$$

$$- 4{,}989{,}600\, R_{[i_1 i_2|f|}{}^a R_{i_3 i_4|a|}{}^b R_{i_5 i_6|b|}{}^c R_{i_7 i_8|c|}{}^d R_{i_9 i_{10}|d|}{}^e R_{i_{11}i_{12}|e|}{}^f R_{i_{13}i_{14}|h|}{}^g R_{i_{15}i_{16}|g|}{}^h R_{i_{17}i_{18}|j|}{}^i R_{i_{19}i_{20}|i|}{}^j V_{i_{21}i_{22}} V_{i_{23}i_{24}]} +$$

$$+ 1{,}663{,}200\, R_{[i_1 i_2|f|}{}^a R_{i_3 i_4|a|}{}^b R_{i_5 i_6|b|}{}^c R_{i_7 i_8|c|}{}^d R_{i_9 i_{10}|d|}{}^e R_{i_{11}i_{12}|e|}{}^f R_{i_{13}i_{14}|h|}{}^g R_{i_{15}i_{16}|g|}{}^h V_{i_{17}i_{18}} V_{i_{19}i_{20}} V_{i_{21}i_{22}} V_{i_{23}i_{24}]} -$$

$$- 110{,}880\, R_{[i_1 i_2|f|}{}^a R_{i_3 i_4|a|}{}^b R_{i_5 i_6|b|}{}^c R_{i_7 i_8|c|}{}^d R_{i_9 i_{10}|d|}{}^e R_{i_{11}i_{12}|e|}{}^f V_{i_{13}i_{14}} V_{i_{15}i_{16}} V_{i_{17}i_{18}} V_{i_{19}i_{20}} V_{i_{21}i_{22}} V_{i_{23}i_{24}]} -$$

$$- 4{,}790{,}016\, R_{[i_1 i_2|e|}{}^a R_{i_3 i_4|a|}{}^b R_{i_5 i_6|b|}{}^c R_{i_7 i_8|c|}{}^d R_{i_9 i_{10}|d|}{}^e R_{i_{11}i_{12}|j|}{}^f R_{i_{13}i_{14}|f|}{}^g R_{i_{15}i_{16}|g|}{}^h R_{i_{17}i_{18}|h|}{}^i R_{i_{19}i_{20}|i|}{}^j R_{i_{21}i_{22}|l|}{}^k R_{i_{23}i_{24}]}{}^l +$$

$$+ 4{,}790{,}016\, R_{[i_1 i_2|e|}{}^a R_{i_3 i_4|a|}{}^b R_{i_5 i_6|b|}{}^c R_{i_7 i_8|c|}{}^d R_{i_9 i_{10}|d|}{}^e R_{i_{11}i_{12}|j|}{}^f R_{i_{13}i_{14}|f|}{}^g R_{i_{15}i_{16}|g|}{}^h R_{i_{17}i_{18}|h|}{}^i R_{i_{19}i_{20}|i|}{}^j V_{i_{21}i_{22}} V_{i_{23}i_{24}]} -$$

$$- 7{,}983{,}360\, R_{[i_1i_2|e|}{}^a R_{i_3i_4|a|}{}^b R_{i_5i_6|b|}{}^c R_{i_7i_8|c|}{}^d R_{i_9i_{10}|d|}{}^e R_{i_{11}i_{12}|i|}{}^f R_{i_{13}i_{14}|f|}{}^g R_{i_{15}i_{16}|g|}{}^h R_{i_{17}i_{18}|h|}{}^i R_{i_{19}i_{20}|l|}{}^j R_{i_{21}i_{22}|j|}{}^k R_{i_{23}i_{24}]k}{}^l +$$

$$+ 11{,}975{,}040\, R_{[i_1i_2|e|}{}^a R_{i_3i_4|a|}{}^b R_{i_5i_6|b|}{}^c R_{i_7i_8|c|}{}^d R_{i_9i_{10}|d|}{}^e R_{i_{11}i_{12}|i|}{}^f R_{i_{13}i_{14}|f|}{}^g R_{i_{15}i_{16}|g|}{}^h R_{i_{17}i_{18}|h|}{}^i R_{i_{19}i_{20}|k|}{}^j R_{i_{21}i_{22}|j|}{}^k V_{i_{23}i_{24}]} -$$

$$- 3{,}991{,}680\, R_{[i_1i_2|e|}{}^a R_{i_3i_4|a|}{}^b R_{i_5i_6|b|}{}^c R_{i_7i_8|c|}{}^d R_{i_9i_{10}|d|}{}^e R_{i_{11}i_{12}|i|}{}^f R_{i_{13}i_{14}|f|}{}^g R_{i_{15}i_{16}|g|}{}^h R_{i_{17}i_{18}|h|}{}^i V_{i_{19}i_{20}} V_{i_{21}i_{22}} V_{i_{23}i_{24}]} +$$

$$+ 5{,}322{,}240\, R_{[i_1i_2|e|}{}^a R_{i_3i_4|a|}{}^b R_{i_5i_6|b|}{}^c R_{i_7i_8|c|}{}^d R_{i_9i_{10}|d|}{}^e R_{i_{11}i_{12}|h|}{}^f R_{i_{13}i_{14}|f|}{}^g R_{i_{15}i_{16}|g|}{}^h R_{i_{17}i_{18}|k|}{}^i R_{i_{19}i_{20}|i|}{}^j R_{i_{21}i_{22}]}{}^k V_{i_{23}i_{24}]} +$$

$$+ 3{,}991{,}680\, R_{[i_1i_2|e|}{}^a R_{i_3i_4|a|}{}^b R_{i_5i_6|b|}{}^c R_{i_7i_8|c|}{}^d R_{i_9i_{10}|d|}{}^e R_{i_{11}i_{12}|h|}{}^f R_{i_{13}i_{14}|f|}{}^g R_{i_{15}i_{16}|g|}{}^h R_{i_{17}i_{18}|j|}{}^i R_{i_{19}i_{20}|i|}{}^j R_{i_{21}i_{22}|l|}{}^k R_{i_{23}i_{24}]k}{}^l -$$

$$- 7{,}983{,}360\, R_{[i_1i_2|e|}{}^a R_{i_3i_4|a|}{}^b R_{i_5i_6|b|}{}^c R_{i_7i_8|c|}{}^d R_{i_9i_{10}|d|}{}^e R_{i_{11}i_{12}|h|}{}^f R_{i_{13}i_{14}|f|}{}^g R_{i_{15}i_{16}|g|}{}^h R_{i_{17}i_{18}|j|}{}^i R_{i_{19}i_{20}|i|}{}^j V_{i_{21}i_{22}} V_{i_{23}i_{24}]} +$$

$$+ 1{,}330{,}560\, R_{[i_1i_2|e|}{}^a R_{i_3i_4|a|}{}^b R_{i_5i_6|b|}{}^c R_{i_7i_8|c|}{}^d R_{i_9i_{10}|d|}{}^e R_{i_{11}i_{12}|f|}{}^f R_{i_{13}i_{14}|f|}{}^g R_{i_{15}i_{16}|g|}{}^h V_{i_{17}i_{18}} V_{i_{19}i_{20}} V_{i_{21}i_{22}} V_{i_{23}i_{24}]} -$$

$$- 1{,}995{,}840\, R_{[i_1i_2|e|}{}^a R_{i_3i_4|a|}{}^b R_{i_5i_6|b|}{}^c R_{i_7i_8|c|}{}^d R_{i_9i_{10}|d|}{}^e R_{i_{11}i_{12}|g|}{}^f R_{i_{13}i_{14}|f|}{}^g R_{i_{15}i_{16}|i|}{}^h R_{i_{17}i_{18}|h|}{}^i R_{i_{19}i_{20}|k|}{}^j R_{i_{21}i_{22}|j|}{}^k V_{i_{23}i_{24}]} +$$

$$+ 1{,}995{,}840\, R_{[i_1i_2|e|}{}^a R_{i_3i_4|a|}{}^b R_{i_5i_6|b|}{}^c R_{i_7i_8|c|}{}^d R_{i_9i_{10}|d|}{}^e R_{i_{11}i_{12}|g|}{}^f R_{i_{13}i_{14}|f|}{}^g R_{i_{15}i_{16}|i|}{}^h R_{i_{17}i_{18}|h|}{}^i V_{i_{19}i_{20}} V_{i_{21}i_{22}} V_{i_{23}i_{24}]} -$$

$$- 399{,}168\, R_{[i_1i_2|e|}{}^a R_{i_3i_4|a|}{}^b R_{i_5i_6|b|}{}^c R_{i_7i_8|c|}{}^d R_{i_9i_{10}|d|}{}^e R_{i_{11}i_{12}|g|}{}^f R_{i_{13}i_{14}|f|}{}^g V_{i_{15}i_{16}} V_{i_{17}i_{18}} V_{i_{19}i_{20}} V_{i_{21}i_{22}} V_{i_{23}i_{24}]} +$$

$$+ 19{,}008\, R_{[i_1i_2|e|}{}^a R_{i_3i_4|a|}{}^b R_{i_5i_6|b|}{}^c R_{i_7i_8|c|}{}^d R_{i_9i_{10}|d|}{}^e V_{i_{11}i_{12}} V_{i_{13}i_{14}} V_{i_{15}i_{16}} V_{i_{17}i_{18}} V_{i_{19}i_{20}} V_{i_{21}i_{22}} V_{i_{23}i_{24}]} -$$

$$- 1{,}247{,}400\, R_{[i_1i_2|d|}{}^a R_{i_3i_4|a|}{}^b R_{i_5i_6|b|}{}^c R_{i_7i_8|c|}{}^d R_{i_9i_{10}|h|}{}^e R_{i_{11}i_{12}|e|}{}^f R_{i_{13}i_{14}|f|}{}^g R_{i_{15}i_{16}|g|}{}^h R_{i_{17}i_{18}|l|}{}^i R_{i_{19}i_{20}|i|}{}^j R_{i_{21}i_{22}|j|}{}^k R_{i_{23}i_{24}]k}{}^l +$$

$$+ 4{,}989{,}600\, R_{[i_1i_2|d|}{}^a R_{i_3i_4|a|}{}^b R_{i_5i_6|b|}{}^c R_{i_7i_8|c|}{}^d R_{i_9i_{10}|h|}{}^e R_{i_{11}i_{12}|e|}{}^f R_{i_{13}i_{14}|f|}{}^g R_{i_{15}i_{16}|g|}{}^h R_{i_{17}i_{18}|k|}{}^i R_{i_{19}i_{20}|i|}{}^j R_{i_{21}i_{22}|j|}{}^k V_{i_{23}i_{24}]} +$$

$$+ 1{,}871{,}100\, R_{[i_1i_2|d|}{}^a R_{i_3i_4|a|}{}^b R_{i_5i_6|b|}{}^c R_{i_7i_8|c|}{}^d R_{i_9i_{10}|h|}{}^e R_{i_{11}i_{12}|e|}{}^f R_{i_{13}i_{14}|f|}{}^g R_{i_{15}i_{16}|g|}{}^h R_{i_{17}i_{18}|j|}{}^i R_{i_{19}i_{20}|i|}{}^j R_{i_{21}i_{22}|l|}{}^k R_{i_{23}i_{24}]k}{}^l -$$

$$- 3{,}742{,}200\, R_{[i_1i_2|d|}{}^a R_{i_3i_4|a|}{}^b R_{i_5i_6|b|}{}^c R_{i_7i_8|c|}{}^d R_{i_9i_{10}|h|}{}^e R_{i_{11}i_{12}|e|}{}^f R_{i_{13}i_{14}|f|}{}^g R_{i_{15}i_{16}|g|}{}^h R_{i_{17}i_{18}|j|}{}^i R_{i_{19}i_{20}|i|}{}^j V_{i_{21}i_{22}} V_{i_{23}i_{24}]} +$$

$$+ 623{,}700\, R_{[i_1i_2|d|}{}^a R_{i_3i_4|a|}{}^b R_{i_5i_6|b|}{}^c R_{i_7i_8|c|}{}^d R_{i_9i_{10}|h|}{}^e R_{i_{11}i_{12}|e|}{}^f R_{i_{13}i_{14}|f|}{}^g R_{i_{15}i_{16}|g|}{}^h V_{i_{17}i_{18}} V_{i_{19}i_{20}} V_{i_{21}i_{22}} V_{i_{23}i_{24}]} +$$

$$+ 3{,}326{,}400\, R_{[i_1i_2|d|}{}^a R_{i_3i_4|a|}{}^b R_{i_5i_6|b|}{}^c R_{i_7i_8|c|}{}^d R_{i_9i_{10}|g|}{}^e R_{i_{11}i_{12}|e|}{}^f R_{i_{13}i_{14}|f|}{}^g R_{i_{15}i_{16}|g|}{}^h R_{i_{17}i_{18}|h|}{}^i R_{i_{19}i_{20}|i|}{}^j R_{i_{21}i_{22}|l|}{}^k R_{i_{23}i_{24}]k}{}^l -$$

$$- 3{,}326{,}400\, R_{[i_1i_2|d|}{}^a R_{i_3i_4|a|}{}^b R_{i_5i_6|b|}{}^c R_{i_7i_8|c|}{}^d R_{i_9i_{10}|g|}{}^e R_{i_{11}i_{12}|e|}{}^f R_{i_{13}i_{14}|f|}{}^g R_{i_{15}i_{16}|g|}{}^h R_{i_{17}i_{18}|h|}{}^i R_{i_{19}i_{20}|i|}{}^j V_{i_{21}i_{22}} V_{i_{23}i_{24}]} -$$

$$- 4{,}989{,}600\, R_{[i_1i_2|d|}{}^a R_{i_3i_4|a|}{}^b R_{i_5i_6|b|}{}^c R_{i_7i_8|c|}{}^d R_{i_9i_{10}|g|}{}^e R_{i_{11}i_{12}|e|}{}^f R_{i_{13}i_{14}|f|}{}^g R_{i_{15}i_{16}|i|}{}^h R_{i_{17}i_{18}|h|}{}^i R_{i_{19}i_{20}|k|}{}^j R_{i_{21}i_{22}|j|}{}^k V_{i_{23}i_{24}]} +$$

$$+ 3{,}326{,}400\, R_{[i_1i_2|d|}{}^a R_{i_3i_4|a|}{}^b R_{i_5i_6|b|}{}^c R_{i_7i_8|c|}{}^d R_{i_9i_{10}|g|}{}^e R_{i_{11}i_{12}|e|}{}^f R_{i_{13}i_{14}|f|}{}^g R_{i_{15}i_{16}|i|}{}^h R_{i_{17}i_{18}|h|}{}^i V_{i_{19}i_{20}} V_{i_{21}i_{22}} V_{i_{23}i_{24}]} -$$

$$- 332{,}640\, R_{[i_1i_2|d|}{}^a R_{i_3i_4|a|}{}^b R_{i_5i_6|b|}{}^c R_{i_7i_8|c|}{}^d R_{i_9i_{10}|g|}{}^e R_{i_{11}i_{12}|e|}{}^f R_{i_{13}i_{14}|f|}{}^g V_{i_{15}i_{16}} V_{i_{17}i_{18}} V_{i_{19}i_{20}} V_{i_{21}i_{22}} V_{i_{23}i_{24}]} -$$

$$- 311{,}850\, R_{[i_1i_2|d|}{}^a R_{i_3i_4|a|}{}^b R_{i_5i_6|b|}{}^c R_{i_7i_8|c|}{}^d R_{i_9i_{10}|f|}{}^e R_{i_{11}i_{12}|e|}{}^f R_{i_{13}i_{14}|h|}{}^g R_{i_{15}i_{16}|g|}{}^h R_{i_{17}i_{18}|j|}{}^i R_{i_{19}i_{20}|i|}{}^j R_{i_{21}i_{22}|l|}{}^k R_{i_{23}i_{24}]k}{}^l +$$

$$+ 1{,}247{,}400\, R_{[i_1i_2|d|}{}^a R_{i_3i_4|a|}{}^b R_{i_5i_6|b|}{}^c R_{i_7i_8|c|}{}^d R_{i_9i_{10}|f|}{}^e R_{i_{11}i_{12}|e|}{}^f R_{i_{13}i_{14}|h|}{}^g R_{i_{15}i_{16}|g|}{}^h R_{i_{17}i_{18}|j|}{}^i R_{i_{19}i_{20}|i|}{}^j V_{i_{21}i_{22}} V_{i_{23}i_{24}]} -$$

$$- 623{,}700\, R_{[i_1i_2|d|}{}^a R_{i_3i_4|a|}{}^b R_{i_5i_6|b|}{}^c R_{i_7i_8|c|}{}^d R_{i_9i_{10}|f|}{}^e R_{i_{11}i_{12}|e|}{}^f R_{i_{13}i_{14}|h|}{}^g R_{i_{15}i_{16}|g|}{}^h V_{i_{17}i_{18}} V_{i_{19}i_{20}} V_{i_{21}i_{22}} V_{i_{23}i_{24}]} +$$

$$+ 83{,}160\, R_{[i_1i_2|d|}{}^a R_{i_3i_4|a|}{}^b R_{i_5i_6|b|}{}^c R_{i_7i_8|c|}{}^d R_{i_9i_{10}|f|}{}^e R_{i_{11}i_{12}|e|}{}^f V_{i_{13}i_{14}} V_{i_{15}i_{16}} V_{i_{17}i_{18}} V_{i_{19}i_{20}} V_{i_{21}i_{22}} V_{i_{23}i_{24}]} -$$

$$- 2970\, R_{[i_1i_2|d|}{}^a R_{i_3i_4|a|}{}^b R_{i_5i_6|b|}{}^c R_{i_7i_8|c|}{}^d V_{i_9i_{10}} V_{i_{11}i_{12}} V_{i_{13}i_{14}} V_{i_{15}i_{16}} V_{i_{17}i_{18}} V_{i_{19}i_{20}} V_{i_{21}i_{22}} V_{i_{23}i_{24}]} +$$

$$+ 246{,}400\, R_{[i_1i_2|c|}{}^a R_{i_3i_4|a|}{}^b R_{i_5i_6|b|}{}^c R_{i_7i_8|f|}{}^d R_{i_9i_{10}|d|}{}^e R_{i_{11}i_{12}|e|}{}^f R_{i_{13}i_{14}|i|}{}^g R_{i_{15}i_{16}|g|}{}^h R_{i_{17}i_{18}|h|}{}^i R_{i_{19}i_{20}|l|}{}^j R_{i_{21}i_{22}|j|}{}^k R_{i_{23}i_{24}]k}{}^l -$$

$$- 1{,}478{,}400\, R_{[i_1i_2|c|}{}^a R_{i_3i_4|a|}{}^b R_{i_5i_6|b|}{}^c R_{i_7i_8|f|}{}^d R_{i_9i_{10}|d|}{}^e R_{i_{11}i_{12}|e|}{}^f R_{i_{13}i_{14}|i|}{}^g R_{i_{15}i_{16}|g|}{}^h R_{i_{17}i_{18}|h|}{}^i R_{i_{19}i_{20}|k|}{}^j R_{i_{21}i_{22}|j|}{}^k V_{i_{23}i_{24}]} +$$

$$+ 492{,}800\, R_{[i_1i_2|c|}{}^a R_{i_3i_4|a|}{}^b R_{i_5i_6|b|}{}^c R_{i_7i_8|f|}{}^d R_{i_9i_{10}|d|}{}^e R_{i_{11}i_{12}|e|}{}^f R_{i_{13}i_{14}|i|}{}^g R_{i_{15}i_{16}|g|}{}^h R_{i_{17}i_{18}|h|}{}^i V_{i_{19}i_{20}} V_{i_{21}i_{22}} V_{i_{23}i_{24}]} -$$

$$- 554{,}400\, R_{[i_1i_2|c|}{}^a R_{i_3i_4|a|}{}^b R_{i_5i_6|b|}{}^c R_{i_7i_8|f|}{}^d R_{i_9i_{10}|d|}{}^e R_{i_{11}i_{12}|e|}{}^f R_{i_{13}i_{14}|h|}{}^g R_{i_{15}i_{16}|g|}{}^h R_{i_{17}i_{18}|j|}{}^i R_{i_{19}i_{20}|i|}{}^j R_{i_{21}i_{22}|l|}{}^k R_{i_{23}i_{24}]k}{}^l +$$

$$+ 1{,}663{,}200\, R_{[i_1i_2|c|}{}^a R_{i_3i_4|a|}{}^b R_{i_5i_6|b|}{}^c R_{i_7i_8|f|}{}^d R_{i_9i_{10}|d|}{}^e R_{i_{11}i_{12}|e|}{}^f R_{i_{13}i_{14}|h|}{}^g R_{i_{15}i_{16}|g|}{}^h R_{i_{17}i_{18}|j|}{}^i R_{i_{19}i_{20}|i|}{}^j V_{i_{21}i_{22}} V_{i_{23}i_{24}]} -$$

$$- 554{,}400\, R_{[i_1i_2|c|}{}^a R_{i_3i_4|a|}{}^b R_{i_5i_6|b|}{}^c R_{i_7i_8|f|}{}^d R_{i_9i_{10}|d|}{}^e R_{i_{11}i_{12}|e|}{}^f R_{i_{13}i_{14}|h|}{}^g R_{i_{15}i_{16}|g|}{}^h V_{i_{17}i_{18}} V_{i_{19}i_{20}} V_{i_{21}i_{22}} V_{i_{23}i_{24}]} +$$

$$+ 36{,}960\, R_{[i_1i_2|c|}{}^a R_{i_3i_4|a|}{}^b R_{i_5i_6|b|}{}^c R_{i_7i_8|f|}{}^d R_{i_9i_{10}|d|}{}^e R_{i_{11}i_{12}|e|}{}^f V_{i_{13}i_{14}} V_{i_{15}i_{16}} V_{i_{17}i_{18}} V_{i_{19}i_{20}} V_{i_{21}i_{22}} V_{i_{23}i_{24}]} +$$

$$+ 415{,}800\, R_{[i_1i_2|c|}{}^a R_{i_3i_4|a|}{}^b R_{i_5i_6|b|}{}^c R_{i_7i_8|e|}{}^d R_{i_9i_{10}|d|}{}^e R_{i_{11}i_{12}|g|}{}^f R_{i_{13}i_{14}|f|}{}^g R_{i_{15}i_{16}|i|}{}^h R_{i_{17}i_{18}|h|}{}^i R_{i_{19}i_{20}|k|}{}^j R_{i_{21}i_{22}|j|}{}^k V_{i_{23}i_{24}]} -$$

$$- 554{,}400\, R_{[i_1i_2|c|}{}^a R_{i_3i_4|a|}{}^b R_{i_5i_6|b|}{}^c R_{i_7i_8|e|}{}^d R_{i_9i_{10}|d|}{}^e R_{i_{11}i_{12}|g|}{}^f R_{i_{13}i_{14}|f|}{}^g R_{i_{15}i_{16}|g|}{}^h R_{i_{17}i_{18}|h|}{}^i V_{i_{19}i_{20}} V_{i_{21}i_{22}} V_{i_{23}i_{24}]} +$$

$$+ 166{,}320\, R_{[i_1i_2|c|}{}^a R_{i_3i_4|a|}{}^b R_{i_5i_6|b|}{}^c R_{i_7i_8|e|}{}^d R_{i_9i_{10}|d|}{}^e R_{i_{11}i_{12}|e|}{}^f R_{i_{13}i_{14}|f|}{}^g V_{i_{15}i_{16}} V_{i_{17}i_{18}} V_{i_{19}i_{20}} V_{i_{21}i_{22}} V_{i_{23}i_{24}]} -$$

$$- 15{,}840\, R_{[i_1i_2|c|}{}^a R_{i_3i_4|a|}{}^b R_{i_5i_6|b|}{}^c R_{i_7i_8|e|}{}^d R_{i_9i_{10}|d|}{}^e V_{i_{11}i_{12}} V_{i_{13}i_{14}} V_{i_{15}i_{16}} V_{i_{17}i_{18}} V_{i_{19}i_{20}} V_{i_{21}i_{22}} V_{i_{23}i_{24}]} +$$

$$+ 440\, R_{[i_1i_2|c|}{}^a R_{i_3i_4|a|}{}^b R_{i_5i_6|b|}{}^c V_{i_7i_8} V_{i_9i_{10}} V_{i_{11}i_{12}} V_{i_{13}i_{14}} V_{i_{15}i_{16}} V_{i_{17}i_{18}} V_{i_{19}i_{20}} V_{i_{21}i_{22}} V_{i_{23}i_{24}]} +$$

$$+ 10{,}395\, R_{[i_1i_2|b|}{}^a R_{i_3i_4|a|}{}^b R_{i_5i_6|d|}{}^c R_{i_7i_8|c|}{}^d R_{i_9i_{10}|f|}{}^e R_{i_{11}i_{12}|e|}{}^f R_{i_{13}i_{14}|h|}{}^g R_{i_{15}i_{16}|g|}{}^h R_{i_{17}i_{18}|j|}{}^i R_{i_{19}i_{20}|i|}{}^j R_{i_{21}i_{22}|l|}{}^k R_{i_{23}i_{24}]k}{}^l -$$

$$- 62{,}370\, R_{[i_1i_2|b|}{}^a R_{i_3i_4|a|}{}^b R_{i_5i_6|d|}{}^c R_{i_7i_8|c|}{}^d R_{i_9i_{10}|f|}{}^e R_{i_{11}i_{12}|e|}{}^f R_{i_{13}i_{14}|h|}{}^g R_{i_{15}i_{16}|g|}{}^h R_{i_{17}i_{18}|j|}{}^i R_{i_{19}i_{20}|i|}{}^j V_{i_{21}i_{22}} V_{i_{23}i_{24}]} +$$

$$+ 51{,}975\, R_{[i_1i_2|b|}{}^a R_{i_3i_4|a|}{}^b R_{i_5i_6|d|}{}^c R_{i_7i_8|c|}{}^d R_{i_9i_{10}|f|}{}^e R_{i_{11}i_{12}|e|}{}^f R_{i_{13}i_{14}|h|}{}^g R_{i_{15}i_{16}|g|}{}^h V_{i_{17}i_{18}} V_{i_{19}i_{20}} V_{i_{21}i_{22}} V_{i_{23}i_{24}]} -$$

$$- 13{,}860\, R_{[i_1i_2|b|}{}^a\, R_{i_3i_4|a|}{}^b\, R_{i_5i_6|d|}{}^c\, R_{i_7i_8|c|}{}^d\, R_{i_9i_{10}|f|}{}^e\, R_{i_{11}i_{12}|e|}{}^f\, V_{i_{13}i_{14}}\, V_{i_{15}i_{16}}\, V_{i_{17}i_{18}}\, V_{i_{19}i_{20}}\, V_{i_{21}i_{22}}\, V_{i_{23}i_{24}}] +$$

$$+ 1485\, R_{[i_1i_2|b|}{}^a\, R_{i_3i_4|a|}{}^b\, R_{i_5i_6|d|}{}^c\, R_{i_7i_8|c|}{}^d\, V_{i_9i_{10}}\, V_{i_{11}i_{12}}\, V_{i_{13}i_{14}}\, V_{i_{15}i_{16}}\, V_{i_{17}i_{18}}\, V_{i_{19}i_{20}}\, V_{i_{21}i_{22}}\, V_{i_{23}i_{24}}] -$$

$$- 66\, R_{[i_1i_2|b|}{}^a\, R_{i_3i_4|a|}{}^b\, V_{i_5i_6}\, V_{i_7i_8}\, V_{i_9i_{10}}\, V_{i_{11}i_{12}}\, V_{i_{13}i_{14}}\, V_{i_{15}i_{16}}\, V_{i_{17}i_{18}}\, V_{i_{19}i_{20}}\, V_{i_{21}i_{22}}\, V_{i_{23}i_{24}}] +$$

$$+ V_{[i_1i_2}\, V_{i_3i_4}\, V_{i_5i_6}\, V_{i_7i_8}\, V_{i_9i_{10}}\, V_{i_{11}i_{12}}\, V_{i_{13}i_{14}}\, V_{i_{15}i_{16}}\, V_{i_{17}i_{18}}\, V_{i_{19}i_{20}}\, V_{i_{21}i_{22}}\, V_{i_{23}i_{24}}])$$

$$= \frac{i^{12}}{2^{12}\pi^{12}12!}\Big( -\, 39{,}916{,}800\, P^{(24)}{}_{i_1i_2i_3i_4i_5i_6i_7i_8i_9i_{10}i_{11}i_{12}i_{13}i_{14}i_{15}i_{16}i_{17}i_{18}i_{19}i_{20}i_{21}i_{22}i_{23}i_{24}} +$$

$$+ 43{,}545{,}600\, P^{(22)}{}_{[i_1i_2i_3i_4i_5i_6i_7i_8i_9i_{10}i_{11}i_{12}i_{13}i_{14}i_{15}i_{16}i_{17}i_{18}i_{19}i_{20}i_{21}i_{22}}\, P^{(2)}{}_{i_{23}i_{24}]} +$$

$$+ 23{,}950{,}080\, P^{(20)}{}_{[i_1i_2i_3i_4i_5i_6i_7i_8i_9i_{10}i_{11}i_{12}i_{13}i_{14}i_{15}i_{16}i_{17}i_{18}i_{19}i_{20}}\, P^{(4)}{}_{i_{21}i_{22}i_{23}i_{24}]} -$$

$$- 23{,}950{,}080\, P^{(20)}{}_{[i_1i_2i_3i_4i_5i_6i_7i_8i_9i_{10}i_{11}i_{12}i_{13}i_{14}i_{15}i_{16}i_{17}i_{18}i_{19}i_{20}}\, P^{(2)}{}_{i_{21}i_{22}}\, P^{(2)}{}_{i_{23}i_{24}]} +$$

$$+ 17{,}740{,}800\, P^{(18)}{}_{[i_1i_2i_3i_4i_5i_6i_7i_8i_9i_{10}i_{11}i_{12}i_{13}i_{14}i_{15}i_{16}i_{17}i_{18}}\, P^{(6)}{}_{i_{19}i_{20}i_{21}i_{22}i_{23}i_{24}]} -$$

$$- 26{,}611{,}200\, P^{(18)}{}_{[i_1i_2i_3i_4i_5i_6i_7i_8i_9i_{10}i_{11}i_{12}i_{13}i_{14}i_{15}i_{16}i_{17}i_{18}}\, P^{(4)}{}_{i_{19}i_{20}i_{21}i_{22}}\, P^{(2)}{}_{i_{23}i_{24}]} +$$

$$+ 8{,}870{,}400\, P^{(18)}{}_{[i_1i_2i_3i_4i_5i_6i_7i_8i_9i_{10}i_{11}i_{12}i_{13}i_{14}i_{15}i_{16}i_{17}i_{18}}\, P^{(2)}{}_{i_{19}i_{20}}\, P^{(2)}{}_{i_{21}i_{22}}\, P^{(2)}{}_{i_{23}i_{24}]} +$$

$$+ 14{,}968{,}800\, P^{(16)}{}_{[i_1i_2i_3i_4i_5i_6i_7i_8i_9i_{10}i_{11}i_{12}i_{13}i_{14}i_{15}i_{16}}\, P^{(8)}{}_{i_{17}i_{18}i_{19}i_{20}i_{21}i_{22}i_{23}i_{24}]} -$$

$$- 19{,}958{,}400\, P^{(16)}{}_{[i_1i_2i_3i_4i_5i_6i_7i_8i_9i_{10}i_{11}i_{12}i_{13}i_{14}i_{15}i_{16}}\, P^{(6)}{}_{i_{17}i_{18}i_{19}i_{20}i_{21}i_{22}}\, P^{(2)}{}_{i_{23}i_{24}]} -$$

$$- 7{,}484{,}400\, P^{(16)}{}_{[i_1i_2i_3i_4i_5i_6i_7i_8i_9i_{10}i_{11}i_{12}i_{13}i_{14}i_{15}i_{16}}\, P^{(4)}{}_{i_{17}i_{18}i_{19}i_{20}}\, P^{(4)}{}_{i_{21}i_{22}i_{23}i_{24}]} +$$

$$+ 14{,}968{,}800\, P^{(16)}{}_{[i_1i_2i_3i_4i_5i_6i_7i_8i_9i_{10}i_{11}i_{12}i_{13}i_{14}i_{15}i_{16}}\, P^{(4)}{}_{i_{17}i_{18}i_{19}i_{20}}\, P^{(2)}{}_{i_{21}i_{22}}\, P^{(2)}{}_{i_{23}i_{24}]} -$$

$$- 2{,}494{,}800\, P^{(16)}{}_{[i_1i_2i_3i_4i_5i_6i_7i_8i_9i_{10}i_{11}i_{12}i_{13}i_{14}i_{15}i_{16}}\, P^{(2)}{}_{i_{17}i_{18}}\, P^{(2)}{}_{i_{19}i_{20}}\, P^{(2)}{}_{i_{21}i_{22}}\, P^{(2)}{}_{i_{23}i_{24}]} +$$

$$+ 13{,}685{,}760\, P^{(14)}{}_{[i_1i_2i_3i_4i_5i_6i_7i_8i_9i_{10}i_{11}i_{12}i_{13}i_{14}}\, P^{(10)}{}_{i_{15}i_{16}i_{17}i_{18}i_{19}i_{20}i_{21}i_{22}i_{23}i_{24}]} -$$

$$- 17{,}107{,}200\, P^{(14)}{}_{[i_1i_2i_3i_4i_5i_6i_7i_8i_9i_{10}i_{11}i_{12}i_{13}i_{14}}\, P^{(8)}{}_{i_{15}i_{16}i_{17}i_{18}i_{19}i_{20}i_{21}i_{22}}\, P^{(2)}{}_{i_{23}i_{24}]} -$$

$$- 11{,}404{,}800\, P^{(14)}{}_{[i_1i_2i_3i_4i_5i_6i_7i_8i_9i_{10}i_{11}i_{12}i_{13}i_{14}}\, P^{(6)}{}_{i_{15}i_{16}i_{17}i_{18}i_{19}i_{20}}\, P^{(4)}{}_{i_{21}i_{22}i_{23}i_{24}]} +$$

$$+ 11{,}404{,}800\, P^{(14)}{}_{[i_1i_2i_3i_4i_5i_6i_7i_8i_9i_{10}i_{11}i_{12}i_{13}i_{14}}\, P^{(6)}{}_{i_{15}i_{16}i_{17}i_{18}i_{19}i_{20}}\, P^{(2)}{}_{i_{21}i_{22}}\, P^{(2)}{}_{i_{23}i_{24}]} +$$

$$+ 8{,}553{,}600\, P^{(14)}{}_{[i_1i_2i_3i_4i_5i_6i_7i_8i_9i_{10}i_{11}i_{12}i_{13}i_{14}}\, P^{(4)}{}_{i_{15}i_{16}i_{17}i_{18}}\, P^{(4)}{}_{i_{19}i_{20}i_{21}i_{22}}\, P^{(2)}{}_{i_{23}i_{24}]} -$$

$$- 5{,}702{,}400\, P^{(14)}{}_{[i_1i_2i_3i_4i_5i_6i_7i_8i_9i_{10}i_{11}i_{12}i_{13}i_{14}}\, P^{(4)}{}_{i_{15}i_{16}i_{17}i_{18}}\, P^{(2)}{}_{i_{19}i_{20}}\, P^{(2)}{}_{i_{21}i_{22}}\, P^{(2)}{}_{i_{23}i_{24}]} +$$

$$+ 570{,}240\, P^{(14)}{}_{[i_1i_2i_3i_4i_5i_6i_7i_8i_9i_{10}i_{11}i_{12}i_{13}i_{14}}\, P^{(2)}{}_{i_{15}i_{16}}\, P^{(2)}{}_{i_{17}i_{18}}\, P^{(2)}{}_{i_{19}i_{20}}\, P^{(2)}{}_{i_{21}i_{22}}\, P^{(2)}{}_{i_{23}i_{24}]} +$$

$$+ 6{,}652{,}800\, P^{(12)}{}_{[i_1i_2i_3i_4i_5i_6i_7i_8i_9i_{10}i_{11}i_{12}}\, P^{(12)}{}_{i_{13}i_{14}i_{15}i_{16}i_{17}i_{18}i_{19}i_{20}i_{21}i_{22}i_{23}i_{24}]} -$$

$$- 15{,}966{,}720\, P^{(12)}{}_{[i_1i_2i_3i_4i_5i_6i_7i_8i_9i_{10}i_{11}i_{12}}\, P^{(10)}{}_{i_{13}i_{14}i_{15}i_{16}i_{17}i_{18}i_{19}i_{20}i_{21}i_{22}}\, P^{(2)}{}_{i_{23}i_{24}]} -$$

$$- 9{,}979{,}200\, P^{(12)}{}_{[i_1i_2i_3i_4i_5i_6i_7i_8i_9i_{10}i_{11}i_{12}}\, P^{(8)}{}_{i_{13}i_{14}i_{15}i_{16}i_{17}i_{18}i_{19}i_{20}}\, P^{(4)}{}_{i_{21}i_{22}i_{23}i_{24}]} +$$

$$+ 9{,}979{,}200\, P^{(12)}{}_{[i_1i_2i_3i_4i_5i_6i_7i_8i_9i_{10}i_{11}i_{12}}\, P^{(8)}{}_{i_{13}i_{14}i_{15}i_{16}i_{17}i_{18}i_{19}i_{20}}\, P^{(2)}{}_{i_{21}i_{22}}\, P^{(2)}{}_{i_{23}i_{24}]} -$$

$$- 4{,}435{,}200\, P^{(12)}{}_{[i_1i_2i_3i_4i_5i_6i_7i_8i_9i_{10}i_{11}i_{12}}\, P^{(6)}{}_{i_{13}i_{14}i_{15}i_{16}i_{17}i_{18}}\, P^{(6)}{}_{i_{19}i_{20}i_{21}i_{22}i_{23}i_{24}]} +$$

$$+ 13{,}305{,}600\, P^{(12)}{}_{[i_1i_2i_3i_4i_5i_6i_7i_8i_9i_{10}i_{11}i_{12}}\, P^{(6)}{}_{i_{13}i_{14}i_{15}i_{16}i_{17}i_{18}}\, P^{(4)}{}_{i_{19}i_{20}i_{21}i_{22}}\, P^{(2)}{}_{i_{23}i_{24}]} -$$

$$- 4{,}435{,}200\, P^{(12)}{}_{[i_1i_2i_3i_4i_5i_6i_7i_8i_9i_{10}i_{11}i_{12}}\, P^{(6)}{}_{i_{13}i_{14}i_{15}i_{16}i_{17}i_{18}}\, P^{(2)}{}_{i_{19}i_{20}}\, P^{(2)}{}_{i_{21}i_{22}}\, P^{(2)}{}_{i_{23}i_{24}]} +$$

$$+ 1{,}663{,}200\, P^{(12)}{}_{[i_1i_2i_3i_4i_5i_6i_7i_8i_9i_{10}i_{11}i_{12}}\, P^{(4)}{}_{i_{13}i_{14}i_{15}i_{16}}\, P^{(4)}{}_{i_{17}i_{18}i_{19}i_{20}}\, P^{(4)}{}_{i_{21}i_{22}i_{23}i_{24}]} -$$

$$- 4{,}989{,}600\, P^{(12)}{}_{[i_1i_2i_3i_4i_5i_6i_7i_8i_9i_{10}i_{11}i_{12}}\, P^{(4)}{}_{i_{13}i_{14}i_{15}i_{16}}\, P^{(4)}{}_{i_{17}i_{18}i_{19}i_{20}}\, P^{(2)}{}_{i_{21}i_{22}}\, P^{(2)}{}_{i_{23}i_{24}]} +$$

$$+ 1{,}663{,}200\, P^{(12)}{}_{[i_1i_2i_3i_4i_5i_6i_7i_8i_9i_{10}i_{11}i_{12}}\, P^{(4)}{}_{i_{13}i_{14}i_{15}i_{16}}\, P^{(2)}{}_{i_{17}i_{18}}\, P^{(2)}{}_{i_{19}i_{20}}\, P^{(2)}{}_{i_{21}i_{22}}\, P^{(2)}{}_{i_{23}i_{24}]} -$$

$$- 110{,}880\, P^{(12)}{}_{[i_1i_2i_3i_4i_5i_6i_7i_8i_9i_{10}i_{11}i_{12}}\, P^{(2)}{}_{i_{13}i_{14}}\, P^{(2)}{}_{i_{15}i_{16}}\, P^{(2)}{}_{i_{17}i_{18}}\, P^{(2)}{}_{i_{19}i_{20}}\, P^{(2)}{}_{i_{21}i_{22}}\, P^{(2)}{}_{i_{23}i_{24}]} -$$

$$- 4{,}790{,}016\, P^{(10)}{}_{[i_1i_2i_3i_4i_5i_6i_7i_8i_9i_{10}}\, P^{(10)}{}_{i_{11}i_{12}i_{13}i_{14}i_{15}i_{16}i_{17}i_{18}i_{19}i_{20}}\, P^{(4)}{}_{i_{21}i_{22}i_{23}i_{24}]} +$$

$$+ 4{,}790{,}016\, P^{(10)}{}_{[i_1i_2i_3i_4i_5i_6i_7i_8i_9i_{10}}\, P^{(10)}{}_{i_{11}i_{12}i_{13}i_{14}i_{15}i_{16}i_{17}i_{18}i_{19}i_{20}}\, P^{(2)}{}_{i_{21}i_{22}}\, P^{(2)}{}_{i_{23}i_{24}]} -$$

$$- 7{,}983{,}360\, P^{(10)}{}_{[i_1i_2i_3i_4i_5i_6i_7i_8i_9i_{10}}\, P^{(8)}{}_{i_{11}i_{12}i_{13}i_{14}i_{15}i_{16}i_{17}i_{18}}\, P^{(6)}{}_{i_{19}i_{20}i_{21}i_{22}i_{23}i_{24}]} +$$

$$+ 11{,}975{,}040\, P^{(10)}{}_{[i_1i_2i_3i_4i_5i_6i_7i_8i_9i_{10}}\, P^{(8)}{}_{i_{11}i_{12}i_{13}i_{14}i_{15}i_{16}i_{17}i_{18}}\, P^{(4)}{}_{i_{19}i_{20}i_{21}i_{22}}\, P^{(2)}{}_{i_{23}i_{24}]} -$$

$$- 3{,}991{,}680\, P^{(10)}{}_{[i_1i_2i_3i_4i_5i_6i_7i_8i_9i_{10}}\, P^{(8)}{}_{i_{11}i_{12}i_{13}i_{14}i_{15}i_{16}i_{17}i_{18}}\, P^{(2)}{}_{i_{19}i_{20}}\, P^{(2)}{}_{i_{21}i_{22}}\, P^{(2)}{}_{i_{23}i_{24}]} +$$

$$+ 5{,}322{,}240\, P^{(10)}{}_{[i_1i_2i_3i_4i_5i_6i_7i_8i_9i_{10}}\, P^{(6)}{}_{i_{11}i_{12}i_{13}i_{14}i_{15}i_{16}}\, P^{(6)}{}_{i_{17}i_{18}i_{19}i_{20}i_{21}i_{22}}\, P^{(2)}{}_{i_{23}i_{24}]} +$$

$$+ 3{,}991{,}680\, P^{(10)}{}_{[i_1i_2i_3i_4i_5i_6i_7i_8i_9i_{10}}\, P^{(6)}{}_{i_{11}i_{12}i_{13}i_{14}i_{15}i_{16}}\, P^{(4)}{}_{i_{17}i_{18}i_{19}i_{20}}\, P^{(4)}{}_{i_{21}i_{22}i_{23}i_{24}]} -$$

$$\begin{aligned}
&- 7{,}983{,}360\ P^{(10)}_{[i_1 i_2 i_3 i_4 i_5 i_6 i_7 i_8 i_9 i_{10}}\ P^{(6)}_{i_{11} i_{12} i_{13} i_{14} i_{15} i_{16}}\ P^{(4)}_{i_{17} i_{18} i_{19} i_{20}}\ P^{(2)}_{i_{21} i_{22}}\ P^{(2)}_{i_{23} i_{24}]} + \\
&+ 1{,}330{,}560\ P^{(10)}_{[i_1 i_2 i_3 i_4 i_5 i_6 i_7 i_8 i_9 i_{10}}\ P^{(6)}_{i_{11} i_{12} i_{13} i_{14} i_{15} i_{16}}\ P^{(2)}_{i_{17} i_{18}}\ P^{(2)}_{i_{19} i_{20}}\ P^{(2)}_{i_{21} i_{22}}\ P^{(2)}_{i_{23} i_{24}]} - \\
&- 1{,}995{,}840\ P^{(10)}_{[i_1 i_2 i_3 i_4 i_5 i_6 i_7 i_8 i_9 i_{10}}\ P^{(4)}_{i_{11} i_{12} i_{13} i_{14}}\ P^{(4)}_{i_{15} i_{16} i_{17} i_{18}}\ P^{(4)}_{i_{19} i_{20} i_{21} i_{22}}\ P^{(2)}_{i_{23} i_{24}]} + \\
&+ 1{,}995{,}840\ P^{(10)}_{[i_1 i_2 i_3 i_4 i_5 i_6 i_7 i_8 i_9 i_{10}}\ P^{(4)}_{i_{11} i_{12} i_{13} i_{14}}\ P^{(4)}_{i_{15} i_{16} i_{17} i_{18}}\ P^{(2)}_{i_{19} i_{20}}\ P^{(2)}_{i_{21} i_{22}}\ P^{(2)}_{i_{23} i_{24}]} - \\
&- 399{,}168\ P^{(10)}_{[i_1 i_2 i_3 i_4 i_5 i_6 i_7 i_8 i_9 i_{10}}\ P^{(4)}_{i_{11} i_{12} i_{13} i_{14}}\ P^{(2)}_{i_{15} i_{16}}\ P^{(2)}_{i_{17} i_{18}}\ P^{(2)}_{i_{19} i_{20}}\ P^{(2)}_{i_{21} i_{22}}\ P^{(2)}_{i_{23} i_{24}]} + \\
&+ 19{,}008\ P^{(10)}_{[i_1 i_2 i_3 i_4 i_5 i_6 i_7 i_8 i_9 i_{10}}\ P^{(2)}_{i_{11} i_{12}}\ P^{(2)}_{i_{13} i_{14}}\ P^{(2)}_{i_{15} i_{16}}\ P^{(2)}_{i_{17} i_{18}}\ P^{(2)}_{i_{19} i_{20}}\ P^{(2)}_{i_{21} i_{22}}\ P^{(2)}_{i_{23} i_{24}]} - \\
&- 1{,}247{,}400\ P^{(8)}_{[i_1 i_2 i_3 i_4 i_5 i_6 i_7 i_8}\ P^{(8)}_{i_9 i_{10} i_{11} i_{12} i_{13} i_{14} i_{15} i_{16}}\ P^{(8)}_{i_{17} i_{18} i_{19} i_{20} i_{21} i_{22} i_{23} i_{24}]} + \\
&+ 4{,}989{,}600\ P^{(8)}_{[i_1 i_2 i_3 i_4 i_5 i_6 i_7 i_8}\ P^{(8)}_{i_9 i_{10} i_{11} i_{12} i_{13} i_{14} i_{15} i_{16}}\ P^{(6)}_{i_{17} i_{18} i_{19} i_{20} i_{21} i_{22}}\ P^{(2)}_{i_{23} i_{24}]} + \\
&+ 1{,}871{,}100\ P^{(8)}_{[i_1 i_2 i_3 i_4 i_5 i_6 i_7 i_8}\ P^{(8)}_{i_9 i_{10} i_{11} i_{12} i_{13} i_{14} i_{15} i_{16}}\ P^{(4)}_{i_{17} i_{18} i_{19} i_{20}}\ P^{(4)}_{i_{21} i_{22} i_{23} i_{24}]} - \\
&- 3{,}742{,}200\ P^{(8)}_{[i_1 i_2 i_3 i_4 i_5 i_6 i_7 i_8}\ P^{(8)}_{i_9 i_{10} i_{11} i_{12} i_{13} i_{14} i_{15} i_{16}}\ P^{(4)}_{i_{17} i_{18} i_{19} i_{20}}\ P^{(2)}_{i_{21} i_{22}}\ P^{(2)}_{i_{23} i_{24}]} + \\
&+ 623{,}700\ P^{(8)}_{[i_1 i_2 i_3 i_4 i_5 i_6 i_7 i_8}\ P^{(8)}_{i_9 i_{10} i_{11} i_{12} i_{13} i_{14} i_{15} i_{16}}\ P^{(2)}_{i_{17} i_{18}}\ P^{(2)}_{i_{19} i_{20}}\ P^{(2)}_{i_{21} i_{22}}\ P^{(2)}_{i_{23} i_{24}]} + \\
&+ 3{,}326{,}400\ P^{(8)}_{[i_1 i_2 i_3 i_4 i_5 i_6 i_7 i_8}\ P^{(6)}_{i_9 i_{10} i_{11} i_{12} i_{13} i_{14}}\ P^{(6)}_{i_{15} i_{16} i_{17} i_{18} i_{19} i_{20}}\ P^{(4)}_{i_{21} i_{22} i_{23} i_{24}]} - \\
&- 3{,}326{,}400\ P^{(8)}_{[i_1 i_2 i_3 i_4 i_5 i_6 i_7 i_8}\ P^{(6)}_{i_9 i_{10} i_{11} i_{12} i_{13} i_{14}}\ P^{(6)}_{i_{15} i_{16} i_{17} i_{18} i_{19} i_{20}}\ P^{(2)}_{i_{21} i_{22}}\ P^{(2)}_{i_{23} i_{24}]} - \\
&- 4{,}989{,}600\ P^{(8)}_{[i_1 i_2 i_3 i_4 i_5 i_6 i_7 i_8}\ P^{(6)}_{i_9 i_{10} i_{11} i_{12} i_{13} i_{14}}\ P^{(4)}_{i_{15} i_{16} i_{17} i_{18}}\ P^{(4)}_{i_{19} i_{20} i_{21} i_{22}}\ P^{(2)}_{i_{23} i_{24}]} + \\
&+ 3{,}326{,}400\ P^{(8)}_{[i_1 i_2 i_3 i_4 i_5 i_6 i_7 i_8}\ P^{(6)}_{i_9 i_{10} i_{11} i_{12} i_{13} i_{14}}\ P^{(4)}_{i_{15} i_{16} i_{17} i_{18}}\ P^{(2)}_{i_{19} i_{20}}\ P^{(2)}_{i_{21} i_{22}}\ P^{(2)}_{i_{23} i_{24}]} - \\
&- 332{,}640\ P^{(8)}_{[i_1 i_2 i_3 i_4 i_5 i_6 i_7 i_8}\ P^{(6)}_{i_9 i_{10} i_{11} i_{12} i_{13} i_{14}}\ P^{(2)}_{i_{15} i_{16}}\ P^{(2)}_{i_{17} i_{18}}\ P^{(2)}_{i_{19} i_{20}}\ P^{(2)}_{i_{21} i_{22}}\ P^{(2)}_{i_{23} i_{24}]} - \\
&- 311{,}850\ P^{(8)}_{[i_1 i_2 i_3 i_4 i_5 i_6 i_7 i_8}\ P^{(4)}_{i_9 i_{10} i_{11} i_{12}}\ P^{(4)}_{i_{13} i_{14} i_{15} i_{16}}\ P^{(4)}_{i_{17} i_{18} i_{19} i_{20}}\ P^{(4)}_{i_{21} i_{22} i_{23} i_{24}]} + \\
&+ 1{,}247{,}400\ P^{(8)}_{[i_1 i_2 i_3 i_4 i_5 i_6 i_7 i_8}\ P^{(4)}_{i_9 i_{10} i_{11} i_{12}}\ P^{(4)}_{i_{13} i_{14} i_{15} i_{16}}\ P^{(4)}_{i_{17} i_{18} i_{19} i_{20}}\ P^{(2)}_{i_{21} i_{22}}\ P^{(2)}_{i_{23} i_{24}]} - \\
&- 623{,}700\ P^{(8)}_{[i_1 i_2 i_3 i_4 i_5 i_6 i_7 i_8}\ P^{(4)}_{i_9 i_{10} i_{11} i_{12}}\ P^{(4)}_{i_{13} i_{14} i_{15} i_{16}}\ P^{(2)}_{i_{17} i_{18}}\ P^{(2)}_{i_{19} i_{20}}\ P^{(2)}_{i_{21} i_{22}}\ P^{(2)}_{i_{23} i_{24}]} + \\
&+ 83{,}160\ P^{(8)}_{[i_1 i_2 i_3 i_4 i_5 i_6 i_7 i_8}\ P^{(4)}_{i_9 i_{10} i_{11} i_{12}}\ P^{(2)}_{i_{13} i_{14}}\ P^{(2)}_{i_{15} i_{16}}\ P^{(2)}_{i_{17} i_{18}}\ P^{(2)}_{i_{19} i_{20}}\ P^{(2)}_{i_{21} i_{22}}\ P^{(2)}_{i_{23} i_{24}]} - \\
&- 2970\ P^{(8)}_{[i_1 i_2 i_3 i_4 i_5 i_6 i_7 i_8}\ P^{(2)}_{i_9 i_{10}}\ P^{(2)}_{i_{11} i_{12}}\ P^{(2)}_{i_{13} i_{14}}\ P^{(2)}_{i_{15} i_{16}}\ P^{(2)}_{i_{17} i_{18}}\ P^{(2)}_{i_{19} i_{20}}\ P^{(2)}_{i_{21} i_{22}}\ P^{(2)}_{i_{23} i_{24}]} + \\
&+ 246{,}400\ P^{(6)}_{[i_1 i_2 i_3 i_4 i_5 i_6}\ P^{(6)}_{i_7 i_8 i_9 i_{10} i_{11} i_{12}}\ P^{(6)}_{i_{13} i_{14} i_{15} i_{16} i_{17} i_{18}}\ P^{(6)}_{i_{19} i_{20} i_{21} i_{22} i_{23} i_{24}]} - \\
&- 1{,}478{,}400\ P^{(6)}_{[i_1 i_2 i_3 i_4 i_5 i_6}\ P^{(6)}_{i_7 i_8 i_9 i_{10} i_{11} i_{12}}\ P^{(6)}_{i_{13} i_{14} i_{15} i_{16} i_{17} i_{18}}\ P^{(4)}_{i_{19} i_{20} i_{21} i_{22}}\ P^{(2)}_{i_{23} i_{24}]} + \\
&+ 492{,}800\ P^{(6)}_{[i_1 i_2 i_3 i_4 i_5 i_6}\ P^{(6)}_{i_7 i_8 i_9 i_{10} i_{11} i_{12}}\ P^{(6)}_{i_{13} i_{14} i_{15} i_{16} i_{17} i_{18}}\ P^{(2)}_{i_{19} i_{20}}\ P^{(2)}_{i_{21} i_{22}}\ P^{(2)}_{i_{23} i_{24}]} - \\
&- 554{,}400\ P^{(6)}_{[i_1 i_2 i_3 i_4 i_5 i_6}\ P^{(6)}_{i_7 i_8 i_9 i_{10} i_{11} i_{12}}\ P^{(4)}_{i_{13} i_{14} i_{15} i_{16}}\ P^{(4)}_{i_{17} i_{18} i_{19} i_{20}}\ P^{(4)}_{i_{21} i_{22} i_{23} i_{24}]} + \\
&+ 1{,}663{,}200\ P^{(6)}_{[i_1 i_2 i_3 i_4 i_5 i_6}\ P^{(6)}_{i_7 i_8 i_9 i_{10} i_{11} i_{12}}\ P^{(4)}_{i_{13} i_{14} i_{15} i_{16}}\ P^{(4)}_{i_{17} i_{18} i_{19} i_{20}}\ P^{(2)}_{i_{21} i_{22}}\ P^{(2)}_{i_{23} i_{24}]} - \\
&- 554{,}400\ P^{(6)}_{[i_1 i_2 i_3 i_4 i_5 i_6}\ P^{(6)}_{i_7 i_8 i_9 i_{10} i_{11} i_{12}}\ P^{(4)}_{i_{13} i_{14} i_{15} i_{16}}\ P^{(2)}_{i_{17} i_{18}}\ P^{(2)}_{i_{19} i_{20}}\ P^{(2)}_{i_{21} i_{22}}\ P^{(2)}_{i_{23} i_{24}]} + \\
&+ 36{,}960\ P^{(6)}_{[i_1 i_2 i_3 i_4 i_5 i_6}\ P^{(6)}_{i_7 i_8 i_9 i_{10} i_{11} i_{12}}\ P^{(2)}_{i_{13} i_{14}}\ P^{(2)}_{i_{15} i_{16}}\ P^{(2)}_{i_{17} i_{18}}\ P^{(2)}_{i_{19} i_{20}}\ P^{(2)}_{i_{21} i_{22}}\ P^{(2)}_{i_{23} i_{24}]} + \\
&+ 415{,}800\ P^{(6)}_{[i_1 i_2 i_3 i_4 i_5 i_6}\ P^{(4)}_{i_7 i_8 i_9 i_{10}}\ P^{(4)}_{i_{11} i_{12} i_{13} i_{14}}\ P^{(4)}_{i_{15} i_{16} i_{17} i_{18}}\ P^{(4)}_{i_{19} i_{20} i_{21} i_{22}}\ P^{(2)}_{i_{23} i_{24}]} - \\
&- 554{,}400\ P^{(6)}_{[i_1 i_2 i_3 i_4 i_5 i_6}\ P^{(4)}_{i_7 i_8 i_9 i_{10}}\ P^{(4)}_{i_{11} i_{12} i_{13} i_{14}}\ P^{(4)}_{i_{15} i_{16} i_{17} i_{18}}\ P^{(2)}_{i_{19} i_{20}}\ P^{(2)}_{i_{21} i_{22}}\ P^{(2)}_{i_{23} i_{24}]} + \\
&+ 166{,}320\ P^{(6)}_{[i_1 i_2 i_3 i_4 i_5 i_6}\ P^{(4)}_{i_7 i_8 i_9 i_{10}}\ P^{(4)}_{i_{11} i_{12} i_{13} i_{14}}\ P^{(2)}_{i_{15} i_{16}}\ P^{(2)}_{i_{17} i_{18}}\ P^{(2)}_{i_{19} i_{20}}\ P^{(2)}_{i_{21} i_{22}}\ P^{(2)}_{i_{23} i_{24}]} - \\
&- 15{,}840\ P^{(6)}_{[i_1 i_2 i_3 i_4 i_5 i_6}\ P^{(4)}_{i_7 i_8 i_9 i_{10}}\ P^{(2)}_{i_{11} i_{12}}\ P^{(2)}_{i_{13} i_{14}}\ P^{(2)}_{i_{15} i_{16}}\ P^{(2)}_{i_{17} i_{18}}\ P^{(2)}_{i_{19} i_{20}}\ P^{(2)}_{i_{21} i_{22}}\ P^{(2)}_{i_{23} i_{24}]} + \\
&+ 440\ P^{(6)}_{[i_1 i_2 i_3 i_4 i_5 i_6}\ P^{(2)}_{i_7 i_8}\ P^{(2)}_{i_9 i_{10}}\ P^{(2)}_{i_{11} i_{12}}\ P^{(2)}_{i_{13} i_{14}}\ P^{(2)}_{i_{15} i_{16}}\ P^{(2)}_{i_{17} i_{18}}\ P^{(2)}_{i_{19} i_{20}}\ P^{(2)}_{i_{21} i_{22}}\ P^{(2)}_{i_{23} i_{24}]} + \\
&+ 10{,}395\ P^{(4)}_{[i_1 i_2 i_3 i_4}\ P^{(4)}_{i_5 i_6 i_7 i_8}\ P^{(4)}_{i_9 i_{10} i_{11} i_{12}}\ P^{(4)}_{i_{13} i_{14} i_{15} i_{16}}\ P^{(4)}_{i_{17} i_{18} i_{19} i_{20}}\ P^{(4)}_{i_{21} i_{22} i_{23} i_{24}]} - \\
&- 62{,}370\ P^{(4)}_{[i_1 i_2 i_3 i_4}\ P^{(4)}_{i_5 i_6 i_7 i_8}\ P^{(4)}_{i_9 i_{10} i_{11} i_{12}}\ P^{(4)}_{i_{13} i_{14} i_{15} i_{16}}\ P^{(4)}_{i_{17} i_{18} i_{19} i_{20}}\ P^{(2)}_{i_{21} i_{22}}\ P^{(2)}_{i_{23} i_{24}]} + \\
&+ 51{,}975\ P^{(4)}_{[i_1 i_2 i_3 i_4}\ P^{(4)}_{i_5 i_6 i_7 i_8}\ P^{(4)}_{i_9 i_{10} i_{11} i_{12}}\ P^{(4)}_{i_{13} i_{14} i_{15} i_{16}}\ P^{(2)}_{i_{17} i_{18}}\ P^{(2)}_{i_{19} i_{20}}\ P^{(2)}_{i_{21} i_{22}}\ P^{(2)}_{i_{23} i_{24}]} - \\
&- 13{,}860\ P^{(4)}_{[i_1 i_2 i_3 i_4}\ P^{(4)}_{i_5 i_6 i_7 i_8}\ P^{(4)}_{i_9 i_{10} i_{11} i_{12}}\ P^{(2)}_{i_{13} i_{14}}\ P^{(2)}_{i_{15} i_{16}}\ P^{(2)}_{i_{17} i_{18}}\ P^{(2)}_{i_{19} i_{20}}\ P^{(2)}_{i_{21} i_{22}}\ P^{(2)}_{i_{23} i_{24}]} + \\
&+ 1485\ P^{(4)}_{[i_1 i_2 i_3 i_4}\ P^{(4)}_{i_5 i_6 i_7 i_8}\ P^{(2)}_{i_9 i_{10}}\ P^{(2)}_{i_{11} i_{12}}\ P^{(2)}_{i_{13} i_{14}}\ P^{(2)}_{i_{15} i_{16}}\ P^{(2)}_{i_{17} i_{18}}\ P^{(2)}_{i_{19} i_{20}}\ P^{(2)}_{i_{21} i_{22}}\ P^{(2)}_{i_{23} i_{24}]} - \\
&- 66\ P^{(4)}_{[i_1 i_2 i_3 i_4}\ P^{(2)}_{i_5 i_6}\ P^{(2)}_{i_7 i_8}\ P^{(2)}_{i_9 i_{10}}\ P^{(2)}_{i_{11} i_{12}}\ P^{(2)}_{i_{13} i_{14}}\ P^{(2)}_{i_{15} i_{16}}\ P^{(2)}_{i_{17} i_{18}}\ P^{(2)}_{i_{19} i_{20}}\ P^{(2)}_{i_{21} i_{22}}\ P^{(2)}_{i_{23} i_{24}]} + \\
&+ P^{(2)}_{[i_1 i_2}\ P^{(2)}_{i_3 i_4}\ P^{(2)}_{i_5 i_6}\ P^{(2)}_{i_7 i_8}\ P^{(2)}_{i_9 i_{10}}\ P^{(2)}_{i_{11} i_{12}}\ P^{(2)}_{i_{13} i_{14}}\ P^{(2)}_{i_{15} i_{16}}\ P^{(2)}_{i_{17} i_{18}}\ P^{(2)}_{i_{19} i_{20}}\ P^{(2)}_{i_{21} i_{22}}\ P^{(2)}_{i_{23} i_{24}]})
\end{aligned}$$

## Coefficient of the $13^{th}$ Chern Form

$$c_{(13)i_1i_2i_3i_4i_5i_6i_7i_8i_9i_{10}i_{11}i_{12}i_{13}i_{14}i_{15}i_{16}i_{17}i_{18}i_{19}i_{20}i_{21}i_{22}i_{23}i_{24}i_{25}i_{26}} = \tag{22}$$

$$= \frac{1}{26!} \langle \mathbf{e}_{i_1} \wedge \mathbf{e}_{i_2} \wedge \mathbf{e}_{i_3} \wedge \mathbf{e}_{i_4} \wedge \mathbf{e}_{i_5} \wedge \mathbf{e}_{i_6} \wedge \mathbf{e}_{i_7} \wedge \mathbf{e}_{i_8} \wedge \mathbf{e}_{i_9} \wedge \mathbf{e}_{i_{10}} \wedge \mathbf{e}_{i_{11}} \wedge \mathbf{e}_{i_{12}} \wedge \mathbf{e}_{i_{13}} \wedge \ldots$$

$$\ldots \wedge \mathbf{e}_{i_{14}} \wedge \mathbf{e}_{i_{15}} \wedge \mathbf{e}_{i_{16}} \wedge \mathbf{e}_{i_{17}} \wedge \mathbf{e}_{i_{18}} \wedge \mathbf{e}_{i_{19}} \wedge \mathbf{e}_{i_{20}} \wedge \mathbf{e}_{i_{21}} \wedge \mathbf{e}_{i_{22}} \wedge \mathbf{e}_{i_{23}} \wedge \mathbf{e}_{i_{24}} \wedge \mathbf{e}_{i_{25}} \wedge \mathbf{e}_{i_{26}}, c_{(13)} \rangle$$

$$= \frac{i^{13}}{2^{26}\pi^{13}13!} \Big(+ 479{,}001{,}600\, R_{[i_1i_2|m|}{}^a\, R_{i_3i_4|a|}{}^b\, R_{i_5i_6|b|}{}^c\, R_{i_7i_8|c|}{}^d\, R_{i_9i_{10}|d|}{}^e\, R_{i_{11}i_{12}|e|}{}^f\, R_{i_{13}i_{14}|f|}{}^g\, R_{i_{15}i_{16}|g|}{}^h\, R_{i_{17}i_{18}|h|}{}^i\, R_{i_{19}i_{20}|i|}{}^j\, R_{i_{21}i_{22}|j|}{}^k\, R_{i_{23}i_{24}|k|}{}^l\, R_{i_{25}i_{26}]l}{}^m -$$

$$- 518{,}918{,}400\, R_{[i_1i_2|l|}{}^a\, R_{i_3i_4|a|}{}^b\, R_{i_5i_6|b|}{}^c\, R_{i_7i_8|c|}{}^d\, R_{i_9i_{10}|d|}{}^e\, R_{i_{11}i_{12}|e|}{}^f\, R_{i_{13}i_{14}|f|}{}^g\, R_{i_{15}i_{16}|g|}{}^h\, R_{i_{17}i_{18}|h|}{}^i\, R_{i_{19}i_{20}|i|}{}^j\, R_{i_{21}i_{22}|j|}{}^k\, R_{i_{23}i_{24}|k|}{}^l\, R_{i_{25}i_{26}]m}{}^m -$$

$$- 283{,}046{,}400\, R_{[i_1i_2|k|}{}^a\, R_{i_3i_4|a|}{}^b\, R_{i_5i_6|b|}{}^c\, R_{i_7i_8|c|}{}^d\, R_{i_9i_{10}|d|}{}^e\, R_{i_{11}i_{12}|e|}{}^f\, R_{i_{13}i_{14}|f|}{}^g\, R_{i_{15}i_{16}|g|}{}^h\, R_{i_{17}i_{18}|h|}{}^i\, R_{i_{19}i_{20}|i|}{}^j\, R_{i_{21}i_{22}|j|}{}^k\, R_{i_{23}i_{24}|m|}{}^l\, R_{i_{25}i_{26}]l}{}^m +$$

$$+ 283{,}046{,}400\, R_{[i_1i_2|k|}{}^a\, R_{i_3i_4|a|}{}^b\, R_{i_5i_6|b|}{}^c\, R_{i_7i_8|c|}{}^d\, R_{i_9i_{10}|d|}{}^e\, R_{i_{11}i_{12}|e|}{}^f\, R_{i_{13}i_{14}|f|}{}^g\, R_{i_{15}i_{16}|g|}{}^h\, R_{i_{17}i_{18}|h|}{}^i\, R_{i_{19}i_{20}|i|}{}^j\, R_{i_{21}i_{22}|j|}{}^k\, R_{i_{23}i_{24}|l|}{}^l\, R_{i_{25}i_{26}]l}{}^m -$$

$$- 207{,}567{,}360\, R_{[i_1i_2|j|}{}^a\, R_{i_3i_4|a|}{}^b\, R_{i_5i_6|b|}{}^c\, R_{i_7i_8|c|}{}^d\, R_{i_9i_{10}|d|}{}^e\, R_{i_{11}i_{12}|e|}{}^f\, R_{i_{13}i_{14}|f|}{}^g\, R_{i_{15}i_{16}|g|}{}^h\, R_{i_{17}i_{18}|h|}{}^i\, R_{i_{19}i_{20}|i|}{}^j\, R_{i_{21}i_{22}|m|}{}^k\, R_{i_{23}i_{24}|k|}{}^l\, R_{i_{25}i_{26}]l}{}^m +$$

$$+ 311{,}351{,}040\, R_{[i_1i_2|j|}{}^a\, R_{i_3i_4|a|}{}^b\, R_{i_5i_6|b|}{}^c\, R_{i_7i_8|c|}{}^d\, R_{i_9i_{10}|d|}{}^e\, R_{i_{11}i_{12}|e|}{}^f\, R_{i_{13}i_{14}|f|}{}^g\, R_{i_{15}i_{16}|g|}{}^h\, R_{i_{17}i_{18}|h|}{}^i\, R_{i_{19}i_{20}|i|}{}^j\, R_{i_{21}i_{22}|l|}{}^k\, R_{i_{23}i_{24}|k|}{}^l\, R_{i_{25}i_{26}]m}{}^m -$$

$$- 103{,}783{,}680\, R_{[i_1i_2|j|}{}^a\, R_{i_3i_4|a|}{}^b\, R_{i_5i_6|b|}{}^c\, R_{i_7i_8|c|}{}^d\, R_{i_9i_{10}|d|}{}^e\, R_{i_{11}i_{12}|e|}{}^f\, R_{i_{13}i_{14}|f|}{}^g\, R_{i_{15}i_{16}|g|}{}^h\, R_{i_{17}i_{18}|h|}{}^i\, R_{i_{19}i_{20}|i|}{}^j\, R_{i_{21}i_{22}|k|}{}^k\, R_{i_{23}i_{24}|l|}{}^l\, R_{i_{25}i_{26}]m}{}^m -$$

$$- 172{,}972{,}800\, R_{[i_1i_2|i|}{}^a\, R_{i_3i_4|a|}{}^b\, R_{i_5i_6|b|}{}^c\, R_{i_7i_8|c|}{}^d\, R_{i_9i_{10}|d|}{}^e\, R_{i_{11}i_{12}|e|}{}^f\, R_{i_{13}i_{14}|f|}{}^g\, R_{i_{15}i_{16}|g|}{}^h\, R_{i_{17}i_{18}|h|}{}^i\, R_{i_{19}i_{20}|m|}{}^j\, R_{i_{21}i_{22}|j|}{}^k\, R_{i_{23}i_{24}|k|}{}^l\, R_{i_{25}i_{26}]l}{}^m +$$

$$+ 230{,}630{,}400\, R_{[i_1i_2|i|}{}^a\, R_{i_3i_4|a|}{}^b\, R_{i_5i_6|b|}{}^c\, R_{i_7i_8|c|}{}^d\, R_{i_9i_{10}|d|}{}^e\, R_{i_{11}i_{12}|e|}{}^f\, R_{i_{13}i_{14}|f|}{}^g\, R_{i_{15}i_{16}|g|}{}^h\, R_{i_{17}i_{18}|h|}{}^i\, R_{i_{19}i_{20}|l|}{}^j\, R_{i_{21}i_{22}|j|}{}^k\, R_{i_{23}i_{24}|k|}{}^l\, R_{i_{25}i_{26}]m}{}^m +$$

$$+ 86{,}486{,}400\, R_{[i_1i_2|i|}{}^a\, R_{i_3i_4|a|}{}^b\, R_{i_5i_6|b|}{}^c\, R_{i_7i_8|c|}{}^d\, R_{i_9i_{10}|d|}{}^e\, R_{i_{11}i_{12}|e|}{}^f\, R_{i_{13}i_{14}|f|}{}^g\, R_{i_{15}i_{16}|g|}{}^h\, R_{i_{17}i_{18}|h|}{}^i\, R_{i_{19}i_{20}|k|}{}^j\, R_{i_{21}i_{22}|j|}{}^k\, R_{i_{23}i_{24}|m|}{}^l\, R_{i_{25}i_{26}]l}{}^m -$$

$$- 172{,}972{,}800\, R_{[i_1i_2|i|}{}^a\, R_{i_3i_4|a|}{}^b\, R_{i_5i_6|b|}{}^c\, R_{i_7i_8|c|}{}^d\, R_{i_9i_{10}|d|}{}^e\, R_{i_{11}i_{12}|e|}{}^f\, R_{i_{13}i_{14}|f|}{}^g\, R_{i_{15}i_{16}|g|}{}^h\, R_{i_{17}i_{18}|h|}{}^i\, R_{i_{19}i_{20}|k|}{}^j\, R_{i_{21}i_{22}|j|}{}^k\, R_{i_{23}i_{24}|l|}{}^l\, R_{i_{25}i_{26}]m}{}^m +$$

$$+ 28{,}828{,}800\, R_{[i_1i_2|i|}{}^a\, R_{i_3i_4|a|}{}^b\, R_{i_5i_6|b|}{}^c\, R_{i_7i_8|c|}{}^d\, R_{i_9i_{10}|d|}{}^e\, R_{i_{11}i_{12}|e|}{}^f\, R_{i_{13}i_{14}|f|}{}^g\, R_{i_{15}i_{16}|g|}{}^h\, R_{i_{17}i_{18}|h|}{}^i\, R_{i_{19}i_{20}|j|}{}^j\, R_{i_{21}i_{22}|j|}{}^k\, R_{i_{23}i_{24}|k|}{}^l\, R_{i_{25}i_{26}]l}{}^m -$$

$$- 155{,}675{,}520\, R_{[i_1i_2|h|}{}^a\, R_{i_3i_4|a|}{}^b\, R_{i_5i_6|b|}{}^c\, R_{i_7i_8|c|}{}^d\, R_{i_9i_{10}|d|}{}^e\, R_{i_{11}i_{12}|e|}{}^f\, R_{i_{13}i_{14}|f|}{}^g\, R_{i_{15}i_{16}|g|}{}^h\, R_{i_{17}i_{18}|m|}{}^i\, R_{i_{19}i_{20}|i|}{}^j\, R_{i_{21}i_{22}|j|}{}^k\, R_{i_{23}i_{24}|k|}{}^l\, R_{i_{25}i_{26}]l}{}^m +$$

$$+ 194{,}594{,}400\, R_{[i_1i_2|h|}{}^a\, R_{i_3i_4|a|}{}^b\, R_{i_5i_6|b|}{}^c\, R_{i_7i_8|c|}{}^d\, R_{i_9i_{10}|d|}{}^e\, R_{i_{11}i_{12}|e|}{}^f\, R_{i_{13}i_{14}|f|}{}^g\, R_{i_{15}i_{16}|g|}{}^h\, R_{i_{17}i_{18}|l|}{}^i\, R_{i_{19}i_{20}|i|}{}^j\, R_{i_{21}i_{22}|j|}{}^k\, R_{i_{23}i_{24}|k|}{}^l\, R_{i_{25}i_{26}]m}{}^m +$$

$$+ 129{,}729{,}600\, R_{[i_1i_2|h|}{}^a\, R_{i_3i_4|a|}{}^b\, R_{i_5i_6|b|}{}^c\, R_{i_7i_8|c|}{}^d\, R_{i_9i_{10}|d|}{}^e\, R_{i_{11}i_{12}|e|}{}^f\, R_{i_{13}i_{14}|f|}{}^g\, R_{i_{15}i_{16}|g|}{}^h\, R_{i_{17}i_{18}|k|}{}^i\, R_{i_{19}i_{20}|i|}{}^j\, R_{i_{21}i_{22}|j|}{}^k\, R_{i_{23}i_{24}|m|}{}^l\, R_{i_{25}i_{26}]l}{}^m -$$

$$- 129{,}729{,}600\, R_{[i_1i_2|h|}{}^a\, R_{i_3i_4|a|}{}^b\, R_{i_5i_6|b|}{}^c\, R_{i_7i_8|c|}{}^d\, R_{i_9i_{10}|d|}{}^e\, R_{i_{11}i_{12}|e|}{}^f\, R_{i_{13}i_{14}|f|}{}^g\, R_{i_{15}i_{16}|g|}{}^h\, R_{i_{17}i_{18}|k|}{}^i\, R_{i_{19}i_{20}|i|}{}^j\, R_{i_{21}i_{22}|j|}{}^k\, R_{i_{23}i_{24}|l|}{}^l\, R_{i_{25}i_{26}]m}{}^m -$$

$$- 97{,}297{,}200\, R_{[i_1i_2|h|}{}^a\, R_{i_3i_4|a|}{}^b\, R_{i_5i_6|b|}{}^c\, R_{i_7i_8|c|}{}^d\, R_{i_9i_{10}|d|}{}^e\, R_{i_{11}i_{12}|e|}{}^f\, R_{i_{13}i_{14}|f|}{}^g\, R_{i_{15}i_{16}|g|}{}^h\, R_{i_{17}i_{18}|j|}{}^i\, R_{i_{19}i_{20}|i|}{}^j\, R_{i_{21}i_{22}|l|}{}^k\, R_{i_{23}i_{24}|k|}{}^l\, R_{i_{25}i_{26}]m}{}^m +$$

$$+ 64{,}864{,}800\, R_{[i_1i_2|h|}{}^a\, R_{i_3i_4|a|}{}^b\, R_{i_5i_6|b|}{}^c\, R_{i_7i_8|c|}{}^d\, R_{i_9i_{10}|d|}{}^e\, R_{i_{11}i_{12}|e|}{}^f\, R_{i_{13}i_{14}|f|}{}^g\, R_{i_{15}i_{16}|g|}{}^h\, R_{i_{17}i_{18}|j|}{}^i\, R_{i_{19}i_{20}|i|}{}^j\, R_{i_{21}i_{22}|k|}{}^k\, R_{i_{23}i_{24}|l|}{}^l\, R_{i_{25}i_{26}]m}{}^m -$$

$$- 6{,}486{,}480\, R_{[i_1i_2|h|}{}^a\, R_{i_3i_4|a|}{}^b\, R_{i_5i_6|b|}{}^c\, R_{i_7i_8|c|}{}^d\, R_{i_9i_{10}|d|}{}^e\, R_{i_{11}i_{12}|e|}{}^f\, R_{i_{13}i_{14}|f|}{}^g\, R_{i_{15}i_{16}|g|}{}^h\, R_{i_{17}i_{18}|i|}{}^i\, R_{i_{19}i_{20}|j|}{}^j\, R_{i_{21}i_{22}|k|}{}^k\, R_{i_{23}i_{24}|l|}{}^l\, R_{i_{25}i_{26}]m}{}^m -$$

$$- 148{,}262{,}400\, R_{[i_1i_2|g|}{}^a\, R_{i_3i_4|a|}{}^b\, R_{i_5i_6|b|}{}^c\, R_{i_7i_8|c|}{}^d\, R_{i_9i_{10}|d|}{}^e\, R_{i_{11}i_{12}|e|}{}^f\, R_{i_{13}i_{14}|f|}{}^g\, R_{i_{15}i_{16}|m|}{}^h\, R_{i_{17}i_{18}|h|}{}^i\, R_{i_{19}i_{20}|i|}{}^j\, R_{i_{21}i_{22}|j|}{}^k\, R_{i_{23}i_{24}|k|}{}^l\, R_{i_{25}i_{26}]l}{}^m +$$

$$+ 177{,}914{,}880\, R_{[i_1i_2|g|}{}^a\, R_{i_3i_4|a|}{}^b\, R_{i_5i_6|b|}{}^c\, R_{i_7i_8|c|}{}^d\, R_{i_9i_{10}|d|}{}^e\, R_{i_{11}i_{12}|e|}{}^f\, R_{i_{13}i_{14}|f|}{}^g\, R_{i_{15}i_{16}|l|}{}^h\, R_{i_{17}i_{18}|h|}{}^i\, R_{i_{19}i_{20}|i|}{}^j\, R_{i_{21}i_{22}|j|}{}^k\, R_{i_{23}i_{24}|k|}{}^l\, R_{i_{25}i_{26}]m}{}^m +$$

$$+ 111{,}196{,}800\, R_{[i_1i_2|g|}{}^a\, R_{i_3i_4|a|}{}^b\, R_{i_5i_6|b|}{}^c\, R_{i_7i_8|c|}{}^d\, R_{i_9i_{10}|d|}{}^e\, R_{i_{11}i_{12}|e|}{}^f\, R_{i_{13}i_{14}|f|}{}^g\, R_{i_{15}i_{16}|k|}{}^h\, R_{i_{17}i_{18}|h|}{}^i\, R_{i_{19}i_{20}|i|}{}^j\, R_{i_{21}i_{22}|j|}{}^k\, R_{i_{23}i_{24}|m|}{}^l\, R_{i_{25}i_{26}]l}{}^m -$$

$$- 111{,}196{,}800\, R_{[i_1i_2|g|}{}^a\, R_{i_3i_4|a|}{}^b\, R_{i_5i_6|b|}{}^c\, R_{i_7i_8|c|}{}^d\, R_{i_9i_{10}|d|}{}^e\, R_{i_{11}i_{12}|e|}{}^f\, R_{i_{13}i_{14}|f|}{}^g\, R_{i_{15}i_{16}|k|}{}^h\, R_{i_{17}i_{18}|h|}{}^i\, R_{i_{19}i_{20}|i|}{}^j\, R_{i_{21}i_{22}|j|}{}^k\, R_{i_{23}i_{24}|l|}{}^l\, R_{i_{25}i_{26}]m}{}^m +$$

$$+ 49{,}420{,}800\, R_{[i_1i_2|g|}{}^a\, R_{i_3i_4|a|}{}^b\, R_{i_5i_6|b|}{}^c\, R_{i_7i_8|c|}{}^d\, R_{i_9i_{10}|d|}{}^e\, R_{i_{11}i_{12}|e|}{}^f\, R_{i_{13}i_{14}|f|}{}^g\, R_{i_{15}i_{16}|j|}{}^h\, R_{i_{17}i_{18}|h|}{}^i\, R_{i_{19}i_{20}|i|}{}^j\, R_{i_{21}i_{22}|m|}{}^k\, R_{i_{23}i_{24}|k|}{}^l\, R_{i_{25}i_{26}]l}{}^m -$$

$$- 148{,}262{,}400\, R_{[i_1i_2|g|}{}^a\, R_{i_3i_4|a|}{}^b\, R_{i_5i_6|b|}{}^c\, R_{i_7i_8|c|}{}^d\, R_{i_9i_{10}|d|}{}^e\, R_{i_{11}i_{12}|e|}{}^f\, R_{i_{13}i_{14}|f|}{}^g\, R_{i_{15}i_{16}|j|}{}^h\, R_{i_{17}i_{18}|h|}{}^i\, R_{i_{19}i_{20}|i|}{}^j\, R_{i_{21}i_{22}|l|}{}^k\, R_{i_{23}i_{24}|k|}{}^l\, R_{i_{25}i_{26}]m}{}^m +$$

$$+ 49{,}420{,}800\, R_{[i_1i_2|g|}{}^a\, R_{i_3i_4|a|}{}^b\, R_{i_5i_6|b|}{}^c\, R_{i_7i_8|c|}{}^d\, R_{i_9i_{10}|d|}{}^e\, R_{i_{11}i_{12}|e|}{}^f\, R_{i_{13}i_{14}|f|}{}^g\, R_{i_{15}i_{16}|j|}{}^h\, R_{i_{17}i_{18}|h|}{}^i\, R_{i_{19}i_{20}|i|}{}^j\, R_{i_{21}i_{22}|k|}{}^k\, R_{i_{23}i_{24}|l|}{}^l\, R_{i_{25}i_{26}]m}{}^m -$$

$$- 18{,}532{,}800\, R_{[i_1i_2|g|}{}^a\, R_{i_3i_4|a|}{}^b\, R_{i_5i_6|b|}{}^c\, R_{i_7i_8|c|}{}^d\, R_{i_9i_{10}|d|}{}^e\, R_{i_{11}i_{12}|e|}{}^f\, R_{i_{13}i_{14}|f|}{}^g\, R_{i_{15}i_{16}|i|}{}^h\, R_{i_{17}i_{18}|h|}{}^i\, R_{i_{19}i_{20}|k|}{}^j\, R_{i_{21}i_{22}|j|}{}^k\, R_{i_{23}i_{24}|m|}{}^l\, R_{i_{25}i_{26}]l}{}^m +$$

$$+ 55{,}598{,}400\, R_{[i_1i_2|g|}{}^a\, R_{i_3i_4|a|}{}^b\, R_{i_5i_6|b|}{}^c\, R_{i_7i_8|c|}{}^d\, R_{i_9i_{10}|d|}{}^e\, R_{i_{11}i_{12}|e|}{}^f\, R_{i_{13}i_{14}|f|}{}^g\, R_{i_{15}i_{16}|i|}{}^h\, R_{i_{17}i_{18}|h|}{}^i\, R_{i_{19}i_{20}|k|}{}^j\, R_{i_{21}i_{22}|j|}{}^k\, R_{i_{23}i_{24}|l|}{}^l\, R_{i_{25}i_{26}]m}{}^m -$$

$$- 18{,}532{,}800\, R_{[i_1i_2|g|}{}^a\, R_{i_3i_4|a|}{}^b\, R_{i_5i_6|b|}{}^c\, R_{i_7i_8|c|}{}^d\, R_{i_9i_{10}|d|}{}^e\, R_{i_{11}i_{12}|e|}{}^f\, R_{i_{13}i_{14}|f|}{}^g\, R_{i_{15}i_{16}|i|}{}^h\, R_{i_{17}i_{18}|h|}{}^i\, R_{i_{19}i_{20}|j|}{}^j\, R_{i_{21}i_{22}|j|}{}^k\, R_{i_{23}i_{24}|l|}{}^l\, R_{i_{25}i_{26}]m}{}^m +$$

$$+ 1{,}235{,}520\, R_{[i_1i_2|g|}{}^a\, R_{i_3i_4|a|}{}^b\, R_{i_5i_6|b|}{}^c\, R_{i_7i_8|c|}{}^d\, R_{i_9i_{10}|d|}{}^e\, R_{i_{11}i_{12}|e|}{}^f\, R_{i_{13}i_{14}|f|}{}^g\, R_{i_{15}i_{16}|h|}{}^h\, R_{i_{17}i_{18}|i|}{}^i\, R_{i_{19}i_{20}|j|}{}^j\, R_{i_{21}i_{22}|k|}{}^k\, R_{i_{23}i_{24}|l|}{}^l\, R_{i_{25}i_{26}]m}{}^m +$$

$$+ 86{,}486{,}400\, R_{[i_1i_2|f|}{}^a\, R_{i_3i_4|a|}{}^b\, R_{i_5i_6|b|}{}^c\, R_{i_7i_8|c|}{}^d\, R_{i_9i_{10}|d|}{}^e\, R_{i_{11}i_{12}|e|}{}^f\, R_{i_{13}i_{14}|l|}{}^g\, R_{i_{15}i_{16}|g|}{}^h\, R_{i_{17}i_{18}|h|}{}^i\, R_{i_{19}i_{20}|i|}{}^j\, R_{i_{21}i_{22}|j|}{}^k\, R_{i_{23}i_{24}|k|}{}^l\, R_{i_{25}i_{26}]m}{}^m +$$

$$+ 103{,}783{,}680\, R_{[i_1i_2|f|}{}^a\, R_{i_3i_4|a|}{}^b\, R_{i_5i_6|b|}{}^c\, R_{i_7i_8|c|}{}^d\, R_{i_9i_{10}|d|}{}^e\, R_{i_{11}i_{12}|e|}{}^f\, R_{i_{13}i_{14}|k|}{}^g\, R_{i_{15}i_{16}|g|}{}^h\, R_{i_{17}i_{18}|h|}{}^i\, R_{i_{19}i_{20}|i|}{}^j\, R_{i_{21}i_{22}|j|}{}^k\, R_{i_{23}i_{24}|m|}{}^l\, R_{i_{25}i_{26}]l}{}^m -$$

$$- 103{,}783{,}680\, R_{[i_1i_2|f|}{}^a\, R_{i_3i_4|a|}{}^b\, R_{i_5i_6|b|}{}^c\, R_{i_7i_8|c|}{}^d\, R_{i_9i_{10}|d|}{}^e\, R_{i_{11}i_{12}|e|}{}^f\, R_{i_{13}i_{14}|k|}{}^g\, R_{i_{15}i_{16}|g|}{}^h\, R_{i_{17}i_{18}|h|}{}^i\, R_{i_{19}i_{20}|i|}{}^j\, R_{i_{21}i_{22}|j|}{}^k\, R_{i_{23}i_{24}|l|}{}^l\, R_{i_{25}i_{26}]m}{}^m +$$

$$+ 86{,}486{,}400\, R_{[i_1i_2|f|}{}^a\, R_{i_3i_4|a|}{}^b\, R_{i_5i_6|b|}{}^c\, R_{i_7i_8|c|}{}^d\, R_{i_9i_{10}|d|}{}^e\, R_{i_{11}i_{12}|e|}{}^f\, R_{i_{13}i_{14}|j|}{}^g\, R_{i_{15}i_{16}|g|}{}^h\, R_{i_{17}i_{18}|h|}{}^i\, R_{i_{19}i_{20}|i|}{}^j\, R_{i_{21}i_{22}|m|}{}^k\, R_{i_{23}i_{24}|k|}{}^l\, R_{i_{25}i_{26}]l}{}^m -$$

$$- 129{,}729{,}600\, R_{[i_1i_2|f|}{}^a\, R_{i_3i_4|a|}{}^b\, R_{i_5i_6|b|}{}^c\, R_{i_7i_8|c|}{}^d\, R_{i_9i_{10}|d|}{}^e\, R_{i_{11}i_{12}|e|}{}^f\, R_{i_{13}i_{14}|j|}{}^g\, R_{i_{15}i_{16}|g|}{}^h\, R_{i_{17}i_{18}|h|}{}^i\, R_{i_{19}i_{20}|i|}{}^j\, R_{i_{21}i_{22}|l|}{}^k\, R_{i_{23}i_{24}|k|}{}^l\, R_{i_{25}i_{26}]m}{}^m +$$

$$+ 43{,}243{,}200\, R_{[i_1i_2|f|}{}^a\, R_{i_3i_4|a|}{}^b\, R_{i_5i_6|b|}{}^c\, R_{i_7i_8|c|}{}^d\, R_{i_9i_{10}|d|}{}^e\, R_{i_{11}i_{12}|e|}{}^f\, R_{i_{13}i_{14}|j|}{}^g\, R_{i_{15}i_{16}|g|}{}^h\, R_{i_{17}i_{18}|h|}{}^i\, R_{i_{19}i_{20}|i|}{}^j\, R_{i_{21}i_{22}|k|}{}^k\, R_{i_{23}i_{24}|l|}{}^l\, R_{i_{25}i_{26}]m}{}^m -$$

$$- 57{,}657{,}600\, R_{[i_1i_2|f|}{}^a\, R_{i_3i_4|a|}{}^b\, R_{i_5i_6|b|}{}^c\, R_{i_7i_8|c|}{}^d\, R_{i_9i_{10}|d|}{}^e\, R_{i_{11}i_{12}|e|}{}^f\, R_{i_{13}i_{14}|i|}{}^g\, R_{i_{15}i_{16}|g|}{}^h\, R_{i_{17}i_{18}|h|}{}^i\, R_{i_{19}i_{20}|l|}{}^j\, R_{i_{21}i_{22}|j|}{}^k\, R_{i_{23}i_{24}|k|}{}^l\, R_{i_{25}i_{26}]m}{}^m -$$

$$- 43{,}243{,}200\, R_{[i_1i_2|f|}{}^a R_{i_3i_4|a|}{}^b R_{i_5i_6|b|}{}^c R_{i_7i_8|c|}{}^d R_{i_9i_{10}|d|}{}^e R_{i_{11}i_{12}|e|}{}^f R_{i_{13}i_{14}|i|}{}^g R_{i_{15}i_{16}|g|}{}^h R_{i_{17}i_{18}|h|}{}^i R_{i_{19}i_{20}|k|}{}^j R_{i_{21}i_{22}|j|}{}^k R_{i_{23}i_{24}|m|}{}^l R_{i_{25}i_{26}]l}{}^m +$$

$$+ 86{,}486{,}400\, R_{[i_1i_2|f|}{}^a R_{i_3i_4|a|}{}^b R_{i_5i_6|b|}{}^c R_{i_7i_8|c|}{}^d R_{i_9i_{10}|d|}{}^e R_{i_{11}i_{12}|e|}{}^f R_{i_{13}i_{14}|i|}{}^g R_{i_{15}i_{16}|g|}{}^h R_{i_{17}i_{18}|h|}{}^i R_{i_{19}i_{20}|k|}{}^j R_{i_{21}i_{22}|j|}{}^k R_{i_{23}i_{24}|l|}{}^l R_{i_{25}i_{26}]m}{}^m -$$

$$- 14{,}414{,}400\, R_{[i_1i_2|f|}{}^a R_{i_3i_4|a|}{}^b R_{i_5i_6|b|}{}^c R_{i_7i_8|c|}{}^d R_{i_9i_{10}|d|}{}^e R_{i_{11}i_{12}|e|}{}^f R_{i_{13}i_{14}|i|}{}^g R_{i_{15}i_{16}|g|}{}^h R_{i_{17}i_{18}|h|}{}^i R_{i_{19}i_{20}|k|}{}^j R_{i_{21}i_{22}|j|}{}^k R_{i_{23}i_{24}|l|}{}^l R_{i_{25}i_{26}]m}{}^m +$$

$$+ 21{,}621{,}600\, R_{[i_1i_2|f|}{}^a R_{i_3i_4|a|}{}^b R_{i_5i_6|b|}{}^c R_{i_7i_8|c|}{}^d R_{i_9i_{10}|d|}{}^e R_{i_{11}i_{12}|e|}{}^f R_{i_{13}i_{14}|h|}{}^g R_{i_{15}i_{16}|g|}{}^h R_{i_{17}i_{18}|j|}{}^i R_{i_{19}i_{20}|i|}{}^j R_{i_{21}i_{22}|k|}{}^k R_{i_{23}i_{24}|k|}{}^l R_{i_{25}i_{26}]m}{}^m -$$

$$- 21{,}621{,}600\, R_{[i_1i_2|f|}{}^a R_{i_3i_4|a|}{}^b R_{i_5i_6|b|}{}^c R_{i_7i_8|c|}{}^d R_{i_9i_{10}|d|}{}^e R_{i_{11}i_{12}|e|}{}^f R_{i_{13}i_{14}|h|}{}^g R_{i_{15}i_{16}|g|}{}^h R_{i_{17}i_{18}|j|}{}^i R_{i_{19}i_{20}|i|}{}^j R_{i_{21}i_{22}|k|}{}^k R_{i_{23}i_{24}|l|}{}^l R_{i_{25}i_{26}]m}{}^m +$$

$$+ 4{,}324{,}320\, R_{[i_1i_2|f|}{}^a R_{i_3i_4|a|}{}^b R_{i_5i_6|b|}{}^c R_{i_7i_8|c|}{}^d R_{i_9i_{10}|d|}{}^e R_{i_{11}i_{12}|e|}{}^f R_{i_{13}i_{14}|h|}{}^g R_{i_{15}i_{16}|g|}{}^h R_{i_{17}i_{18}|i|}{}^i R_{i_{19}i_{20}|j|}{}^j R_{i_{21}i_{22}|k|}{}^k R_{i_{23}i_{24}|l|}{}^l R_{i_{25}i_{26}]m}{}^m -$$

$$- 205{,}920\, R_{[i_1i_2|f|}{}^a R_{i_3i_4|a|}{}^b R_{i_5i_6|b|}{}^c R_{i_7i_8|c|}{}^d R_{i_9i_{10}|d|}{}^e R_{i_{11}i_{12}|e|}{}^f R_{i_{13}i_{14}|g|}{}^g R_{i_{15}i_{16}|h|}{}^h R_{i_{17}i_{18}|i|}{}^i R_{i_{19}i_{20}|j|}{}^j R_{i_{21}i_{22}|k|}{}^k R_{i_{23}i_{24}|l|}{}^l R_{i_{25}i_{26}]m}{}^m +$$

$$+ 41{,}513{,}472\, R_{[i_1i_2|e|}{}^a R_{i_3i_4|a|}{}^b R_{i_5i_6|b|}{}^c R_{i_7i_8|c|}{}^d R_{i_9i_{10}|d|}{}^e R_{i_{11}i_{12}|j|}{}^f R_{i_{13}i_{14}|f|}{}^g R_{i_{15}i_{16}|g|}{}^h R_{i_{17}i_{18}|h|}{}^i R_{i_{19}i_{20}|i|}{}^j R_{i_{21}i_{22}|m|}{}^k R_{i_{23}i_{24}|k|}{}^l R_{i_{25}i_{26}]l}{}^m -$$

$$- 62{,}270{,}208\, R_{[i_1i_2|e|}{}^a R_{i_3i_4|a|}{}^b R_{i_5i_6|b|}{}^c R_{i_7i_8|c|}{}^d R_{i_9i_{10}|d|}{}^e R_{i_{11}i_{12}|j|}{}^f R_{i_{13}i_{14}|f|}{}^g R_{i_{15}i_{16}|g|}{}^h R_{i_{17}i_{18}|h|}{}^i R_{i_{19}i_{20}|i|}{}^j R_{i_{21}i_{22}|k|}{}^k R_{i_{23}i_{24}|l|}{}^l R_{i_{25}i_{26}]m}{}^m +$$

$$+ 20{,}756{,}736\, R_{[i_1i_2|e|}{}^a R_{i_3i_4|a|}{}^b R_{i_5i_6|b|}{}^c R_{i_7i_8|c|}{}^d R_{i_9i_{10}|d|}{}^e R_{i_{11}i_{12}|i|}{}^f R_{i_{13}i_{14}|f|}{}^g R_{i_{15}i_{16}|g|}{}^h R_{i_{17}i_{18}|h|}{}^i R_{i_{19}i_{20}|j|}{}^j R_{i_{21}i_{22}|k|}{}^k R_{i_{23}i_{24}|l|}{}^l R_{i_{25}i_{26}]m}{}^m +$$

$$+ 38{,}918{,}880\, R_{[i_1i_2|e|}{}^a R_{i_3i_4|a|}{}^b R_{i_5i_6|b|}{}^c R_{i_7i_8|c|}{}^d R_{i_9i_{10}|d|}{}^e R_{i_{11}i_{12}|i|}{}^f R_{i_{13}i_{14}|f|}{}^g R_{i_{15}i_{16}|g|}{}^h R_{i_{17}i_{18}|h|}{}^i R_{i_{19}i_{20}|m|}{}^j R_{i_{21}i_{22}|j|}{}^k R_{i_{23}i_{24}|k|}{}^l R_{i_{25}i_{26}]l}{}^m -$$

$$- 103{,}783{,}680\, R_{[i_1i_2|e|}{}^a R_{i_3i_4|a|}{}^b R_{i_5i_6|b|}{}^c R_{i_7i_8|c|}{}^d R_{i_9i_{10}|d|}{}^e R_{i_{11}i_{12}|i|}{}^f R_{i_{13}i_{14}|f|}{}^g R_{i_{15}i_{16}|g|}{}^h R_{i_{17}i_{18}|h|}{}^i R_{i_{19}i_{20}|l|}{}^j R_{i_{21}i_{22}|j|}{}^k R_{i_{23}i_{24}|k|}{}^l R_{i_{25}i_{26}]m}{}^m -$$

$$- 38{,}918{,}880\, R_{[i_1i_2|e|}{}^a R_{i_3i_4|a|}{}^b R_{i_5i_6|b|}{}^c R_{i_7i_8|c|}{}^d R_{i_9i_{10}|d|}{}^e R_{i_{11}i_{12}|i|}{}^f R_{i_{13}i_{14}|f|}{}^g R_{i_{15}i_{16}|g|}{}^h R_{i_{17}i_{18}|h|}{}^i R_{i_{19}i_{20}|k|}{}^j R_{i_{21}i_{22}|j|}{}^k R_{i_{23}i_{24}|l|}{}^l R_{i_{25}i_{26}]m}{}^m +$$

$$+ 77{,}837{,}760\, R_{[i_1i_2|e|}{}^a R_{i_3i_4|a|}{}^b R_{i_5i_6|b|}{}^c R_{i_7i_8|c|}{}^d R_{i_9i_{10}|d|}{}^e R_{i_{11}i_{12}|i|}{}^f R_{i_{13}i_{14}|f|}{}^g R_{i_{15}i_{16}|g|}{}^h R_{i_{17}i_{18}|h|}{}^i R_{i_{19}i_{20}|j|}{}^j R_{i_{21}i_{22}|k|}{}^k R_{i_{23}i_{24}|l|}{}^l R_{i_{25}i_{26}]m}{}^m -$$

$$- 12{,}972{,}960\, R_{[i_1i_2|e|}{}^a R_{i_3i_4|a|}{}^b R_{i_5i_6|b|}{}^c R_{i_7i_8|c|}{}^d R_{i_9i_{10}|d|}{}^e R_{i_{11}i_{12}|i|}{}^f R_{i_{13}i_{14}|f|}{}^g R_{i_{15}i_{16}|g|}{}^h R_{i_{17}i_{18}|h|}{}^i R_{i_{19}i_{20}|j|}{}^j R_{i_{21}i_{22}|k|}{}^k R_{i_{23}i_{24}|l|}{}^l R_{i_{25}i_{26}]m}{}^m -$$

$$- 34{,}594{,}560\, R_{[i_1i_2|e|}{}^a R_{i_3i_4|a|}{}^b R_{i_5i_6|b|}{}^c R_{i_7i_8|c|}{}^d R_{i_9i_{10}|d|}{}^e R_{i_{11}i_{12}|h|}{}^f R_{i_{13}i_{14}|f|}{}^g R_{i_{15}i_{16}|g|}{}^h R_{i_{17}i_{18}|k|}{}^i R_{i_{19}i_{20}|i|}{}^j R_{i_{21}i_{22}|j|}{}^k R_{i_{23}i_{24}|m|}{}^l R_{i_{25}i_{26}]l}{}^m +$$

$$+ 34{,}594{,}560\, R_{[i_1i_2|e|}{}^a R_{i_3i_4|a|}{}^b R_{i_5i_6|b|}{}^c R_{i_7i_8|c|}{}^d R_{i_9i_{10}|d|}{}^e R_{i_{11}i_{12}|h|}{}^f R_{i_{13}i_{14}|f|}{}^g R_{i_{15}i_{16}|g|}{}^h R_{i_{17}i_{18}|k|}{}^i R_{i_{19}i_{20}|i|}{}^j R_{i_{21}i_{22}|j|}{}^k R_{i_{23}i_{24}|l|}{}^l R_{i_{25}i_{26}]m}{}^m +$$

$$+ 51{,}891{,}840\, R_{[i_1i_2|e|}{}^a R_{i_3i_4|a|}{}^b R_{i_5i_6|b|}{}^c R_{i_7i_8|c|}{}^d R_{i_9i_{10}|d|}{}^e R_{i_{11}i_{12}|h|}{}^f R_{i_{13}i_{14}|f|}{}^g R_{i_{15}i_{16}|g|}{}^h R_{i_{17}i_{18}|j|}{}^i R_{i_{19}i_{20}|i|}{}^j R_{i_{21}i_{22}|k|}{}^k R_{i_{23}i_{24}|l|}{}^l R_{i_{25}i_{26}]m}{}^m -$$

$$- 34{,}594{,}560\, R_{[i_1i_2|e|}{}^a R_{i_3i_4|a|}{}^b R_{i_5i_6|b|}{}^c R_{i_7i_8|c|}{}^d R_{i_9i_{10}|d|}{}^e R_{i_{11}i_{12}|h|}{}^f R_{i_{13}i_{14}|f|}{}^g R_{i_{15}i_{16}|g|}{}^h R_{i_{17}i_{18}|j|}{}^i R_{i_{19}i_{20}|i|}{}^j R_{i_{21}i_{22}|k|}{}^k R_{i_{23}i_{24}|l|}{}^l R_{i_{25}i_{26}]m}{}^m +$$

$$+ 3{,}459{,}456\, R_{[i_1i_2|e|}{}^a R_{i_3i_4|a|}{}^b R_{i_5i_6|b|}{}^c R_{i_7i_8|c|}{}^d R_{i_9i_{10}|d|}{}^e R_{i_{11}i_{12}|f|}{}^f R_{i_{13}i_{14}|f|}{}^g R_{i_{15}i_{16}|g|}{}^h R_{i_{17}i_{18}|i|}{}^i R_{i_{19}i_{20}|j|}{}^j R_{i_{21}i_{22}|k|}{}^k R_{i_{23}i_{24}|l|}{}^l R_{i_{25}i_{26}]m}{}^m +$$

$$+ 3{,}243{,}240\, R_{[i_1i_2|e|}{}^a R_{i_3i_4|a|}{}^b R_{i_5i_6|b|}{}^c R_{i_7i_8|c|}{}^d R_{i_9i_{10}|d|}{}^e R_{i_{11}i_{12}|g|}{}^f R_{i_{13}i_{14}|f|}{}^g R_{i_{15}i_{16}|i|}{}^h R_{i_{17}i_{18}|h|}{}^i R_{i_{19}i_{20}|k|}{}^j R_{i_{21}i_{22}|j|}{}^k R_{i_{23}i_{24}|m|}{}^l R_{i_{25}i_{26}]l}{}^m -$$

$$- 12{,}972{,}960\, R_{[i_1i_2|e|}{}^a R_{i_3i_4|a|}{}^b R_{i_5i_6|b|}{}^c R_{i_7i_8|c|}{}^d R_{i_9i_{10}|d|}{}^e R_{i_{11}i_{12}|g|}{}^f R_{i_{13}i_{14}|f|}{}^g R_{i_{15}i_{16}|i|}{}^h R_{i_{17}i_{18}|h|}{}^i R_{i_{19}i_{20}|k|}{}^j R_{i_{21}i_{22}|j|}{}^k R_{i_{23}i_{24}|l|}{}^l R_{i_{25}i_{26}]m}{}^m +$$

$$+ 6{,}486{,}480\, R_{[i_1i_2|e|}{}^a R_{i_3i_4|a|}{}^b R_{i_5i_6|b|}{}^c R_{i_7i_8|c|}{}^d R_{i_9i_{10}|d|}{}^e R_{i_{11}i_{12}|g|}{}^f R_{i_{13}i_{14}|f|}{}^g R_{i_{15}i_{16}|i|}{}^h R_{i_{17}i_{18}|h|}{}^i R_{i_{19}i_{20}|j|}{}^j R_{i_{21}i_{22}|k|}{}^k R_{i_{23}i_{24}|l|}{}^l R_{i_{25}i_{26}]m}{}^m -$$

$$- 864{,}864\, R_{[i_1i_2|e|}{}^a R_{i_3i_4|a|}{}^b R_{i_5i_6|b|}{}^c R_{i_7i_8|c|}{}^d R_{i_9i_{10}|d|}{}^e R_{i_{11}i_{12}|g|}{}^f R_{i_{13}i_{14}|f|}{}^g R_{i_{15}i_{16}|h|}{}^h R_{i_{17}i_{18}|i|}{}^i R_{i_{19}i_{20}|j|}{}^j R_{i_{21}i_{22}|k|}{}^k R_{i_{23}i_{24}|l|}{}^l R_{i_{25}i_{26}]m}{}^m +$$

$$+ 30{,}888\, R_{[i_1i_2|e|}{}^a R_{i_3i_4|a|}{}^b R_{i_5i_6|b|}{}^c R_{i_7i_8|c|}{}^d R_{i_9i_{10}|d|}{}^e R_{i_{11}i_{12}|f|}{}^f R_{i_{13}i_{14}|g|}{}^g R_{i_{15}i_{16}|h|}{}^h R_{i_{17}i_{18}|i|}{}^i R_{i_{19}i_{20}|j|}{}^j R_{i_{21}i_{22}|k|}{}^k R_{i_{23}i_{24}|l|}{}^l R_{i_{25}i_{26}]m}{}^m -$$

$$- 16{,}216{,}200\, R_{[i_1i_2|d|}{}^a R_{i_3i_4|a|}{}^b R_{i_5i_6|b|}{}^c R_{i_7i_8|c|}{}^d R_{i_9i_{10}|h|}{}^e R_{i_{11}i_{12}|e|}{}^f R_{i_{13}i_{14}|f|}{}^g R_{i_{15}i_{16}|g|}{}^h R_{i_{17}i_{18}|l|}{}^i R_{i_{19}i_{20}|i|}{}^j R_{i_{21}i_{22}|j|}{}^k R_{i_{23}i_{24}|k|}{}^l R_{i_{25}i_{26}]m}{}^m -$$

$$- 32{,}432{,}400\, R_{[i_1i_2|d|}{}^a R_{i_3i_4|a|}{}^b R_{i_5i_6|b|}{}^c R_{i_7i_8|c|}{}^d R_{i_9i_{10}|h|}{}^e R_{i_{11}i_{12}|e|}{}^f R_{i_{13}i_{14}|f|}{}^g R_{i_{15}i_{16}|g|}{}^h R_{i_{17}i_{18}|k|}{}^i R_{i_{19}i_{20}|i|}{}^j R_{i_{21}i_{22}|j|}{}^k R_{i_{23}i_{24}|m|}{}^l R_{i_{25}i_{26}]l}{}^m +$$

$$+ 32{,}432{,}400\, R_{[i_1i_2|d|}{}^a R_{i_3i_4|a|}{}^b R_{i_5i_6|b|}{}^c R_{i_7i_8|c|}{}^d R_{i_9i_{10}|h|}{}^e R_{i_{11}i_{12}|e|}{}^f R_{i_{13}i_{14}|f|}{}^g R_{i_{15}i_{16}|g|}{}^h R_{i_{17}i_{18}|k|}{}^i R_{i_{19}i_{20}|i|}{}^j R_{i_{21}i_{22}|j|}{}^k R_{i_{23}i_{24}|l|}{}^l R_{i_{25}i_{26}]m}{}^m +$$

$$+ 24{,}324{,}300\, R_{[i_1i_2|d|}{}^a R_{i_3i_4|a|}{}^b R_{i_5i_6|b|}{}^c R_{i_7i_8|c|}{}^d R_{i_9i_{10}|h|}{}^e R_{i_{11}i_{12}|e|}{}^f R_{i_{13}i_{14}|f|}{}^g R_{i_{15}i_{16}|g|}{}^h R_{i_{17}i_{18}|i|}{}^i R_{i_{19}i_{20}|j|}{}^j R_{i_{21}i_{22}|k|}{}^k R_{i_{23}i_{24}|l|}{}^l R_{i_{25}i_{26}]m}{}^m -$$

$$- 16{,}216{,}200\, R_{[i_1i_2|d|}{}^a R_{i_3i_4|a|}{}^b R_{i_5i_6|b|}{}^c R_{i_7i_8|c|}{}^d R_{i_9i_{10}|h|}{}^e R_{i_{11}i_{12}|e|}{}^f R_{i_{13}i_{14}|f|}{}^g R_{i_{15}i_{16}|g|}{}^h R_{i_{17}i_{18}|i|}{}^i R_{i_{19}i_{20}|j|}{}^j R_{i_{21}i_{22}|k|}{}^k R_{i_{23}i_{24}|l|}{}^l R_{i_{25}i_{26}]m}{}^m +$$

$$+ 1{,}621{,}620\, R_{[i_1i_2|d|}{}^a R_{i_3i_4|a|}{}^b R_{i_5i_6|b|}{}^c R_{i_7i_8|c|}{}^d R_{i_9i_{10}|h|}{}^e R_{i_{11}i_{12}|e|}{}^f R_{i_{13}i_{14}|f|}{}^g R_{i_{15}i_{16}|g|}{}^h R_{i_{17}i_{18}|i|}{}^i R_{i_{19}i_{20}|j|}{}^j R_{i_{21}i_{22}|k|}{}^k R_{i_{23}i_{24}|l|}{}^l R_{i_{25}i_{26}]m}{}^m -$$

$$- 9{,}609{,}600\, R_{[i_1i_2|d|}{}^a R_{i_3i_4|a|}{}^b R_{i_5i_6|b|}{}^c R_{i_7i_8|c|}{}^d R_{i_9i_{10}|h|}{}^e R_{i_{11}i_{12}|e|}{}^f R_{i_{13}i_{14}|g|}{}^g R_{i_{15}i_{16}|j|}{}^h R_{i_{17}i_{18}|h|}{}^i R_{i_{19}i_{20}|i|}{}^j R_{i_{21}i_{22}|m|}{}^k R_{i_{23}i_{24}|k|}{}^l R_{i_{25}i_{26}]l}{}^m +$$

$$+ 43{,}243{,}200\, R_{[i_1i_2|d|}{}^a R_{i_3i_4|a|}{}^b R_{i_5i_6|b|}{}^c R_{i_7i_8|c|}{}^d R_{i_9i_{10}|g|}{}^e R_{i_{11}i_{12}|e|}{}^f R_{i_{13}i_{14}|f|}{}^g R_{i_{15}i_{16}|j|}{}^h R_{i_{17}i_{18}|h|}{}^i R_{i_{19}i_{20}|i|}{}^j R_{i_{21}i_{22}|k|}{}^k R_{i_{23}i_{24}|l|}{}^l R_{i_{25}i_{26}]m}{}^m -$$

$$- 14{,}414{,}400\, R_{[i_1i_2|d|}{}^a R_{i_3i_4|a|}{}^b R_{i_5i_6|b|}{}^c R_{i_7i_8|c|}{}^d R_{i_9i_{10}|g|}{}^e R_{i_{11}i_{12}|e|}{}^f R_{i_{13}i_{14}|f|}{}^g R_{i_{15}i_{16}|j|}{}^h R_{i_{17}i_{18}|h|}{}^i R_{i_{19}i_{20}|i|}{}^j R_{i_{21}i_{22}|k|}{}^k R_{i_{23}i_{24}|l|}{}^l R_{i_{25}i_{26}]m}{}^m +$$

$$+ 10{,}810{,}800\, R_{[i_1i_2|d|}{}^a R_{i_3i_4|a|}{}^b R_{i_5i_6|b|}{}^c R_{i_7i_8|c|}{}^d R_{i_9i_{10}|g|}{}^e R_{i_{11}i_{12}|e|}{}^f R_{i_{13}i_{14}|f|}{}^g R_{i_{15}i_{16}|i|}{}^h R_{i_{17}i_{18}|h|}{}^i R_{i_{19}i_{20}|j|}{}^j R_{i_{21}i_{22}|k|}{}^k R_{i_{23}i_{24}|m|}{}^l R_{i_{25}i_{26}]l}{}^m -$$

$$- 32{,}432{,}400\, R_{[i_1i_2|d|}{}^a R_{i_3i_4|a|}{}^b R_{i_5i_6|b|}{}^c R_{i_7i_8|c|}{}^d R_{i_9i_{10}|g|}{}^e R_{i_{11}i_{12}|e|}{}^f R_{i_{13}i_{14}|f|}{}^g R_{i_{15}i_{16}|i|}{}^h R_{i_{17}i_{18}|h|}{}^i R_{i_{19}i_{20}|j|}{}^j R_{i_{21}i_{22}|k|}{}^k R_{i_{23}i_{24}|l|}{}^l R_{i_{25}i_{26}]m}{}^m +$$

$$+ 10{,}810{,}800\, R_{[i_1i_2|d|}{}^a R_{i_3i_4|a|}{}^b R_{i_5i_6|b|}{}^c R_{i_7i_8|c|}{}^d R_{i_9i_{10}|g|}{}^e R_{i_{11}i_{12}|e|}{}^f R_{i_{13}i_{14}|f|}{}^g R_{i_{15}i_{16}|i|}{}^h R_{i_{17}i_{18}|h|}{}^i R_{i_{19}i_{20}|j|}{}^j R_{i_{21}i_{22}|k|}{}^k R_{i_{23}i_{24}|l|}{}^l R_{i_{25}i_{26}]m}{}^m -$$

$$- 720{,}720\, R_{[i_1i_2|d|}{}^a R_{i_3i_4|a|}{}^b R_{i_5i_6|b|}{}^c R_{i_7i_8|c|}{}^d R_{i_9i_{10}|g|}{}^e R_{i_{11}i_{12}|e|}{}^f R_{i_{13}i_{14}|f|}{}^g R_{i_{15}i_{16}|g|}{}^h R_{i_{17}i_{18}|i|}{}^i R_{i_{19}i_{20}|j|}{}^j R_{i_{21}i_{22}|k|}{}^k R_{i_{23}i_{24}|l|}{}^l R_{i_{25}i_{26}]m}{}^m -$$

$$- 4{,}054{,}050\, R_{[i_1i_2|d|}{}^a R_{i_3i_4|a|}{}^b R_{i_5i_6|b|}{}^c R_{i_7i_8|c|}{}^d R_{i_9i_{10}|f|}{}^e R_{i_{11}i_{12}|e|}{}^f R_{i_{13}i_{14}|g|}{}^g R_{i_{15}i_{16}|h|}{}^h R_{i_{17}i_{18}|i|}{}^i R_{i_{19}i_{20}|j|}{}^j R_{i_{21}i_{22}|k|}{}^k R_{i_{23}i_{24}|l|}{}^l R_{i_{25}i_{26}]m}{}^m +$$

$$+ 5{,}405{,}400\, R_{[i_1i_2|d|}{}^a R_{i_3i_4|a|}{}^b R_{i_5i_6|b|}{}^c R_{i_7i_8|c|}{}^d R_{i_9i_{10}|e|}{}^e R_{i_{11}i_{12}|e|}{}^f R_{i_{13}i_{14}|h|}{}^g R_{i_{15}i_{16}|g|}{}^h R_{i_{17}i_{18}|j|}{}^i R_{i_{19}i_{20}|i|}{}^j R_{i_{21}i_{22}|k|}{}^k R_{i_{23}i_{24}|l|}{}^l R_{i_{25}i_{26}]m}{}^m -$$

$$- 1{,}621{,}620\, R_{[i_1i_2|d|}{}^a R_{i_3i_4|a|}{}^b R_{i_5i_6|b|}{}^c R_{i_7i_8|c|}{}^d R_{i_9i_{10}|e|}{}^e R_{i_{11}i_{12}|e|}{}^f R_{i_{13}i_{14}|h|}{}^g R_{i_{15}i_{16}|g|}{}^h R_{i_{17}i_{18}|i|}{}^i R_{i_{19}i_{20}|j|}{}^j R_{i_{21}i_{22}|k|}{}^k R_{i_{23}i_{24}|l|}{}^l R_{i_{25}i_{26}]m}{}^m +$$

$$+ 154{,}440\, R_{[i_1i_2|d|}{}^a R_{i_3i_4|a|}{}^b R_{i_5i_6|b|}{}^c R_{i_7i_8|c|}{}^d R_{i_9i_{10}|f|}{}^e R_{i_{11}i_{12}|e|}{}^f R_{i_{13}i_{14}|g|}{}^g R_{i_{15}i_{16}|h|}{}^h R_{i_{17}i_{18}|i|}{}^i R_{i_{19}i_{20}|j|}{}^j R_{i_{21}i_{22}|k|}{}^k R_{i_{23}i_{24}|l|}{}^l R_{i_{25}i_{26}]m}{}^m -$$

$$- 4290\, R_{[i_1i_2|d|}{}^a R_{i_3i_4|a|}{}^b R_{i_5i_6|b|}{}^c R_{i_7i_8|c|}{}^d R_{i_9i_{10}|e|}{}^e R_{i_{11}i_{12}|f|}{}^f R_{i_{13}i_{14}|g|}{}^g R_{i_{15}i_{16}|h|}{}^h R_{i_{17}i_{18}|i|}{}^i R_{i_{19}i_{20}|j|}{}^j R_{i_{21}i_{22}|k|}{}^k R_{i_{23}i_{24}|l|}{}^l R_{i_{25}i_{26}]m}{}^m +$$

$$+ 3{,}203{,}200\, R_{[i_1i_2|c|}{}^a R_{i_3i_4|a|}{}^b R_{i_5i_6|b|}{}^c R_{i_7i_8|f|}{}^d R_{i_9i_{10}|d|}{}^e R_{i_{11}i_{12}|e|}{}^f R_{i_{13}i_{14}|i|}{}^g R_{i_{15}i_{16}|g|}{}^h R_{i_{17}i_{18}|h|}{}^i R_{i_{19}i_{20}|l|}{}^j R_{i_{21}i_{22}|j|}{}^k R_{i_{23}i_{24}|k|}{}^l R_{i_{25}i_{26}]m}{}^m +$$

$$+ 4{,}804{,}800\, R_{[i_1i_2|c|}{}^a R_{i_3i_4|a|}{}^b R_{i_5i_6|b|}{}^c R_{i_7i_8|f|}{}^d R_{i_9i_{10}|d|}{}^e R_{i_{11}i_{12}|e|}{}^f R_{i_{13}i_{14}|i|}{}^g R_{i_{15}i_{16}|g|}{}^h R_{i_{17}i_{18}|h|}{}^i R_{i_{19}i_{20}|k|}{}^j R_{i_{21}i_{22}|j|}{}^k R_{i_{23}i_{24}|m|}{}^l R_{i_{25}i_{26}]l}{}^m -$$

$$- 9{,}609{,}600\, R_{[i_1i_2|c|}{}^a R_{i_3i_4|a|}{}^b R_{i_5i_6|b|}{}^c R_{i_7i_8|f|}{}^d R_{i_9i_{10}|d|}{}^e R_{i_{11}i_{12}|e|}{}^f R_{i_{13}i_{14}|i|}{}^g R_{i_{15}i_{16}|g|}{}^h R_{i_{17}i_{18}|h|}{}^i R_{i_{19}i_{20}|k|}{}^j R_{i_{21}i_{22}|j|}{}^k R_{i_{23}i_{24}|l|}{}^l R_{i_{25}i_{26}]m}{}^m +$$

$$+ 1{,}601{,}600\, R_{[i_1i_2|c|}{}^a R_{i_3i_4|a|}{}^b R_{i_5i_6|b|}{}^c R_{i_7i_8|f|}{}^d R_{i_9i_{10}|d|}{}^e R_{i_{11}i_{12}|e|}{}^f R_{i_{13}i_{14}|i|}{}^g R_{i_{15}i_{16}|g|}{}^h R_{i_{17}i_{18}|h|}{}^i R_{i_{19}i_{20}|j|}{}^j R_{i_{21}i_{22}|k|}{}^k R_{i_{23}i_{24}|l|}{}^l R_{i_{25}i_{26}]m}{}^m -$$

$$- 7{,}207{,}200\, R_{[i_1i_2|c|}{}^a R_{i_3i_4|a|}{}^b R_{i_5i_6|b|}{}^c R_{i_7i_8|f|}{}^d R_{i_9i_{10}|d|}{}^e R_{i_{11}i_{12}|e|}{}^f R_{i_{13}i_{14}|h|}{}^g R_{i_{15}i_{16}|g|}{}^h R_{i_{17}i_{18}|j|}{}^i R_{i_{19}i_{20}|i|}{}^j R_{i_{21}i_{22}|k|}{}^k R_{i_{23}i_{24}|l|}{}^l R_{i_{25}i_{26}]m}{}^m +$$

$$+ 7{,}207{,}200\, R_{[i_1i_2|c|}{}^a R_{i_3i_4|a|}{}^b R_{i_5i_6|b|}{}^c R_{i_7i_8|f|}{}^d R_{i_9i_{10}|d|}{}^e R_{i_{11}i_{12}|e|}{}^f R_{i_{13}i_{14}|h|}{}^g R_{i_{15}i_{16}|g|}{}^h R_{i_{17}i_{18}|i|}{}^i R_{i_{19}i_{20}|j|}{}^j R_{i_{21}i_{22}|k|}{}^k R_{i_{23}i_{24}|l|}{}^l R_{i_{25}i_{26}]m}{}^m -$$

$$- 1{,}441{,}440\, R_{[i_1i_2|c|}{}^a R_{i_3i_4|a|}{}^b R_{i_5i_6|b|}{}^c R_{i_7i_8|f|}{}^d R_{i_9i_{10}|d|}{}^e R_{i_{11}i_{12}|e|}{}^f R_{i_{13}i_{14}|h|}{}^g R_{i_{15}i_{16}|g|}{}^h R_{i_{17}i_{18}|i|}{}^i R_{i_{19}i_{20}|j|}{}^j R_{i_{21}i_{22}|k|}{}^k R_{i_{23}i_{24}|l|}{}^l R_{i_{25}i_{26}]m}{}^m +$$

$$+ 68{,}640\, R_{[i_1i_2|c|}{}^a R_{i_3i_4|a|}{}^b R_{i_5i_6|b|}{}^c R_{i_7i_8|f|}{}^d R_{i_9i_{10}|d|}{}^e R_{i_{11}i_{12}|e|}{}^f R_{i_{13}i_{14}|g|}{}^g R_{i_{15}i_{16}|h|}{}^h R_{i_{17}i_{18}|i|}{}^i R_{i_{19}i_{20}|j|}{}^j R_{i_{21}i_{22}|k|}{}^k R_{i_{23}i_{24}|l|}{}^l R_{i_{25}i_{26}]m}{}^m -$$

$$- 540{,}540\, R_{[i_1i_2|c|}{}^a R_{i_3i_4|a|}{}^b R_{i_5i_6|b|}{}^c R_{i_7i_8|e|}{}^d R_{i_9i_{10}|d|}{}^e R_{i_{11}i_{12}|f|}{}^f R_{i_{13}i_{14}|g|}{}^g R_{i_{15}i_{16}|h|}{}^h R_{i_{17}i_{18}|i|}{}^i R_{i_{19}i_{20}|j|}{}^j R_{i_{21}i_{22}|k|}{}^k R_{i_{23}i_{24}|m|}{}^l R_{i_{25}i_{26}]l}{}^m +$$

$$+ 2{,}702{,}700\, R_{[i_1i_2|c|}{}^a R_{i_3i_4|a|}{}^b R_{i_5i_6|b|}{}^c R_{i_7i_8|e|}{}^d R_{i_9i_{10}|d|}{}^e R_{i_{11}i_{12}|g|}{}^f R_{i_{13}i_{14}|f|}{}^g R_{i_{15}i_{16}|i|}{}^h R_{i_{17}i_{18}|h|}{}^i R_{i_{19}i_{20}|k|}{}^j R_{i_{21}i_{22}|j|}{}^k R_{i_{23}i_{24}|l|}{}^l R_{i_{25}i_{26}]m}{}^m -$$

$$- 1{,}801{,}800\, R_{[i_1i_2|c|}{}^a R_{i_3i_4|a|}{}^b R_{i_5i_6|b|}{}^c R_{i_7i_8|e|}{}^d R_{i_9i_{10}|d|}{}^e R_{i_{11}i_{12}|f|}{}^f R_{i_{13}i_{14}|g|}{}^g R_{i_{15}i_{16}|h|}{}^h R_{i_{17}i_{18}|i|}{}^i R_{i_{19}i_{20}|j|}{}^j R_{i_{21}i_{22}|k|}{}^k R_{i_{23}i_{24}|l|}{}^l R_{i_{25}i_{26}]m}{}^m +$$

$$+ 360{,}360\, R_{[i_1i_2|c|}{}^a R_{i_3i_4|a|}{}^b R_{i_5i_6|b|}{}^c R_{i_7i_8|e|}{}^d R_{i_9i_{10}|d|}{}^e R_{i_{11}i_{12}|f|}{}^f R_{i_{13}i_{14}|g|}{}^g R_{i_{15}i_{16}|h|}{}^h R_{i_{17}i_{18}|i|}{}^i R_{i_{19}i_{20}|j|}{}^j R_{i_{21}i_{22}|k|}{}^k R_{i_{23}i_{24}|l|}{}^l R_{i_{25}i_{26}]m}{}^m -$$

$$- 25{,}740\, R_{[i_1i_2|c|}{}^a R_{i_3i_4|a|}{}^b R_{i_5i_6|b|}{}^c R_{i_7i_8|e|}{}^d R_{i_9i_{10}|d|}{}^e R_{i_{11}i_{12}|f|}{}^f R_{i_{13}i_{14}|g|}{}^g R_{i_{15}i_{16}|h|}{}^h R_{i_{17}i_{18}|i|}{}^i R_{i_{19}i_{20}|j|}{}^j R_{i_{21}i_{22}|k|}{}^k R_{i_{23}i_{24}|l|}{}^l R_{i_{25}i_{26}]m}{}^m +$$

$$+ 572\, R_{[i_1i_2|c|}{}^a R_{i_3i_4|a|}{}^b R_{i_5i_6|b|}{}^c R_{i_7i_8|d|}{}^d R_{i_9i_{10}|e|}{}^e R_{i_{11}i_{12}|f|}{}^f R_{i_{13}i_{14}|g|}{}^g R_{i_{15}i_{16}|h|}{}^h R_{i_{17}i_{18}|i|}{}^i R_{i_{19}i_{20}|j|}{}^j R_{i_{21}i_{22}|k|}{}^k R_{i_{23}i_{24}|l|}{}^l R_{i_{25}i_{26}]m}{}^m +$$

$$+ 135{,}135\, R_{[i_1i_2|b|}{}^a R_{i_3i_4|a|}{}^b R_{i_5i_6|d|}{}^c R_{i_7i_8|c|}{}^d R_{i_9i_{10}|f|}{}^e R_{i_{11}i_{12}|e|}{}^f R_{i_{13}i_{14}|h|}{}^g R_{i_{15}i_{16}|g|}{}^h R_{i_{17}i_{18}|j|}{}^i R_{i_{19}i_{20}|i|}{}^j R_{i_{21}i_{22}|k|}{}^k R_{i_{23}i_{24}|l|}{}^l R_{i_{25}i_{26}]m}{}^m -$$

$$- 270{,}270\, R_{[i_1i_2|b|}{}^a R_{i_3i_4|a|}{}^b R_{i_5i_6|d|}{}^c R_{i_7i_8|c|}{}^d R_{i_9i_{10}|f|}{}^e R_{i_{11}i_{12}|e|}{}^f R_{i_{13}i_{14}|h|}{}^g R_{i_{15}i_{16}|g|}{}^h R_{i_{17}i_{18}|i|}{}^i R_{i_{19}i_{20}|j|}{}^j R_{i_{21}i_{22}|k|}{}^k R_{i_{23}i_{24}|l|}{}^l R_{i_{25}i_{26}]m}{}^m +$$

$$+ 135{,}135\, R_{[i_1i_2|b|}{}^a R_{i_3i_4|a|}{}^b R_{i_5i_6|d|}{}^c R_{i_7i_8|c|}{}^d R_{i_9i_{10}|f|}{}^e R_{i_{11}i_{12}|e|}{}^f R_{i_{13}i_{14}|h|}{}^g R_{i_{15}i_{16}|g|}{}^h R_{i_{17}i_{18}|i|}{}^i R_{i_{19}i_{20}|j|}{}^j R_{i_{21}i_{22}|k|}{}^k R_{i_{23}i_{24}|l|}{}^l R_{i_{25}i_{26}]m}{}^m -$$

$$- 25{,}740\, R_{[i_1i_2|b|}{}^a R_{i_3i_4|a|}{}^b R_{i_5i_6|d|}{}^c R_{i_7i_8|c|}{}^d R_{i_9i_{10}|f|}{}^e R_{i_{11}i_{12}|e|}{}^f R_{i_{13}i_{14}|g|}{}^g R_{i_{15}i_{16}|h|}{}^h R_{i_{17}i_{18}|i|}{}^i R_{i_{19}i_{20}|j|}{}^j R_{i_{21}i_{22}|k|}{}^k R_{i_{23}i_{24}|l|}{}^l R_{i_{25}i_{26}]m}{}^m +$$

$$+ 2145\, R_{[i_1i_2|b|}{}^a R_{i_3i_4|a|}{}^b R_{i_5i_6|d|}{}^c R_{i_7i_8|c|}{}^d R_{i_9i_{10}|e|}{}^e R_{i_{11}i_{12}|f|}{}^f R_{i_{13}i_{14}|g|}{}^g R_{i_{15}i_{16}|h|}{}^h R_{i_{17}i_{18}|i|}{}^i R_{i_{19}i_{20}|j|}{}^j R_{i_{21}i_{22}|k|}{}^k R_{i_{23}i_{24}|l|}{}^l R_{i_{25}i_{26}]m}{}^m -$$

$$- 78\, R_{[i_1i_2|b|}{}^a R_{i_3i_4|a|}{}^b R_{i_5i_6|c|}{}^c R_{i_7i_8|d|}{}^d R_{i_9i_{10}|e|}{}^e R_{i_{11}i_{12}|f|}{}^f R_{i_{13}i_{14}|g|}{}^g R_{i_{15}i_{16}|h|}{}^h R_{i_{17}i_{18}|i|}{}^i R_{i_{19}i_{20}|j|}{}^j R_{i_{21}i_{22}|k|}{}^k R_{i_{23}i_{24}|l|}{}^l R_{i_{25}i_{26}]m}{}^m +$$

$$+ R_{[i_1i_2|a|}{}^a R_{i_3i_4|b|}{}^b R_{i_5i_6|c|}{}^c R_{i_7i_8|d|}{}^d R_{i_9i_{10}|e|}{}^e R_{i_{11}i_{12}|f|}{}^f R_{i_{13}i_{14}|g|}{}^g R_{i_{15}i_{16}|h|}{}^h R_{i_{17}i_{18}|i|}{}^i R_{i_{19}i_{20}|j|}{}^j R_{i_{21}i_{22}|k|}{}^k R_{i_{23}i_{24}|l|}{}^l R_{i_{25}i_{26}]m}{}^m )$$

$$= \frac{i^{13}}{2^{26} \pi^{13} 13!} \Big( + 479{,}001{,}600\, R_{[i_1i_2|m|}{}^a R_{i_3i_4|a|}{}^b R_{i_5i_6|b|}{}^c R_{i_7i_8|c|}{}^d R_{i_9i_{10}|d|}{}^e R_{i_{11}i_{12}|e|}{}^f R_{i_{13}i_{14}|f|}{}^g R_{i_{15}i_{16}|g|}{}^h R_{i_{17}i_{18}|h|}{}^i R_{i_{19}i_{20}|i|}{}^j R_{i_{21}i_{22}|j|}{}^k R_{i_{23}i_{24}|k|}{}^l R_{i_{25}i_{26}]l}{}^m -$$

$$- 518{,}918{,}400\, R_{[i_1i_2|l|}{}^a R_{i_3i_4|a|}{}^b R_{i_5i_6|b|}{}^c R_{i_7i_8|c|}{}^d R_{i_9i_{10}|d|}{}^e R_{i_{11}i_{12}|e|}{}^f R_{i_{13}i_{14}|f|}{}^g R_{i_{15}i_{16}|g|}{}^h R_{i_{17}i_{18}|h|}{}^i R_{i_{19}i_{20}|i|}{}^j R_{i_{21}i_{22}|j|}{}^k R_{i_{23}i_{24}|k|}{}^l V_{i_{25}i_{26}]} -$$

$$- 283{,}046{,}400\, R_{[i_1i_2|k|}{}^a R_{i_3i_4|a|}{}^b R_{i_5i_6|b|}{}^c R_{i_7i_8|c|}{}^d R_{i_9i_{10}|d|}{}^e R_{i_{11}i_{12}|e|}{}^f R_{i_{13}i_{14}|f|}{}^g R_{i_{15}i_{16}|g|}{}^h R_{i_{17}i_{18}|h|}{}^i R_{i_{19}i_{20}|i|}{}^j R_{i_{21}i_{22}|j|}{}^k R_{i_{23}i_{24}|m|}{}^l R_{i_{25}i_{26}]l}{}^m +$$

$$+ 283{,}046{,}400\, R_{[i_1i_2|k|}{}^a R_{i_3i_4|a|}{}^b R_{i_5i_6|b|}{}^c R_{i_7i_8|c|}{}^d R_{i_9i_{10}|d|}{}^e R_{i_{11}i_{12}|e|}{}^f R_{i_{13}i_{14}|f|}{}^g R_{i_{15}i_{16}|g|}{}^h R_{i_{17}i_{18}|h|}{}^i R_{i_{19}i_{20}|i|}{}^j R_{i_{21}i_{22}|j|}{}^k V_{i_{23}i_{24}} V_{i_{25}i_{26}]} -$$

$$- 207{,}567{,}360\, R_{[i_1i_2|j|}{}^a R_{i_3i_4|a|}{}^b R_{i_5i_6|b|}{}^c R_{i_7i_8|c|}{}^d R_{i_9i_{10}|d|}{}^e R_{i_{11}i_{12}|e|}{}^f R_{i_{13}i_{14}|f|}{}^g R_{i_{15}i_{16}|g|}{}^h R_{i_{17}i_{18}|h|}{}^i R_{i_{19}i_{20}|i|}{}^j R_{i_{21}i_{22}|k|}{}^k R_{i_{23}i_{24}|l|}{}^l R_{i_{25}i_{26}]m}{}^m +$$

$$+ 311{,}351{,}040\, R_{[i_1i_2|j|}{}^a R_{i_3i_4|a|}{}^b R_{i_5i_6|b|}{}^c R_{i_7i_8|c|}{}^d R_{i_9i_{10}|d|}{}^e R_{i_{11}i_{12}|e|}{}^f R_{i_{13}i_{14}|f|}{}^g R_{i_{15}i_{16}|g|}{}^h R_{i_{17}i_{18}|h|}{}^i R_{i_{19}i_{20}|i|}{}^j R_{i_{21}i_{22}|l|}{}^k R_{i_{23}i_{24}|k|}{}^l V_{i_{25}i_{26}]} -$$

$$- 103{,}783{,}680\, R_{[i_1i_2|j|}{}^a R_{i_3i_4|a|}{}^b R_{i_5i_6|b|}{}^c R_{i_7i_8|c|}{}^d R_{i_9i_{10}|d|}{}^e R_{i_{11}i_{12}|e|}{}^f R_{i_{13}i_{14}|f|}{}^g R_{i_{15}i_{16}|g|}{}^h R_{i_{17}i_{18}|h|}{}^i R_{i_{19}i_{20}|i|}{}^j V_{i_{21}i_{22}} V_{i_{23}i_{24}} V_{i_{25}i_{26}]} -$$

$$- 172{,}972{,}800\, R_{[i_1i_2|i|}{}^a R_{i_3i_4|a|}{}^b R_{i_5i_6|b|}{}^c R_{i_7i_8|c|}{}^d R_{i_9i_{10}|d|}{}^e R_{i_{11}i_{12}|e|}{}^f R_{i_{13}i_{14}|f|}{}^g R_{i_{15}i_{16}|g|}{}^h R_{i_{17}i_{18}|h|}{}^i R_{i_{19}i_{20}|m|}{}^j R_{i_{21}i_{22}|j|}{}^k R_{i_{23}i_{24}|k|}{}^l R_{i_{25}i_{26}]l}{}^m +$$

$$+ 230{,}630{,}400\, R_{[i_1i_2|i|}{}^a R_{i_3i_4|a|}{}^b R_{i_5i_6|b|}{}^c R_{i_7i_8|c|}{}^d R_{i_9i_{10}|d|}{}^e R_{i_{11}i_{12}|e|}{}^f R_{i_{13}i_{14}|f|}{}^g R_{i_{15}i_{16}|g|}{}^h R_{i_{17}i_{18}|h|}{}^i R_{i_{19}i_{20}|l|}{}^j R_{i_{21}i_{22}|j|}{}^k R_{i_{23}i_{24}|k|}{}^l V_{i_{25}i_{26}]} +$$

$$+ 86{,}486{,}400\, R_{[i_1i_2|i|}{}^a R_{i_3i_4|a|}{}^b R_{i_5i_6|b|}{}^c R_{i_7i_8|c|}{}^d R_{i_9i_{10}|d|}{}^e R_{i_{11}i_{12}|e|}{}^f R_{i_{13}i_{14}|f|}{}^g R_{i_{15}i_{16}|g|}{}^h R_{i_{17}i_{18}|h|}{}^i R_{i_{19}i_{20}|k|}{}^j R_{i_{21}i_{22}|j|}{}^k R_{i_{23}i_{24}|m|}{}^l R_{i_{25}i_{26}]l}{}^m -$$

$$- 172{,}972{,}800\, R_{[i_1i_2|i|}{}^a R_{i_3i_4|a|}{}^b R_{i_5i_6|b|}{}^c R_{i_7i_8|c|}{}^d R_{i_9i_{10}|d|}{}^e R_{i_{11}i_{12}|e|}{}^f R_{i_{13}i_{14}|f|}{}^g R_{i_{15}i_{16}|g|}{}^h R_{i_{17}i_{18}|h|}{}^i R_{i_{19}i_{20}|k|}{}^j R_{i_{21}i_{22}|j|}{}^k V_{i_{23}i_{24}} V_{i_{25}i_{26}]} +$$

$$+ 28{,}828{,}800\, R_{[i_1i_2|i|}{}^a R_{i_3i_4|a|}{}^b R_{i_5i_6|b|}{}^c R_{i_7i_8|c|}{}^d R_{i_9i_{10}|d|}{}^e R_{i_{11}i_{12}|e|}{}^f R_{i_{13}i_{14}|f|}{}^g R_{i_{15}i_{16}|g|}{}^h R_{i_{17}i_{18}|h|}{}^i V_{i_{19}i_{20}} V_{i_{21}i_{22}} V_{i_{23}i_{24}} V_{i_{25}i_{26}]} -$$

$$- 155{,}675{,}520\, R_{[i_1i_2|h|}{}^a R_{i_3i_4|a|}{}^b R_{i_5i_6|b|}{}^c R_{i_7i_8|c|}{}^d R_{i_9i_{10}|d|}{}^e R_{i_{11}i_{12}|e|}{}^f R_{i_{13}i_{14}|f|}{}^g R_{i_{15}i_{16}|g|}{}^h R_{i_{17}i_{18}|m|}{}^i R_{i_{19}i_{20}|i|}{}^j R_{i_{21}i_{22}|j|}{}^k R_{i_{23}i_{24}|k|}{}^l R_{i_{25}i_{26}]l}{}^m +$$

$$+ 194{,}594{,}400\, R_{[i_1i_2|h|}{}^a R_{i_3i_4|a|}{}^b R_{i_5i_6|b|}{}^c R_{i_7i_8|c|}{}^d R_{i_9i_{10}|d|}{}^e R_{i_{11}i_{12}|e|}{}^f R_{i_{13}i_{14}|f|}{}^g R_{i_{15}i_{16}|g|}{}^h R_{i_{17}i_{18}|l|}{}^i R_{i_{19}i_{20}|i|}{}^j R_{i_{21}i_{22}|j|}{}^k R_{i_{23}i_{24}|k|}{}^l V_{i_{25}i_{26}]} +$$

$$+ 129729600\, R_{[i_1i_2|h|}{}^a R_{i_3i_4|a|}{}^b R_{i_5i_6|b|}{}^c R_{i_7i_8|c|}{}^d R_{i_9i_{10}|d|}{}^e R_{i_{11}i_{12}|e|}{}^f R_{i_{13}i_{14}|f|}{}^g R_{i_{15}i_{16}|g|}{}^h R_{i_{17}i_{18}|k|}{}^i R_{i_{19}i_{20}|i|}{}^j R_{i_{21}i_{22}|j|}{}^k R_{i_{23}i_{24}|m|}{}^l R_{i_{25}i_{26}]l}{}^m -$$

$$- 129729600\, R_{[i_1i_2|h|}{}^a R_{i_3i_4|a|}{}^b R_{i_5i_6|b|}{}^c R_{i_7i_8|c|}{}^d R_{i_9i_{10}|d|}{}^e R_{i_{11}i_{12}|e|}{}^f R_{i_{13}i_{14}|f|}{}^g R_{i_{15}i_{16}|g|}{}^h R_{i_{17}i_{18}|k|}{}^i R_{i_{19}i_{20}|i|}{}^j R_{i_{21}i_{22}|j|}{}^k V_{i_{23}i_{24}} V_{i_{25}i_{26}]} -$$

$$- 97297200\, R_{[i_1i_2|h|}{}^a R_{i_3i_4|a|}{}^b R_{i_5i_6|b|}{}^c R_{i_7i_8|c|}{}^d R_{i_9i_{10}|d|}{}^e R_{i_{11}i_{12}|e|}{}^f R_{i_{13}i_{14}|f|}{}^g R_{i_{15}i_{16}|g|}{}^h R_{i_{17}i_{18}|j|}{}^i R_{i_{19}i_{20}|i|}{}^j R_{i_{21}i_{22}|l|}{}^k R_{i_{23}i_{24}|k|}{}^l V_{i_{25}i_{26}]} +$$

$$+ 64{,}864{,}800\, R_{[i_1i_2|h|}{}^a R_{i_3i_4|a|}{}^b R_{i_5i_6|b|}{}^c R_{i_7i_8|c|}{}^d R_{i_9i_{10}|d|}{}^e R_{i_{11}i_{12}|e|}{}^f R_{i_{13}i_{14}|f|}{}^g R_{i_{15}i_{16}|g|}{}^h R_{i_{17}i_{18}|j|}{}^i R_{i_{19}i_{20}|i|}{}^j V_{i_{21}i_{22}} V_{i_{23}i_{24}} V_{i_{25}i_{26}]} -$$

$$- 6{,}486{,}480\ R_{[i_1i_2|h|}{}^a R_{i_3i_4|a|}{}^b R_{i_5i_6|b|}{}^c R_{i_7i_8|c|}{}^d R_{i_9i_{10}|d|}{}^e R_{i_{11}i_{12}|e|}{}^f R_{i_{13}i_{14}|f|}{}^g R_{i_{15}i_{16}|g]}{}^h V_{i_{17}i_{18}} V_{i_{19}i_{20}} V_{i_{21}i_{22}} V_{i_{23}i_{24}} V_{i_{25}i_{26}]} -$$

$$- 148{,}262{,}400\ R_{[i_1i_2|g|}{}^a R_{i_3i_4|a|}{}^b R_{i_5i_6|b|}{}^c R_{i_7i_8|c|}{}^d R_{i_9i_{10}|d|}{}^e R_{i_{11}i_{12}|e|}{}^f R_{i_{13}i_{14}|f|}{}^g R_{i_{15}i_{16}|m|}{}^h R_{i_{17}i_{18}|h|}{}^i R_{i_{19}i_{20}|i|}{}^j R_{i_{21}i_{22}|j|}{}^k R_{i_{23}i_{24}|k|}{}^l R_{i_{25}i_{26}]l}{}^m +$$

$$+ 177{,}914{,}880\ R_{[i_1i_2|g|}{}^a R_{i_3i_4|a|}{}^b R_{i_5i_6|b|}{}^c R_{i_7i_8|c|}{}^d R_{i_9i_{10}|d|}{}^e R_{i_{11}i_{12}|e|}{}^f R_{i_{13}i_{14}|f|}{}^g R_{i_{15}i_{16}|l|}{}^h R_{i_{17}i_{18}|h|}{}^i R_{i_{19}i_{20}|i|}{}^j R_{i_{21}i_{22}|j|}{}^k R_{i_{23}i_{24}|k|}{}^l V_{i_{25}i_{26}]} +$$

$$+ 111{,}196{,}800\ R_{[i_1i_2|g|}{}^a R_{i_3i_4|a|}{}^b R_{i_5i_6|b|}{}^c R_{i_7i_8|c|}{}^d R_{i_9i_{10}|d|}{}^e R_{i_{11}i_{12}|e|}{}^f R_{i_{13}i_{14}|f|}{}^g R_{i_{15}i_{16}|k|}{}^h R_{i_{17}i_{18}|h|}{}^i R_{i_{19}i_{20}|i|}{}^j R_{i_{21}i_{22}|j|}{}^k R_{i_{23}i_{24}|m|}{}^l R_{i_{25}i_{26}]l}{}^m -$$

$$- 111{,}196{,}800\ R_{[i_1i_2|g|}{}^a R_{i_3i_4|a|}{}^b R_{i_5i_6|b|}{}^c R_{i_7i_8|c|}{}^d R_{i_9i_{10}|d|}{}^e R_{i_{11}i_{12}|e|}{}^f R_{i_{13}i_{14}|f|}{}^g R_{i_{15}i_{16}|k|}{}^h R_{i_{17}i_{18}|h|}{}^i R_{i_{19}i_{20}|i|}{}^j R_{i_{21}i_{22}|j|}{}^k V_{i_{23}i_{24}} V_{i_{25}i_{26}]} +$$

$$+ 49{,}420{,}800\ R_{[i_1i_2|g|}{}^a R_{i_3i_4|a|}{}^b R_{i_5i_6|b|}{}^c R_{i_7i_8|c|}{}^d R_{i_9i_{10}|d|}{}^e R_{i_{11}i_{12}|e|}{}^f R_{i_{13}i_{14}|f|}{}^g R_{i_{15}i_{16}|j|}{}^h R_{i_{17}i_{18}|h|}{}^i R_{i_{19}i_{20}|i|}{}^j R_{i_{21}i_{22}|m|}{}^k R_{i_{23}i_{24}|k|}{}^l R_{i_{25}i_{26}]l}{}^m -$$

$$- 148{,}262{,}400\ R_{[i_1i_2|g|}{}^a R_{i_3i_4|a|}{}^b R_{i_5i_6|b|}{}^c R_{i_7i_8|c|}{}^d R_{i_9i_{10}|d|}{}^e R_{i_{11}i_{12}|e|}{}^f R_{i_{13}i_{14}|f|}{}^g R_{i_{15}i_{16}|j|}{}^h R_{i_{17}i_{18}|h|}{}^i R_{i_{19}i_{20}|i|}{}^j R_{i_{21}i_{22}|l|}{}^k R_{i_{23}i_{24}|k|}{}^l V_{i_{25}i_{26}]} +$$

$$+ 49{,}420{,}800\ R_{[i_1i_2|g|}{}^a R_{i_3i_4|a|}{}^b R_{i_5i_6|b|}{}^c R_{i_7i_8|c|}{}^d R_{i_9i_{10}|d|}{}^e R_{i_{11}i_{12}|e|}{}^f R_{i_{13}i_{14}|f|}{}^g R_{i_{15}i_{16}|j|}{}^h R_{i_{17}i_{18}|h|}{}^i R_{i_{19}i_{20}|i|}{}^j V_{i_{21}i_{22}} V_{i_{23}i_{24}} V_{i_{25}i_{26}]} -$$

$$- 18{,}532{,}800\ R_{[i_1i_2|g|}{}^a R_{i_3i_4|a|}{}^b R_{i_5i_6|b|}{}^c R_{i_7i_8|c|}{}^d R_{i_9i_{10}|d|}{}^e R_{i_{11}i_{12}|e|}{}^f R_{i_{13}i_{14}|f|}{}^g R_{i_{15}i_{16}|i|}{}^h R_{i_{17}i_{18}|h|}{}^i R_{i_{19}i_{20}|k|}{}^j R_{i_{21}i_{22}|j|}{}^k R_{i_{23}i_{24}|m|}{}^l R_{i_{25}i_{26}]l}{}^m +$$

$$+ 55{,}598{,}400\ R_{[i_1i_2|g|}{}^a R_{i_3i_4|a|}{}^b R_{i_5i_6|b|}{}^c R_{i_7i_8|c|}{}^d R_{i_9i_{10}|d|}{}^e R_{i_{11}i_{12}|e|}{}^f R_{i_{13}i_{14}|f|}{}^g R_{i_{15}i_{16}|i|}{}^h R_{i_{17}i_{18}|h|}{}^i R_{i_{19}i_{20}|k|}{}^j R_{i_{21}i_{22}|j|}{}^k V_{i_{23}i_{24}} V_{i_{25}i_{26}]} -$$

$$- 18{,}532{,}800\ R_{[i_1i_2|g|}{}^a R_{i_3i_4|a|}{}^b R_{i_5i_6|b|}{}^c R_{i_7i_8|c|}{}^d R_{i_9i_{10}|d|}{}^e R_{i_{11}i_{12}|e|}{}^f R_{i_{13}i_{14}|f|}{}^g R_{i_{15}i_{16}|i|}{}^h R_{i_{17}i_{18}|h|}{}^i V_{i_{19}i_{20}} V_{i_{21}i_{22}} V_{i_{23}i_{24}} V_{i_{25}i_{26}]} +$$

$$+ 1{,}235{,}520\ R_{[i_1i_2|g|}{}^a R_{i_3i_4|a|}{}^b R_{i_5i_6|b|}{}^c R_{i_7i_8|c|}{}^d R_{i_9i_{10}|d|}{}^e R_{i_{11}i_{12}|e|}{}^f R_{i_{13}i_{14}|f|}{}^g V_{i_{15}i_{16}} V_{i_{17}i_{18}} V_{i_{19}i_{20}} V_{i_{21}i_{22}} V_{i_{23}i_{24}} V_{i_{25}i_{26}]} +$$

$$+ 86{,}486{,}400\ R_{[i_1i_2|f|}{}^a R_{i_3i_4|a|}{}^b R_{i_5i_6|b|}{}^c R_{i_7i_8|c|}{}^d R_{i_9i_{10}|d|}{}^e R_{i_{11}i_{12}|e|}{}^f R_{i_{13}i_{14}|l|}{}^g R_{i_{15}i_{16}|g|}{}^h R_{i_{17}i_{18}|h|}{}^i R_{i_{19}i_{20}|i|}{}^j R_{i_{21}i_{22}|j|}{}^k R_{i_{23}i_{24}|k|}{}^l V_{i_{25}i_{26}]} +$$

$$+ 103{,}783{,}680\ R_{[i_1i_2|f|}{}^a R_{i_3i_4|a|}{}^b R_{i_5i_6|b|}{}^c R_{i_7i_8|c|}{}^d R_{i_9i_{10}|d|}{}^e R_{i_{11}i_{12}|e|}{}^f R_{i_{13}i_{14}|k|}{}^g R_{i_{15}i_{16}|g|}{}^h R_{i_{17}i_{18}|h|}{}^i R_{i_{19}i_{20}|i|}{}^j R_{i_{21}i_{22}|j|}{}^k R_{i_{23}i_{24}|m|}{}^l R_{i_{25}i_{26}]l}{}^m -$$

$$- 103{,}783{,}680\ R_{[i_1i_2|f|}{}^a R_{i_3i_4|a|}{}^b R_{i_5i_6|b|}{}^c R_{i_7i_8|c|}{}^d R_{i_9i_{10}|d|}{}^e R_{i_{11}i_{12}|e|}{}^f R_{i_{13}i_{14}|k|}{}^g R_{i_{15}i_{16}|g|}{}^h R_{i_{17}i_{18}|h|}{}^i R_{i_{19}i_{20}|i|}{}^j R_{i_{21}i_{22}|j|}{}^k V_{i_{23}i_{24}} V_{i_{25}i_{26}]} +$$

$$+ 86{,}486{,}400\ R_{[i_1i_2|f|}{}^a R_{i_3i_4|a|}{}^b R_{i_5i_6|b|}{}^c R_{i_7i_8|c|}{}^d R_{i_9i_{10}|d|}{}^e R_{i_{11}i_{12}|e|}{}^f R_{i_{13}i_{14}|j|}{}^g R_{i_{15}i_{16}|g|}{}^h R_{i_{17}i_{18}|h|}{}^i R_{i_{19}i_{20}|i|}{}^j R_{i_{21}i_{22}|m|}{}^k R_{i_{23}i_{24}|k|}{}^l R_{i_{25}i_{26}]l}{}^m -$$

$$- 129{,}729{,}600\ R_{[i_1i_2|f|}{}^a R_{i_3i_4|a|}{}^b R_{i_5i_6|b|}{}^c R_{i_7i_8|c|}{}^d R_{i_9i_{10}|d|}{}^e R_{i_{11}i_{12}|e|}{}^f R_{i_{13}i_{14}|j|}{}^g R_{i_{15}i_{16}|g|}{}^h R_{i_{17}i_{18}|h|}{}^i R_{i_{19}i_{20}|i|}{}^j R_{i_{21}i_{22}|l|}{}^k R_{i_{23}i_{24}|k|}{}^l V_{i_{25}i_{26}]} +$$

$$+ 43{,}243{,}200\ R_{[i_1i_2|f|}{}^a R_{i_3i_4|a|}{}^b R_{i_5i_6|b|}{}^c R_{i_7i_8|c|}{}^d R_{i_9i_{10}|d|}{}^e R_{i_{11}i_{12}|e|}{}^f R_{i_{13}i_{14}|j|}{}^g R_{i_{15}i_{16}|g|}{}^h R_{i_{17}i_{18}|h|}{}^i R_{i_{19}i_{20}|i|}{}^j V_{i_{21}i_{22}} V_{i_{23}i_{24}} V_{i_{25}i_{26}]} -$$

$$- 57{,}657{,}600\ R_{[i_1i_2|f|}{}^a R_{i_3i_4|a|}{}^b R_{i_5i_6|b|}{}^c R_{i_7i_8|c|}{}^d R_{i_9i_{10}|d|}{}^e R_{i_{11}i_{12}|e|}{}^f R_{i_{13}i_{14}|i|}{}^g R_{i_{15}i_{16}|g|}{}^h R_{i_{17}i_{18}|h|}{}^i R_{i_{19}i_{20}|l|}{}^j R_{i_{21}i_{22}|j|}{}^k R_{i_{23}i_{24}|k|}{}^l V_{i_{25}i_{26}]} -$$

$$- 43{,}243{,}200\ R_{[i_1i_2|f|}{}^a R_{i_3i_4|a|}{}^b R_{i_5i_6|b|}{}^c R_{i_7i_8|c|}{}^d R_{i_9i_{10}|d|}{}^e R_{i_{11}i_{12}|e|}{}^f R_{i_{13}i_{14}|i|}{}^g R_{i_{15}i_{16}|g|}{}^h R_{i_{17}i_{18}|h|}{}^i R_{i_{19}i_{20}|k|}{}^j R_{i_{21}i_{22}|j|}{}^k R_{i_{23}i_{24}|m|}{}^l R_{i_{25}i_{26}]l}{}^m +$$

$$+ 86{,}486{,}400\ R_{[i_1i_2|f|}{}^a R_{i_3i_4|a|}{}^b R_{i_5i_6|b|}{}^c R_{i_7i_8|c|}{}^d R_{i_9i_{10}|d|}{}^e R_{i_{11}i_{12}|e|}{}^f R_{i_{13}i_{14}|i|}{}^g R_{i_{15}i_{16}|g|}{}^h R_{i_{17}i_{18}|h|}{}^i R_{i_{19}i_{20}|k|}{}^j R_{i_{21}i_{22}|j|}{}^k V_{i_{23}i_{24}} V_{i_{25}i_{26}]} -$$

$$- 14{,}414{,}400\ R_{[i_1i_2|f|}{}^a R_{i_3i_4|a|}{}^b R_{i_5i_6|b|}{}^c R_{i_7i_8|c|}{}^d R_{i_9i_{10}|d|}{}^e R_{i_{11}i_{12}|e|}{}^f R_{i_{13}i_{14}|i|}{}^g R_{i_{15}i_{16}|g|}{}^h R_{i_{17}i_{18}|h|}{}^i V_{i_{19}i_{20}} V_{i_{21}i_{22}} V_{i_{23}i_{24}} V_{i_{25}i_{26}]} +$$

$$+ 21{,}621{,}600\ R_{[i_1i_2|f|}{}^a R_{i_3i_4|a|}{}^b R_{i_5i_6|b|}{}^c R_{i_7i_8|c|}{}^d R_{i_9i_{10}|d|}{}^e R_{i_{11}i_{12}|e|}{}^f R_{i_{13}i_{14}|h|}{}^g R_{i_{15}i_{16}|g|}{}^h R_{i_{17}i_{18}|j|}{}^i R_{i_{19}i_{20}|i|}{}^j R_{i_{21}i_{22}|l|}{}^k R_{i_{23}i_{24}|k|}{}^l V_{i_{25}i_{26}]} -$$

$$- 21{,}621{,}600\ R_{[i_1i_2|f|}{}^a R_{i_3i_4|a|}{}^b R_{i_5i_6|b|}{}^c R_{i_7i_8|c|}{}^d R_{i_9i_{10}|d|}{}^e R_{i_{11}i_{12}|e|}{}^f R_{i_{13}i_{14}|h|}{}^g R_{i_{15}i_{16}|g|}{}^h R_{i_{17}i_{18}|j|}{}^i R_{i_{19}i_{20}|i|}{}^j V_{i_{21}i_{22}} V_{i_{23}i_{24}} V_{i_{25}i_{26}]} +$$

$$+ 4{,}324{,}320\ R_{[i_1i_2|f|}{}^a R_{i_3i_4|a|}{}^b R_{i_5i_6|b|}{}^c R_{i_7i_8|c|}{}^d R_{i_9i_{10}|d|}{}^e R_{i_{11}i_{12}|e|}{}^f R_{i_{13}i_{14}|h|}{}^g R_{i_{15}i_{16}|g|}{}^h V_{i_{17}i_{18}} V_{i_{19}i_{20}} V_{i_{21}i_{22}} V_{i_{23}i_{24}} V_{i_{25}i_{26}]} -$$

$$- 205{,}920\ R_{[i_1i_2|f|}{}^a R_{i_3i_4|a|}{}^b R_{i_5i_6|b|}{}^c R_{i_7i_8|c|}{}^d R_{i_9i_{10}|d|}{}^e R_{i_{11}i_{12}|e]}{}^f V_{i_{13}i_{14}} V_{i_{15}i_{16}} V_{i_{17}i_{18}} V_{i_{19}i_{20}} V_{i_{21}i_{22}} V_{i_{23}i_{24}} V_{i_{25}i_{26}]} +$$

$$+ 41{,}513{,}472\ R_{[i_1i_2|e|}{}^a R_{i_3i_4|a|}{}^b R_{i_5i_6|b|}{}^c R_{i_7i_8|c|}{}^d R_{i_9i_{10}|d|}{}^e R_{i_{11}i_{12}|j|}{}^f R_{i_{13}i_{14}|f|}{}^g R_{i_{15}i_{16}|g|}{}^h R_{i_{17}i_{18}|h|}{}^i R_{i_{19}i_{20}|i|}{}^j R_{i_{21}i_{22}|m|}{}^k R_{i_{23}i_{24}|k|}{}^l R_{i_{25}i_{26}]l}{}^m -$$

$$- 62{,}270{,}208\ R_{[i_1i_2|e|}{}^a R_{i_3i_4|a|}{}^b R_{i_5i_6|b|}{}^c R_{i_7i_8|c|}{}^d R_{i_9i_{10}|d|}{}^e R_{i_{11}i_{12}|j|}{}^f R_{i_{13}i_{14}|f|}{}^g R_{i_{15}i_{16}|g|}{}^h R_{i_{17}i_{18}|h|}{}^i R_{i_{19}i_{20}|i|}{}^j R_{i_{21}i_{22}|l|}{}^k R_{i_{23}i_{24}|k|}{}^l V_{i_{25}i_{26}]} +$$

$$+ 20{,}756{,}736\ R_{[i_1i_2|e|}{}^a R_{i_3i_4|a|}{}^b R_{i_5i_6|b|}{}^c R_{i_7i_8|c|}{}^d R_{i_9i_{10}|d|}{}^e R_{i_{11}i_{12}|j|}{}^f R_{i_{13}i_{14}|f|}{}^g R_{i_{15}i_{16}|g|}{}^h R_{i_{17}i_{18}|h|}{}^i R_{i_{19}i_{20}|i|}{}^j V_{i_{21}i_{22}} V_{i_{23}i_{24}} V_{i_{25}i_{26}]} +$$

$$+ 38{,}918{,}880\ R_{[i_1i_2|e|}{}^a R_{i_3i_4|a|}{}^b R_{i_5i_6|b|}{}^c R_{i_7i_8|c|}{}^d R_{i_9i_{10}|d|}{}^e R_{i_{11}i_{12}|i|}{}^f R_{i_{13}i_{14}|f|}{}^g R_{i_{15}i_{16}|g|}{}^h R_{i_{17}i_{18}|h|}{}^i R_{i_{19}i_{20}|m|}{}^j R_{i_{21}i_{22}|j|}{}^k R_{i_{23}i_{24}|k|}{}^l R_{i_{25}i_{26}]l}{}^m -$$

$$- 103{,}783{,}680\ R_{[i_1i_2|e|}{}^a R_{i_3i_4|a|}{}^b R_{i_5i_6|b|}{}^c R_{i_7i_8|c|}{}^d R_{i_9i_{10}|d|}{}^e R_{i_{11}i_{12}|i|}{}^f R_{i_{13}i_{14}|f|}{}^g R_{i_{15}i_{16}|g|}{}^h R_{i_{17}i_{18}|h|}{}^i R_{i_{19}i_{20}|l|}{}^j R_{i_{21}i_{22}|j|}{}^k R_{i_{23}i_{24}|k|}{}^l V_{i_{25}i_{26}]} -$$

$$- 38{,}918{,}880\ R_{[i_1i_2|e|}{}^a R_{i_3i_4|a|}{}^b R_{i_5i_6|b|}{}^c R_{i_7i_8|c|}{}^d R_{i_9i_{10}|d|}{}^e R_{i_{11}i_{12}|i|}{}^f R_{i_{13}i_{14}|f|}{}^g R_{i_{15}i_{16}|g|}{}^h R_{i_{17}i_{18}|h|}{}^i R_{i_{19}i_{20}|k|}{}^j R_{i_{21}i_{22}|j|}{}^k R_{i_{23}i_{24}|m|}{}^l R_{i_{25}i_{26}]l}{}^m +$$

$$+ 77{,}837{,}760\ R_{[i_1i_2|e|}{}^a R_{i_3i_4|a|}{}^b R_{i_5i_6|b|}{}^c R_{i_7i_8|c|}{}^d R_{i_9i_{10}|d|}{}^e R_{i_{11}i_{12}|i|}{}^f R_{i_{13}i_{14}|f|}{}^g R_{i_{15}i_{16}|g|}{}^h R_{i_{17}i_{18}|h|}{}^i R_{i_{19}i_{20}|k|}{}^j R_{i_{21}i_{22}|j|}{}^k V_{i_{23}i_{24}} V_{i_{25}i_{26}]} -$$

$$- 12{,}972{,}960\ R_{[i_1i_2|e|}{}^a R_{i_3i_4|a|}{}^b R_{i_5i_6|b|}{}^c R_{i_7i_8|c|}{}^d R_{i_9i_{10}|d|}{}^e R_{i_{11}i_{12}|i|}{}^f R_{i_{13}i_{14}|f|}{}^g R_{i_{15}i_{16}|g|}{}^h R_{i_{17}i_{18}|h|}{}^i V_{i_{19}i_{20}} V_{i_{21}i_{22}} V_{i_{23}i_{24}} V_{i_{25}i_{26}]} -$$

$$- 34{,}594{,}560\ R_{[i_1i_2|e|}{}^a R_{i_3i_4|a|}{}^b R_{i_5i_6|b|}{}^c R_{i_7i_8|c|}{}^d R_{i_9i_{10}|d|}{}^e R_{i_{11}i_{12}|h|}{}^f R_{i_{13}i_{14}|f|}{}^g R_{i_{15}i_{16}|g|}{}^h R_{i_{17}i_{18}|k|}{}^i R_{i_{19}i_{20}|j|}{}^j R_{i_{21}i_{22}|j|}{}^k R_{i_{23}i_{24}|m|}{}^l R_{i_{25}i_{26}]l}{}^m +$$

$$+ 34{,}594{,}560\ R_{[i_1i_2|e|}{}^a R_{i_3i_4|a|}{}^b R_{i_5i_6|b|}{}^c R_{i_7i_8|c|}{}^d R_{i_9i_{10}|d|}{}^e R_{i_{11}i_{12}|h|}{}^f R_{i_{13}i_{14}|f|}{}^g R_{i_{15}i_{16}|g|}{}^h R_{i_{17}i_{18}|k|}{}^i R_{i_{19}i_{20}|j|}{}^j R_{i_{21}i_{22}|j|}{}^k V_{i_{23}i_{24}} V_{i_{25}i_{26}]} +$$

$$+ 51{,}891{,}840\ R_{[i_1i_2|e|}{}^a R_{i_3i_4|a|}{}^b R_{i_5i_6|b|}{}^c R_{i_7i_8|c|}{}^d R_{i_9i_{10}|d|}{}^e R_{i_{11}i_{12}|h|}{}^f R_{i_{13}i_{14}|f|}{}^g R_{i_{15}i_{16}|g|}{}^h R_{i_{17}i_{18}|j|}{}^i R_{i_{19}i_{20}|i|}{}^j R_{i_{21}i_{22}|j|}{}^k R_{i_{23}i_{24}|k|}{}^l V_{i_{25}i_{26}]} -$$

$$- 34{,}594{,}560\ R_{[i_1i_2|e|}{}^a R_{i_3i_4|a|}{}^b R_{i_5i_6|b|}{}^c R_{i_7i_8|c|}{}^d R_{i_9i_{10}|d|}{}^e R_{i_{11}i_{12}|h|}{}^f R_{i_{13}i_{14}|f|}{}^g R_{i_{15}i_{16}|g|}{}^h R_{i_{17}i_{18}|j|}{}^i R_{i_{19}i_{20}|i|}{}^j V_{i_{21}i_{22}} V_{i_{23}i_{24}} V_{i_{25}i_{26}]} +$$

$$+ 3{,}459{,}456\ R_{[i_1i_2|e|}{}^a R_{i_3i_4|a|}{}^b R_{i_5i_6|b|}{}^c R_{i_7i_8|c|}{}^d R_{i_9i_{10}|d|}{}^e R_{i_{11}i_{12}|h|}{}^f R_{i_{13}i_{14}|f|}{}^g R_{i_{15}i_{16}|g|}{}^h V_{i_{17}i_{18}} V_{i_{19}i_{20}} V_{i_{21}i_{22}} V_{i_{23}i_{24}} V_{i_{25}i_{26}]} +$$

$$+ 3{,}243{,}240\ R_{[i_1i_2|e|}{}^a R_{i_3i_4|a|}{}^b R_{i_5i_6|b|}{}^c R_{i_7i_8|c|}{}^d R_{i_9i_{10}|d|}{}^e R_{i_{11}i_{12}|g|}{}^f R_{i_{13}i_{14}|f|}{}^g R_{i_{15}i_{16}|i|}{}^h R_{i_{17}i_{18}|h|}{}^i R_{i_{19}i_{20}|k|}{}^j R_{i_{21}i_{22}|j|}{}^k R_{i_{23}i_{24}|m|}{}^l R_{i_{25}i_{26}]l}{}^m -$$

$$- 12{,}972{,}960\ R_{[i_1i_2|e|}{}^a R_{i_3i_4|a|}{}^b R_{i_5i_6|b|}{}^c R_{i_7i_8|c|}{}^d R_{i_9i_{10}|d|}{}^e R_{i_{11}i_{12}|g|}{}^f R_{i_{13}i_{14}|f|}{}^g R_{i_{15}i_{16}|i|}{}^h R_{i_{17}i_{18}|h|}{}^i R_{i_{19}i_{20}|k|}{}^j R_{i_{21}i_{22}|j|}{}^k V_{i_{23}i_{24}} V_{i_{25}i_{26}]} +$$

$$+ 6{,}486{,}480\, R_{[i_1 i_2|e|}{}^a R_{i_3 i_4|a|}{}^b R_{i_5 i_6|b|}{}^c R_{i_7 i_8|c|}{}^d R_{i_9 i_{10}|d|}{}^e R_{i_{11} i_{12}|g|}{}^f R_{i_{13} i_{14}|f|}{}^g R_{i_{15} i_{16}|i|}{}^h R_{i_{17} i_{18}|h|}{}^i V_{i_{19} i_{20}} V_{i_{21} i_{22}} V_{i_{23} i_{24}} V_{i_{25} i_{26}]} -$$

$$- 864{,}864\, R_{[i_1 i_2|e|}{}^a R_{i_3 i_4|a|}{}^b R_{i_5 i_6|b|}{}^c R_{i_7 i_8|c|}{}^d R_{i_9 i_{10}|d|}{}^e R_{i_{11} i_{12}|g|}{}^f R_{i_{13} i_{14}|f|}{}^g V_{i_{15} i_{16}} V_{i_{17} i_{18}} V_{i_{19} i_{20}} V_{i_{21} i_{22}} V_{i_{23} i_{24}} V_{i_{25} i_{26}]} +$$

$$+ 30{,}888\, R_{[i_1 i_2|e|}{}^a R_{i_3 i_4|a|}{}^b R_{i_5 i_6|b|}{}^c R_{i_7 i_8|c|}{}^d R_{i_9 i_{10}|d|}{}^e V_{i_{11} i_{12}} V_{i_{13} i_{14}} V_{i_{15} i_{16}} V_{i_{17} i_{18}} V_{i_{19} i_{20}} V_{i_{21} i_{22}} V_{i_{23} i_{24}} V_{i_{25} i_{26}]} -$$

$$- 16{,}216{,}200\, R_{[i_1 i_2|d|}{}^a R_{i_3 i_4|a|}{}^b R_{i_5 i_6|b|}{}^c R_{i_7 i_8|c|}{}^d R_{i_9 i_{10}|h|}{}^e R_{i_{11} i_{12}|e|}{}^f R_{i_{13} i_{14}|f|}{}^g R_{i_{15} i_{16}|g|}{}^h R_{i_{17} i_{18}|l|}{}^i R_{i_{19} i_{20}|i|}{}^j R_{i_{21} i_{22}|j|}{}^k R_{i_{23} i_{24}|k|}{}^l V_{i_{25} i_{26}]} -$$

$$- 32{,}432{,}400\, R_{[i_1 i_2|d|}{}^a R_{i_3 i_4|a|}{}^b R_{i_5 i_6|b|}{}^c R_{i_7 i_8|c|}{}^d R_{i_9 i_{10}|h|}{}^e R_{i_{11} i_{12}|e|}{}^f R_{i_{13} i_{14}|f|}{}^g R_{i_{15} i_{16}|g|}{}^h R_{i_{17} i_{18}|k|}{}^i R_{i_{19} i_{20}|i|}{}^j R_{i_{21} i_{22}|j|}{}^k R_{i_{23} i_{24}|m|}{}^l R_{i_{25} i_{26}|l|}{}^m +$$

$$+ 32{,}432{,}400\, R_{[i_1 i_2|d|}{}^a R_{i_3 i_4|a|}{}^b R_{i_5 i_6|b|}{}^c R_{i_7 i_8|c|}{}^d R_{i_9 i_{10}|h|}{}^e R_{i_{11} i_{12}|e|}{}^f R_{i_{13} i_{14}|f|}{}^g R_{i_{15} i_{16}|g|}{}^h R_{i_{17} i_{18}|k|}{}^i R_{i_{19} i_{20}|i|}{}^j R_{i_{21} i_{22}|j|}{}^k V_{i_{23} i_{24}} V_{i_{25} i_{26}]} +$$

$$+ 24{,}324{,}300\, R_{[i_1 i_2|d|}{}^a R_{i_3 i_4|a|}{}^b R_{i_5 i_6|b|}{}^c R_{i_7 i_8|c|}{}^d R_{i_9 i_{10}|h|}{}^e R_{i_{11} i_{12}|e|}{}^f R_{i_{13} i_{14}|f|}{}^g R_{i_{15} i_{16}|g|}{}^h R_{i_{17} i_{18}|j|}{}^i R_{i_{19} i_{20}|i|}{}^j R_{i_{21} i_{22}|l|}{}^k R_{i_{23} i_{24}|k|}{}^l V_{i_{25} i_{26}]} -$$

$$- 16{,}216{,}200\, R_{[i_1 i_2|d|}{}^a R_{i_3 i_4|a|}{}^b R_{i_5 i_6|b|}{}^c R_{i_7 i_8|c|}{}^d R_{i_9 i_{10}|h|}{}^e R_{i_{11} i_{12}|e|}{}^f R_{i_{13} i_{14}|f|}{}^g R_{i_{15} i_{16}|g|}{}^h R_{i_{17} i_{18}|j|}{}^i R_{i_{19} i_{20}|i|}{}^j V_{i_{21} i_{22}} V_{i_{23} i_{24}} V_{i_{25} i_{26}]} +$$

$$+ 1{,}621{,}620\, R_{[i_1 i_2|d|}{}^a R_{i_3 i_4|a|}{}^b R_{i_5 i_6|b|}{}^c R_{i_7 i_8|c|}{}^d R_{i_9 i_{10}|h|}{}^e R_{i_{11} i_{12}|e|}{}^f R_{i_{13} i_{14}|f|}{}^g R_{i_{15} i_{16}|g|}{}^h V_{i_{17} i_{18}} V_{i_{19} i_{20}} V_{i_{21} i_{22}} V_{i_{23} i_{24}} V_{i_{25} i_{26}]} -$$

$$- 9{,}609{,}600\, R_{[i_1 i_2|d|}{}^a R_{i_3 i_4|a|}{}^b R_{i_5 i_6|b|}{}^c R_{i_7 i_8|c|}{}^d R_{i_9 i_{10}|g|}{}^e R_{i_{11} i_{12}|e|}{}^f R_{i_{13} i_{14}|f|}{}^g R_{i_{15} i_{16}|j|}{}^h R_{i_{17} i_{18}|h|}{}^i R_{i_{19} i_{20}|i|}{}^j R_{i_{21} i_{22}|m|}{}^k R_{i_{23} i_{24}|k|}{}^l R_{i_{25} i_{26}|l|}{}^m +$$

$$+ 43{,}243{,}200\, R_{[i_1 i_2|d|}{}^a R_{i_3 i_4|a|}{}^b R_{i_5 i_6|b|}{}^c R_{i_7 i_8|c|}{}^d R_{i_9 i_{10}|g|}{}^e R_{i_{11} i_{12}|e|}{}^f R_{i_{13} i_{14}|f|}{}^g R_{i_{15} i_{16}|j|}{}^h R_{i_{17} i_{18}|h|}{}^i R_{i_{19} i_{20}|i|}{}^j R_{i_{21} i_{22}|l|}{}^k R_{i_{23} i_{24}|k|}{}^l V_{i_{25} i_{26}]} -$$

$$- 14{,}414{,}400\, R_{[i_1 i_2|d|}{}^a R_{i_3 i_4|a|}{}^b R_{i_5 i_6|b|}{}^c R_{i_7 i_8|c|}{}^d R_{i_9 i_{10}|g|}{}^e R_{i_{11} i_{12}|e|}{}^f R_{i_{13} i_{14}|f|}{}^g R_{i_{15} i_{16}|j|}{}^h R_{i_{17} i_{18}|h|}{}^i R_{i_{19} i_{20}|i|}{}^j V_{i_{21} i_{22}} V_{i_{23} i_{24}} V_{i_{25} i_{26}]} +$$

$$+ 10{,}810{,}800\, R_{[i_1 i_2|d|}{}^a R_{i_3 i_4|a|}{}^b R_{i_5 i_6|b|}{}^c R_{i_7 i_8|c|}{}^d R_{i_9 i_{10}|g|}{}^e R_{i_{11} i_{12}|e|}{}^f R_{i_{13} i_{14}|f|}{}^g R_{i_{15} i_{16}|i|}{}^h R_{i_{17} i_{18}|h|}{}^i R_{i_{19} i_{20}|k|}{}^j R_{i_{21} i_{22}|j|}{}^k R_{i_{23} i_{24}|m|}{}^l R_{i_{25} i_{26}|l|}{}^m -$$

$$- 32{,}432{,}400\, R_{[i_1 i_2|d|}{}^a R_{i_3 i_4|a|}{}^b R_{i_5 i_6|b|}{}^c R_{i_7 i_8|c|}{}^d R_{i_9 i_{10}|g|}{}^e R_{i_{11} i_{12}|e|}{}^f R_{i_{13} i_{14}|f|}{}^g R_{i_{15} i_{16}|i|}{}^h R_{i_{17} i_{18}|h|}{}^i R_{i_{19} i_{20}|k|}{}^j R_{i_{21} i_{22}|j|}{}^k V_{i_{23} i_{24}} V_{i_{25} i_{26}]} +$$

$$+ 10{,}810{,}800\, R_{[i_1 i_2|d|}{}^a R_{i_3 i_4|a|}{}^b R_{i_5 i_6|b|}{}^c R_{i_7 i_8|c|}{}^d R_{i_9 i_{10}|g|}{}^e R_{i_{11} i_{12}|e|}{}^f R_{i_{13} i_{14}|f|}{}^g R_{i_{15} i_{16}|i|}{}^h R_{i_{17} i_{18}|h|}{}^i V_{i_{19} i_{20}} V_{i_{21} i_{22}} V_{i_{23} i_{24}} V_{i_{25} i_{26}]} -$$

$$- 720{,}720\, R_{[i_1 i_2|d|}{}^a R_{i_3 i_4|a|}{}^b R_{i_5 i_6|b|}{}^c R_{i_7 i_8|c|}{}^d R_{i_9 i_{10}|g|}{}^e R_{i_{11} i_{12}|e|}{}^f R_{i_{13} i_{14}|f|}{}^g V_{i_{15} i_{16}} V_{i_{17} i_{18}} V_{i_{19} i_{20}} V_{i_{21} i_{22}} V_{i_{23} i_{24}} V_{i_{25} i_{26}]} -$$

$$- 4{,}054{,}050\, R_{[i_1 i_2|d|}{}^a R_{i_3 i_4|a|}{}^b R_{i_5 i_6|b|}{}^c R_{i_7 i_8|c|}{}^d R_{i_9 i_{10}|e|}{}^e R_{i_{11} i_{12}|e|}{}^f R_{i_{13} i_{14}|h|}{}^g R_{i_{15} i_{16}|g|}{}^h R_{i_{17} i_{18}|j|}{}^i R_{i_{19} i_{20}|i|}{}^j R_{i_{21} i_{22}|l|}{}^k R_{i_{23} i_{24}|k|}{}^l V_{i_{25} i_{26}]} +$$

$$+ 5{,}405{,}400\, R_{[i_1 i_2|d|}{}^a R_{i_3 i_4|a|}{}^b R_{i_5 i_6|b|}{}^c R_{i_7 i_8|c|}{}^d R_{i_9 i_{10}|f|}{}^e R_{i_{11} i_{12}|e|}{}^f R_{i_{13} i_{14}|h|}{}^g R_{i_{15} i_{16}|g|}{}^h R_{i_{17} i_{18}|j|}{}^i R_{i_{19} i_{20}|i|}{}^j V_{i_{21} i_{22}} V_{i_{23} i_{24}} V_{i_{25} i_{26}]} -$$

$$- 1{,}621{,}620\, R_{[i_1 i_2|d|}{}^a R_{i_3 i_4|a|}{}^b R_{i_5 i_6|b|}{}^c R_{i_7 i_8|c|}{}^d R_{i_9 i_{10}|f|}{}^e R_{i_{11} i_{12}|e|}{}^f R_{i_{13} i_{14}|h|}{}^g R_{i_{15} i_{16}|g|}{}^h V_{i_{17} i_{18}} V_{i_{19} i_{20}} V_{i_{21} i_{22}} V_{i_{23} i_{24}} V_{i_{25} i_{26}]} +$$

$$+ 154{,}440\, R_{[i_1 i_2|d|}{}^a R_{i_3 i_4|a|}{}^b R_{i_5 i_6|b|}{}^c R_{i_7 i_8|c|}{}^d R_{i_9 i_{10}|e|}{}^e R_{i_{11} i_{12}|e|}{}^f V_{i_{13} i_{14}} V_{i_{15} i_{16}} V_{i_{17} i_{18}} V_{i_{19} i_{20}} V_{i_{21} i_{22}} V_{i_{23} i_{24}} V_{i_{25} i_{26}]} -$$

$$- 4290\, R_{[i_1 i_2|d|}{}^a R_{i_3 i_4|a|}{}^b R_{i_5 i_6|b|}{}^c R_{i_7 i_8|c|}{}^d V_{i_9 i_{10}} V_{i_{11} i_{12}} V_{i_{13} i_{14}} V_{i_{15} i_{16}} V_{i_{17} i_{18}} V_{i_{19} i_{20}} V_{i_{21} i_{22}} V_{i_{23} i_{24}} V_{i_{25} i_{26}]} +$$

$$+ 3{,}203{,}200\, R_{[i_1 i_2|c|}{}^a R_{i_3 i_4|a|}{}^b R_{i_5 i_6|b|}{}^c R_{i_7 i_8|f|}{}^d R_{i_9 i_{10}|d|}{}^e R_{i_{11} i_{12}|e|}{}^f R_{i_{13} i_{14}|i|}{}^g R_{i_{15} i_{16}|g|}{}^h R_{i_{17} i_{18}|h|}{}^i R_{i_{19} i_{20}|l|}{}^j R_{i_{21} i_{22}|j|}{}^k R_{i_{23} i_{24}|k|}{}^l V_{i_{25} i_{26}]} +$$

$$+ 4{,}804{,}800\, R_{[i_1 i_2|c|}{}^a R_{i_3 i_4|a|}{}^b R_{i_5 i_6|b|}{}^c R_{i_7 i_8|f|}{}^d R_{i_9 i_{10}|d|}{}^e R_{i_{11} i_{12}|e|}{}^f R_{i_{13} i_{14}|i|}{}^g R_{i_{15} i_{16}|g|}{}^h R_{i_{17} i_{18}|h|}{}^i R_{i_{19} i_{20}|k|}{}^j R_{i_{21} i_{22}|j|}{}^k R_{i_{23} i_{24}|m|}{}^l R_{i_{25} i_{26}|l|}{}^m -$$

$$- 9{,}609{,}600\, R_{[i_1 i_2|c|}{}^a R_{i_3 i_4|a|}{}^b R_{i_5 i_6|b|}{}^c R_{i_7 i_8|f|}{}^d R_{i_9 i_{10}|d|}{}^e R_{i_{11} i_{12}|e|}{}^f R_{i_{13} i_{14}|i|}{}^g R_{i_{15} i_{16}|g|}{}^h R_{i_{17} i_{18}|h|}{}^i R_{i_{19} i_{20}|k|}{}^j R_{i_{21} i_{22}|j|}{}^k V_{i_{23} i_{24}} V_{i_{25} i_{26}]} +$$

$$+ 1{,}601{,}600\, R_{[i_1 i_2|c|}{}^a R_{i_3 i_4|a|}{}^b R_{i_5 i_6|b|}{}^c R_{i_7 i_8|f|}{}^d R_{i_9 i_{10}|d|}{}^e R_{i_{11} i_{12}|e|}{}^f R_{i_{13} i_{14}|i|}{}^g R_{i_{15} i_{16}|g|}{}^h R_{i_{17} i_{18}|h|}{}^i V_{i_{19} i_{20}} V_{i_{21} i_{22}} V_{i_{23} i_{24}} V_{i_{25} i_{26}]} -$$

$$- 7{,}207{,}200\, R_{[i_1 i_2|c|}{}^a R_{i_3 i_4|a|}{}^b R_{i_5 i_6|b|}{}^c R_{i_7 i_8|f|}{}^d R_{i_9 i_{10}|d|}{}^e R_{i_{11} i_{12}|e|}{}^f R_{i_{13} i_{14}|h|}{}^g R_{i_{15} i_{16}|g|}{}^h R_{i_{17} i_{18}|j|}{}^i R_{i_{19} i_{20}|i|}{}^j R_{i_{21} i_{22}|l|}{}^k R_{i_{23} i_{24}|k|}{}^l V_{i_{25} i_{26}]} +$$

$$+ 7{,}207{,}200\, R_{[i_1 i_2|c|}{}^a R_{i_3 i_4|a|}{}^b R_{i_5 i_6|b|}{}^c R_{i_7 i_8|f|}{}^d R_{i_9 i_{10}|d|}{}^e R_{i_{11} i_{12}|e|}{}^f R_{i_{13} i_{14}|h|}{}^g R_{i_{15} i_{16}|g|}{}^h R_{i_{17} i_{18}|j|}{}^i R_{i_{19} i_{20}|i|}{}^j V_{i_{21} i_{22}} V_{i_{23} i_{24}} V_{i_{25} i_{26}]} -$$

$$- 1{,}441{,}440\, R_{[i_1 i_2|c|}{}^a R_{i_3 i_4|a|}{}^b R_{i_5 i_6|b|}{}^c R_{i_7 i_8|f|}{}^d R_{i_9 i_{10}|d|}{}^e R_{i_{11} i_{12}|e|}{}^f R_{i_{13} i_{14}|h|}{}^g R_{i_{15} i_{16}|g|}{}^h V_{i_{17} i_{18}} V_{i_{19} i_{20}} V_{i_{21} i_{22}} V_{i_{23} i_{24}} V_{i_{25} i_{26}]} +$$

$$+ 68{,}640\, R_{[i_1 i_2|c|}{}^a R_{i_3 i_4|a|}{}^b R_{i_5 i_6|b|}{}^c R_{i_7 i_8|f|}{}^d R_{i_9 i_{10}|d|}{}^e R_{i_{11} i_{12}|e|}{}^f V_{i_{13} i_{14}} V_{i_{15} i_{16}} V_{i_{17} i_{18}} V_{i_{19} i_{20}} V_{i_{21} i_{22}} V_{i_{23} i_{24}} V_{i_{25} i_{26}]} -$$

$$- 540{,}540\, R_{[i_1 i_2|c|}{}^a R_{i_3 i_4|a|}{}^b R_{i_5 i_6|b|}{}^c R_{i_7 i_8|e|}{}^d R_{i_9 i_{10}|d|}{}^e R_{i_{11} i_{12}|g|}{}^f R_{i_{13} i_{14}|f|}{}^g R_{i_{15} i_{16}|i|}{}^h R_{i_{17} i_{18}|h|}{}^i R_{i_{19} i_{20}|k|}{}^j R_{i_{21} i_{22}|j|}{}^k R_{i_{23} i_{24}|m|}{}^l R_{i_{25} i_{26}|l|}{}^m +$$

$$+ 2{,}702{,}700\, R_{[i_1 i_2|c|}{}^a R_{i_3 i_4|a|}{}^b R_{i_5 i_6|b|}{}^c R_{i_7 i_8|e|}{}^d R_{i_9 i_{10}|d|}{}^e R_{i_{11} i_{12}|g|}{}^f R_{i_{13} i_{14}|f|}{}^g R_{i_{15} i_{16}|i|}{}^h R_{i_{17} i_{18}|h|}{}^i R_{i_{19} i_{20}|k|}{}^j R_{i_{21} i_{22}|j|}{}^k V_{i_{23} i_{24}} V_{i_{25} i_{26}]} -$$

$$- 1{,}801{,}800\, R_{[i_1 i_2|c|}{}^a R_{i_3 i_4|a|}{}^b R_{i_5 i_6|b|}{}^c R_{i_7 i_8|e|}{}^d R_{i_9 i_{10}|d|}{}^e R_{i_{11} i_{12}|g|}{}^f R_{i_{13} i_{14}|f|}{}^g R_{i_{15} i_{16}|i|}{}^h R_{i_{17} i_{18}|h|}{}^i V_{i_{19} i_{20}} V_{i_{21} i_{22}} V_{i_{23} i_{24}} V_{i_{25} i_{26}]} +$$

$$+ 360{,}360\, R_{[i_1 i_2|c|}{}^a R_{i_3 i_4|a|}{}^b R_{i_5 i_6|b|}{}^c R_{i_7 i_8|e|}{}^d R_{i_9 i_{10}|d|}{}^e R_{i_{11} i_{12}|g|}{}^f R_{i_{13} i_{14}|f|}{}^g V_{i_{15} i_{16}} V_{i_{17} i_{18}} V_{i_{19} i_{20}} V_{i_{21} i_{22}} V_{i_{23} i_{24}} V_{i_{25} i_{26}]} -$$

$$- 25{,}740\, R_{[i_1 i_2|c|}{}^a R_{i_3 i_4|a|}{}^b R_{i_5 i_6|b|}{}^c R_{i_7 i_8|e|}{}^d R_{i_9 i_{10}|d|}{}^e V_{i_{11} i_{12}} V_{i_{13} i_{14}} V_{i_{15} i_{16}} V_{i_{17} i_{18}} V_{i_{19} i_{20}} V_{i_{21} i_{22}} V_{i_{23} i_{24}} V_{i_{25} i_{26}]} +$$

$$+ 572\, R_{[i_1 i_2|c|}{}^a R_{i_3 i_4|a|}{}^b R_{i_5 i_6|b|}{}^c V_{i_7 i_8} V_{i_9 i_{10}} V_{i_{11} i_{12}} V_{i_{13} i_{14}} V_{i_{15} i_{16}} V_{i_{17} i_{18}} V_{i_{19} i_{20}} V_{i_{21} i_{22}} V_{i_{23} i_{24}} V_{i_{25} i_{26}]} +$$

$$+ 135{,}135\, R_{[i_1 i_2|b|}{}^a R_{i_3 i_4|a|}{}^b R_{i_5 i_6|d|}{}^c R_{i_7 i_8|c|}{}^d R_{i_9 i_{10}|f|}{}^e R_{i_{11} i_{12}|e|}{}^f R_{i_{13} i_{14}|h|}{}^g R_{i_{15} i_{16}|g|}{}^h R_{i_{17} i_{18}|j|}{}^i R_{i_{19} i_{20}|i|}{}^j R_{i_{21} i_{22}|l|}{}^k R_{i_{23} i_{24}|k|}{}^l V_{i_{25} i_{26}]} -$$

$$- 270{,}270\, R_{[i_1 i_2|b|}{}^a R_{i_3 i_4|a|}{}^b R_{i_5 i_6|d|}{}^c R_{i_7 i_8|c|}{}^d R_{i_9 i_{10}|f|}{}^e R_{i_{11} i_{12}|e|}{}^f R_{i_{13} i_{14}|h|}{}^g R_{i_{15} i_{16}|g|}{}^h R_{i_{17} i_{18}|j|}{}^i R_{i_{19} i_{20}|i|}{}^j V_{i_{21} i_{22}} V_{i_{23} i_{24}} V_{i_{25} i_{26}]} +$$

$$+ 135{,}135\, R_{[i_1 i_2|b|}{}^a R_{i_3 i_4|a|}{}^b R_{i_5 i_6|d|}{}^c R_{i_7 i_8|c|}{}^d R_{i_9 i_{10}|f|}{}^e R_{i_{11} i_{12}|e|}{}^f R_{i_{13} i_{14}|h|}{}^g R_{i_{15} i_{16}|g|}{}^h V_{i_{17} i_{18}} V_{i_{19} i_{20}} V_{i_{21} i_{22}} V_{i_{23} i_{24}} V_{i_{25} i_{26}]} -$$

$$- 25{,}740\, R_{[i_1 i_2|b|}{}^a R_{i_3 i_4|a|}{}^b R_{i_5 i_6|d|}{}^c R_{i_7 i_8|c|}{}^d R_{i_9 i_{10}|f|}{}^e R_{i_{11} i_{12}|e|}{}^f V_{i_{13} i_{14}} V_{i_{15} i_{16}} V_{i_{17} i_{18}} V_{i_{19} i_{20}} V_{i_{21} i_{22}} V_{i_{23} i_{24}} V_{i_{25} i_{26}]} +$$

$$+ 2145\, R_{[i_1 i_2|b|}{}^a R_{i_3 i_4|a|}{}^b R_{i_5 i_6|d|}{}^c R_{i_7 i_8|c|}{}^d V_{i_9 i_{10}} V_{i_{11} i_{12}} V_{i_{13} i_{14}} V_{i_{15} i_{16}} V_{i_{17} i_{18}} V_{i_{19} i_{20}} V_{i_{21} i_{22}} V_{i_{23} i_{24}} V_{i_{25} i_{26}]} -$$

$$- 78\, R_{[i_1 i_2|b|}{}^a R_{i_3 i_4|a|}{}^b V_{i_5 i_6} V_{i_7 i_8} V_{i_9 i_{10}} V_{i_{11} i_{12}} V_{i_{13} i_{14}} V_{i_{15} i_{16}} V_{i_{17} i_{18}} V_{i_{19} i_{20}} V_{i_{21} i_{22}} V_{i_{23} i_{24}} V_{i_{25} i_{26}]} +$$

$$+ V_{[i_1 i_2} V_{i_3 i_4} V_{i_5 i_6} V_{i_7 i_8} V_{i_9 i_{10}} V_{i_{11} i_{12}} V_{i_{13} i_{14}} V_{i_{15} i_{16}} V_{i_{17} i_{18}} V_{i_{19} i_{20}} V_{i_{21} i_{22}} V_{i_{23} i_{24}} V_{i_{25} i_{26}]})$$

$$= \frac{i^{13}}{2^{13} \pi^{13} 13!} (+ 479{,}001{,}600 \, P^{(26)}{}_{i_1 i_2 i_3 i_4 i_5 i_6 i_7 i_8 i_9 i_{10} i_{11} i_{12} i_{13} i_{14} i_{15} i_{16} i_{17} i_{18} i_{19} i_{20} i_{21} i_{22} i_{23} i_{24} i_{25} i_{26}} +$$

$$- 518{,}918{,}400 \, P^{(24)}{}_{[i_1 i_2 i_3 i_4 i_5 i_6 i_7 i_8 i_9 i_{10} i_{11} i_{12} i_{13} i_{14} i_{15} i_{16} i_{17} i_{18} i_{19} i_{20} i_{21} i_{22} i_{23} i_{24}} P^{(2)}{}_{i_{25} i_{26}]} -$$

$$- 283{,}046{,}400 \, P^{(22)}{}_{[i_1 i_2 i_3 i_4 i_5 i_6 i_7 i_8 i_9 i_{10} i_{11} i_{12} i_{13} i_{14} i_{15} i_{16} i_{17} i_{18} i_{19} i_{20} i_{21} i_{22}} P^{(4)}{}_{i_{23} i_{24} i_{25} i_{26}]} +$$

$$+ 283{,}046{,}400 \, P^{(22)}{}_{[i_1 i_2 i_3 i_4 i_5 i_6 i_7 i_8 i_9 i_{10} i_{11} i_{12} i_{13} i_{14} i_{15} i_{16} i_{17} i_{18} i_{19} i_{20} i_{21} i_{22}} P^{(2)}{}_{i_{23} i_{24}} P^{(2)}{}_{i_{25} i_{26}]} -$$

$$- 207{,}567{,}360 \, P^{(20)}{}_{[i_1 i_2 i_3 i_4 i_5 i_6 i_7 i_8 i_9 i_{10} i_{11} i_{12} i_{13} i_{14} i_{15} i_{16} i_{17} i_{18} i_{19} i_{20}} P^{(6)}{}_{i_{21} i_{22} i_{23} i_{24} i_{25} i_{26}]} +$$

$$+ 311{,}351{,}040 \, P^{(20)}{}_{[i_1 i_2 i_3 i_4 i_5 i_6 i_7 i_8 i_9 i_{10} i_{11} i_{12} i_{13} i_{14} i_{15} i_{16} i_{17} i_{18} i_{19} i_{20}} P^{(4)}{}_{i_{21} i_{22} i_{23} i_{24}} P^{(2)}{}_{i_{25} i_{26}]} +$$

$$- 103{,}783{,}680 \, P^{(20)}{}_{[i_1 i_2 i_3 i_4 i_5 i_6 i_7 i_8 i_9 i_{10} i_{11} i_{12} i_{13} i_{14} i_{15} i_{16} i_{17} i_{18} i_{19} i_{20}} P^{(2)}{}_{i_{21} i_{22}} P^{(2)}{}_{i_{23} i_{24}} P^{(2)}{}_{i_{25} i_{26}]} -$$

$$- 172{,}972{,}800 \, P^{(18)}{}_{[i_1 i_2 i_3 i_4 i_5 i_6 i_7 i_8 i_9 i_{10} i_{11} i_{12} i_{13} i_{14} i_{15} i_{16} i_{17} i_{18}} P^{(8)}{}_{i_{19} i_{20} i_{21} i_{22} i_{23} i_{24} i_{25} i_{26}]} +$$

$$+ 230{,}630{,}400 \, P^{(18)}{}_{[i_1 i_2 i_3 i_4 i_5 i_6 i_7 i_8 i_9 i_{10} i_{11} i_{12} i_{13} i_{14} i_{15} i_{16} i_{17} i_{18}} P^{(6)}{}_{i_{19} i_{20} i_{21} i_{22} i_{23} i_{24}} P^{(2)}{}_{i_{25} i_{26}]} +$$

$$+ 86{,}486{,}400 \, P^{(18)}{}_{[i_1 i_2 i_3 i_4 i_5 i_6 i_7 i_8 i_9 i_{10} i_{11} i_{12} i_{13} i_{14} i_{15} i_{16} i_{17} i_{18}} P^{(4)}{}_{i_{19} i_{20} i_{21} i_{22}} P^{(4)}{}_{i_{23} i_{24} i_{25} i_{26}]} -$$

$$- 172{,}972{,}800 \, P^{(18)}{}_{[i_1 i_2 i_3 i_4 i_5 i_6 i_7 i_8 i_9 i_{10} i_{11} i_{12} i_{13} i_{14} i_{15} i_{16} i_{17} i_{18}} P^{(4)}{}_{i_{19} i_{20} i_{21} i_{22}} P^{(2)}{}_{i_{23} i_{24}} P^{(2)}{}_{i_{25} i_{26}]} +$$

$$+ 28{,}828{,}800 \, P^{(18)}{}_{[i_1 i_2 i_3 i_4 i_5 i_6 i_7 i_8 i_9 i_{10} i_{11} i_{12} i_{13} i_{14} i_{15} i_{16} i_{17} i_{18}} P^{(2)}{}_{i_{19} i_{20}} P^{(2)}{}_{i_{21} i_{22}} P^{(2)}{}_{i_{23} i_{24}} P^{(2)}{}_{i_{25} i_{26}]} -$$

$$- 155{,}675{,}520 \, P^{(16)}{}_{[i_1 i_2 i_3 i_4 i_5 i_6 i_7 i_8 i_9 i_{10} i_{11} i_{12} i_{13} i_{14} i_{15} i_{16}} P^{(10)}{}_{i_{17} i_{18} i_{19} i_{20} i_{21} i_{22} i_{23} i_{24} i_{25} i_{26}]} +$$

$$+ 194{,}594{,}400 \, P^{(16)}{}_{[i_1 i_2 i_3 i_4 i_5 i_6 i_7 i_8 i_9 i_{10} i_{11} i_{12} i_{13} i_{14} i_{15} i_{16}} P^{(8)}{}_{i_{17} i_{18} i_{19} i_{20} i_{21} i_{22} i_{23} i_{24}} P^{(2)}{}_{i_{25} i_{26}]} +$$

$$+ 129{,}729{,}600 \, P^{(16)}{}_{[i_1 i_2 i_3 i_4 i_5 i_6 i_7 i_8 i_9 i_{10} i_{11} i_{12} i_{13} i_{14} i_{15} i_{16}} P^{(6)}{}_{i_{17} i_{18} i_{19} i_{20} i_{21} i_{22}} P^{(4)}{}_{i_{23} i_{24} i_{25} i_{26}]} -$$

$$- 129{,}729{,}600 \, P^{(16)}{}_{[i_1 i_2 i_3 i_4 i_5 i_6 i_7 i_8 i_9 i_{10} i_{11} i_{12} i_{13} i_{14} i_{15} i_{16}} P^{(6)}{}_{i_{17} i_{18} i_{19} i_{20} i_{21} i_{22}} P^{(2)}{}_{i_{23} i_{24}} P^{(2)}{}_{i_{25} i_{26}]} -$$

$$- 97{,}297{,}200 \, P^{(16)}{}_{[i_1 i_2 i_3 i_4 i_5 i_6 i_7 i_8 i_9 i_{10} i_{11} i_{12} i_{13} i_{14} i_{15} i_{16}} P^{(4)}{}_{i_{17} i_{18} i_{19} i_{20}} P^{(4)}{}_{i_{21} i_{22} i_{23} i_{24}} P^{(2)}{}_{i_{25} i_{26}]} +$$

$$+ 64{,}864{,}800 \, P^{(16)}{}_{[i_1 i_2 i_3 i_4 i_5 i_6 i_7 i_8 i_9 i_{10} i_{11} i_{12} i_{13} i_{14} i_{15} i_{16}} P^{(4)}{}_{i_{17} i_{18} i_{19} i_{20}} P^{(2)}{}_{i_{21} i_{22}} P^{(2)}{}_{i_{23} i_{24}} P^{(2)}{}_{i_{25} i_{26}]} -$$

$$- 6{,}486{,}480 \, P^{(16)}{}_{[i_1 i_2 i_3 i_4 i_5 i_6 i_7 i_8 i_9 i_{10} i_{11} i_{12} i_{13} i_{14} i_{15} i_{16}} P^{(2)}{}_{i_{17} i_{18}} P^{(2)}{}_{i_{19} i_{20}} P^{(2)}{}_{i_{21} i_{22}} P^{(2)}{}_{i_{23} i_{24}} P^{(2)}{}_{i_{25} i_{26}]} -$$

$$- 148{,}262{,}400 \, P^{(14)}{}_{[i_1 i_2 i_3 i_4 i_5 i_6 i_7 i_8 i_9 i_{10} i_{11} i_{12} i_{13} i_{14}} P^{(12)}{}_{i_{15} i_{16} i_{17} i_{18} i_{19} i_{20} i_{21} i_{22} i_{23} i_{24} i_{25} i_{26}]} +$$

$$+ 177{,}914{,}880 \, P^{(14)}{}_{[i_1 i_2 i_3 i_4 i_5 i_6 i_7 i_8 i_9 i_{10} i_{11} i_{12} i_{13} i_{14}} P^{(10)}{}_{i_{15} i_{16} i_{17} i_{18} i_{19} i_{20} i_{21} i_{22} i_{23} i_{24}} P^{(2)}{}_{i_{25} i_{26}]} +$$

$$+ 111{,}196{,}800 \, P^{(14)}{}_{[i_1 i_2 i_3 i_4 i_5 i_6 i_7 i_8 i_9 i_{10} i_{11} i_{12} i_{13} i_{14}} P^{(8)}{}_{i_{15} i_{16} i_{17} i_{18} i_{19} i_{20} i_{21} i_{22}} P^{(4)}{}_{i_{23} i_{24} i_{25} i_{26}]} -$$

$$- 111{,}196{,}800 \, P^{(14)}{}_{[i_1 i_2 i_3 i_4 i_5 i_6 i_7 i_8 i_9 i_{10} i_{11} i_{12} i_{13} i_{14}} P^{(8)}{}_{i_{15} i_{16} i_{17} i_{18} i_{19} i_{20} i_{21} i_{22}} P^{(2)}{}_{i_{23} i_{24}} P^{(2)}{}_{i_{25} i_{26}]} +$$

$$+ 49{,}420{,}800 \, P^{(14)}{}_{[i_1 i_2 i_3 i_4 i_5 i_6 i_7 i_8 i_9 i_{10} i_{11} i_{12} i_{13} i_{14}} P^{(6)}{}_{i_{15} i_{16} i_{17} i_{18} i_{19} i_{20}} P^{(6)}{}_{i_{21} i_{22} i_{23} i_{24} i_{25} i_{26}]} +$$

$$- 148{,}262{,}400 \, P^{(14)}{}_{[i_1 i_2 i_3 i_4 i_5 i_6 i_7 i_8 i_9 i_{10} i_{11} i_{12} i_{13} i_{14}} P^{(6)}{}_{i_{15} i_{16} i_{17} i_{18} i_{19} i_{20}} P^{(4)}{}_{i_{21} i_{22} i_{23} i_{24}} P^{(2)}{}_{i_{25} i_{26}]} +$$

$$+ 49{,}420{,}800 \, P^{(14)}{}_{[i_1 i_2 i_3 i_4 i_5 i_6 i_7 i_8 i_9 i_{10} i_{11} i_{12} i_{13} i_{14}} P^{(6)}{}_{i_{15} i_{16} i_{17} i_{18} i_{19} i_{20}} P^{(2)}{}_{i_{21} i_{22}} P^{(2)}{}_{i_{23} i_{24}} P^{(2)}{}_{i_{25} i_{26}]} -$$

$$- 18{,}532{,}800 \, P^{(14)}{}_{[i_1 i_2 i_3 i_4 i_5 i_6 i_7 i_8 i_9 i_{10} i_{11} i_{12} i_{13} i_{14}} P^{(4)}{}_{i_{15} i_{16} i_{17} i_{18}} P^{(4)}{}_{i_{19} i_{20} i_{21} i_{22}} P^{(4)}{}_{i_{23} i_{24} i_{25} i_{26}]} +$$

$$+ 55{,}598{,}400 \, P^{(14)}{}_{[i_1 i_2 i_3 i_4 i_5 i_6 i_7 i_8 i_9 i_{10} i_{11} i_{12} i_{13} i_{14}} P^{(4)}{}_{i_{15} i_{16} i_{17} i_{18}} P^{(4)}{}_{i_{19} i_{20} i_{21} i_{22}} P^{(2)}{}_{i_{23} i_{24}} P^{(2)}{}_{i_{25} i_{26}]} -$$

$$- 18{,}532{,}800 \, P^{(14)}{}_{[i_1 i_2 i_3 i_4 i_5 i_6 i_7 i_8 i_9 i_{10} i_{11} i_{12} i_{13} i_{14}} P^{(4)}{}_{i_{15} i_{16} i_{17} i_{18}} P^{(2)}{}_{i_{19} i_{20}} P^{(2)}{}_{i_{21} i_{22}} P^{(2)}{}_{i_{23} i_{24}} P^{(2)}{}_{i_{25} i_{26}]} +$$

$$+ 1{,}235{,}520 \, P^{(14)}{}_{[i_1 i_2 i_3 i_4 i_5 i_6 i_7 i_8 i_9 i_{10} i_{11} i_{12} i_{13} i_{14}} P^{(2)}{}_{i_{15} i_{16}} P^{(2)}{}_{i_{17} i_{18}} P^{(2)}{}_{i_{19} i_{20}} P^{(2)}{}_{i_{21} i_{22}} P^{(2)}{}_{i_{23} i_{24}} P^{(2)}{}_{i_{25} i_{26}]} +$$

$$+ 86{,}486{,}400 \, P^{(12)}{}_{[i_1 i_2 i_3 i_4 i_5 i_6 i_7 i_8 i_9 i_{10} i_{11} i_{12}} P^{(12)}{}_{i_{13} i_{14} i_{15} i_{16} i_{17} i_{18} i_{19} i_{20} i_{21} i_{22} i_{23} i_{24}} P^{(2)}{}_{i_{25} i_{26}]} +$$

$$+ 103{,}783{,}680 \, P^{(12)}{}_{[i_1 i_2 i_3 i_4 i_5 i_6 i_7 i_8 i_9 i_{10} i_{11} i_{12}} P^{(10)}{}_{i_{13} i_{14} i_{15} i_{16} i_{17} i_{18} i_{19} i_{20} i_{21} i_{22}} P^{(4)}{}_{i_{23} i_{24} i_{25} i_{26}]} -$$

$$- 103{,}783{,}680 \, P^{(12)}{}_{[i_1 i_2 i_3 i_4 i_5 i_6 i_7 i_8 i_9 i_{10} i_{11} i_{12}} P^{(10)}{}_{i_{13} i_{14} i_{15} i_{16} i_{17} i_{18} i_{19} i_{20} i_{21} i_{22}} P^{(2)}{}_{i_{23} i_{24}} P^{(2)}{}_{i_{25} i_{26}]} +$$

$$+ 86{,}486{,}400 \, P^{(12)}{}_{[i_1 i_2 i_3 i_4 i_5 i_6 i_7 i_8 i_9 i_{10} i_{11} i_{12}} P^{(8)}{}_{i_{13} i_{14} i_{15} i_{16} i_{17} i_{18} i_{19} i_{20}} P^{(6)}{}_{i_{21} i_{22} i_{23} i_{24} i_{25} i_{26}]} -$$

$$- 129{,}729{,}600 \, P^{(12)}{}_{[i_1 i_2 i_3 i_4 i_5 i_6 i_7 i_8 i_9 i_{10} i_{11} i_{12}} P^{(8)}{}_{i_{13} i_{14} i_{15} i_{16} i_{17} i_{18} i_{19} i_{20}} P^{(4)}{}_{i_{21} i_{22} i_{23} i_{24}} P^{(2)}{}_{i_{25} i_{26}]} +$$

$$+ 43{,}243{,}200 \, P^{(12)}{}_{[i_1 i_2 i_3 i_4 i_5 i_6 i_7 i_8 i_9 i_{10} i_{11} i_{12}} P^{(8)}{}_{i_{13} i_{14} i_{15} i_{16} i_{17} i_{18} i_{19} i_{20}} P^{(2)}{}_{i_{21} i_{22}} P^{(2)}{}_{i_{23} i_{24}} P^{(2)}{}_{i_{25} i_{26}]} -$$

$$- 57{,}657{,}600 \, P^{(12)}{}_{[i_1 i_2 i_3 i_4 i_5 i_6 i_7 i_8 i_9 i_{10} i_{11} i_{12}} P^{(6)}{}_{i_{13} i_{14} i_{15} i_{16} i_{17} i_{18}} P^{(6)}{}_{i_{19} i_{20} i_{21} i_{22} i_{23} i_{24}} P^{(2)}{}_{i_{25} i_{26}]} -$$

$$- 43{,}243{,}200 \, P^{(12)}{}_{[i_1 i_2 i_3 i_4 i_5 i_6 i_7 i_8 i_9 i_{10} i_{11} i_{12}} P^{(6)}{}_{i_{13} i_{14} i_{15} i_{16} i_{17} i_{18}} P^{(4)}{}_{i_{19} i_{20} i_{21} i_{22}} P^{(4)}{}_{i_{23} i_{24} i_{25} i_{26}]} +$$

$$+ 86{,}486{,}400 \, P^{(12)}{}_{[i_1 i_2 i_3 i_4 i_5 i_6 i_7 i_8 i_9 i_{10} i_{11} i_{12}} P^{(6)}{}_{i_{13} i_{14} i_{15} i_{16} i_{17} i_{18}} P^{(4)}{}_{i_{19} i_{20} i_{21} i_{22}} P^{(2)}{}_{i_{23} i_{24}} P^{(2)}{}_{i_{25} i_{26}]} -$$

$$- 14{,}414{,}400 \, P^{(12)}{}_{[i_1 i_2 i_3 i_4 i_5 i_6 i_7 i_8 i_9 i_{10} i_{11} i_{12}} P^{(6)}{}_{i_{13} i_{14} i_{15} i_{16} i_{17} i_{18}} P^{(2)}{}_{i_{19} i_{20}} P^{(2)}{}_{i_{21} i_{22}} P^{(2)}{}_{i_{23} i_{24}} P^{(2)}{}_{i_{25} i_{26}]} +$$

$$+ 21{,}621{,}600 \; P^{(12)}{}_{[i_1 i_2 i_3 i_4 i_5 i_6 i_7 i_8 i_9 i_{10} i_{11} i_{12}} \; P^{(4)}{}_{i_{13} i_{14} i_{15} i_{16}} \; P^{(4)}{}_{i_{17} i_{18} i_{19} i_{20}} \; P^{(4)}{}_{i_{21} i_{22} i_{23} i_{24}} \; P^{(2)}{}_{i_{25} i_{26}]} -$$

$$- 21{,}621{,}600 \; P^{(12)}{}_{[i_1 i_2 i_3 i_4 i_5 i_6 i_7 i_8 i_9 i_{10} i_{11} i_{12}} \; P^{(4)}{}_{i_{13} i_{14} i_{15} i_{16}} \; P^{(4)}{}_{i_{17} i_{18} i_{19} i_{20}} \; P^{(2)}{}_{i_{21} i_{22}} \; P^{(2)}{}_{i_{23} i_{24}} \; P^{(2)}{}_{i_{25} i_{26}]} +$$

$$+ 4{,}324{,}320 \; P^{(12)}{}_{[i_1 i_2 i_3 i_4 i_5 i_6 i_7 i_8 i_9 i_{10} i_{11} i_{12}} \; P^{(4)}{}_{i_{13} i_{14} i_{15} i_{16}} \; P^{(2)}{}_{i_{17} i_{18}} \; P^{(2)}{}_{i_{19} i_{20}} \; P^{(2)}{}_{i_{21} i_{22}} \; P^{(2)}{}_{i_{23} i_{24}} \; P^{(2)}{}_{i_{25} i_{26}]} -$$

$$- 205{,}920 \; P^{(12)}{}_{[i_1 i_2 i_3 i_4 i_5 i_6 i_7 i_8 i_9 i_{10} i_{11} i_{12}} \; P^{(2)}{}_{i_{13} i_{14}} \; P^{(2)}{}_{i_{15} i_{16}} \; P^{(2)}{}_{i_{17} i_{18}} \; P^{(2)}{}_{i_{19} i_{20}} \; P^{(2)}{}_{i_{21} i_{22}} \; P^{(2)}{}_{i_{23} i_{24}} \; P^{(2)}{}_{i_{25} i_{26}]} +$$

$$+ 41{,}513{,}472 \; P^{(10)}{}_{[i_1 i_2 i_3 i_4 i_5 i_6 i_7 i_8 i_9 i_{10}} \; P^{(10)}{}_{i_{11} i_{12} i_{13} i_{14} i_{15} i_{16} i_{17} i_{18} i_{19} i_{20}} \; P^{(6)}{}_{i_{21} i_{22} i_{23} i_{24} i_{25} i_{26}]} -$$

$$- 62{,}270{,}208 \; P^{(10)}{}_{[i_1 i_2 i_3 i_4 i_5 i_6 i_7 i_8 i_9 i_{10}} \; P^{(10)}{}_{i_{11} i_{12} i_{13} i_{14} i_{15} i_{16} i_{17} i_{18} i_{19} i_{20}} \; P^{(4)}{}_{i_{21} i_{22} i_{23} i_{24}} \; P^{(2)}{}_{i_{25} i_{26}]} +$$

$$+ 20{,}756{,}736 \; P^{(10)}{}_{[i_1 i_2 i_3 i_4 i_5 i_6 i_7 i_8 i_9 i_{10}} \; P^{(10)}{}_{i_{11} i_{12} i_{13} i_{14} i_{15} i_{16} i_{17} i_{18} i_{19} i_{20}} \; P^{(2)}{}_{i_{21} i_{22}} \; P^{(2)}{}_{i_{23} i_{24}} \; P^{(2)}{}_{i_{25} i_{26}]} +$$

$$+ 38{,}918{,}880 \; P^{(10)}{}_{[i_1 i_2 i_3 i_4 i_5 i_6 i_7 i_8 i_9 i_{10}} \; P^{(8)}{}_{i_{11} i_{12} i_{13} i_{14} i_{15} i_{16} i_{17} i_{18}} \; P^{(8)}{}_{i_{19} i_{20} i_{21} i_{22} i_{23} i_{24} i_{25} i_{26}]} -$$

$$- 103{,}783{,}680 \; P^{(10)}{}_{[i_1 i_2 i_3 i_4 i_5 i_6 i_7 i_8 i_9 i_{10}} \; P^{(8)}{}_{i_{11} i_{12} i_{13} i_{14} i_{15} i_{16} i_{17} i_{18}} \; P^{(6)}{}_{i_{19} i_{20} i_{21} i_{22} i_{23} i_{24}} \; P^{(2)}{}_{i_{25} i_{26}]} -$$

$$- 38{,}918{,}880 \; P^{(10)}{}_{[i_1 i_2 i_3 i_4 i_5 i_6 i_7 i_8 i_9 i_{10}} \; P^{(8)}{}_{i_{11} i_{12} i_{13} i_{14} i_{15} i_{16} i_{17} i_{18}} \; P^{(4)}{}_{i_{19} i_{20} i_{21} i_{22}} \; P^{(4)}{}_{i_{23} i_{24} i_{25} i_{26}]} +$$

$$+ 77{,}837{,}760 \; P^{(10)}{}_{[i_1 i_2 i_3 i_4 i_5 i_6 i_7 i_8 i_9 i_{10}} \; P^{(8)}{}_{i_{11} i_{12} i_{13} i_{14} i_{15} i_{16} i_{17} i_{18}} \; P^{(4)}{}_{i_{19} i_{20} i_{21} i_{22}} \; P^{(2)}{}_{i_{23} i_{24}} \; P^{(2)}{}_{i_{25} i_{26}]} -$$

$$- 12{,}972{,}960 \; P^{(10)}{}_{[i_1 i_2 i_3 i_4 i_5 i_6 i_7 i_8 i_9 i_{10}} \; P^{(8)}{}_{i_{11} i_{12} i_{13} i_{14} i_{15} i_{16} i_{17} i_{18}} \; P^{(2)}{}_{i_{19} i_{20}} \; P^{(2)}{}_{i_{21} i_{22}} \; P^{(2)}{}_{i_{23} i_{24}} \; P^{(2)}{}_{i_{25} i_{26}]} -$$

$$- 34{,}594{,}560 \; P^{(10)}{}_{[i_1 i_2 i_3 i_4 i_5 i_6 i_7 i_8 i_9 i_{10}} \; P^{(6)}{}_{i_{11} i_{12} i_{13} i_{14} i_{15} i_{16}} \; P^{(6)}{}_{i_{17} i_{18} i_{19} i_{20} i_{21} i_{22}} \; P^{(4)}{}_{i_{23} i_{24} i_{25} i_{26}]} +$$

$$+ 34{,}594{,}560 \; P^{(10)}{}_{[i_1 i_2 i_3 i_4 i_5 i_6 i_7 i_8 i_9 i_{10}} \; P^{(6)}{}_{i_{11} i_{12} i_{13} i_{14} i_{15} i_{16}} \; P^{(6)}{}_{i_{17} i_{18} i_{19} i_{20} i_{21} i_{22}} \; P^{(2)}{}_{i_{23} i_{24}} \; P^{(2)}{}_{i_{25} i_{26}]} +$$

$$+ 51{,}891{,}840 \; P^{(10)}{}_{[i_1 i_2 i_3 i_4 i_5 i_6 i_7 i_8 i_9 i_{10}} \; P^{(6)}{}_{i_{11} i_{12} i_{13} i_{14} i_{15} i_{16}} \; P^{(4)}{}_{i_{17} i_{18} i_{19} i_{20}} \; P^{(4)}{}_{i_{21} i_{22} i_{23} i_{24}} \; P^{(2)}{}_{i_{25} i_{26}]} -$$

$$- 34{,}594{,}560 \; P^{(10)}{}_{[i_1 i_2 i_3 i_4 i_5 i_6 i_7 i_8 i_9 i_{10}} \; P^{(6)}{}_{i_{11} i_{12} i_{13} i_{14} i_{15} i_{16}} \; P^{(4)}{}_{i_{17} i_{18} i_{19} i_{20}} \; P^{(2)}{}_{i_{21} i_{22}} \; P^{(2)}{}_{i_{23} i_{24}} \; P^{(2)}{}_{i_{25} i_{26}]} +$$

$$+ 3{,}459{,}456 \; P^{(10)}{}_{[i_1 i_2 i_3 i_4 i_5 i_6 i_7 i_8 i_9 i_{10}} \; P^{(6)}{}_{i_{11} i_{12} i_{13} i_{14} i_{15} i_{16}} \; P^{(2)}{}_{i_{17} i_{18}} \; P^{(2)}{}_{i_{19} i_{20}} \; P^{(2)}{}_{i_{21} i_{22}} \; P^{(2)}{}_{i_{23} i_{24}} \; P^{(2)}{}_{i_{25} i_{26}]} +$$

$$+ 3{,}243{,}240 \; P^{(10)}{}_{[i_1 i_2 i_3 i_4 i_5 i_6 i_7 i_8 i_9 i_{10}} \; P^{(4)}{}_{i_{11} i_{12} i_{13} i_{14}} \; P^{(4)}{}_{i_{15} i_{16} i_{17} i_{18}} \; P^{(4)}{}_{i_{19} i_{20} i_{21} i_{22}} \; P^{(4)}{}_{i_{23} i_{24} i_{25} i_{26}]} -$$

$$- 12{,}972{,}960 \; P^{(10)}{}_{[i_1 i_2 i_3 i_4 i_5 i_6 i_7 i_8 i_9 i_{10}} \; P^{(4)}{}_{i_{11} i_{12} i_{13} i_{14}} \; P^{(4)}{}_{i_{15} i_{16} i_{17} i_{18}} \; P^{(4)}{}_{i_{19} i_{20} i_{21} i_{22}} \; P^{(2)}{}_{i_{23} i_{24}} \; P^{(2)}{}_{i_{25} i_{26}]} +$$

$$+ 6{,}486{,}480 \; P^{(10)}{}_{[i_1 i_2 i_3 i_4 i_5 i_6 i_7 i_8 i_9 i_{10}} \; P^{(4)}{}_{i_{11} i_{12} i_{13} i_{14}} \; P^{(4)}{}_{i_{15} i_{16} i_{17} i_{18}} \; P^{(2)}{}_{i_{19} i_{20}} \; P^{(2)}{}_{i_{21} i_{22}} \; P^{(2)}{}_{i_{23} i_{24}} \; P^{(2)}{}_{i_{25} i_{26}]} -$$

$$- 864{,}864 \; P^{(10)}{}_{[i_1 i_2 i_3 i_4 i_5 i_6 i_7 i_8 i_9 i_{10}} \; P^{(4)}{}_{i_{11} i_{12} i_{13} i_{14}} \; P^{(2)}{}_{i_{15} i_{16}} \; P^{(2)}{}_{i_{17} i_{18}} \; P^{(2)}{}_{i_{19} i_{20}} \; P^{(2)}{}_{i_{21} i_{22}} \; P^{(2)}{}_{i_{23} i_{24}} \; P^{(2)}{}_{i_{25} i_{26}]} +$$

$$+ 30{,}888 \; P^{(10)}{}_{[i_1 i_2 i_3 i_4 i_5 i_6 i_7 i_8 i_9 i_{10}} \; P^{(2)}{}_{i_{11} i_{12}} \; P^{(2)}{}_{i_{13} i_{14}} \; P^{(2)}{}_{i_{15} i_{16}} \; P^{(2)}{}_{i_{17} i_{18}} \; P^{(2)}{}_{i_{19} i_{20}} \; P^{(2)}{}_{i_{21} i_{22}} \; P^{(2)}{}_{i_{23} i_{24}} \; P^{(2)}{}_{i_{25} i_{26}]} -$$

$$- 16{,}216{,}200 \; P^{(8)}{}_{[i_1 i_2 i_3 i_4 i_5 i_6 i_7 i_8} \; P^{(8)}{}_{i_9 i_{10} i_{11} i_{12} i_{13} i_{14} i_{15} i_{16}} \; P^{(8)}{}_{i_{17} i_{18} i_{19} i_{20} i_{21} i_{22} i_{23} i_{24}} \; P^{(2)}{}_{i_{25} i_{26}]} -$$

$$- 32{,}432{,}400 \; P^{(8)}{}_{[i_1 i_2 i_3 i_4 i_5 i_6 i_7 i_8} \; P^{(8)}{}_{i_9 i_{10} i_{11} i_{12} i_{13} i_{14} i_{15} i_{16}} \; P^{(6)}{}_{i_{17} i_{18} i_{19} i_{20} i_{21} i_{22}} \; P^{(4)}{}_{i_{23} i_{24} i_{25} i_{26}]} +$$

$$+ 32{,}432{,}400 \; P^{(8)}{}_{[i_1 i_2 i_3 i_4 i_5 i_6 i_7 i_8} \; P^{(8)}{}_{i_9 i_{10} i_{11} i_{12} i_{13} i_{14} i_{15} i_{16}} \; P^{(6)}{}_{i_{17} i_{18} i_{19} i_{20} i_{21} i_{22}} \; P^{(2)}{}_{i_{23} i_{24}} \; P^{(2)}{}_{i_{25} i_{26}]} +$$

$$+ 24{,}324{,}300 \; P^{(8)}{}_{[i_1 i_2 i_3 i_4 i_5 i_6 i_7 i_8} \; P^{(8)}{}_{i_9 i_{10} i_{11} i_{12} i_{13} i_{14} i_{15} i_{16}} \; P^{(4)}{}_{i_{17} i_{18} i_{19} i_{20}} \; P^{(4)}{}_{i_{21} i_{22} i_{23} i_{24}} \; P^{(2)}{}_{i_{25} i_{26}]} -$$

$$- 16{,}216{,}200 \; P^{(8)}{}_{[i_1 i_2 i_3 i_4 i_5 i_6 i_7 i_8} \; P^{(8)}{}_{i_9 i_{10} i_{11} i_{12} i_{13} i_{14} i_{15} i_{16}} \; P^{(4)}{}_{i_{17} i_{18} i_{19} i_{20}} \; P^{(2)}{}_{i_{21} i_{22}} \; P^{(2)}{}_{i_{23} i_{24}} \; P^{(2)}{}_{i_{25} i_{26}]} +$$

$$+ 1{,}621{,}620 \; P^{(8)}{}_{[i_1 i_2 i_3 i_4 i_5 i_6 i_7 i_8} \; P^{(8)}{}_{i_9 i_{10} i_{11} i_{12} i_{13} i_{14} i_{15} i_{16}} \; P^{(2)}{}_{i_{17} i_{18}} \; P^{(2)}{}_{i_{19} i_{20}} \; P^{(2)}{}_{i_{21} i_{22}} \; P^{(2)}{}_{i_{23} i_{24}} \; P^{(2)}{}_{i_{25} i_{26}]} -$$

$$- 9{,}609{,}600 \; P^{(8)}{}_{[i_1 i_2 i_3 i_4 i_5 i_6 i_7 i_8} \; P^{(6)}{}_{i_9 i_{10} i_{11} i_{12} i_{13} i_{14}} \; P^{(6)}{}_{i_{15} i_{16} i_{17} i_{18} i_{19} i_{20}} \; P^{(6)}{}_{i_{21} i_{22} i_{23} i_{24} i_{25} i_{26}]} +$$

$$+ 43{,}243{,}200 \; P^{(8)}{}_{[i_1 i_2 i_3 i_4 i_5 i_6 i_7 i_8} \; P^{(6)}{}_{i_9 i_{10} i_{11} i_{12} i_{13} i_{14}} \; P^{(6)}{}_{i_{15} i_{16} i_{17} i_{18} i_{19} i_{20}} \; P^{(4)}{}_{i_{21} i_{22} i_{23} i_{24}} \; P^{(2)}{}_{i_{25} i_{26}]} -$$

$$- 14{,}414{,}400 \; P^{(8)}{}_{[i_1 i_2 i_3 i_4 i_5 i_6 i_7 i_8} \; P^{(6)}{}_{i_9 i_{10} i_{11} i_{12} i_{13} i_{14}} \; P^{(6)}{}_{i_{15} i_{16} i_{17} i_{18} i_{19} i_{20}} \; P^{(2)}{}_{i_{21} i_{22}} \; P^{(2)}{}_{i_{23} i_{24}} \; P^{(2)}{}_{i_{25} i_{26}]} +$$

$$+ 10{,}810{,}800 \; P^{(8)}{}_{[i_1 i_2 i_3 i_4 i_5 i_6 i_7 i_8} \; P^{(6)}{}_{i_9 i_{10} i_{11} i_{12} i_{13} i_{14}} \; P^{(4)}{}_{i_{15} i_{16} i_{17} i_{18}} \; P^{(4)}{}_{i_{19} i_{20} i_{21} i_{22}} \; P^{(4)}{}_{i_{23} i_{24} i_{25} i_{26}]} -$$

$$- 32{,}432{,}400 \; P^{(8)}{}_{[i_1 i_2 i_3 i_4 i_5 i_6 i_7 i_8} \; P^{(6)}{}_{i_9 i_{10} i_{11} i_{12} i_{13} i_{14}} \; P^{(4)}{}_{i_{15} i_{16} i_{17} i_{18}} \; P^{(4)}{}_{i_{19} i_{20} i_{21} i_{22}} \; P^{(2)}{}_{i_{23} i_{24}} \; P^{(2)}{}_{i_{25} i_{26}]} +$$

$$+ 10{,}810{,}800 \; P^{(8)}{}_{[i_1 i_2 i_3 i_4 i_5 i_6 i_7 i_8} \; P^{(6)}{}_{i_9 i_{10} i_{11} i_{12} i_{13} i_{14}} \; P^{(4)}{}_{i_{15} i_{16} i_{17} i_{18}} \; P^{(2)}{}_{i_{19} i_{20}} \; P^{(2)}{}_{i_{21} i_{22}} \; P^{(2)}{}_{i_{23} i_{24}} \; P^{(2)}{}_{i_{25} i_{26}]} -$$

$$- 720{,}720 \; P^{(8)}{}_{[i_1 i_2 i_3 i_4 i_5 i_6 i_7 i_8} \; P^{(6)}{}_{i_9 i_{10} i_{11} i_{12} i_{13} i_{14}} \; P^{(2)}{}_{i_{15} i_{16}} \; P^{(2)}{}_{i_{17} i_{18}} \; P^{(2)}{}_{i_{19} i_{20}} \; P^{(2)}{}_{i_{21} i_{22}} \; P^{(2)}{}_{i_{23} i_{24}} \; P^{(2)}{}_{i_{25} i_{26}]} -$$

$$- 4{,}054{,}050 \; P^{(8)}{}_{[i_1 i_2 i_3 i_4 i_5 i_6 i_7 i_8} \; P^{(4)}{}_{i_9 i_{10} i_{11} i_{12}} \; P^{(4)}{}_{i_{13} i_{14} i_{15} i_{16}} \; P^{(4)}{}_{i_{17} i_{18} i_{19} i_{20}} \; P^{(4)}{}_{i_{21} i_{22} i_{23} i_{24}} \; P^{(2)}{}_{i_{25} i_{26}]} +$$

$$+ 5{,}405{,}400 \; P^{(8)}{}_{[i_1 i_2 i_3 i_4 i_5 i_6 i_7 i_8} \; P^{(4)}{}_{i_9 i_{10} i_{11} i_{12}} \; P^{(4)}{}_{i_{13} i_{14} i_{15} i_{16}} \; P^{(4)}{}_{i_{17} i_{18} i_{19} i_{20}} \; P^{(2)}{}_{i_{21} i_{22}} \; P^{(2)}{}_{i_{23} i_{24}} \; P^{(2)}{}_{i_{25} i_{26}]} -$$

$$- 1{,}621{,}620 \; P^{(8)}{}_{[i_1 i_2 i_3 i_4 i_5 i_6 i_7 i_8} \; P^{(4)}{}_{i_9 i_{10} i_{11} i_{12}} \; P^{(4)}{}_{i_{13} i_{14} i_{15} i_{16}} \; P^{(2)}{}_{i_{17} i_{18}} \; P^{(2)}{}_{i_{19} i_{20}} \; P^{(2)}{}_{i_{21} i_{22}} \; P^{(2)}{}_{i_{23} i_{24}} \; P^{(2)}{}_{i_{25} i_{26}]} +$$

$$+ 154{,}440 \; P^{(8)}{}_{[i_1 i_2 i_3 i_4 i_5 i_6 i_7 i_8} \; P^{(4)}{}_{i_9 i_{10} i_{11} i_{12}} \; P^{(2)}{}_{i_{13} i_{14}} \; P^{(2)}{}_{i_{15} i_{16}} \; P^{(2)}{}_{i_{17} i_{18}} \; P^{(2)}{}_{i_{19} i_{20}} \; P^{(2)}{}_{i_{21} i_{22}} \; P^{(2)}{}_{i_{23} i_{24}} \; P^{(2)}{}_{i_{25} i_{26}]} -$$

$$- 4290 \; P^{(8)}{}_{[i_1 i_2 i_3 i_4 i_5 i_6 i_7 i_8} \; P^{(2)}{}_{i_9 i_{10}} \; P^{(2)}{}_{i_{11} i_{12}} \; P^{(2)}{}_{i_{13} i_{14}} \; P^{(2)}{}_{i_{15} i_{16}} \; P^{(2)}{}_{i_{17} i_{18}} \; P^{(2)}{}_{i_{19} i_{20}} \; P^{(2)}{}_{i_{21} i_{22}} \; P^{(2)}{}_{i_{23} i_{24}} \; P^{(2)}{}_{i_{25} i_{26}]} +$$

$$+ 3{,}203{,}200 \; P^{(6)}{}_{[i_1 i_2 i_3 i_4 i_5 i_6} \; P^{(6)}{}_{i_7 i_8 i_9 i_{10} i_{11} i_{12}} \; P^{(6)}{}_{i_{13} i_{14} i_{15} i_{16} i_{17} i_{18}} \; P^{(6)}{}_{i_{19} i_{20} i_{21} i_{22} i_{23} i_{24}} \; P^{(2)}{}_{i_{25} i_{26}]} +$$

$$+ 4{,}804{,}800\, P^{(6)}_{[i_1i_2i_3i_4i_5i_6}\, P^{(6)}_{i_7i_8i_9i_{10}i_{11}i_{12}}\, P^{(6)}_{i_{13}i_{14}i_{15}i_{16}i_{17}i_{18}}\, P^{(4)}_{i_{19}i_{20}i_{21}i_{22}}\, P^{(4)}_{i_{23}i_{24}i_{25}i_{26}]} -$$

$$- 9{,}609{,}600\, P^{(6)}_{[i_1i_2i_3i_4i_5i_6}\, P^{(6)}_{i_7i_8i_9i_{10}i_{11}i_{12}}\, P^{(6)}_{i_{13}i_{14}i_{15}i_{16}i_{17}i_{18}}\, P^{(4)}_{i_{19}i_{20}i_{21}i_{22}}\, P^{(2)}_{i_{23}i_{24}}\, P^{(2)}_{i_{25}i_{26}]} +$$

$$+ 1{,}601{,}600\, P^{(6)}_{[i_1i_2i_3i_4i_5i_6}\, P^{(6)}_{i_7i_8i_9i_{10}i_{11}i_{12}}\, P^{(6)}_{i_{13}i_{14}i_{15}i_{16}i_{17}i_{18}}\, P^{(2)}_{i_{19}i_{20}}\, P^{(2)}_{i_{21}i_{22}}\, P^{(2)}_{i_{23}i_{24}}\, P^{(2)}_{i_{25}i_{26}]} -$$

$$- 7{,}207{,}200\, P^{(6)}_{[i_1i_2i_3i_4i_5i_6}\, P^{(6)}_{i_7i_8i_9i_{10}i_{11}i_{12}}\, P^{(4)}_{i_{13}i_{14}i_{15}i_{16}}\, P^{(4)}_{i_{17}i_{18}i_{19}i_{20}}\, P^{(4)}_{i_{21}i_{22}i_{23}i_{24}}\, P^{(2)}_{i_{25}i_{26}]} +$$

$$+ 7{,}207{,}200\, P^{(6)}_{[i_1i_2i_3i_4i_5i_6}\, P^{(6)}_{i_7i_8i_9i_{10}i_{11}i_{12}}\, P^{(4)}_{i_{13}i_{14}i_{15}i_{16}}\, P^{(4)}_{i_{17}i_{18}i_{19}i_{20}}\, P^{(2)}_{i_{21}i_{22}}\, P^{(2)}_{i_{23}i_{24}}\, P^{(2)}_{i_{25}i_{26}]} -$$

$$- 1{,}441{,}440\, P^{(6)}_{[i_1i_2i_3i_4i_5i_6}\, P^{(6)}_{i_7i_8i_9i_{10}i_{11}i_{12}}\, P^{(4)}_{i_{13}i_{14}i_{15}i_{16}}\, P^{(2)}_{i_{17}i_{18}}\, P^{(2)}_{i_{19}i_{20}}\, P^{(2)}_{i_{21}i_{22}}\, P^{(2)}_{i_{23}i_{24}}\, P^{(2)}_{i_{25}i_{26}]} +$$

$$+ 68{,}640\, P^{(6)}_{[i_1i_2i_3i_4i_5i_6}\, P^{(6)}_{i_7i_8i_9i_{10}i_{11}i_{12}}\, P^{(2)}_{i_{13}i_{14}}\, P^{(2)}_{i_{15}i_{16}}\, P^{(2)}_{i_{17}i_{18}}\, P^{(2)}_{i_{19}i_{20}}\, P^{(2)}_{i_{21}i_{22}}\, P^{(2)}_{i_{23}i_{24}}\, P^{(2)}_{i_{25}i_{26}]} -$$

$$- 540{,}540\, P^{(6)}_{[i_1i_2i_3i_4i_5i_6}\, P^{(4)}_{i_7i_8i_9i_{10}}\, P^{(4)}_{i_{11}i_{12}i_{13}i_{14}}\, P^{(4)}_{i_{15}i_{16}i_{17}i_{18}}\, P^{(4)}_{i_{19}i_{20}i_{21}i_{22}}\, P^{(4)}_{i_{23}i_{24}i_{25}i_{26}]} +$$

$$+ 2{,}702{,}700\, P^{(6)}_{[i_1i_2i_3i_4i_5i_6}\, P^{(4)}_{i_7i_8i_9i_{10}}\, P^{(4)}_{i_{11}i_{12}i_{13}i_{14}}\, P^{(4)}_{i_{15}i_{16}i_{17}i_{18}}\, P^{(4)}_{i_{19}i_{20}i_{21}i_{22}}\, P^{(2)}_{i_{23}i_{24}}\, P^{(2)}_{i_{25}i_{26}]} -$$

$$- 1{,}801{,}800\, P^{(6)}_{[i_1i_2i_3i_4i_5i_6}\, P^{(4)}_{i_7i_8i_9i_{10}}\, P^{(4)}_{i_{11}i_{12}i_{13}i_{14}}\, P^{(4)}_{i_{15}i_{16}i_{17}i_{18}}\, P^{(2)}_{i_{19}i_{20}}\, P^{(2)}_{i_{21}i_{22}}\, P^{(2)}_{i_{23}i_{24}}\, P^{(2)}_{i_{25}i_{26}]} +$$

$$+ 360{,}360\, P^{(6)}_{[i_1i_2i_3i_4i_5i_6}\, P^{(4)}_{i_7i_8i_9i_{10}}\, P^{(4)}_{i_{11}i_{12}i_{13}i_{14}}\, P^{(2)}_{i_{15}i_{16}}\, P^{(2)}_{i_{17}i_{18}}\, P^{(2)}_{i_{19}i_{20}}\, P^{(2)}_{i_{21}i_{22}}\, P^{(2)}_{i_{23}i_{24}}\, P^{(2)}_{i_{25}i_{26}]} -$$

$$- 25{,}740\, P^{(6)}_{[i_1i_2i_3i_4i_5i_6}\, P^{(4)}_{i_7i_8i_9i_{10}}\, P^{(2)}_{i_{11}i_{12}}\, P^{(2)}_{i_{13}i_{14}}\, P^{(2)}_{i_{15}i_{16}}\, P^{(2)}_{i_{17}i_{18}}\, P^{(2)}_{i_{19}i_{20}}\, P^{(2)}_{i_{21}i_{22}}\, P^{(2)}_{i_{23}i_{24}}\, P^{(2)}_{i_{25}i_{26}]} +$$

$$+ 572\, P^{(6)}_{[i_1i_2i_3i_4i_5i_6}\, P^{(2)}_{i_7i_8}\, P^{(2)}_{i_9i_{10}}\, P^{(2)}_{i_{11}i_{12}}\, P^{(2)}_{i_{13}i_{14}}\, P^{(2)}_{i_{15}i_{16}}\, P^{(2)}_{i_{17}i_{18}}\, P^{(2)}_{i_{19}i_{20}}\, P^{(2)}_{i_{21}i_{22}}\, P^{(2)}_{i_{23}i_{24}}\, P^{(2)}_{i_{25}i_{26}]} +$$

$$+ 135{,}135\, P^{(4)}_{[i_1i_2i_3i_4}\, P^{(4)}_{i_5i_6i_7i_8}\, P^{(4)}_{i_9i_{10}i_{11}i_{12}}\, P^{(4)}_{i_{13}i_{14}i_{15}i_{16}}\, P^{(4)}_{i_{17}i_{18}i_{19}i_{20}}\, P^{(4)}_{i_{21}i_{22}i_{23}i_{24}}\, P^{(2)}_{i_{25}i_{26}]} -$$

$$- 270{,}270\, P^{(4)}_{[i_1i_2i_3i_4}\, P^{(4)}_{i_5i_6i_7i_8}\, P^{(4)}_{i_9i_{10}i_{11}i_{12}}\, P^{(4)}_{i_{13}i_{14}i_{15}i_{16}}\, P^{(4)}_{i_{17}i_{18}i_{19}i_{20}}\, P^{(2)}_{i_{21}i_{22}}\, P^{(2)}_{i_{23}i_{24}}\, P^{(2)}_{i_{25}i_{26}]} +$$

$$+ 135{,}135\, P^{(4)}_{[i_1i_2i_3i_4}\, P^{(4)}_{i_5i_6i_7i_8}\, P^{(4)}_{i_9i_{10}i_{11}i_{12}}\, P^{(4)}_{i_{13}i_{14}i_{15}i_{16}}\, P^{(2)}_{i_{17}i_{18}}\, P^{(2)}_{i_{19}i_{20}}\, P^{(2)}_{i_{21}i_{22}}\, P^{(2)}_{i_{23}i_{24}}\, P^{(2)}_{i_{25}i_{26}]} -$$

$$- 25{,}740\, P^{(4)}_{[i_1i_2i_3i_4}\, P^{(4)}_{i_5i_6i_7i_8}\, P^{(4)}_{i_9i_{10}i_{11}i_{12}}\, P^{(2)}_{i_{13}i_{14}}\, P^{(2)}_{i_{15}i_{16}}\, P^{(2)}_{i_{17}i_{18}}\, P^{(2)}_{i_{19}i_{20}}\, P^{(2)}_{i_{21}i_{22}}\, P^{(2)}_{i_{23}i_{24}}\, P^{(2)}_{i_{25}i_{26}]} +$$

$$+ 2145\, P^{(4)}_{[i_1i_2i_3i_4}\, P^{(4)}_{i_5i_6i_7i_8}\, P^{(2)}_{i_9i_{10}}\, P^{(2)}_{i_{11}i_{12}}\, P^{(2)}_{i_{13}i_{14}}\, P^{(2)}_{i_{15}i_{16}}\, P^{(2)}_{i_{17}i_{18}}\, P^{(2)}_{i_{19}i_{20}}\, P^{(2)}_{i_{21}i_{22}}\, P^{(2)}_{i_{23}i_{24}}\, P^{(2)}_{i_{25}i_{26}]} -$$

$$- 78\, P^{(4)}_{[i_1i_2i_3i_4}\, P^{(2)}_{i_5i_6}\, P^{(2)}_{i_7i_8}\, P^{(2)}_{i_9i_{10}}\, P^{(2)}_{i_{11}i_{12}}\, P^{(2)}_{i_{13}i_{14}}\, P^{(2)}_{i_{15}i_{16}}\, P^{(2)}_{i_{17}i_{18}}\, P^{(2)}_{i_{19}i_{20}}\, P^{(2)}_{i_{21}i_{22}}\, P^{(2)}_{i_{23}i_{24}}\, P^{(2)}_{i_{25}i_{26}]} +$$

$$+ P^{(2)}_{[i_1i_2}\, P^{(2)}_{i_3i_4}\, P^{(2)}_{i_5i_6}\, P^{(2)}_{i_7i_8}\, P^{(2)}_{i_9i_{10}}\, P^{(2)}_{i_{11}i_{12}}\, P^{(2)}_{i_{13}i_{14}}\, P^{(2)}_{i_{15}i_{16}}\, P^{(2)}_{i_{17}i_{18}}\, P^{(2)}_{i_{19}i_{20}}\, P^{(2)}_{i_{21}i_{22}}\, P^{(2)}_{i_{23}i_{24}}\, P^{(2)}_{i_{25}i_{26}]})$$

## CONCLUDING REMARK

For a check, note that the magnitudes of the numerical factors in the preceding expressions for $c_{(p)i_1i_2\ldots i_{2p}}$ for $0 \le p \le 13$ in Eqs. (9) through (22) add up—aside from the respective overall numerical factors—to the corresponding numbers $p!$, values for which appear in Table 2 (see below).

TABLE 2. SUMS OF THE MAGNITUDES OF THE NUMERICAL FACTORS IN THE GENERAL EXPRESSIONS FOR $c_{(p)i_1i_2\ldots i_{2p}}$ FOR $0 \le p \le 13$

| ORDER | SUM OF MAGNITUDES OF NUMERICAL FACTORS |
|---|---|
| $p$ | $p!$ |
| 0 | 1 |
| 1 | 1 |
| 2 | 2 |
| 3 | 6 |
| 4 | 24 |
| 5 | 120 |
| 6 | 720 |
| 7 | 5040 |
| 8 | 40,320 |
| 9 | 362,880 |
| 10 | 3,628,800 |
| 11 | 39,916,800 |
| 12 | 479,001,600 |
| 13 | 6,227,020,800 |